\documentclass[letterpaper,11pt]{article}
\pdfoutput=1 

\usepackage{jheppub} 

\usepackage[T1]{fontenc} 

\usepackage{amsmath,amsthm,amssymb,amsfonts,mathrsfs}
\usepackage[normalem]{ulem}
\usepackage{hyperref}
\hypersetup{
    pdftex,
    pdfstartview=FitH,
    bookmarksnumbered=true,
    pagebackref,
    colorlinks,
    breaklinks,
    urlcolor=Maroon,
    linkcolor=Maroon,
    citecolor=Bittersweet
}
\usepackage[dvipsnames]{xcolor}
\usepackage{tikz}
\usetikzlibrary{calc,matrix,decorations.pathmorphing,decorations.markings,arrows,positioning,intersections,mindmap,backgrounds}
\usepackage{booktabs}
\usepackage[utf8]{inputenc}
\usepackage[yyyymmdd,hhmmss]{datetime}
\usepackage{cancel}
\usepackage{caption}
\usepackage{enumitem}
\usepackage{mathtools}
\usepackage{booktabs}

\newcommand{\bfn}[2]{{\mathrm{B}\!\!\left[\begin{matrix}{#1}\\{#2}\end{matrix}\right]}}
\newcommand{\residue}[1]{{\underset{#1}{\text{Res}}}}
\newcommand{\checked}{{}}
\newcommand{\temp}[1]{{}}

\newtheorem{proposition}{Proposition}

\newcommand{\thistheoremname}{}
\newtheorem*{genericthm*}{\thistheoremname}
\newenvironment{conjecture}[1]
  {\renewcommand{\thistheoremname}{#1}%
   \begin{genericthm*}}
  {\end{genericthm*}}

\title{Simplicity in AdS Perturbative Dynamics}

\author{Ellis Ye Yuan}
\affiliation{Institute for Advanced Study,\\Einstein Drive, Princeton, NJ 08540, USA}\emailAdd{yyuan@ias.edu}

\abstract{We investigate analytic properties of loop-level perturbative dynamics in pure AdS, with the scalar effective theories with non-derivative couplings  as a prototype.  Explicit computations reveal certain (perhaps unexpected) simplicity regarding the pole structure of the results, in both the Mellin amplitude and a closely related object that we call Mellin pre-amplitude.  Correspondingly we propose a pair of conjectures for arbitrary diagrams at all loops, based on non-trivial evidence up to two loops (and higher loops in a special class of diagrams). We also inspect the structure of residues at poles in the physical channels for several one-loop examples up to a 4-point box, as well as a two-loop double-triangle diagram. These analyses are performed using the recursive construction of Mellin (pre-)amplitudes recently prescribed in \cite{Yuan:2017vgp}, for which we provide detailed derivation and generalization in this paper. Along the way we derive a set of alternative diagrammatic rules for tree (pre-)amplitudes, which are better suited to our loop construction. On the mathematical aspect we share some new thoughts on improving the contour analysis of multi-dimensional Mellin integrals, which are the essential ingredients that make our approach practical.}

\begin{document} 
\maketitle
\flushbottom

\newpage

\section{Introduction and Summary}

In this paper we develop in detail a systematized analysis of scattering problem in Anti de Sitter space in the perturbative regime, i.e., Witten diagrams, to all loop levels, with a focus on scalar effective theory with the simplest class of interactions, $\sum_{p=3}^\infty g_p\phi^p$.

Perturbation theory in AdS is not only the simplest playground for studying weakly coupled QFT dynamics beyond flat spacetime, but also relevant for precision checks of AdS/CFT correspondence \cite{Maldacena:1997re,Gubser:1998bc,Witten:1998qj} (for a useful review on this aspect, e.g., \cite{DHoker:2002nbb}). One promising advancement in this study arises from the observation that correlation functions enjoying conformal symmetries are more naturally represented in Mellin space (where they are commonly called \emph{Mellin amplitudes}) than in the original spacetime \cite{Mack:2009mi,Mack:2009gy}, in much the same way as the momentum space for scattering in flat spacetime. As was first postulated in \cite{Penedones:2010ue}, this is particularly so for holographic CFTs, since there the Mellin amplitudes manifest the bulk scattering process in its poles and residues and may even admit of diagrammatic rules. This was firmly verified at tree level and confirmed in certain special class of loop diagrams \cite{Penedones:2010ue,Fitzpatrick:2011ia,Paulos:2011ie,Nandan:2011wc}.

Despite of the tremendous efforts in the past (see \cite{Cardona:2017tsw,Giombi:2017hpr} for some direct computations more recently), most of the established knowledge about AdS perturbation is restricted to tree level, and for quite some time the results at loop level were only available for bubble diagrams and in principle also for their natural generalizations (as will be reviewed in Section \ref{sec:bubblereview}). It was only very recently that conformal bootstrap method was designed to determine CFT data of multi-trace operators due to bulk loop corrections, order by order in $1/N$, from the assumed known data of single-trace operators \cite{Alday:2016njk,Aharony:2016dwx,Alday:2017xua,Aprile:2017bgs,Aprile:2017xsp}; consistency in the large $N$ expansion was also utilized to directly determine analytic data of Mellin amplitudes for some specific loop diagrams \cite{Aharony:2016dwx}.  In a sense we are at a primitive stage in the understanding of dynamics in AdS (in comparison to that in Minkowski\footnote{For the flat spacetime scattering, convenient all-loop results for the loop integrand are known at least for special theories such as the maximal super Yang--Mills in 4d and various novel geometric structures have been discovered therein; understanding of the analytic structure at one loop is pretty mature; and explicit analytic results are also available for many processes at higher loops.}). This is mainly because most of the existing calculations more or less has to rely on specific spacetime integrations at some stage, during which Schwinger parameters have to be introduced, and the technical difficulty of this grows rapidly with the number of loops.

Nevertheless, the fact that the Mellin amplitudes maintain to be \emph{meromorphic} to all loop levels suggests that, perhaps a little bit counter-intuitive at first sight, the perturbative dynamics in AdS should enjoy certain simplicity that is rather obscure in its flat space counterpart. Meromorphy has proven to be a useful property time and again in various branches of studies; even restricting to perturbation theory, this was the main source of idea leading to the Britto--Cachazo--Feng--Witten recursion relation \cite{Britto:2004ap,Britto:2005fq} and its various geometric implications. Analogous approach was recently attempted in the Witten diagrams in \cite{Raju:2012zr,Raju:2012zs} as well. In this paper we would like to exploit this property at a different angle: meromophy indicates that a Witten diagram at arbitrary loops is effectively similar to a tree diagram. We are going to make this analogy precise and develop it further in Section \ref{sec:arbitraryloops}. In fact, this is quite naturally expected from the boundary point of view, given that any correlation function arising from a bulk perturbative calculation should ultimately allow an OPE expansion, which is structurally tree-like, and the loop dynamics only affects higher-order corrections in $1/N$ to various CFT data.

Moreover, given the practical convenience of Mellin space representation for (holographic) CFT correlators, it should be helpful to make full use of its power by developing machinery that computes \emph{solely} in Mellin space. To the author's knowledge though, this was not thoroughly explored in the previous literature. This has the potential advantage of avoiding the difficulty that overwhelms the previous investigations to loop-level dynamics as mentioned before. To set up such analysis, it is useful to work with the so-called split representation for the bulk-to-bulk propagators \cite{Penedones:2010ue,Costa:2014kfa}, as will be reviewed shortly. 

These two intuitions combined together naturally leads to (integral) recursion relations among Mellin amplitudes that are applicable to the construction of Witten diagrams at arbitrary loop level.

Before we move on, let us point out that apart from the expected applications in holography, Mellin amplitudes for loop-level Witten diagrams could potentially be very useful from a completely different point of view, as a probe for the analytic structure of dynamics in flat space via the connection between the large radius limit of AdS scattering and its flat space counterpart \cite{Penedones:2010ue,Paulos:2016fap}. While this is not explored in this paper, we it leave for the future.

In the rest of this introduction, we first quickly review necessary existing techniques for computations in AdS and set up conventions for later discussions. We then provide a review on some qualitative features of loop-level dynamics that can be expected from existing results. Readers already familiar with these can skip to Section \ref{sec:summary}, which is a comprehensive summary of results in this paper. Outline of the rest of the paper is drawn in Section \ref{sec:guidance}.

\subsection{Review and conventions}

\temp{All are fixed.}

In this paper we set $R_{\rm AdS}=1$ and work with the convention that the bulk-to-boundary propagator of a scalar field is normalized to
\begin{equation}
\checked{}
G_{\rm{b}\partial}^\Delta[X,P]=\frac{\mathcal{C}_\Delta}{(-2P\cdot X)^{\Delta}},\qquad \mathcal{C}_\Delta=\frac{1}{2\pi^h}\frac{\Gamma[\Delta]}{\Gamma[\Delta-h+1]},
\end{equation}
where $h\equiv\frac{d}{2}$ is half of the boundary spacetime dimensions. We always use $P,Q,\ldots$ for boundary coordinates and $X,Y,\ldots$ for bulk cooridnates, using the embedding formalism
\begin{equation}
P\cdot P\equiv0,\quad P\sim\lambda P\;\;(\lambda\in\mathbb{R}_+),\qquad X\cdot X\equiv-1.
\end{equation} 
We use the \emph{split representation} for the bulk-to-bulk propagator \cite{Penedones:2010ue}
\begin{equation}\label{eq:bbpropagatordef}
\checked{}
G_{\rm bb}^{\underline\Delta}[X_1,X_2]=\int[\mathrm{d}c]_{\underline\Delta}\underset{\partial\rm{AdS}}{\int}\!\!\mathrm{d}P\frac{\mathcal{N}_c}{(-2P\cdot X_1)^{h-c}(-2P\cdot X_2)^{h+c}},\quad\mathcal{N}_c=\frac{\Gamma[h\pm c]}{2\pi^{2h}\Gamma[\pm c]}.
\end{equation}
We always underline the conformal dimension of a bulk-to-bulk propagator in order to distinguish it from that of a bulk-to-boundary propagator. The explicit definition of the \emph{spectrum integral} is
\begin{equation}\label{eq:spectrummeasure}
\checked{}
\int[\mathrm{d}c]_{\underline\Delta}\cdots\equiv\int_{-i\infty}^{+i\infty}\frac{\mathrm{d}c}{2\pi i}\frac{1}{(\underline\Delta-h)^2-c^2}\cdots,
\end{equation}
where the $c$ contour is chosen such as to separate the left poles and the right poles, which will be explained in more detail in Section \ref{sec:MtoM}.  In \eqref{eq:bbpropagatordef} we also abbreviate $\Gamma[a\pm b]\equiv\Gamma[a+b]\Gamma[a-b]$, and more generally we always employ the notation
\begin{equation}\label{eq:pmabbreviation}
\Gamma[a_0\pm a_1\pm \cdots\pm a_m]\equiv\prod_{\{\sigma\}}\Gamma[a_0+\sum_{i=1}^m\sigma_ia_a],
\end{equation}
where the product scans over all $2^m$ possible combinations of the signs $\{\sigma_1,\ldots,\sigma_m\}$. This is going to be frequently used in the paper.

We also have the \emph{Symanzik star formula} \cite{Symanzik:1972wj,Paulos:2011ie}
\begin{equation}\label{eq:symanzik}
\checked{}
\int_0^\infty\prod_{i=1}^n\mathrm{d}\alpha_i\,\alpha_i^{\Delta_i-1}\exp\left[-\sum_{1\leq i<j\leq n}\alpha_i\alpha_jP_{i\,j}\right]=\frac{1}{2}\int[\mathrm{d}\delta]\prod_{1\leq i<j\leq n}\frac{\Gamma[\delta_{i\,j}]}{P_{i\,j}^{\delta_{i\,j}}}.
\end{equation}
\temp{here I checked the overall coefficient for 3pt and 4pt} where $P_{ij}=-2P_i\cdot P_j$. $\delta_{ij}$'s are not all independent but satisfy
\begin{equation}\label{eq:canonicalconstraints}
\checked{}
\delta_{i\,j}=\delta_{j\,i},\qquad
\delta_{i\,i}=-\Delta_i,\qquad
\sum_{j=1}^n\delta_{i\,j}=0.
\end{equation}
The measure $[\mathrm{d}\delta]$ is then the product of $\frac{\mathrm{d}\delta}{2\pi i}$ for any $\frac{n(n-3)}{2}$ independent $\delta$ variables, while we keep in mind that the remaining variables in the integrand are solved in terms of them using the above constraints. The $\delta$ contours, similar to that of \eqref{eq:spectrummeasure}, all follow the standard prescription for Mellin integrals, which are discussed in Appendix \ref{app:sec:mellinintegrals}, and will be described when we actually work with them later on.

Consequently a generic $n$-point contact diagram with non-derivative coupling is
\begin{equation}\label{eq:contactcorrelator}
\checked{}
\underset{\rm AdS}{\int}\mathrm{d}X\prod_{i=1}^nG_{\rm{b}\partial}^{\Delta_i}[X,P_i]=\frac{\pi^{h}}{2}\prod_{i=1}^n\frac{\mathcal{C}_{\Delta_i}}{\Gamma[\Delta_i]}\Gamma[{\textstyle\frac{\Sigma\Delta}{2}-h}]\int[\mathrm{d}\delta]\prod_{i<j}\frac{\Gamma[\delta_{ij}]}{P_{ij}^{\delta_{ij}}},
\end{equation}
where we abbreviate $\Sigma\Delta\equiv\sum_{i=1}^n\Delta_i$. We preserve the notation $\delta$ for Mellin integration variables that satisfy the above canonical constraints \eqref{eq:canonicalconstraints}.

The \emph{Mellin amplitude} $\mathcal{M}$ of a CFT correlator is defined by
\begin{equation}\label{eq:correlatorMellin}
\checked{}
\langle\mathcal{O}_1\cdots\mathcal{O}_n\rangle
=\int[\mathrm{d}\delta]\,\mathcal{M}\prod_{i<j}\frac{\Gamma[\delta_{ij}]}{P_{ij}^{\delta_{ij}}}.
\end{equation}
Here the measure $[\mathrm{d}\delta]$ follows the same definition as that appearing in the Symanzik formula \eqref{eq:symanzik} and the set of Mellin variables undergo the same constraints \eqref{eq:canonicalconstraints}. Note here and in \eqref{eq:symanzik} we have assumed that the spacetime dimension $d$ (or $h$) is sufficiently large, because otherwise the Mellin variables receive extra linear constraints from vanishing Gram determinant (similar to the flat space Mandelstam variables, see \cite{Penedones:2016voo}). However, in this paper we do not explore the latter case.

With this setup the Mellin amplitude of a contact diagram is as simple as
\begin{equation}\label{eq:contactMellin}
\checked{}
\mathcal{M}_{\text{contact}}=\frac{\pi^h}{2}\prod_{i=1}^n\frac{\mathcal{C}_{\Delta_i}}{\Gamma[\Delta_i]}\times\Gamma[{\textstyle\frac{\Sigma\Delta}{2}-h}].
\end{equation}
Beyond the contact diagram the Mellin amplitude will in general have non-trivial dependence on the Mellin variables. In fact, as we will also explicitly see later on, it depends on these variables only via their Mandelstam-like combinations, e.g.,
\begin{equation}
s_A=\sum_{i\in A}\Delta_i-2\sum_{i<j\in A}\delta_{i\,j}.
\end{equation}
For simplicity we will just call these Mandelstam variables; by \eqref{eq:canonicalconstraints} obviously they satisfy the same relations as those for flat space scattering.

In this paper we will equally focus on a slightly different quantity, $M$, which is prior to all the spectrum integrals, i.e.,
\begin{equation}\label{eq:Mellinnormalization}
\checked{}
\mathcal{M}[\{\Delta\},\{\underline\Delta\},h]=\int\mathcal{N}\,M[\{\Delta\},\{c\},h],\qquad
\mathcal{N}=\frac{\pi^{(1-L)h}}{2^{2V+L-1}}
\prod_{i=1}^n\frac{\mathcal{C}_{\Delta_i}}{\Gamma[\Delta_i]}\prod_{a}\frac{[\mathrm{d}c_a]_{\underline\Delta_a}}{\Gamma[\pm c_a]}.
\end{equation}
We call these \emph{(Mellin) pre-amplitude}. The reason why this quantity is emphasized will be clear later in Section \ref{sec:polestructure}. 
Here we use $\mathcal{N}$ to denote our choice of overall normalization constant, and the label $a$ runs over all bulk-to-bulk propagators in the Witten diagram. $V$ and $L$ denote the number of bulk vertices and the number of loops, respectively. 
In the special case of a contact diagram we then have
\begin{equation}
\checked{}
M_{\rm contact}=\Gamma[{\textstyle\frac{\Sigma\Delta}{2}-h}].
\end{equation}

Finally we also collect here  useful formulas for bulk and boundary spacetime integrals
\begin{align}
\checked{}\underset{\rm AdS}{\int}\mathrm{d}X\exp[-2X\cdot Y]&=\pi^h\int_0^\infty\mathrm{d}z\,z^{-h-1}\exp[-z-Y^2/z],\\
\checked{}\underset{\partial\rm{AdS}}{\int}\mathrm{d}Q\frac{1}{(-2Q\cdot Y)^{2h}}&=\pi^h\frac{\Gamma[h]}{\Gamma[2h]}\frac{1}{(-Y^2)^h}.
\end{align}

\subsection{Qualitative estimations at loop level}\label{sec:bubblereview}

Before going to detailed analysis of Witten diagrams, let us draw a rough estimation on some qualitative features of Mellin amplitudes at generic loop level.

Two known properties of bulk-to-bulk propagators are needed for this estimation. For our purpose it suffices to describe the qualitative result. Quantitative details are reviewed in Appendix \ref{app:sec:bbpropagators}. 
\begin{enumerate}[noitemsep,nolistsep]
\item Consider two bulk-to-bulk propagators anchored at the same pair of bulk points. Their product allows a linear expansion onto single bulk-to-bulk propagators \cite{Penedones:2010ue}
\begin{equation}\label{eq:Gbbid1}
G_{\rm bb}^{\underline\Delta_1}[X,Y]\,G_{\rm bb}^{\underline\Delta_2}[X,Y]=\sum_{\underline\Delta'}\alpha^{\underline\Delta'}_{\underline\Delta_1\,\underline\Delta_1}\,G_{\rm bb}^{\underline\Delta'}[X,Y],\qquad
\underline\Delta'=\underline\Delta_1+\underline\Delta_2+2m,
\end{equation}
where $m$ is some non-negative integer and $\alpha$ denotes the coefficients in the expansion. This is exactly what make one-loop bubble diagrams accessible in previous literature, by reducing the problem to that of exchange diagrams at tree level.
\item Consider two bulk-to-bulk propagators which now share only one end point, and we further assume this common point is not connected to any other boundary or bulk points in the diagram. In this case the integration of this common point over the AdS bulk induces another linear expansion \cite{Hijano:2015zsa}
\begin{equation}\label{eq:Gbbid2}
\underset{\rm AdS}{\int}\mathrm{d}Z\,G_{\rm bb}^{\underline\Delta_1}[X,Z]\,G_{\rm bb}^{\underline\Delta_2}[Z,Y]=\beta_{\underline\Delta_1\,\underline\Delta_2}\,G_{\rm bb}^{\underline\Delta_1}[X,Y]+\beta_{\underline\Delta_2\,\underline\Delta_1}\,G_{\rm bb}^{\underline\Delta_2}[X,Y],
\end{equation}
where the $\beta$'s are some other coefficients.
\end{enumerate}

\begin{figure}[ht]
\captionsetup{margin=2em}
\begin{center}
\begin{tikzpicture}
\draw [black,ultra thick] (0,0) circle [radius=2.5];
\draw [black,thick] (-1.5,0) -- (150:2.5);
\draw [black,thick] (-1.5,0) -- (180:2.5);
\draw [black,thick] (-1.5,0) -- (210:2.5);
\draw [black,thick] (1.5,0) -- (30:2.5);
\draw [black,thick] (1.5,0) -- (10:2.5);
\draw [black,thick] (1.5,0) -- (-10:2.5);
\draw [black,thick] (1.5,0) -- (-30:2.5);
\draw [black,thick] (-1.5,0) -- (1.5,0);
\draw [black,thick] (180:1.5) arc [start angle=180,end angle=360,radius=1.5] arc [start angle=0,end angle=180,radius=1] arc [start angle=180,end angle=360,radius=.5];
\fill [black] (-1.5,0) circle [radius=2pt];
\fill [black] (-.5,0) circle [radius=2pt];
\fill [black] (.5,0) circle [radius=2pt];
\fill [black] (1.5,0) circle [radius=2pt];
\draw [red,very thick,dashed] (140:2.5) .. controls (140:1) and (220:1) .. (220:2.5);
\draw [red,very thick,dashed] (85:2.5) .. controls (85:1) and (275:1) .. (275:2.5);
\draw [red,very thick,dashed] (50:2.5) .. controls (50:1) and (-50:1) .. (-50:2.5);
\node [anchor=north] at (0,-1.5) {$1$};
\node [anchor=south] at (-1,0) {$2$};
\node [anchor=south] at (.5,1) {$3$};
\node [anchor=north] at (1,0) {$4$};
\node [anchor=north] at (0,-.5) {$5$};
\node [anchor=south] at (0,0) {$6$};
\node [anchor=south east] at (-1.5,0) {$X$};
\node [anchor=north east] at (-.5,0) {$Z$};
\node [anchor=south] at (.5,0) {$W$};
\node [anchor=south east] at (1.5,0) {$Y$};
\end{tikzpicture}
\end{center}
\vspace{-.5em}\caption{An example of generalized bubble diagram. The red dashed curves serve as a convenient book-keeping of the operators potentially present in a given OPE of the boundary correlator (which are not the multi-trace operators formed by the boundary operators), or equivalently, potential poles of the Mellin amplitude in the corresponding Mandelstam variable.}
\label{fig:bubblegeneralization}
\end{figure}
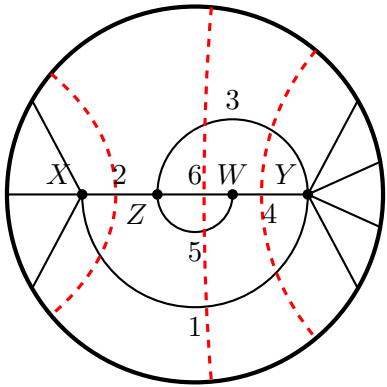

\subsubsection{Generalized bubble diagrams}\label{sec:generalizedubbles}

The two identities \eqref{eq:Gbbid1} and \eqref{eq:Gbbid2} allows us to easily analyze a special class of diagrams, all of which can be ultimately reduced to the tree-level exchange by nested applications of them. Let us call them \emph{generalized bubble diagrams}.
 
A simple yet non-trivial example of such diagrams is shown in Figure \ref{fig:bubblegeneralization}, which consists of six bulk-to-bulk propagators and three loops. To deal with this diagram, it is obvious that we can first apply \eqref{eq:Gbbid1} to propagator 5 and 6, obtaining a linear sum of $G_{\rm bb}[Z,W]$ with conformal dimensions valued in $\underline\Delta_5+\underline\Delta_6+2m_1$ ($m_1\in\mathbb{N}$); and then apply \eqref{eq:Gbbid2} for the $W$ integration, so that both these and the original propagator 4 are now anchored at $Z$ and $Y$; then a second application of \eqref{eq:Gbbid1} on these propagators with propagator 3 results in a linear combination of $G_{\rm bb}[Z,Y]$, with dimensions valued in $\underline\Delta_3+\underline\Delta_5+\underline\Delta_6+2m_2$ and in $\underline\Delta_3+\underline\Delta_4+2m_3$ ($m_2,m_3\in\mathbb{N}$). A further \eqref{eq:Gbbid2} gets rid of the $Z$ integration and by another \eqref{eq:Gbbid1} we finally turn this diagram into a linear combination of exchange diagrams, schematically of the form
\begin{equation}\label{eq:generalizedbubbledecomposition}
\sum_{n_1}G_{\rm bb}^{\underline\Delta_{12}+2n_1}[X,Y]+\sum_{n_2}G_{\rm bb}^{\underline\Delta_{1356}+2n_2}[X,Y]+\sum_{n_3}G_{\rm bb}^{\underline\Delta_{134}+2n_3}[X,Y],
\end{equation}
where all the summations are over non-negative integers, and we abbreviate, e.g., $\underline\Delta_{12}=\underline\Delta_1+\underline\Delta_2$. We also ignored the coefficients. As a result, the Mellin amplitude for this diagram is in principle explicitly known, though we do not further work out the details here.

Once the correlator associated to this diagram is obtained, its OPE expansion and its contribution to the spectrum as seen in the corresponding channel are of great interest. There are relatively two different types of operators that we may observe in a chosen expansion. 
\begin{enumerate}[noitemsep,nolistsep]
\item In any OPE channel we should always be able to observe multi-trace primary operators (and their descendents) formed by the group of boundary operators on either side. From the Mellin amplitude point of view this is taken care of by the poles of the $\Gamma$ functions in its dressing factor $\prod_{i<j}\frac{\Gamma[\delta_{i\,j}]}{P_{i\,j}^{\delta_{i\,j}}}$ in \eqref{eq:correlatorMellin}. They do not manifest as poles of the Mellin amplitude $\mathcal{M}$.
\item In the special case of single exchange diagram, in its direct channel we will in addition observe the very single-trace operator (and its descendents) associated to the bulk field in the exchange. This can be worked out, e.g., by directly expanding the diagram into geodesic Witten diagrams as prescribed in \cite{Aharony:2016dwx}. Combined with qualitative result for the bubble diagrams and their generalization discussed above, this immediately teaches us that for the generalized bubble diagrams we will in principle also observe multi-trace operators formed by the single-trace ones associated to the original bulk-to-bulk propagators, and the pattern is tightly tied to the topology of the diagram. To be explicit, for our example in Figure \ref{fig:bubblegeneralization}, the expansion \eqref{eq:generalizedbubbledecomposition} that the OPE in the direct channel should contain the double-, triple- and quadruple-trace primary operators of the schematic form
\begin{equation}\label{eq:multitraceexample}
[\mathcal{O}_{\underline1}\mathcal{O}_{\underline2}],\qquad
[\mathcal{O}_{\underline1}\mathcal{O}_{\underline3}\mathcal{O}_{\underline4}],\qquad
[\mathcal{O}_{\underline1}\mathcal{O}_{\underline3}\mathcal{O}_{\underline5}\mathcal{O}_{\underline6}].
\end{equation}
As is implied by the result for Mellin amplitudes of exchange diagrams \eqref{eq:Mexchange}, these operators together with their descendants are in principle manifest in the Mellin amplitude as simple poles of the Mandelstam variable at their corresponding twists.
\end{enumerate}

The operators of the second type above come with a simple visualization in the diagrams: for a given OPE channel we draw an auxiliary curve/hypersurface (marked red in Figure \ref{fig:bubblegeneralization}) that separates the group of points in the OPE limit from the rest, and require that it ``cuts'' the original diagram into two pieces, each of which is connected, and avoids interaction vertices. Generically there can be several inequivalent choices of cuts in a specified channel. For each choice we collect the set of bulk-to-bulk propagators it cuts across, and the multi-trace operators formed by the corresponding single-trace operators is expected to be present in the spectrum in this OPE expansion. Applying this rule to Figure \ref{fig:bubblegeneralization} we exactly recover the three families concluded in \eqref{eq:multitraceexample}.

The validity of the above rule is a direct consequence of the identities \eqref{eq:Gbbid1} and \eqref{eq:Gbbid2} among bulk-to-bulk propagators. The notation is in close analogy with cutting Feynman diagrams, and many diagrammatic intuitions there can be directly borrowed. In flat spacetime a cut diagram admits of further physical interpretation as discontinuities across branch cuts of the amplitude, which in physical region reveals information about the spectrum as well. To avoid confusion, however, we emphasize that in this paper ``cut'' in AdS merely refers to a convenient graphical book-keeping and we do not mean to do any specific operation on the bulk-to-bulk propagators that are cut by the auxiliary curve.

\subsubsection{Generic Witten diagrams}

For generic Witten diagrams beyond the generalized bubbles, although the above discussions do not directly apply, they can still help to perform preliminary estimations. Intuitively this is because for an arbitrary loop diagram together with a specified OPE channel the bulk integrations always contains a sub-region where the original Witten diagram degenerates into a generalized bubble consistent with the channel (apart from possibly additional tadpoles, which are however irrelevant). 
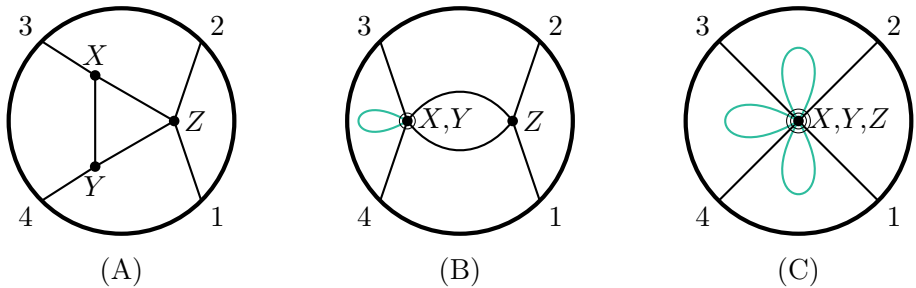
\begin{figure}[ht]
\captionsetup{margin=2em}
\begin{center}
\begin{tikzpicture}
\begin{scope}[xshift=-4.5cm]
\node [anchor=center] at (0,-2) {(A)};
\draw [black,ultra thick] (0,0) circle [radius=1.5];
\draw [black,thick] (120:.7) -- (240:.7) -- (.7,0) -- cycle;
\draw [black,thick] (120:.7) -- (135:1.5);
\draw [black,thick] (240:.7) -- (225:1.5);
\draw [black,thick] (45:1.5) -- (0:.7) -- (-45:1.5);
\fill [black] (0:.7) circle [radius=2pt];
\fill [black] (120:.7) circle [radius=2pt];
\fill [black] (240:.7) circle [radius=2pt];
\node [anchor=center] at (-45:1.8) {$1$};
\node [anchor=center] at (45:1.8) {$2$};
\node [anchor=center] at (135:1.8) {$3$};
\node [anchor=center] at (225:1.8) {$4$};
\node [anchor=south] at (120:.7) {$X$};
\node [anchor=north] at (240:.7) {$Y$};
\node [anchor=west] at (0:.7) {$Z$};
\end{scope}
\begin{scope}
\node [anchor=center] at (0,-2) {(B)};
\draw [black,ultra thick] (0,0) circle [radius=1.5];
\draw [black,thick] (135:1.5) -- (180:.7) -- (-135:1.5);
\draw [black,thick] (45:1.5) -- (0:.7) -- (-45:1.5);
\draw [black,thick] (0:.7) .. controls (60:.6) and (120:.6) .. (180:.7) .. controls (-120:.6) and (-60:.6) .. (0:.7);
\draw [SeaGreen,thick] (180:.7) .. controls ($(180:.7)+(150:1)$) and ($(180:.7)+(-150:1)$) .. (180:.7);
\fill [black] (0:.7) circle [radius=2pt];
\fill [black] (180:.7) circle [radius=2pt];
\draw [black] (180:.7) circle [radius=3pt];
\node [anchor=center] at (-45:1.8) {$1$};
\node [anchor=center] at (45:1.8) {$2$};
\node [anchor=center] at (135:1.8) {$3$};
\node [anchor=center] at (225:1.8) {$4$};
\node [anchor=west] at (180:.7) {$X,\!Y$};
\node [anchor=west] at (0:.7) {$Z$};
\end{scope}
\begin{scope}[xshift=4.5cm]
\node [anchor=center] at (0,-2) {(C)};
\draw [black,ultra thick] (0,0) circle [radius=1.5];
\draw [black,thick] (-45:1.5) -- (135:1.5);
\draw [black,thick] (45:1.5) -- (225:1.5);
\draw [SeaGreen,thick] (0,0) .. controls (60:1.5) and (120:1.5) .. (0,0);
\draw [SeaGreen,thick] (0,0) .. controls (150:1.5) and (-150:1.5) .. (0,0);
\draw [SeaGreen,thick] (0,0) .. controls (-60:1.5) and (-120:1.5) .. (0,0);
\fill [black] (0,0) circle [radius=2pt];
\draw [black] (0,0) circle [radius=3pt];
\draw [black] (0,0) circle [radius=4.5pt];
\node [anchor=center] at (-45:1.8) {$1$};
\node [anchor=center] at (45:1.8) {$2$};
\node [anchor=center] at (135:1.8) {$3$};
\node [anchor=center] at (225:1.8) {$4$};
\node [anchor=west] at (0,0) {$X,\!Y,\!Z$};
\end{scope}
\end{tikzpicture}
\end{center}
\vspace{-1.5em}\caption{Degeneration of the 4-point triangle into generalized bubbles. (A) original diagram; (B) degeneration in the $S$ channel; (C) degeneration in the $T$ channel. There can be additional tadpoles in the degenerate diagrams (marked green), but these are not relevant for our estimation.}
\label{fig:triangledegeneration}
\end{figure}
For instance in the simplest case let us consider a 4-point triangle diagram as in Figure \ref{fig:triangledegeneration}, for which neither \eqref{eq:Gbbid1} nor \eqref{eq:Gbbid2} is available. But in the $S$ channel as we perform the bulk integrations there is a sub-region where $X=Y$ and the diagram degenerates into a bubble in this channel. So we would conclude a double-trace operator corresponding to the two remaining propagators, if contribution from all other regions are absent. On the other hand, in the $T$ channel it is obvious the diagram has to degenerate into a contact diagram in order to fit into the generalized bubbles, and correspondingly we can see no operators of the second type in this part of the contribution. At the level of the Mellin amplitude, all such operators that are present in the OPE should again manifest as poles in the amplitude.

More straightforwardly, we can apply the cutting rules described above to the original diagram and reach the same qualitative predictions. Of course, this does not plainly mean that the original diagram itself is identical to the summation of these various degenerate ones. Hence a rigorous verification is still in need.



\subsection{Summary of results}\label{sec:summary}

Given the qualitative estimations in the previous subsection, it may appear a bit surprising that the actual Mellin amplitudes for scalar diagrams are much simpler, in the sense that a fair amount of poles expected from these intuitions turn out to be absent.

Let us take this opportunity to collect some of our main results.

\subsubsection{Mellin (pre-)amplitudes}

First of all, as we mentioned previously, in this paper we are interested not only in the Mellin amplitude $\mathcal{M}$ (as commonly defined in literature) but also equally in a closely related object $M$ that we call Mellin pre-amplitude, as defined in \eqref{eq:Mellinnormalization}.   Apart from practical concern in actual computations, a main reason and advantage in introducing this notion is that, while it is a certain integrand to the Mellin amplitude and thus simpler, it still manages to capture many features of the amplitude manifestly and is itself elegantly organized in structure.

To be precise, let us first restrict to diagrams at tree level. Here the Mellin pre-amplitude always  factorizes into elementary pieces, each of which is tied to either a bulk-to-bulk propagator or a bulk vertex of the given diagram. This is in strict analogy with the Feynman rules for flat space scattering as studied in the momentum space. Explicitly, as will be derived in Section \ref{sec:treediagrammaticrules},
\begin{itemize}[noitemsep,nolistsep]
\item For each bulk-to-bulk propagator $a$, with $c_a$ and $s_a$ denoting its corresponding spectrum variable and Mandelstam variable, we obtain a factor
\begin{equation}\label{intro:treepropagatorcontribution}
\Gamma[{\textstyle\frac{h\pm c_a-s_a}{2}}]
\equiv\Gamma[{\textstyle\frac{h+ c_a-s_a}{2}}]\,\Gamma[{\textstyle\frac{h- c_a-s_a}{2}}].
\end{equation}
\item For each bulk vertex $A$, which is incident to $r$ bulk-to-bulk propagators (label them by $\{1,2,\ldots,r\}$; we call this vertex to have valency $r$), and for which the total sum of conformal dimensions of boundary points attached to it being $\Delta_A$, we obtain a factor
\begin{equation}\label{intro:polecontribution}
\int[\mathrm{d}w]\,\Gamma[{\textstyle\frac{\Delta_{A}+(r-2)h+\sum_{a=1}^r(c_a+2w_a)}{2}}]\prod_{a=1}^r\frac{\Gamma[-w_a]\Gamma[-c_a-w_a]\Gamma[\frac{h+c_a-s_{a}}{2}+w_a]}{\Gamma[\frac{h\pm c_a-s_{a}}{2}]},
\end{equation}
where $[\mathrm{d}w]\equiv\prod_{a=1}^r\frac{\mathrm{d}w_a}{2\pi i}$, whose integral contours follow the standard Mellin integrals.  In particular, the poles of this function always follow a simple pattern, such that we also frequently re-write this contribution into
\begin{equation}\label{intro:polecorrection}
\Gamma[{\textstyle\frac{\Delta_{A}+(r-2)h\pm c_1\pm\cdots\pm c_r}{2}}]\,C_A[\Delta_a;\substack{c_1,\ldots,c_r\\s_1,\ldots,s_r}],
\end{equation}
where $C_A$ is a pole-free function determined by identifying \eqref{intro:polecontribution} and \eqref{intro:polecorrection}, and we call it the \emph{correction function} associated to the vertex $A$.
\end{itemize}
Note that in \eqref{intro:treepropagatorcontribution} and \eqref{intro:polecorrection} we employed the abbreviation \eqref{eq:pmabbreviation}. The tree pre-amplitude is then simply the ordinary product of all these contributions from elements in the diagram
\begin{equation}
M_{\rm tree}=\prod_a\eqref{intro:treepropagatorcontribution}\prod_{A}\eqref{intro:polecorrection}.
\end{equation}
All the poles of $M_{\rm tree}$ are straightforwardly indicated by the explicit $\Gamma$ functions thus obtained. Note that at the level of the Mellin pre-amplitude there is completely no dependence on the actual conformal dimensions $\underline\Delta$ of the bulk-to-bulk propagators, as they only enter via the spectral integrals \eqref{eq:bbpropagatordef} that convey $M$ to $\mathcal{M}$. In $M$ their roles are played instead by the spectrum variables $c$.  The equivalence between the diagrammatic rules presented above and the existing knowledge from previous literature can be easily understood, as will be discussed in Appendix \ref{app:sec:proofofequivalence}.

\vspace{1em}

Given this simple diagrammatic structure of the expression for pre-amplitudes at tree level, it is interesting to inquire to what extent this holds at loop level.   It can be confirmed in general (as in Section \ref{sec:localityofinsertion}) that the pre-amplitude of a loop-level diagram again factorizes, into individual contribution from: (1) each tree bulk-to-bulk propagator (following the same rule \eqref{intro:treepropagatorcontribution}), and (2) each one-particle irreducible (1PI) part of the diagram.  However, the contribution from each 1PI sub-diagram has to be a complicated function that does not directly factorize on its own.

Despite this, if we restrict our scope to the pole structure only, then it turns out there equally exist a set of diagrammatic rules that (we conjecture to) hold for arbitrary diagrams, as will be observed in Section \ref{sec:polediagrammaticrules}. To set up notation, whenever we write down a $\Gamma$ function we also equally refer to the family of poles in it. In general we perform the following analysis:
\begin{itemize}[noitemsep,nolistsep]
\item \textbf{Vertex rule.} For each bulk vertex (with the total conformal dimension of its attached boundary points being $\Delta_A$ and incident to bulk-to-bulk propagators $\{1,2,\ldots,r\}$) we read off the poles
\begin{equation}
\Gamma[{\textstyle\frac{\Delta_A+(r-2)h\pm c_1\pm \cdots\pm c_r}{2}}],
\end{equation}
which are the same poles as those in the tree-level rule \eqref{intro:polecorrection}.
\item \textbf{Channel rule.} For each physical channel (specified by Mandelstam variable $S$) we collect the set of possible cuts as discussed in Section \ref{sec:generalizedubbles}. For each specific cut (which runs across bulk-to-bulk propagators $\{1,2,\ldots,r\}$), we read off the poles
\begin{equation}\label{intro:channelrule}
\Gamma[{\textstyle\frac{rh\pm c_1\pm\cdots\pm c_r-S}{2}}].
\end{equation}
This is a generalization to the tree-level rule \eqref{intro:treepropagatorcontribution}. Recall that an allowed cut should divide the original diagram into two parts, each being a \emph{connected} diagram. And in particular we also take into consideration trivial channels such that only one boundary point is on one side of the cut (in which case $S$ reduces to the conformal dimension of this point).
\item \textbf{Loop contraction rule.} At loop level we also consider degenerate diagrams obtained from the given diagram by contracting existing loops into bulk vertices in all possible ways. For every new vertex thus emerged, we apply the vertex rule above as well.
\item \textbf{Generalized bubble rule.} If the given diagram itself or any of its sub-diagram is a generalized bubble as defined in Section \ref{sec:generalizedubbles}, then for any explicit bulk-to-bulk propagator $a$ inside this generalized bubble we further read off the poles
\begin{equation}
\Gamma[h\pm c_a].
\end{equation}
\end{itemize}
For a given diagram we apply the above four rules and collect all the poles thus obtained, and the genuine poles of the Mellin amplitude have to belong to this set.  However, in general this set of families of poles can be redundant.  In order to precisely locate the genuine poles, we need to introduce a notion of ``\emph{compositeness}''.  Roughly speaking, based on the result at this stage we imagine further doing integrals of the spectrum variables $c$ as well as the boundary point conformal dimensions $\Delta$, whose contour follows the standard Mellin integrals.  If it turns out that a family of poles in our collection can equally emerge from other families of poles in this same collection via (a subset of) these integrals, then we call them composite poles.  Otherwise the poles are non-composite. With this notion we then propose our first conjecture (in Section \ref{sec:enhancedrules}), regarding the precise pole structure of pre-amplitudes:
\begin{conjecture}{Conjecture on $M$}
For an arbitrary scalar Witten diagram, we work out all the poles following the four rules prescribed above. After eliminating the composite poles, the remaining families exactly constitute all the genuine poles of the pre-amplitude of the given diagram.
\end{conjecture}
While being a conjecture, we have confirmed its validity in many non-trivial examples up to two loops (Section \ref{sec:polestructure} and Appendix \ref{app:sec:nonplanar2loop}) as well as at higher loops for the generalized bubbles (Appendix \ref{app:sec:polegbubbles}). In practice this allows us to estimate the pole structure of pre-amplitudes for any diagram by directly carrying out the above procedure.

To illustrate this in an explicit example, let us consider the following 4-point non-planar diagram at two loops.
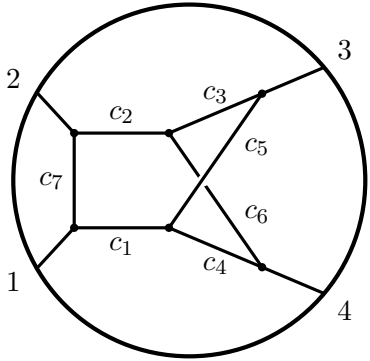
\begin{center}
\begin{tikzpicture}
\begin{scope}[scale=.9]
\draw [black,ultra thick] (0,0) circle [radius=2.6];
\coordinate (up) at (-.3,.7);
\coordinate (down) at (-.3,-.7);
\draw [black,very thick] (150:2.6) -- ($(up)+(-1.4,0)$) -- ($(down)+(-1.4,0)$) -- (-150:2.6);
\draw [black,very thick] (40:2.6) -- (up) -- ($(up)+(-1.4,0)$);
\draw [black,very thick] ($(down)+(-1.4,0)$) -- (down) -- (-40:2.6);
\draw [black,very thick] (up) -- ($(down)!.6!(-40:2.6)$);
\fill [white] (.21,0) circle [radius=2.5pt];
\draw [black,very thick] (down) -- ($(up)!.6!(40:2.6)$);
\fill [black] (up) circle [radius=1.75pt];
\fill [black] (down) circle [radius=1.75pt];
\fill [black] ($(up)+(-1.4,0)$) circle [radius=1.75pt];
\fill [black] ($(down)+(-1.4,0)$) circle [radius=1.75pt];
\fill [black] ($(up)!.6!(40:2.6)$) circle [radius=1.75pt];
\fill [black] ($(down)!.6!(-40:2.6)$) circle [radius=1.75pt];
\node [anchor=center] at (-150:3) {$1$};
\node [anchor=center] at (150:3) {$2$};
\node [anchor=center] at (40:3) {$3$};
\node [anchor=center] at (-40:3) {$4$};
\node [anchor=north] at ($(down)+(-.7,0)$) {$c_1$};
\node [anchor=south] at ($(up)+(-.7,0)$) {$c_2$};
\node [anchor=north] at ($(down)!.3!(-40:2.6)$) {$c_4$};
\node [anchor=south] at ($(up)!.3!(40:2.6)$) {$c_3$};
\node [anchor=center] at (1,.5) {$c_5$};
\node [anchor=center] at (1,-.5) {$c_6$};
\node [anchor=east] at (-1.7,0) {$c_7$};
\end{scope}
\end{tikzpicture}
\end{center}
From the vertex rule we have poles
\begin{equation}\label{intro:vertexrule}
\Gamma[{\textstyle\frac{\Delta_1\pm c_1\pm c_7}{2}}]\,
\Gamma[{\textstyle\frac{\Delta_2\pm c_2\pm c_7}{2}}]\,
\Gamma[{\textstyle\frac{h\pm c_1\pm c_4\pm c_5}{2}}]\,
\Gamma[{\textstyle\frac{h\pm c_2\pm c_3\pm c_6}{2}}]\,
\Gamma[{\textstyle\frac{\Delta_3\pm c_3\pm c_5}{2}}]\,
\Gamma[{\textstyle\frac{\Delta_4\pm c_4\pm c_6}{2}}],
\end{equation}
one for each of the six bulk vertices. From the channel rule we have poles
\begin{equation}\label{intro:channelruletrivial}
\begin{split}
\Delta_1:\quad&\Gamma[{\textstyle\frac{2h\pm c_1\pm c_7-\Delta_1}{2}}]\,
{\color{ForestGreen}\Gamma[{\textstyle\frac{3h\pm c_4\pm c_5\pm c_7-\Delta_1}{2}}]},\\
\Delta_2:\quad&\Gamma[{\textstyle\frac{2h\pm c_2\pm c_7-\Delta_2}{2}}]\,
{\color{ForestGreen}\Gamma[{\textstyle\frac{3h\pm c_3\pm c_6\pm c_7-\Delta_2}{2}}]},\\
\Delta_3:\quad&\Gamma[{\textstyle\frac{2h\pm c_3\pm c_5-\Delta_3}{2}}]\,
{\color{ForestGreen}\Gamma[{\textstyle\frac{3h\pm c_1\pm c_3\pm c_4-\Delta_3}{2}}]\,
\Gamma[{\textstyle\frac{3h\pm c_2\pm c_5\pm c_6-\Delta_3}{2}}]},\\
\Delta_4:\quad&\Gamma[{\textstyle\frac{2h\pm c_4\pm c_6-\Delta_4}{2}}]\,
{\color{ForestGreen}\Gamma[{\textstyle\frac{3h\pm c_1\pm c_5\pm c_6-\Delta_4}{2}}]\,
\Gamma[{\textstyle\frac{3h\pm c_2\pm c_3\pm c_4-\Delta_4}{2}}]},
\end{split}
\end{equation}
from the four trivial channels, as well as poles
\begin{equation}\label{intro:channelrulenontrivial}
\begin{split}
S:\quad&\Gamma[{\textstyle\frac{2h\pm c_1\pm c_2-S}{2}}]\,
{\color{ForestGreen}\Gamma[{\textstyle\frac{3h\pm c_1\pm c_3\pm c_6-S}{2}}]\,
\Gamma[{\textstyle\frac{3h\pm c_2\pm c_4\pm c_5-S}{2}}]},\\
T:\quad&\Gamma[{\textstyle\frac{3h\pm c_5\pm c_6\pm c_7-T}{2}}],\\
U:\quad&\Gamma[{\textstyle\frac{3h\pm c_3\pm c_4\pm c_7}{2}}],
\end{split}
\end{equation}
from the three non-trivial physical channels. Furthermore, this two loop diagram can induce four degenerate diagrams via loop contraction, three by contracting one of the loops, and one by contracting all existing loops. Each of these diagrams contains one new emergent bulk vertex, and so from the loop contraction rule we also have poles
\begin{equation}\label{intro:loopcontractionrule}
{\color{ForestGreen}\Gamma[{\textstyle\frac{\Delta_{123}\pm c_4\pm c_6}{2}}]}\,
{\color{ForestGreen}\Gamma[{\textstyle\frac{\Delta_{124}\pm c_3\pm c_5}{2}}]}\,
\Gamma[{\textstyle\frac{\Delta_{34}\pm c_1\pm c_2}{2}}]\,
\Gamma[{\textstyle\frac{\Delta_{1234}}{2}-h}].
\end{equation}
Since neither this diagram nor any of its sub-diagram is a generalized bubble, here we do not have to worry about the generalized bubble rule.  Now collecting all the poles in \eqref{intro:vertexrule}, \eqref{intro:channelruletrivial}, \eqref{intro:channelrulenontrivial} and \eqref{intro:loopcontractionrule}, we observe that the poles marked green are composite ones. To name some examples,
\begin{align}
\label{intro:compositeness}\Gamma[{\textstyle\frac{2h+c_1+c_2-S}{2}}]\,\Gamma[{\textstyle\frac{h-c_2+c_3+c_6}{2}}]&\xrightarrow{\int\mathrm{d}c_2}{\color{ForestGreen}\Gamma[{\textstyle\frac{3h+c_1+c_3+c_6-S}{2}}]},\\
\Gamma[{\textstyle\frac{2h+c_4+c_6-\Delta_4}{2}}]\,\Gamma[{\textstyle\frac{\Delta_{1234}}{2}-h}]&\xrightarrow{\int\mathrm{d}\Delta_4}{\color{ForestGreen}\Gamma[{\textstyle\frac{\Delta_{123}+c_4+c_6}{2}}]}.
\end{align}
In the above formulas the expression on LHS does not necessarily mean a specific integrand but just indicates poles that are present before the specified Mellin integral, while that on RHS indicates poles that can potentially emerge from the poles on LHS via the integral (they emerge due to the pinching of the integration contour following the general discussion in Appendix \ref{app:sec:poleemergence}, but in practice this can be simply treated as eliminating the integrated variable(s) in the arguments of the $\Gamma$ functions). Here we explicitly observe the necessity of considering both $c$ and $\Delta$ integrals in the identification of composite poles, which is plausible since in the split representation \eqref{eq:bbpropagatordef} $h\pm c$ plays the same role as the conformal dimension of an ordinary boundary point. 

Hence according to our conjecture above, after ruling out the families of poles marked green, the rest are all the genuine poles of the pre-amplitude for our example. This turns out to be in compete agreement with results from explicit computation. 

\vspace{1em}

Next we move on to the Mellin amplitude $\mathcal{M}$, via the spectrum integrals. Recalling \eqref{eq:Mellinnormalization}, in these integrals the normalization $\mathcal{N}$ brings  to the integrand extra poles in the $c_a$ variable
\begin{equation}\label{intro:polesfromnormalization}
\frac{1}{(\underline\Delta_a-h)^2-c_a^2}
\end{equation}
for each bulk-to-bulk propagator $a$. This is the only means by which the actual conformal dimensions $\underline\Delta$ of bulk-to-bulk propagators enter Mellin amplitudes. Hence these poles are crucial for the emergence of poles of the amplitude at the end.   In this paper we do not dive into a complete discussion regarding the generic pattern of the entire pole structure for Mellin amplitudes\footnote{Nevertheless, we still provide in Appendix \ref{app:sec:resultpolesofamplitudes} a summary of the complete pole structures of the one-loop Mellin amplitudes considered in this paper worked out from our analysis, for the readers' reference.}.  Instead, we will solely focus on the pattern of the poles in the Mandelstam variables, which captures information relevant for the spectrum observed in OPE expansions of the corresponding boundary correlators.

In fact, from our discussions so far we have enough intuitions to expect a correspondence between poles in the Mandelstam variables for both the Mellin pre-amplitude $M$ and the Mellin amplitude $\mathcal{M}$
\begin{equation}\label{intro:poleMMcorrespondence}
\parbox{12cm}{\centering\tikz{
\node [anchor=east] (M) at (-1.8,0) {$\Gamma[\frac{rh\pm c_1\pm\cdots\pm c_r-S}{2}]$};
\node [anchor=west] (calM) at (1.8,0) {$\Gamma[\frac{\underline\Delta_1+\cdots+\underline\Delta_r-S}{2}]$};
\node [anchor=center] (cut) at (0,1.6) {$\underset{(1,2,\ldots,r)}{r\text{-cut}}$};
\draw [black,thick,<->] (M.north east) -- (cut.south west);
\node [anchor=south,rotate=35] at (-1.3,.7) {\scriptsize $\substack{\text{ channel rule}\\\eqref{intro:channelrule}}$};
\draw [black,thick,<->] (calM.north west) -- (cut.south east);
\node [anchor=south,rotate=-35] at (1.35,.75) {\scriptsize $\substack{\text{diagrammatic}\\\text{intuition}\\\text{Section \ref{sec:generalizedubbles}}}$};
\draw [black,thick,->] (M.east) -- (calM.west);
\node [anchor=north] at (0,0) {$\tiny \int[\mathrm{d}c_1]_{\underline\Delta_1}\cdots[\mathrm{d}c_r]_{\underline\Delta_r}$};
}}
\end{equation}
This correspondence is conveyed by the allowed cuts of the given diagram, which dictate the poles in $M$ by the channel rule \eqref{intro:channelrule} and meanwhile suggests the presence of poles in $\mathcal{M}$ by the discussion in Section \ref{sec:generalizedubbles}. Furthermore, the poles \eqref{intro:polesfromnormalization} also tend to imply a simple connection between the two families of poles via pinching the contours for the spectrum integrals of $\{c_1,c_2,\ldots,c_r\}$. 

However, there appear to be some tension, because by our conjecture on $M$ it may occur that poles of $M$ implied by a certain cut are absent due to compositeness, in which case \eqref{intro:poleMMcorrespondence} seems to break down. This suggests that when analyzing the poles $\Gamma[\frac{\underline\Delta_1+\cdots+\underline\Delta_r-S}{2}]$ that potentially appear in $\mathcal{M}$ we have to distinguish between two situations, which we call minimal and non-minimal respectively:
\begin{itemize}[noitemsep,nolistsep]
\item \textbf{Minimal poles.} Firstly, there are poles of $\mathcal{M}$ whose counterparts in $M$ are indeed present (i.e., non-composite). We call such poles \emph{minimal poles}, and their corresponding cuts \emph{minimal cuts}. These poles emerge from the spectrum integrals in a ``minimal'' way. Taking the above example for instance, we have
\begin{equation}
\begin{rcases}\quad\quad\;\;\Gamma[{\textstyle\frac{2h+c_1+c_2-S}{2}}]\\{\textstyle\frac{1}{(\underline\Delta_1-h)-c_1}}\,{\textstyle\frac{1}{(\underline\Delta_2-h)-c_2}}\end{rcases}\xrightarrow{\int\mathrm{d}c_1\mathrm{d}c_2}\Gamma[{\textstyle\frac{\underline\Delta_{12}-S}{2}}].
\end{equation}
In other words, they emerge immediately after the integrals tied to the cut propagators are performed, while all the rest spectrum integrals are irrelevant.
\item \textbf{Non-minimal poles.} Poles of $\mathcal{M}$ whose counterparts in $M$ are ruled out by compositeness are called \emph{non-minimal}. In principle such poles can still emerge from the spectrum integrals, exactly due to the compositeness of the corresponding ones in $M$, but this has to involve integrals of extra $c$ variables for propagators not encountered in the cut. For example, following \eqref{intro:compositeness} we have
\begin{equation}
\begin{rcases}
\begin{rcases}
\Gamma[{\textstyle\frac{2h+c_1+c_2-S}{2}}]\\
\,\Gamma[{\textstyle\frac{h-c_2+c_3+c_6}{2}}]
\end{rcases}
\xrightarrow{\int\mathrm{d}c_2}{\color{ForestGreen}\Gamma[{\textstyle\frac{3h+ c_1+c_3+c_6-S}{2}}]}\\
\quad\quad\quad\quad\;{\textstyle\frac{1}{(\underline\Delta_1-h)-c_1}}\,{\textstyle\frac{1}{(\underline\Delta_3-h)-c_3}}\,{\textstyle\frac{1}{(\underline\Delta_6-h)-c_6}}
\end{rcases}
\xrightarrow{\int\mathrm{d}c_1\mathrm{d}c_3\mathrm{d}c_6}\Gamma[{\textstyle\frac{\underline\Delta_{136}-S}{2}}].
\end{equation}
Here propagator 2 is not cut, but the integral $c_2$ is still needed and in the intermediate step it exactly reproduces the composite poles $\Gamma[\frac{3h+c_1+c_3+c_5-S}{2}]$ that we ruled out in $M$ previously.
\end{itemize}
More details about the emergence of the non-minimal poles are discussed in Section \ref{sec:nonminimalpoles}. We note here that in order for mathematical rigor one has to perform a careful analysis on the detailed geometry associated to the pinching of the spectrum integral contours when understanding the emergence of these poles in the physical channels, but in practice it turns out that they always follow the patterns described above.  

Although these two types of poles arise in relatively different ways, their local properties can nevertheless be analyzed under the same general framework of multi-dimensional Mellin integrals, as systematically prescribed in Appendix \ref{app:sec:mellinintegrals}.  Very amusingly, and perhaps a bit counter-intuitively, for all the explicit examples that we have studied (except for those containing generalized bubbles) all the non-minimal poles are confirmed to be absent by explicit residue computation.  This leads us to propose another conjecture (in Section \ref{sec:conjectureonamplitude}) regarding the poles of Mellin amplitudes $\mathcal{M}$ in the physical channels:
\begin{conjecture}{Conjecture on $\mathcal{M}$}
For diagrams that do not contain any sub-diagram (including itself) as a generalized bubble, the only families of poles of the Mandelstam variables present in the Mellin amplitude are in one-to-one correspondence to its minimal cuts.
\end{conjecture}
In particular, this means that in our two-loop 1PI diagram above there is only a single family of double-trace poles $\Gamma[\frac{\underline\Delta_{12}-S}{2}]$ in the $S$ channel, despite the presence of two three-cuts.  We are not aware of a straightforward physics argument in favor of this phenomenon.  It would be interesting to gain a better understanding and a complete proof for it and further learn about its consequences. We leave this exploration for the future.  

Also we emphasis here that the above conjecture does not apply to diagrams that contain generalized bubbles as sub-diagrams, as will be explicitly verified in many examples in Appendix \ref{app:sec:polegbubbles} (in particular \eqref{app:eq:resgbsubdiagram}). For these more general diagrams, it is tempting to conjecture that the absence of the non-minimal poles still holds for any cuts that do not run across any generalized bubble sub-diagram, but fails otherwise.  We leave this for future verification. This also urges for a better understanding about properties of effective bulk-to-bulk propagators in general.

\vspace{1em}

Apart from the investigations into the pole structure of both the Mellin amplitude and the Mellin pre-amplitude as summarized so far, we also draw some preliminary studies on the structure of residues at poles of $\mathcal{M}$ in the physical channel, in Section \ref{sec:residuecomputation}.  
Our computation explicitly verifies the factorization into sub-diagrams as required by unitarity.
\begin{table}[h]\centering
\caption{List of diagrams analyzed in this paper.}\label{tab:examples}
\begin{tabular}{@{}c|cc|c|cc@{}}
\toprule
diagram & construction & labeling of & $M$ pole & $\mathcal{M}$ pole & $\mathcal{M}$ leading\\
&& integrand poles & structure & structure & residues\\
\midrule
\parbox{1.6cm}{\centering\tikz{\begin{scope}[scale=.3]
\draw [white] (0,-2.9) -- (0,2.9);
\draw [black,ultra thick] (0,0) circle [radius=2.6];
\coordinate (up) at (-.3,.7);
\coordinate (down) at (-.3,-.7);
\draw [black,thick] (150:2.6) -- (-1.4,0) -- (-150:2.6);
\draw [black,thick] (30:2.6) -- (1.4,0) -- (-30:2.6);
\draw [black,thick] (-1.4,0) .. controls ($(-1.4,0)+(60:1)$) and ($(1.4,0)+(120:1)$) .. (1.4,0) .. controls ($(1.4,0)+(-120:1)$) and ($(-1.4,0)+(-60:1)$) .. (-1.4,0);
\end{scope}}}
&\begin{tabular}{c}Fig.~\ref{fig:bubble4pt}\\\eqref{eq:M0bubble4pt},\eqref{eq:Kbubble4pt}\end{tabular}&\begin{tabular}{c}\eqref{eq:bubblelabeledintegrand},\eqref{app:eq:bubble4tremainpoles}\end{tabular}&\begin{tabular}{c}\eqref{eq:bubble4ptpoles}\\\eqref{eq:bubble4ptMpoleS},\eqref{eq:bubble4ptMpolerest}\end{tabular}&\eqref{app:eq:bubble4ptcalM}&\eqref{eq:bubble4ptleadingresidue}\\
\parbox{1.6cm}{\centering\tikz{\begin{scope}[scale=.3]
\draw [white] (0,-2.9) -- (0,2.9);
\draw [black,ultra thick] (0,0) circle [radius=2.6];
\draw [black,thick] (60:1.4) -- (-1.4,0) -- (-60:1.4) -- cycle;
\draw [black,thick] (-1.4,0) -- (-2.6,0);
\draw [black,thick] (60:1.4) -- (60:2.6);
\draw [black,thick] (-60:1.4) -- (-60:2.6);
\end{scope}}}
&\begin{tabular}{c}Fig.~\ref{fig:triangle} (A)\\\eqref{eq:M0triangle3pt},\eqref{eq:Ktriangle3pt}\end{tabular}&\begin{tabular}{c}\eqref{eq:labeledpolesintegrandtriangle3pt},\eqref{app:eq:triangl3ptremainpoles}\end{tabular}&\eqref{eq:triangle3ptMpolesemerge},\eqref{eq:triangle3ptpolesall}&\eqref{app:eq:triangle3ptcalMpoles}&\\
\parbox{1.6cm}{\centering\tikz{\begin{scope}[scale=.3]
\draw [white] (0,-2.9) -- (0,2.9);
\draw [black,ultra thick] (0,0) circle [radius=2.6];
\draw [black,thick] (45:1.4) -- (-1,0) -- (-45:1.4) -- cycle;
\draw [black,thick] (150:2.6) -- (-1,0) -- (-150:2.6);
\draw [black,thick] (45:1.4) -- (45:2.6);
\draw [black,thick] (-45:1.4) -- (-45:2.6);
\end{scope}}}
&\begin{tabular}{c}Fig.~\ref{fig:triangle} (B)\\\eqref{eq:M0triangle4ptquartic},\eqref{eq:Ktriangle4ptquartic}\end{tabular}&\begin{tabular}{c}\eqref{eq:labeledpolesintegrandtriangle4pt},\eqref{eq:triangle4ptremainpoles}\end{tabular}&\eqref{eq:preMpolestriangle4pt},\eqref{eq:preMtriangle4ptpolesall}&\eqref{app:eq:triangle4ptQpolesS},\eqref{app:eq:triangle4ptQpolesall}&\eqref{eq:quartictriangleleadingresidue}\\
\parbox{1.6cm}{\centering\tikz{\begin{scope}[scale=.3]
\draw [white] (0,-2.9) -- (0,2.9);
\draw [black,ultra thick] (0,0) circle [radius=2.6];
\draw [black,thick] (30:1.4) -- (-.3,0) -- (-30:1.4) -- cycle;
\draw [black,thick] (150:2.6) -- (-1.4,0) -- (-150:2.6);
\draw [black,thick] (-.3,0) -- (-1.4,0);
\draw [black,thick] (30:1.4) -- (45:2.6);
\draw [black,thick] (-30:1.4) -- (-45:2.6);
\end{scope}}}
&\begin{tabular}{c}Fig.~\ref{fig:triangle4ptC}\\\eqref{eq:M0triangle4ptcubic},\eqref{eq:Ktriangle4ptcubic}\end{tabular}&\begin{tabular}{c}\eqref{eq:labeledpolesintegrandtriangle4ptC},\eqref{app:eq:triangle4ptCremainpoles}\end{tabular}&\eqref{eq:poleresult4pttrianglecubic},\eqref{eq:poleresult4pttrianglecubicshort}&\begin{tabular}{c}\eqref{app:eq:triangle4ptCpoleS},\eqref{app:eq:triangle4ptCpolesall}\\\checkmark \end{tabular}&\eqref{eq:triangle4ptCleadingminimalres}\\
\parbox{1.6cm}{\centering\tikz{\begin{scope}[scale=.3]
\draw [white] (0,-2.9) -- (0,2.9);
\draw [black,ultra thick] (0,0) circle [radius=2.6];
\draw [black,thick] (45:1.4) -- (135:1.4) -- (-135:1.4) -- (-45:1.4) -- cycle;
\draw [black,thick] (45:1.4) -- (45:2.6);
\draw [black,thick] (135:1.4) -- (135:2.6);
\draw [black,thick] (-135:1.4) -- (-135:2.6);
\draw [black,thick] (-45:1.4) -- (-45:2.6);
\end{scope}}}
&\begin{tabular}{c}Fig.~\ref{fig:box4pt}\\\eqref{eq:M0box4pt},\eqref{eq:Kbox4pt}\end{tabular}&\begin{tabular}{c}\eqref{eq:labeledpolesintgrandbox},\eqref{app:eq:box4ptremainpoles}\end{tabular}&\eqref{eq:resultpolesM4pt},\eqref{eq:box4ptMpolesshort}&\eqref{app:eq:box4ptpoleS},\eqref{app:eq:box4ptpolesall}&\begin{tabular}{c}\eqref{eq:box4ptconsistencycheck}\\\eqref{eq:leadingTresidue}\\\eqref{eq:box4ptSTres}\end{tabular}\\
\midrule
\parbox{1.6cm}{\centering\tikz{\begin{scope}[scale=.3]
\draw [white] (0,-2.9) -- (0,2.9);
\draw [black,ultra thick] (0,0) circle [radius=2.6];
\coordinate (up) at (-.3,.7);
\coordinate (down) at (-.3,-.7);
\draw [black,thick] (150:2.6) -- (-1.4,0) -- (-150:2.6);
\draw [black,thick] (30:2.6) -- (1.4,0) -- (-30:2.6);
\draw [black,thick] (-1.4,0) -- (0,-1) -- (1.4,0) -- (0,1) -- cycle;
\draw [black,thick] (0,-1) -- (0,1);
\end{scope}}}
&\begin{tabular}{c}Fig.~\ref{fig:triangle2doubletriangle}\\\eqref{eq:doubletrianglekernel}\\\eqref{eq:doubletriangletreeM0},\eqref{eq:doubletriangletreekernel}\end{tabular}&\begin{tabular}{c}\eqref{eq:relevantpolesinM0fortwoloop},\eqref{eq:polesinMp0Kprelevant},\\\eqref{eq:polesoneloopemerge},\eqref{eq:polesoneloopdirect},\\\eqref{eq:polestwoloopnormalization}\end{tabular}&\eqref{eq:resultdoubletrianglepolepreM},\eqref{eq:doubletriangleMpolesshort}&\checkmark &\eqref{eq:doubletriangleres12expr}\\
\parbox{1.6cm}{\centering\tikz{\begin{scope}[scale=.3]
\draw [white] (0,-2.9) -- (0,2.9);
\draw [black,ultra thick] (0,0) circle [radius=2.6];
\coordinate (up) at (-.3,.7);
\coordinate (down) at (-.3,-.7);
\draw [black,thick] (150:2.6) -- (-1.4,0) -- (-150:2.6);
\draw [black,thick] (40:2.6) -- (up) -- (-1.4,0) -- (down) -- (-40:2.6);
\draw [black,thick] (up) -- ($(down)!.6!(-40:2.6)$);
\fill [white] (.21,0) circle [radius=5pt];
\draw [black,thick] (down) -- ($(up)!.6!(40:2.6)$);
\end{scope}}}
&Fig.~\ref{app:fig:constructnonplanarA}&\begin{tabular}{c}\eqref{app:eq:nonplanarAintegrandpoles1},\eqref{app:eq:nonplanarArestpolesoneloop}\\\eqref{app:eq:nonplanarAintegrandpoles2},\eqref{app:eq:nonplanarAintegrandpoles3}\end{tabular}&\eqref{app:eq:nonplanarAprepoles}&\checkmark &\\
\parbox{1.6cm}{\centering\tikz{\begin{scope}[scale=.3]
\draw [white] (0,-2.9) -- (0,2.9);
\draw [black,ultra thick] (0,0) circle [radius=2.6];
\coordinate (up) at (-.3,.7);
\coordinate (down) at (-.3,-.7);
\draw [black,thick] (150:2.6) -- ($(up)+(-1.4,0)$) -- ($(down)+(-1.4,0)$) -- (-150:2.6);
\draw [black,thick] (40:2.6) -- (up) -- ($(up)+(-1.4,0)$);
\draw [black,thick] ($(down)+(-1.4,0)$) -- (down) -- (-40:2.6);
\draw [black,thick] (up) -- ($(down)!.6!(-40:2.6)$);
\fill [white] (.21,0) circle [radius=5pt];
\draw [black,thick] (down) -- ($(up)!.6!(40:2.6)$);
\end{scope}}}
&&&\eqref{app:eq:nonplanarBMpoleSTU},\eqref{app:eq:nonplanarBpolesrest}&\checkmark &\\
\midrule
\parbox{1.6cm}{\centering\tikz{\begin{scope}[scale=.3]
\draw [white] (0,-2.9) -- (0,2.9);
\draw [black,ultra thick] (0,0) circle [radius=2.6];
\node [anchor=center,align=center] at (0,0) {\scriptsize generalized\\[-.5em]\scriptsize bubbles};
\end{scope}}}
&&&App.~\ref{app:sec:polegbubbles}&&\begin{tabular}{c}App.~\ref{app:sec:polegbubbles}\\non-min.\\poles\end{tabular}\\
\bottomrule
\end{tabular}
\end{table}
In this paper we work on many explicit diagrams (assuming completely generic data for the conformal dimensions) to illustrate various points, but the discussion for each individual diagram is divided into different parts and contained in different sections, due to the need in the organization of the paper. In Table \ref{tab:examples} we list out these diagrams together with pointers to places at various stages of the analyses on them.
In this table, the third column collects our choice of labeling for different families of poles in the integrand, which are useful in analyzing the local properties of the poles of the (pre-)amplitudes. In the fifth column the diagrams with a check mark allows non-minimal cuts, and we have verified that their corresponding non-minimal poles are absent from $\mathcal{M}$.

\subsubsection{Methodology}

Next let us switch to methodology, which is another main theme of this paper.  As pointed out at the beginning, in general the computation of the diagram can be simplified by exploiting the power of Mellin (pre-)amplitude being a meromorphic function in Mellin space. While this has largely been prescribed in our previous letter \cite{Yuan:2017vgp}, let us very quickly summarize here the recursive construction of arbitrary scalar diagrams thus resulted.

The general idea is to construct a target diagram by assuming the complete knowledge of a lower-loop diagram from which it is obtained by gluing two boundary points via the split representation \eqref{eq:bbpropagatordef}.  Let us name these two points by $\{0,n+1\}$ and the remaining boundary points are labeled as $\{1,2,\ldots,n\}$.   We start by collecting a maximal set of OPE channels of the lower-loop diagram perceived by its Mellin amplitude (i.e, those with corresponding poles of the Mandelstam variables), such that these OPE channels obey tree-level consistency (i.e., we are able to do OPEs one after another sequentially). As we regard each of these OPE channels as an effective tree propagator, the original lower-loop diagram induces an effective tree diagram. There is thus a unique chain of propagators linking point $0$ and point $n+1$, as exemplified in the figure below.
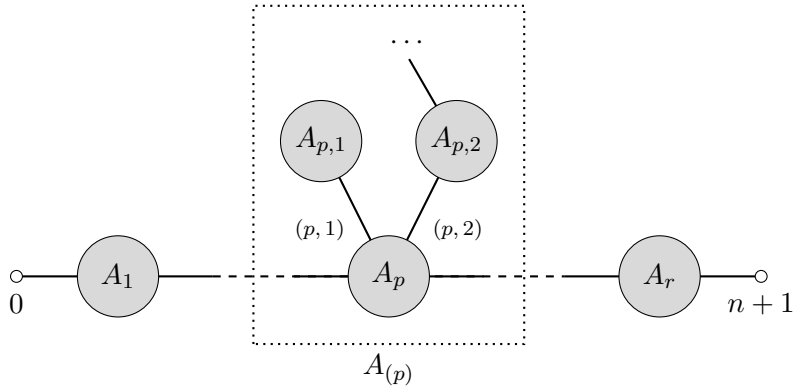
\begin{center}
\begin{tikzpicture}
\begin{scope}[scale=.9]
\draw [black,thick] (-5.5,0) -- (-2.6,0);
\draw [black,thick,dashed] (-2.6,0) -- (2.6,0);
\draw [black,thick] (5.5,0) -- (2.6,0);
\draw [black,thick] (-1.4,0) -- (1.4,0);
\draw [black,thick] (-1,2) -- (0,0) -- (1,2) -- +(120:1.4);
\draw [black,fill=black!15!white] (-4,0) circle [radius=.6];
\draw [black,fill=black!15!white] (0,0) circle [radius=.6];
\draw [black,fill=black!15!white] (-1,2) circle [radius=.6];
\draw [black,fill=black!15!white] (1,2) circle [radius=.6];
\draw [black,fill=black!15!white] (4,0) circle [radius=.6];
\node [anchor=center] at (-4,0) {$A_1$};
\node [anchor=center] at (0,0) {$A_p$};
\node [anchor=center] at (-1,2) {$A_{p,1}$};
\node [anchor=center] at (1,2) {$A_{p,2}$};
\node [anchor=center] at (4,0) {$A_r$};
\node [anchor=south] at ($(1,2)+(120:1.4)$) {$\cdots$};
\node [anchor=north east] at (-.5,1) {\scriptsize $(p,1)$};
\node [anchor=north west] at (.5,1) {\scriptsize $(p,2)$};
\draw [black,thick,dotted] (-2,-1) rectangle (2,4);
\node [anchor=north] at (0,-1) {$A_{(p)}$};
\draw [black,fill=white] (-5.5,0) circle [radius=2.5pt];
\draw [black,fill=white] (5.5,0) circle [radius=2.5pt];
\node [anchor=center] at (-5.5,-.4) {$0$};
\node [anchor=center] at (5.5,-.4) {$n+1$};
\end{scope}
\end{tikzpicture}
\end{center}
Here each grey blob represents a bulk vertex together with any possible bulk-to-boundary propagators incident to it. We assume altogether $r$ vertices on the chain, labeled as $\{A_1,\ldots,A_r\}$ in sequence. In general there can be extra effective tree structure grown out of some $A_p$, and we denote this entire tree sub-diagram as $A_{(p)}$. The effective vertices therein are labeled according to their depth with respect to the chain, as can be easily understood from the above figure. Now we denote (1) the total conformal dimension of boundary points attached to vertex $A_a$ as $\Delta_{A_a}$, (2) the Mandelstam variable for each propagator $(a)$ in the additional tree structure as $\Xi_{(a)}$, (3) the Mandelstam variable for the propagator in between $A_{a-1}$ and $A_a$ as $\Xi_a$, and (4) Mandelstam variable associated to the channel that cuts the sub-diagram from some $A_{(a)}$ to some $A_{(b)}$ out of the rest as $\xi_{a,b}$. 

With these notations, the Mellin pre-amplitude of the new diagram thus constructed can be obtained as an integral operation acting on that of the original lower-loop diagram
\begin{equation}\label{intro:recursionschematic}
M_{\rm new}[s]=(\text{constant})\int[\mathrm{d}\Xi,\mathrm{d}\xi]\,M_{\rm old}[\Xi]\,K[\Xi,\xi;s],
\end{equation}
where the integral kernel $K[\Xi,\xi;s]$ is a product of Beta functions directly determined by the above effective tree diagram
\begin{equation}
\begin{split}
K=&\prod_{a=1}^{r}\bfn{\frac{(\xi_{a,r}-\xi_{a+1,r})-(\Xi_a-\Xi_{a+1})}{2}}{\frac{(\xi_{a,r}-\xi_{a+1,r})-(\xi_{1,a-1}-\xi_{1,a})}{2}}\prod_{p\in\{A_{(a)}\text{ vertices}\}}\!\!\!\bfn{\frac{\Delta_{A_p}-\Xi_{(p)}+\sum_m \Xi_{(p,m)}}{2}}{\frac{\Delta_{A_p}-s_{(p)}+\sum_{m}s_{(p,m)}}{2}}\\
&\times\prod_{1\leq a<b\leq r}\!\!\bfn{\frac{\xi_{a,b-1}+\xi_{a+1,b}-\xi_{a+1,b-1}-\xi_{a,b}}{2}}{\frac{s_{a,b-1}+s_{a+1,b}-s_{a+1,b-1}-s_{a,b}}{2}},
\end{split}
\end{equation}
with the abbreviation $B[\substack{x\\y}]\equiv\frac{\Gamma[x]\Gamma[y-x]}{\Gamma[y]}$. The $s$ variables are the same as the $\xi$ variables but constructed from the Mellin variables of the new diagram. The integrals involved in \eqref{intro:recursionschematic} consists of all the independent $\Xi$ and $\xi$ variables associated to the original Mellin space. With this, an arbitrary scalar diagram at any loops can be obtained recursively from a certain tree diagram, for which the pre-amplitude is generically known (and is simple) by the result presented at the beginning of this subsection.  We leave the rigorous derivation of this integral relation as well as its detailed subtleties to Section \ref{sec:oneloopdiagrams} and \ref{sec:arbitraryloops}.

The virtue of this recursive construction method involves several aspects:
\begin{itemize}[noitemsep,nolistsep]
\item All the integrals encountered in the resulting representation for pre-amplitudes are standard Mellin integrals, while the integrand to start with are as simple as can be directly read off from the diagram under study.  These integrals are of the same type as the spectrum integrals that further connect the pre-amplitude to the amplitude, so that we are able to treat these integrals altogether at the same footing when analyzing local properties of poles of the Mellin amplitude. This will be illustrated in various explicit examples in Section \ref{sec:residuecomputation} and subsequent sections.
\item This construction of loop diagrams is highly asymmetric.  Even if the target diagram itself may enjoy symmetries (under reflections, rotations, etc), these are usually not manifest at all in the integral representation resulting from the construction.  While this may not be very appealing for some specific purposes, this appearing asymmetry is actually very powerful computational-wise, because any possible symmetry of the target diagram has to correctly emerge from the Mellin integrals non-trivially.  On the one hand, frequently such symmetry argument allows us to very quickly land on a final conclusion, e.g., regarding the pole structure of the pre-amplitude, without the need of checking hard the absence of certain potential families of poles (as we will see in many examples later on).  On the other hand, the proper emergence of symmetries also provides a strong self-consistency check for the computation (e.g., \eqref{eq:boxtotalres}).
\item In general the same diagram can be constructed in many different ways, all of which leading to Mellin integral representations of the same (pre-)amplitude but are not at all obviously seen to be equivalent to each other before the remaining integrals are actually performed.  This fact also provides great convenience in cross-checking the results.  In addition, some specific constructions may turn out to be more convenient for specific applications than others, and in practice one has the freedom of choosing that best suited to the purpose. This is going to be illustrated in the computation of residues for a box diagram in Section \ref{eq:boxresidues}.
\end{itemize}

Of course, all what the above construction itself gives us is only some specific Mellin integral representation of (pre-)amplitudes. Ultimately we are interested in the ``final result''.  Numerically there is in principle no difficulty in performing these integrals upon any specified values of the data \cite{Czakon:2005rk,Hahn:2004fe,Hahn:2014fua}.  If instead analytic result is needed, usually one may ask to expand the amplitude onto some familiar/standard functions at the end.   However, this expansion itself can potentially be troublesome as it will almost always introduce extra redundant information that are not at all physical (such as some spurious poles in individual pieces, as can be easily encountered by deforming the Mellin integral contours and turn them into summation of residues). Hence it is not wise to perform any such expansion if the expansion basis themselves are not physically relevant enough.

It is not our purpose in this paper to provide answers to any such expansion. Instead, for meromorphic functions all what we care about are their pole structures together with the local properties of each pole.  It is of our main interest here to seek for a systematic and efficeint method 
\begin{itemize}[noitemsep,nolistsep]
\item to determine the entire pole structure,
\item to estimate the order of each specific pole,
\item to work out a representation (still belonging to the Mellin integrals) for the residue at each pole,
\end{itemize}
\emph{directly} using the given Mellin integral representation for an arbitrary meromorphic function.  In this paper we provide answers to all these generic questions.  In short, the basic underlying idea is that poles of the Mellin integrals always emerge by pinching the contour by a certain \emph{simplex} formed by the singularities of the integrand, or a composition of simplices. It can be foreseen that the resulting integral representations for residues have to be (far) simpler that the represention for the Mellin amplitude that we start with, since many integrals have to be localized at the pinching\footnote{As we will see in explicit examples, in special situations the remaining integrals will also be as simple as can be straightforwardly performed.}. A brief overview that is sufficient for most of the applications in this paper is presented in Section \ref{sec:polealgorithm}, and we leave a complete account on this systematic treatment to the entire Appendix \ref{app:sec:mellinintegrals}.  This involves ideas that, to our knowledge, are not found in previous literature.

\subsection{Outline of the paper}\label{sec:guidance}

Here we outline the main contents of the paper.  Each section/appendix is written in as much a self-contained way as possible, so that the reader can feel free to move around different parts. In particular, for those who are mainly interested in actual applications or are already familiar with the construction procedure, we suggest them to directly focus on Section \ref{sec:polestructure}, \ref{sec:residuecomputation}, \ref{sec:doubletrianglediagram}, \ref{sec:conjectureonamplitude}, and the appendices pointed out therein.

In Section  \ref{sec:elementaryoperations} we introduce two elementary operations on Witten diagrams. Diagrammatically, one inserts an extra bulk-vertex to an existing diagram (Section \ref{sec:vertexinsertiongeneral}) while the other takes two of the original boundary points and creates an extra loop (Section \ref{sec:loopformation}). Arbitrary scalar diagrams can thus be constructed by a recursive application of these two operations starting from the contact diagrams.  We translate these operations into integral operations directly acting in Mellin space, which turns out to possess a compact expression whose integral kernel is completely independent of the original diagram.

In Section \ref{sec:treediagramsrevisited} we restrict our scope to tree level. On the one hand, we show how the operation of vertex insertion can be simplified, such that the integrals involved are only tied to the Mandelstam variables naturally appearing in the original Mellin (pre-)amplitude (Section \ref{sec:kernelgenerictrees}). While the resulting new kernel now depends on the original diagram, it still maintains to be simple and follows diagrammatic rules.  On the other hand, we show that this recursive construction directly implies that the pre-amplitude has to factorize into contributions locally from each tree-propagator and each 1PI part of the diagram (Section \ref{sec:localityofinsertion}). At tree level this leads to a set of diagrammatic rules directly for the pre-amplitudes (Section \ref{sec:treediagrammaticrules}), and their equivalence to the previous results in literature is further explained in Appendix \ref{app:sec:proofofequivalence}. Further technical details are collected in Appendix \ref{app:sec:forest2tree}.

In Section \ref{sec:oneloopdiagrams} we borrow the insight from Section \ref{sec:treediagramsrevisited} and seek for simplified integral formulas for loop formation in the construction of one-loop diagrams. This is first done for the special case of necklace diagrams in Section \ref{sec:necklace} and then further generalized to arbitrary one-loop diagrams in Section \ref{sec:generaloneloop}.  In the resulting formula the integrals are again tied to the Mandelstam variables naturally associated to the original diagram, and the new kernel satisfies another set of diagrammatic rules. We provide various explicit examples to illustrate this construction, including: (1) 4-point and 3-point bubble diagrams, (2) 3-point triangle diagram, (3) 4-point triangle diagram that involves a quartic vertex, (4) 4-point triangle diagram with only cubic vertices, (5) 4-point box diagram.

In Section \ref{sec:polestructure} and \ref{sec:residuecomputation} we explore the consequence of our construction in actual applications at one loop.  In Section \ref{sec:polestructure} we present how to obtain the complete knowledge of the pole structure of the pre-amplitude. Firstly we summarize a general algorithm for the detection of poles in Mellin integrals (Section \ref{sec:polealgorithm}), and comment on a strategy that helps further simplify the analysis (Section \ref{sec:strategy}). Then we apply this method to the explicit examples in the previous section and determine all the poles in their pre-amplitudes.  As a result, the pole structure of pre-amplitudes turns out to obey diagrammatic rules (partly similar to that at tree level), and at the end we propose a conjecture on this structure which we expect to hold for arbitrary Witten diagrams (Section \ref{sec:enhancedrules}). Further evidence for this conjecture from the generalized bubble diagrams are presented in Appendix \ref{app:sec:polegbubbles}.  Some subtle technical details in the check of absence of poles are collected in Appendix \ref{sec:checkfakepoles}.

In Section \ref{sec:residuecomputation} we move on to investigate analytic properties of the Mellin amplitudes, using the knowledge we acquired on the pre-amplitudes in the previous section. First of all, in Section \ref{sec:MtoM} we review the detailed definition of the spectrum integrals that bridge these two quantities, and explain how the pole analysis in the previous section can be equally applied here as well as the subtleties involved. We do not present a detailed discussion on the pole structure of the Mellin amplitudes in this paper, but instead summarize the results for our explicit examples in Appendix \ref{app:sec:resultpolesofamplitudes}. We then fully focus on the poles in terms of the Mandelstam variables, whose emergence from our integral representation follows a simple pattern, which we describe in Section \ref{sec:polesinMandelstam}. This includes two types of poles that we name as minimal and non-minimal respectively. Based on this, in the rest of this section we compute the residues at various leading poles of our one-loop examples, which provide some preliminary manifestation of unitarity in AdS scattering process at loop level.

In Section \ref{sec:arbitraryloops} we generalize our construction at one loop (Section \ref{sec:oneloopdiagrams}) to the construction of diagrams at arbitrary loops. This is fulfilled essentially by treating an arbitrary original diagram effectively as a tree-level diagram using a maximal sequence of consistent OPE limits (Section \ref{sec:emergehigherloop}).  After addressing some potential issues tied to the choice of Mandelstam variables at higher loops in Section \ref{sec:Mandelstamindependence}, we present the general prescription for the simplified recursive construction formula for arbitrary loop diagram in Section \ref{sec:generalkernel}.  This is then applied to the construction of a 4-point double triangle diagram at two loops in Section \ref{sec:doubletrianglediagram}. For this explicit example we work out its Mellin integral representation, and use the method discussed in Section \ref{sec:polestructure} and \ref{sec:residuecomputation} to determine its poles structure and to compute the residue at its leading poles in the physical channels. This provides further non-trivial evidence for our conjecture in Section \ref{sec:enhancedrules}. Furthermore, our examples also reveal that the non-minimal poles are actually absent, and correspondingly we propose another conjecture regarding poles of Mellin amplitudes in Section \ref{sec:conjectureonamplitude}. More examples of non-planar diagrams at two-loops are discussed in Appendix \ref{app:sec:nonplanar2loop}.

Since our analysis heavily relies on Mellin integrals, we provide a detailed account on them in Appendix \ref{app:sec:mellinintegrals}, where the main focus is on convenient analytic methods for the identification of their poles and and the computation of the residues.

\newpage

\section{Elementary Operations}\label{sec:elementaryoperations}

\temp{No major revisions further needed. I have double checked the normalizations.}

An arbitrary scalar diagram can be constructed by iterating two elementary operations: (1) vertex insertion, to increase the number of bulk vertices $V$; and (2) loop formation, to increase the number of loops $L$. In this section we start with an arbitrary diagram, which is connected but with no restrictions on its loop level or number of boundary points. We assume the complete knowledge about its Mellin pre-amplitude $M_0$ (in other words any details of its analytic properties), and investigate the actions of these two operations on $M_0$.

Both actions take the form of integrating $M_0$ against some kernel $K$, schematically
\begin{equation}
\checked{}
M[\delta]=\int[\mathrm{d}\tau]_0\,M_0[\tau]\,K[\tau,\delta].
\end{equation}
Without further specification, we always use $\tau$ for the original Mellin variables, and $\delta$ for the new ones. The integration measure $[\mathrm{d}\tau]_0$ is the same as that relating the Mellin amplitude to the correlator in \eqref{eq:correlatorMellin} (for the original diagram).

\subsection{Vertex insertion}\label{sec:vertexinsertiongeneral}

The (bulk) vertex insertion is a special case to the general operation of gluing two existing diagrams by forming a single new propagator, but one sub-diagram is just a contact diagram (thus a single bulk vertex).

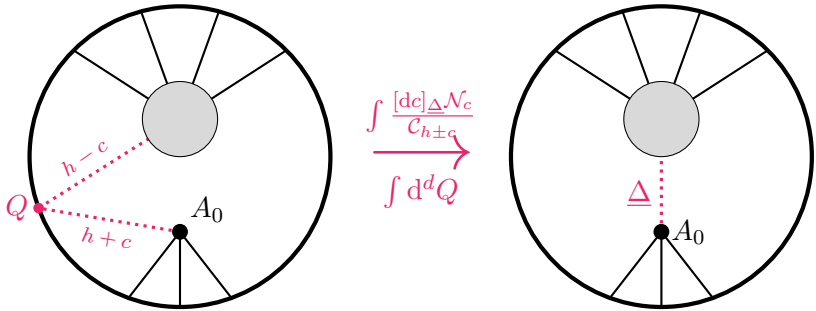
\begin{figure}[ht]
\captionsetup{margin=2em}
\begin{center}
\begin{tikzpicture}
\begin{scope}[xshift=-3.2cm]
\draw [WildStrawberry,very thick,dotted] (-90:1) -- (-160:2) -- (90:.5);
\node [anchor=south west] at (-90:1) {$A_0$};
\node [anchor=south,rotate=35] at ($(-160:2)!.4!(90:.5)$) {\color{WildStrawberry}\scriptsize $h-c$};
\node [anchor=north,rotate=-9] at ($(-160:2)!.5!(-90:1)$) {\color{WildStrawberry}\scriptsize $h+c$};
\draw [black,thick] (-90:1) -- (-110:2);
\draw [black,thick] (-90:1) -- (-90:2);
\draw [black,thick] (-90:1) -- (-70:2);
\fill [black] (-90:1) circle [radius=3pt];
\draw [black,thick] (90:.5) -- (45:2);
\draw [black,thick] (90:.5) -- (75:2);
\draw [black,thick] (90:.5) -- (105:2);
\draw [black,thick] (90:.5) -- (135:2);
\draw [black,fill=black!15!white] (90:.5) circle [radius=.5];
\draw [black,ultra thick] (0,0) circle [radius=2cm];
\fill [WildStrawberry] (-160:2) circle [radius=2pt];
\node [anchor=east] at (-160:2) {\color{WildStrawberry}$Q$};
\end{scope}
\node [anchor=center] at (0,0) {\Huge\color{WildStrawberry} $\longrightarrow$};
\node [anchor=south] at (0,.1) {\color{WildStrawberry} $\int\frac{[\mathrm{d}c]_{\underline\Delta}\mathcal{N}_c}{\mathcal{C}_{h\pm c}}$};
\node [anchor=north] at (0,-.1) {\color{WildStrawberry} $\int\mathrm{d}^dQ$};
\begin{scope}[xshift=3.2cm]
\draw [WildStrawberry,very thick,dotted] (-90:1) -- (90:.5);
\node [anchor=east] at (-90:.5) {\color{WildStrawberry}$\underline\Delta$};
\node [anchor=west] at (-90:1) {$A_0$};
\draw [black,thick] (-90:1) -- (-110:2);
\draw [black,thick] (-90:1) -- (-90:2);
\draw [black,thick] (-90:1) -- (-70:2);
\fill [black] (-90:1) circle [radius=3pt];
\draw [black,thick] (90:.5) -- (45:2);
\draw [black,thick] (90:.5) -- (75:2);
\draw [black,thick] (90:.5) -- (105:2);
\draw [black,thick] (90:.5) -- (135:2);
\draw [black,fill=black!15!white] (90:.5) circle [radius=.5];
\draw [black,ultra thick] (0,0) circle [radius=2cm];
\end{scope}
\end{tikzpicture}
\end{center}
\caption{Vertex insertion to a connected diagram.}
\label{fig:vertexinsertion}
\end{figure}

We assume the original Witten diagram has its boundary points labeled by $\{0,1,\ldots,m\}$, and we insert to it a bulk vertex with boundary points $\{m+1,\ldots,n,n+1\}$. This is done by identifying point $0$ and point $n+1$ and then integrating it over the boundary, thus forming a diagram with $n$ points. We specify the conformal dimensions of the two corresponding bulk-to-boundary propagators as
\begin{equation}\label{eq:Deltahc}
\checked{}
\Delta_0=h-c,\qquad\Delta_{n+1}=h+c.
\end{equation}
Then a further spectrum integration $\int[\mathrm{d}c]_{\underline\Delta}\mathcal{N}_c/\mathcal{C}_{h\pm c}$ lifts the pair into a bulk-to-bulk propagator of dimension $\underline\Delta$. This is shown in Figure \ref{fig:vertexinsertion}.

Explicitly the correlator associated to the Witten diagram thus formed is
\begin{equation}
\begin{split}
\checked{}
\mathcal{I}=&
\int\frac{[\mathrm{d}c]_{\underline\Delta}\mathcal{N}_{c}}{\mathcal{C}_{h\pm c}}\underset{\partial\text{AdS}}{\int}\mathrm{d}Q\left(\mathcal{I}_0\Big|_{\substack{P_0=Q\\\Delta_0=h-c}}\right)\left(\mathcal{I}_{\rm v}\Big|_{\substack{P_{n+1}=Q\\\Delta_{n+1}=h+c}}\right)\\
=&\frac{1}{\pi^h}\int[\mathrm{d}\tau]_0[\mathrm{d}\tau]_{\rm v}\,\mathcal{N}\,M_0[\tau]\,\Gamma[{\textstyle\frac{\sum_{i=m+1}^{n}\Delta_i-h+c}{2}}]\prod_{1\leq i<j\leq m}\frac{\Gamma[\tau_{i\,j}]}{P_{i\,j}^{\tau_{i\,j}}}\prod_{m<i<j\leq n}\frac{\Gamma[\tau_{i\,j}]}{P_{i\,j}^{\tau_{i\,j}}}\\
&\times\underset{\partial\text{AdS}}{\int}\mathrm{d}Q\prod_{k=1}^m\frac{\Gamma[\tau_{0\,k}]}{(-2P_k\cdot Q)^{\tau_{0\,k}}}\,\prod_{k=m+1}^n\frac{\Gamma[\tau_{k\,n+1}]}{(-2P_k\cdot Q)^{\tau_{k\,n+1}}},
\end{split}
\end{equation}
where we have made the identification $P_0=P_{n+1}=Q$. $[\mathrm{d}\tau]_{\rm v}$ is the Mellin measure for the new vertex. $\mathcal{N}=\pi^h[\mathrm{d}c]_{\underline\Delta}\mathcal{N}_{c}\,\mathcal{N}_0\,\mathcal{N}_{\rm v}/\mathcal{C}_{h\pm c}$ is the normalization \eqref{eq:Mellinnormalization} for the new amplitude, with $\mathcal{N}_0$ for $M_0$ and $\mathcal{N}_{\rm v}$ for the vertex. The original Mellin variables $\tau$ satisfy
\begin{align}
\checked{}
\sum_{j=0,\neq i}^m\tau_{i\,j}=\Delta_i,\quad i\leq m;\qquad
\sum_{j=m+1,\neq i}^{n+1}\tau_{i\,j}=\Delta_i,\quad i>m.
\end{align}
Integrating $Q$ yields
\begin{equation}
\begin{split}
\checked{}
\mathcal{I}=&2\int[\mathrm{d}\tau]_0[\mathrm{d}\tau]_{\rm v}\,\mathcal{N}\,M_0\,\Gamma[{\textstyle\frac{\sum_{i=m+1}^{n}\Delta_i-h+c}{2}}]\prod_{1\leq i<j\leq m}\frac{\Gamma[\tau_{i\,j}]}{P_{i\,j}^{\tau_{i\,j}}}\prod_{m<i<j\leq n}\frac{\Gamma[\tau_{i\,j}]}{P_{i\,j}^{\tau_{i\,j}}}\\
&\times\int[\mathrm{d}\alpha]\prod_{k=1}^m\alpha_k^{\tau_{0\,k}-1}\prod_{k=m+1}^n\alpha_k^{\tau_{k\,n+1}-1}\exp\!\left[-\sum_{1\leq i<j\leq n}\alpha_i\alpha_jP_{i\,j}\right],
\end{split}
\end{equation}
where $\int[\mathrm{d}\alpha]=\int_0^\infty\prod_{k=1}^n\mathrm{d}\alpha_k$. Applying Symanzik formula \eqref{eq:symanzik} and further making a proper shift to the resulting Mellin variables we land on an expression for the Mellin pre-amplitude $M$ of the corresponding new diagram
\begin{equation}
\checked{}
M[\delta]=\int[\mathrm{d}\tau]_0[\mathrm{d}\tau]_{\rm v}\,M_0[\tau]\,K'[\tau,\delta],
\end{equation}
with
\begin{equation}
\checked{}
K'[\tau,\delta]=\Gamma[{\textstyle\frac{\sum_{i=m+1}^{n}\Delta_i-h+c}{2}}]\prod_{1\leq i<j\leq m}\!\!\!\frac{\Gamma[\tau_{i\,j}]\Gamma[\delta_{i\,j}-\tau_{i\,j}]}{\Gamma[\delta_{i\,j}]}\prod_{m<i<j\leq n}\!\!\!\frac{\Gamma[\tau_{i\,j}]\Gamma[\delta_{i\,j}-\tau_{i\,j}]}{\Gamma[\delta_{i\,j}]}.
\end{equation}
At this stage $\delta$'s satisfy the canonical constraints \eqref{eq:canonicalconstraints} for the new diagram. In fact, since $M_0$ only depend on $\tau_{i,j}$ with $i,j\leq m$, it is possible to directly complete the $[\mathrm{d}\tau]_{\rm v}$ integrations, which yields
\begin{equation}\label{formula:vertexinsertion}
\checked{}
M[\delta]=\int[\mathrm{d}\tau]_0\,M_0[\tau]\,K_{\rm pre}[\tau,\delta],
\end{equation}
with
\begin{equation}\label{eq:simplifyK2}
\checked{}
K_{\rm pre}[\tau,\delta]=\frac{\Gamma[\frac{\Delta_{A_0}-h\pm c_{(1)}}{2}]\Gamma[\frac{h+c_{(1)}-s_{(1)}}{2}]}{\Gamma[\frac{\Delta_{A_0}-s_{(1)}}{2}]}\prod_{1\leq i<j\leq m}\!\!\!\frac{\Gamma[\tau_{i\,j}]\Gamma[\delta_{i\,j}-\tau_{i\,j}]}{\Gamma[\delta_{i\,j}]}.
\end{equation}
Here we switch the notation $c$ for the spectrum variable of the new propagator to $c_{(1)}$, so as to be consistent with the discussion of general trees later on, and we also abbreviate
\begin{equation}
\checked{}
\Delta_{A_0}=\sum_{i=m+1}^n\Delta_i,\qquad
s_{(1)}=\Delta_{A_0}-2\sum_{m<i<j\leq n}\delta_{i\,j}.
\end{equation}
In other words, these are the total conformal dimension and the Mandelstam variable associated to the group of boundary points attached to the extra vertex (excluding point $n+1$). The formula was also worked out previously in \cite{Fitzpatrick:2011ia}, in the study of factorization of Mellin amplitudes.

The operation of vertex insertions discussed here can be naturally generalized to the situation where we start with a disconnected diagram and add a single vertex to link all its connected components (each can again have arbitrary loops) so as to form a connected diagram, as shown in Figure \ref{app:fig:vertexinsertiontoforest}. We leave this discussion to Appendix \ref{app:sec:forest2tree}, and will need only part of the results there for later discussions.

\subsection{Loop formation}\label{sec:loopformation}

\begin{figure}[ht]
\captionsetup{margin=2em}
\begin{center}
\begin{tikzpicture}
\begin{scope}[xshift=-3.2cm]
\draw [black,thick] (0,0) -- (0:2);
\draw [black,thick] (0,0) -- (60:2);
\draw [black,thick] (0,0) -- (120:2);
\draw [black,thick] (0,0) -- (180:2);
\draw [WildStrawberry,very thick,dotted] (-90:2) .. controls (-1,-1.5) and (-1,-.5) .. (0,0) .. controls (1,-.5) and (1,-1.5) .. (-90:2);
\node [anchor=south,rotate=90] at (-.7,-1.1) {\scriptsize \color{WildStrawberry} $h\!-\!c$};
\node [anchor=south,rotate=-90] at (.7,-1.1) {\scriptsize \color{WildStrawberry} $h\!+\!c$};
\draw [black,fill=black!15!white] (0,0) circle [radius=.85];
\draw [black,ultra thick] (0,0) circle [radius=2];
\fill [WildStrawberry] (-90:2) circle [radius=2pt];
\node [anchor=north] at (-90:2) {\color{WildStrawberry}$Q$};
\end{scope}
\node [anchor=center] at (0,0) {\Huge\color{WildStrawberry} $\longrightarrow$};
\node [anchor=south] at (0,.1) {\color{WildStrawberry} $\int\frac{[\mathrm{d}c]_{\underline\Delta}\mathcal{N}_c}{\mathcal{C}_{h\pm c}}$};
\node [anchor=north] at (0,-.1) {\color{WildStrawberry} $\int\mathrm{d}^dQ$};
\begin{scope}[xshift=3.2cm]
\draw [black,thick] (0,0) -- (0:2);
\draw [black,thick] (0,0) -- (60:2);
\draw [black,thick] (0,0) -- (120:2);
\draw [black,thick] (0,0) -- (180:2);
\draw [WildStrawberry,very thick,dotted] (0,0) .. controls (1,-.5) and (.8,-1.5) .. (-90:1.5) .. controls (-.8,-1.5) and (-1,-.5) .. (0,0);
\node [anchor=south] at (0,-1.5) {\color{WildStrawberry} $\underline\Delta$};
\draw [black,fill=black!15!white] (0,0) circle [radius=.85];
\draw [black,ultra thick] (0,0) circle [radius=2];
\end{scope}
\end{tikzpicture}
\end{center}
\vspace{-1em}\caption{Loop formation.}
\label{fig:loopformation}
\end{figure}
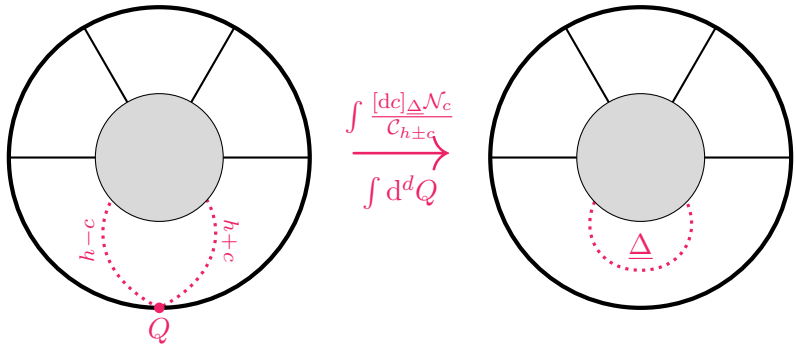

We now look at the second operation, which lifts a given (connected) diagram to one higher loop. This is achieved by taking coincidence limit of two existing boundary points, whose conformal dimensions we assign to be $h\pm c$, integrating over the boundary and further performing the spectrum integration of $c$, thus forming a new bulk-to-bulk propagator that creates an additional loop. This is shown in Figure \ref{fig:loopformation}.

For convenience of later discussions, let us assume that the original amplitude has $n+2$ points, labeled by $\{0,1,\ldots,n+1\}$, and we pick the coincident points to be $\{0,n+1\}$.

Note that na\"ively the identification $P_0=P_{n+1}$ in a representation of the correlator in terms of Mellin amplitude appears to be illegal, due to the presence of the additional factor $P_{0\,n+1}^{-\tau_{0\,n+1}}$ in \eqref{eq:correlatorMellin} as we translate $\mathcal{M}_0$ to the spacetime correlator $\mathcal{I}_0$. This seemingly singularity is merely an artifact of this particular representation, and can be simply resolved by inversely applying the Symanzik formula \eqref{eq:symanzik} so as to turn this factor back into a bulk integration.

In doing this, we require an extra reference point. Without loss of generality we choose this to be point $1$. The Mellin variables associated to a three-point function is fully fixed, and so we are able to treat the following part of the dressing factor $\prod_{i<j}\frac{\Gamma[\tau_{i\,j}]}{P_{i\,j}^{\tau_{i\,j}}}$ as a three-point contact diagram, following \eqref{eq:contactcorrelator}
\begin{equation}\label{eq:backtobulk3pt}
\begin{split}
\checked{}
\frac{\Gamma[\tau_{0\,1}]\Gamma[\tau_{0\,n+1}]\Gamma[\tau_{1\,n+1}]}{P_{0\,1}^{\tau_{0\,1}}P_{0\,n+1}^{\tau_{0\,n+1}}P_{1\,n+1}^{\tau_{1\,n+1}}}=&\frac{2}{\pi^h}\frac{\Gamma[\tilde{\Delta}_0]\Gamma[\tilde{\Delta}_1]\Gamma[\tilde{\Delta}_{n+1}]}{\Gamma[\frac{\tilde{\Delta}_0+\tilde{\Delta}_1+\tilde{\Delta}_{n+1}}{2}-h]}\underbrace{\underset{\partial\text{AdS}}{\int}\mathrm{d}X\!\!\!\!\prod_{i\in\{0,1,n+1\}}\!\!\!\!\!\!(-2P_i\cdot X)^{-\tilde{\Delta}_i}}_{\mathcal{I}_3},
\end{split}
\end{equation}
with
\begin{equation}
\checked{}
\tilde{\Delta}_0=\tau_{0\,1}+\tau_{0\,n+1},\quad
\tilde{\Delta}_1=\tau_{0\,1}+\tau_{1\,n+1},\quad
\tilde{\Delta}_{n+1}=\tau_{0\,n+1}+\tau_{1\,n+1}.
\end{equation}
At this stage it is safe to identify $P_0=P_{n+1}=Q$, after which we can re-integrate $X$, which amounts to
\begin{equation}\label{eq:I3regulated}
\checked{}
\mathcal{I}_3\xrightarrow{P_0=P_{n+1}=Q}\frac{\pi^{h}}{2}\frac{\Gamma[\frac{\tilde{\Delta}_0+\tilde{\Delta}_1+\tilde{\Delta}_{n+1}}{2}-h]\Gamma[\frac{\tilde{\Delta}_0+\tilde{\Delta}_1+\tilde{\Delta}_{n+1}}{2}]}{\Gamma[\tilde{\Delta}_1]\Gamma[\tilde{\Delta}_0+\tilde{\Delta}_{n+1}]}\frac{\Gamma[\frac{\epsilon}{2}+\tau_{0\,n+1}]\Gamma[\frac{\epsilon}{2}-\tau_{0\,n+1}]}{\Gamma[\epsilon](-2P_1\cdot Q)^{\frac{\tilde{\Delta}_0+\tilde{\Delta}_1+\tilde{\Delta}_{n+1}}{2}}}.
\end{equation}
Here we have introduced a regulator $\epsilon$, which is then taken to zero. While for generic value of $\tau$'s this is simply zero, there is subtlety in the neighborhood of $\tau_{0\,n+1}=0$. Specifically, if we take $\tau_{0\,n+1}$ to be an independent Mellin variable and integrate it, note that its integration contour should follow the standard Mellin integral, thus passing between poles from $\Gamma[\frac{\epsilon}{2}+\tau_{0\,n+1}]$ and those from $\Gamma[\frac{\epsilon}{2}-\tau_{0\,n+1}]$, and so the $\tau_{0\,n+1}$ integration in fact produces a pole at $\epsilon=0$ (for a general scenario of the emergence of poles from Mellin integrals, see Appendix \ref{app:sec:mellinintegrals}). This then dictates that
\begin{equation}
\checked{}
\lim_{\epsilon\to0}\int\frac{\mathrm{d}\tau_{0\,n+1}}{2\pi i}\,\mathcal{I}_3\,h[\tau_{0\,n+1}]=\residue{\epsilon=0}\int\frac{\mathrm{d}\tau_{0\,n+1}}{2\pi i}\,\Gamma[\epsilon]\,\mathcal{I}_3\,h[\tau_{0\,n+1}],
\end{equation}
for any function $h[\tau_{0\,n+1}]$ that does not pinch the $\tau_{0\,n+1}$ when $\epsilon\to0$. In computing the residue at this pole, the $\tau_{0\,n+1}$ contour is forced to localized to $|\frac{\epsilon}{2}+\tau_{0\,n+1}|=\epsilon'$. In consequence
\begin{equation}
\begin{split}
\checked{}
\lim_{\epsilon\to0}\int\frac{\mathrm{d}\tau_{0\,n+1}}{2\pi i}\,\mathcal{I}_3\,h[\tau_{0\,n+1}]&=\residue{\epsilon=0}\,\residue{\tau_{0\,n+1}=-\frac{\epsilon}{2}}\left(\Gamma[\epsilon]\,\mathcal{I}_3\,h[\tau_{0\,n+1}]\right)\\
&=\frac{\pi^h}{2}\frac{\Gamma[\tilde\Delta_1-h]}{\Gamma[\tilde\Delta_1]}\frac{h[0]}{(-2P_1\cdot Q)^{\tilde\Delta_1}}.
\end{split}
\end{equation}

Putting every ingredient together we thus have (with $\mathcal{N}=\pi^{h}[\mathrm{d}c]_\Delta\mathcal{N}_c\mathcal{N}_0/\mathcal{C}_{h\pm c}$)
\begin{equation}\label{eq:formingloopintermediate}
\checked{}
I=\frac{1}{\pi^h}\int[\mathrm{d}\tau]_0\,\mathcal{N}\,M_0[\tau]\,\prod_{1\leq i<j\leq n}\frac{\Gamma[\tau_{i\,j}]}{P_{i\,j}^{\tau_{i\,j}}}
\underset{\partial\text{AdS}}{\int}\mathrm{d}Q\,\prod_{k=1}^n\frac{\Gamma[\tau_{0\,k}]\Gamma[\tau_{k\,n+1}]}{(-2P_k\cdot Q)^{\tau_{0\,k}+\tau_{k\,n+1}}},
\end{equation}
where we have set $\tau_{0\,n+1}=0$. Note that although we chose a reference point $1$ when taking the coincidence limit, that choice is completely irrelevant for the formula obtained above, as should be expected. After the $Q$ integration, we further apply a Symanzik transformation, together with a proper shift of the new Mellin variables, similar to what we did in vertex insertions. In the end, we obtain
\begin{equation}\label{formula:loopformation}
\checked{}
M[\delta]=\int[\mathrm{d}\tau]'_0\,M_0[\tau]\,K_{\rm pre}[\tau,\delta],
\end{equation}
with
\begin{equation}
\checked{}
K_{\rm pre}[\tau,\delta]=\prod_{1\leq i<j\leq n}\frac{\Gamma[\tau_{i\,j}]\Gamma[\delta_{i\,j}-\tau_{i\,j}]}{\Gamma[\delta_{i\,j}]}\prod_{k=1}^n\frac{\Gamma[\tau_{0\,k}]\Gamma[\tau_{k\,n+1}]}{\Gamma[\tau_{0\,k}+\tau_{k\,n+1}]}.
\end{equation}
$[\mathrm{d}\tau]'_0$ indicates that the resulting expression here is free of $\tau_{0\,n+1}$. $\delta$'s again satisfy the canonical constraints \eqref{eq:canonicalconstraints} for the new diagram. Note that although we have one less Mellin variable we still have the same amount of original constraints. In particular, those constraints imply a relation between $\tau_{0\,n+1}$ and the others
\begin{equation}
\checked{}
\sum_{1\leq i<j\leq n}\tau_{i\,j}=\frac{\Sigma\Delta}{2}-h+\tau_{0\,n+1}.
\end{equation}
Hence with $\tau_{0\,n+1}=0$ these constraints can be more conveniently written as
\begin{align}
\checked{}
\Delta_k&=\sum_{j=0,\neq k}^{n+1}\tau_{j\,k},\quad k=1,\ldots,n,\\
2c&=\sum_{k=1}^n(\tau_{k\,n+1}-\tau_{0\,k}),\\
\label{eq:extraconstraint}\frac{\Sigma\Delta}{2}-h&=\sum_{1\leq i<j\leq n}\tau_{i\,j}.
\end{align}
In total there are $\frac{(n+2)(n-1)}{2}-1$ independent $\tau$ variables to be integrated.

\newpage

\section{Tree Diagrams Revisited}\label{sec:treediagramsrevisited}

\temp{In this section, apart from the places explicitly marked out, there are no major revisions further needed.}

In the previous section we drew a preliminary step towards obtaining Mellin pre-amplitude of a given diagram by integrating that of certain simpler diagram against an integration kernel $K_{\rm pre}$, which is completely independent of what the original pre-amplitude $M_0$ is.

Tree diagrams can be constructed purely by iterating vertex insertions. In this section we restrict our scope to tree level, and examine in full detail the consequence of the formula \eqref{formula:vertexinsertion} for this operation
\begin{equation}\label{eq:vertexinsertion2}
\checked{}
M[\delta]=\int[\mathrm{d}\tau]_0\,M_0[\tau]\,K_{\rm pre}[\tau,\delta],
\end{equation}
with
\begin{equation}
\checked{}
K_{\rm pre}[\tau,\delta]=\underbrace{\frac{\Gamma[\frac{\Delta_{A_0}-h\pm c_{(1)}}{2}]\Gamma[\frac{h+c_{(1)}-s_{(1)}}{2}]}{\Gamma[\frac{\Delta_{A_0}-s_{(1)}}{2}]}}_{K_2}
\underbrace{\prod_{1\leq i<j\leq m}\frac{\Gamma[\tau_{i\,j}]\Gamma[\delta_{i\,j}-\tau_{i\,j}]}{\Gamma[\delta_{i\,j}]}}_{K_1}.
\end{equation}
For later convenience here we separate the kernel into two parts $K_1$ and $K_2$. 

Scalar tree diagrams were thoroughly explored in previous literature \cite{Penedones:2010ue,Fitzpatrick:2011ia,Paulos:2011ie,Nandan:2011wc,Goncalves:2014rfa}, resulting in a set of diagrammatic rules that determine their Mellin amplitudes $\mathcal{M}$ in terms of summation over explicit poles. 

The main purpose of this section has two folds. One is to derive a representation for the pre-amplitude $M$ instead of the amplitude $\mathcal{M}$, which will be better suited to our later discussions at loop level. For this we explain in Section \ref{sec:localityofinsertion} why $M$ directly factorizes into contributions localized in the diagram, and correspondingly work out a set of diagrammatic rules for $M$ of tree diagrams in Section \ref{sec:treediagrammaticrules}. The other is a simplified recursion formula derived in Section \ref{sec:kernelgenerictrees}, whose role is mainly to motivate our treatment of loop diagrams later on in Section \ref{sec:oneloopdiagrams} and \ref{sec:arbitraryloops}. The equivalence of our results to the ones in previous literature can be easily understood by contour deformations, which are further explained in Appendix \ref{app:sec:proofofequivalence}.

\subsection{Exchange diagrams revisited}

As a warm-up, let us first check how the vertex insertion formula manages to recover the existing result for exchange diagrams.

Note that the kernel $K_{\rm pre}$ depends on $\tau$ only through $K_1$. If $M_0$ itself is a contact diagram, then all the $\tau$ integrations are trivial, leading to the replacement
\begin{equation}\label{eq:contactsimplification}
\checked{}
K_1\longrightarrow\frac{\Gamma[\frac{\Delta_{A_1}-h+ c_{(1)}}{2}]\Gamma[\frac{h-c_{(1)}-s_{(1)}}{2}]}{\Gamma[\frac{\Delta_{A_1}-s_{(1)}}{2}]},\qquad
\Delta_{A_1}=\sum_{i=1}^m\Delta_i.
\end{equation}

Let us check closely how this works. Recall that there are in total $\frac{m(m-1)}{2}$ $\tau$ variables entering this factor; they all come from the original Mellin amplitude of $m+1$ points, which has $\frac{(m+1)(m-2)}{2}$ independent variables. This means we can choose all the explicit variables in $K_1$ to be independent, except for one, which is related to the rest by
\begin{equation}\label{eq:vertexsingleconstraint}
\checked{}
\sum_{1\leq i<j\leq m}\tau_{i\,j}=\frac{\Delta_{A_1}-h+c_{(1)}}{2}.
\end{equation}
Without loss of generality let us assume the variable to be solved by \eqref{eq:vertexsingleconstraint} is $\tau_{i^*j^*}$, then every other $\tau_{i\,j}$ enters $K_1$ exactly four times. Since each individual factors in $K_1$ with fixed $\{i,j\}$ labels is exactly a Beta function, this leads to a special case of Barnes' first lemma
\begin{equation}
\checked{}
\int_{-i\infty}^{+i\infty}\frac{\mathrm{d}z}{2\pi i}\,\Gamma[a_1+z]\Gamma[a_2+z]\Gamma[b_1-z]\Gamma[b_2-z]=\frac{\prod_{i=1}^2\prod_{j=1}^2\Gamma[a_i+b_j]}{\Gamma[a_1+a_2+b_1+b_2]}.
\end{equation}
For instance, the first integration has the form
\begin{equation}
\begin{split}
\checked{}
&\int\frac{\mathrm{d}\tau_{i\,j}}{2\pi i}\frac{\Gamma[\tau_{i\,j}]\Gamma[\delta_{i\,j}\!-\!\tau_{i\,j}]}{\Gamma[\delta_{i\,j}]}\frac{\Gamma[\frac{\Delta_{A_1}-h+c_{(1)}}{2}-\sum'\!\tau-\tau_{i\,j}]\Gamma[\delta_{i^*j^*}-\frac{\Delta_{A_1}-h+c_{(1)}}{2}+\sum'\!\tau+\tau_{i\,j}]}{\Gamma[\delta_{i^*j^*}]}\\
&=\frac{\Gamma[\frac{\Delta_{A_1}-h+c_{(1)}}{2}-\sum'\tau]\Gamma[\delta_{i^*j^*}+\delta_{i\,j}-\frac{\Delta_{A_1}-h+c_{(1)}}{2}+\sum'\tau]}{\Gamma[\delta_{i^*j^*}+\delta_{i\,j}]},
\end{split}
\end{equation}
where $\sum'\tau$ denotes the summation of the remaining $\tau$'s except for $\{\tau_{i^*j^*},\tau_{i\,j}\}$. It is clear that this integration effectively only causes a slight modification to the original factor in $K_1$ labeled by $(i^*j^*)$, by deleting one independent $\tau$ variable while augmenting $\delta_{i^*\,j^*}$ by its corresponding $\delta$. This immediately implies that all the other integrals strictly follow the same pattern. As a result, in the end $K_1$ is replaced by a single Beta function with all the $\delta_{i\,j}$ ($1\leq i<j\leq m$) variables accumulated in the same way as the original $\tau$ summation
\begin{equation}
\checked{}
K_1\longrightarrow\frac{\Gamma[\frac{\Delta_{A_1}-h+c_{(1)}}{2}]\Gamma[\frac{h-c_{(1)}-\Delta_{A_1}}{2}+\sum_{1\leq i<j\leq m}\delta_{i\,j}]}{\Gamma[\sum_{1\leq i<j\leq m}\delta_{i\,j}]}
\equiv\frac{\Gamma[\frac{\Delta_{A_1}-h+c_{(1)}}{2}]\Gamma[\frac{h-c_{(1)}-s_{(1)}}{2}]}{\Gamma[\frac{\Delta_{A_1}-s_{(1)}}{2}]}.
\end{equation}
This $\delta$ summation is then naturally identified with the Mandelstam variable $s_{(1)}$ for the newly formed tree propagator, as in RHS above. Taking into consideration the orignial Mellin integrand $M_0=\Gamma[\frac{\Delta_{A_1}-h-c_{(1)}}{2}]$ for the contact diagram as well as $K_2$, we thus obtain the following well-known formula for a generic scalar exchange diagram \cite{Penedones:2010ue}
\begin{equation}\label{eq:Mexchange}
\begin{split}
\checked{}
M_{\rm exchange}&=\frac{\Gamma[\frac{\Delta_{A_0}-h\pm c_{(1)}}{2}]}{\Gamma[\frac{\Delta_{A_0}-s_{(1)}}{2}]}
\Gamma[{\textstyle\frac{h\pm c_{(1)}-s_{(1)}}{2}}]
\frac{\Gamma[\frac{\Delta_{A_1}-h\pm c_{(1)}}{2}]}{\Gamma[\frac{\Delta_{A_1}-s_{(1)}}{2}]}.
\end{split}
\end{equation}
Here we intentionally organized the result into the above factorized form, such that the first and the third factors appear to come from the two vertices, while the second from the propagator between them. In addition, obviously the convention in the choice of the sign of the spectrum variable $c_{(1)}$ is irrelevant.

\subsection{The integral kernel for generic trees}\label{sec:kernelgenerictrees}

\temp{Correct including the normalization.}

When the original diagram is not a contact diagram $M_0$ has non-trivial dependence on the Mellin variables.  It is both convenient and crucial that Mellin amplitude/pre-amplitude depends on the Mellin variables only via their Mandelstam-like combinations. From the intuition of OPE expansion of the corresponding boundary correlator, it is expected that for scalar tree diagrams the set of Mandelstam variables entering the pre-amplitude should have one-to-one correspondence with the cuts of the tree diagram into two smaller trees. For the exchange diagram \eqref{eq:Mexchange} this is exactly the Mandelstam variable associated to the unqiue tree propagator. For generic trees this can in fact be verified purely from the bulk point of view, as we are going to show in the next subsection. For the time being let us assume this is true for any tree pre-amplitude, and inspect its consequence on the recursion formula.

Note that it is up to our choice how we choose to parametrize the Mellin space using a set of independent variables. As we try to intergrate away the original Mellin variables in the vertex insertion, it is most natural to always take all the Mandelstam variables appearing in $M_0$ to be a subset of the integration variables (it is guaranteed that they are mutually independent as they are associated to a tree), since the remaining independent variables then only explicitly enters the kernel $K_{\rm pre}$. From our experience with the exchange diagrams it is quite probable that this allows us to find out a simplified formula for vertex insertion when restricting our scope to tree diagrams. 

\begin{figure}[ht]
\captionsetup{margin=2em}
\begin{center}
\begin{tikzpicture}
\draw [WildStrawberry,very thick,dotted] (0,0) -- (0,-1.8);
\draw [black,thick] (0,0) -- (-2,1.8) -- (-3,3.6);
\draw [black,thick] (0,0) -- (1,1.8) -- (0,3.6);
\draw [black,thick] (1,1.8) -- (2,3.6);
\draw [black,thick] (0,0) -- (4,1.8);
\draw [black,thick] (-2,1.8) -- (-1.5,2.7);
\draw [black,thick] (0,3.6) -- (.5,4.5);
\draw [black,fill=black!15!white] (0,-1.8) circle [radius=.6];
\draw [black,fill=black!15!white] (0,0) circle [radius=.6];
\draw [black,fill=black!15!white] (-2,1.8) circle [radius=.6];
\draw [black,fill=black!15!white] (1,1.8) circle [radius=.6];
\draw [black,fill=black!15!white] (4,1.8) circle [radius=.6];
\draw [black,fill=black!15!white] (-3,3.6) circle [radius=.6];
\draw [black,fill=black!15!white] (0,3.6) circle [radius=.6];
\draw [black,fill=black!15!white] (2,3.6) circle [radius=.6];
\node [anchor=center] at (0,-1.8) {$A_{0}$};
\node [anchor=center] at (0,0) {$A_{1}$};
\node [anchor=center] at (-2,1.8) {$A_{1,1}$};
\node [anchor=center] at (-3,3.6) {$A_{1,1,1}$};
\node [anchor=center] at (1,1.8) {$A_{1,2}$};
\node [anchor=center] at (0,3.6) {$A_{1,2,1}$};
\node [anchor=center] at (2,3.6) {$A_{1,2,2}$};
\node [anchor=center] at (4,1.8) {$A_{1,3}$};
\node [anchor=south] at (-1.5,2.7) {$\cdots$};
\node [anchor=south] at (.5,4.5) {$\cdots$};
\node [anchor=west] at (0,-.9) {\color{WildStrawberry}\scriptsize $(1)$};
\node [anchor=north east] at (-1,.9) {\scriptsize $(1,1)$};
\node [anchor=north east] at (-2.5,2.7) {\scriptsize $(1,1,1)$};
\node [anchor=south east] at (.5,.9) {\scriptsize $(1,2)$};
\node [anchor=north east] at (.5,2.7) {\scriptsize $(1,2,1)$};
\node [anchor=north west] at (1.5,2.7) {\scriptsize $(1,2,2)$};
\node [anchor=north west] at (2,.9) {\scriptsize $(1,3)$};
\end{tikzpicture}
\end{center}
\caption{General tree diagrams. Each grey blob denotes a bulk interaction vertex, together with possibly any bulk-to-boundary propagators attached to it. The propagators explicitly drawn are bulk-to-bulk propagators only. The newly formed propagator is dotted.}
\label{fig:generaltree}
\end{figure}
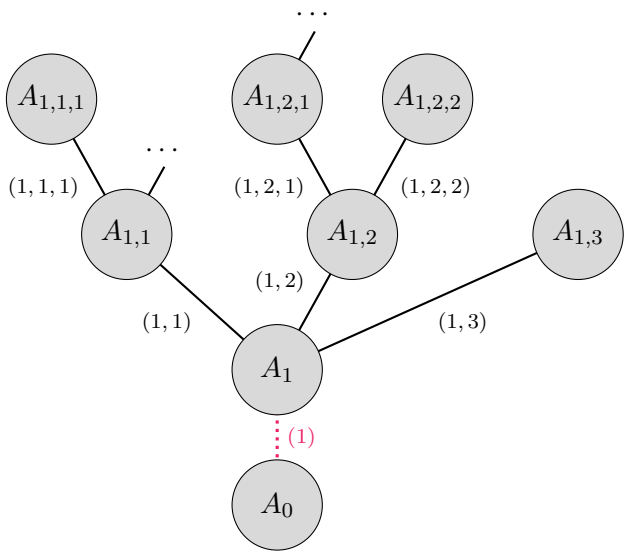
To investigate this possibility for generic tree diagrams, it suffices to just focus on $K_1$. For convenience of subsequent discussions let us first set up some notational conventions. We call the newly inserted vertex $A_0$, and the vertex to which it is inserted $A_1$. We further organize the tree as examplified in Figure \ref{fig:generaltree}, where we name the other vertices by a sequence of subscripts, whose lengths increase with their relative distance to $A_0$ in the diagram. For each vertex the subscripts excluding the last one descends from the unique vertex prior to it, while the last subscript serves to distinguish vertices that descend from the same one. A vertex label is equally used as the set of boundary points attached to it.

We also use similar names for the propagators, and in particular we use these to name the corresponding Mandelstam variable $\Xi$'s, which are (by assumption) those and only those appearing in $M_0$. We use the propagator label in the conformal dimensions to indicate the sum of those boundary points connecting to $A_0$ via this propagator, while we use the vertex label to indicate the sum of boundary points attached to the vertex only. To illustrate these conventions, e.g., we have
\begin{align}
\checked{}\Delta_{A_{1,1}}&=\sum_{i\in A_{1,1}}\Delta_i,\qquad
\Delta_{A_{1,2,1}}=\sum_{i\in A_{1,2,1}}\Delta_i,\\
\checked{}\Delta_{(1,2)}&=\Delta_{A_{1,2}}+\Delta_{(1,2,1)}+\Delta_{(1,2,2)}=\Delta_{A_{1,2}}+\Delta_{A_{1,2,1}}+\Delta_{A_{1,2,2}}+\cdots,
\end{align}
and so on. The Mandelstam variables are defined as, e.g.\footnote{$\sqcup$ denotes disjoint union.},
\begin{equation}
\checked{}
\Xi_{(1,2)}=\Delta_{(1,2)}-2\sum_{i<j\in A_{(1,2)}}\tau_{i,j},\qquad A_{(1,2)}=A_{1,2}\sqcup A_{1,2,1}\sqcup A_{1,2,2}\sqcup\cdots.
\end{equation}
In particular, we have
\begin{equation}\label{eq:overallconstraint}
\checked{}
\sum_{i<j\in A_{(1)}}\tau_{i,j}=\frac{\Delta_{(1)}-h+c}{2},
\end{equation}
so correspondingly $\Xi_{(1)}=h-c$, although this is just a specific value.

As we commented previously, it is preferrable to take all the Mandelstam variables manifestly shown in $M_0$ to be independent integration variables. When $M_0$ arises from a contact diagram there is only the single constraint \eqref{eq:overallconstraint} (recall that any $\tau_{0\,k}$ is absent from the kernel), and by solving arbitrary choice of $\tau$ we easily integrate the other independent ones by Barnes' first lemma. Each time we take one more Mandelstam variable to be independent, we need to pick up one more ordinary $\tau$ variable to solve; hence the number of $\tau$ variables to be solved is always the same as the number of vertices in $M_0$.

There is a natural choice of ordinary $\tau$'s to be solved such that the later integrations are made manifest and easy. Let us start from the bottom of the tree in Figure \ref{fig:generaltree} and focus on vertex $A_1$. Associated to it is the constraint \eqref{eq:overallconstraint}. Note that all the existing $\tau$'s in $K_1$ enter this constraint, but there are certain subset which never appear in other upcoming constraints. To be explicit, let us denote the set of all $\tau$ variables with labels only in $A_{(a)}$ as $T_{(a)}$ (when there is no confusion we also use this to denote the sum of them). Consequently we have the disjoint decomposition
\begin{equation}\label{eq:Taudisjointunion}
\checked{}
T_{(1)}=T_{(1,1)}\sqcup T_{(1,2)}\sqcup T_{(1,3)}\sqcup T_{(1)}^{\rm c},
\end{equation}
where $T_{(1)}^{\rm c}$ is defined to be the complement of the disjoint union $T_{(1,1)}\sqcup T_{(1,2)}\sqcup T_{(1,3)}$ in $T_{(1)}$. These are the variables that never enters any other constraints to be imposed. 

The disjoint decomposition \eqref{eq:Taudisjointunion} allows us to rewrite the constraint \eqref{eq:overallconstraint} as
\begin{equation}
\checked{}
T_{(1)}^{\rm c}=\frac{\Delta_{(1)}-\Xi_{(1)}}{2}-\sum_{m=1}^3\frac{\Delta_{(1,m)}-\Xi_{(1,m)}}{2}=\frac{\Delta_{A_1}-\Xi_{(1)}+\sum_m \Xi_{(1,m)}}{2},
\end{equation}
so that the variables entering other constraints do not make an appearance. It is then convenient to pick up an arbitrary variable from $T_{(1)}^{\rm c}$ and solve it using this constraint. Consequently, this makes it possible to extract part of $K_1$
\begin{equation}
\checked{}
\prod_{\tau\in T_{(1)}^{\rm c}}\frac{\Gamma[\tau_{i\,j}]\Gamma[\delta_{i\,j}-\tau_{i\,j}]}{\Gamma[\delta_{i\,j}]}
\end{equation}
and integrate away the remaining variables within the set $T_{(1)}^{\rm c}$. This yields a single Beta function, which explicitly is
\begin{equation}
\checked{}
\bfn{\frac{\Delta_{A_1}-\Xi_{(1)}+\sum_m \Xi_{(1,m)}}{2}}{D_{(1)}^{\rm c}}\equiv
\bfn{\frac{\Delta_{A_1}-\Xi_{(1)}+\sum_m \Xi_{(1,m)}}{2}}{\frac{\Delta_{A_1}-s_{(1)}+\sum_{m}s_{(1,m)}}{2}},
\end{equation}
where $D_{(1)}^{\rm c}$ refers to $T_{(1)}^{\rm c}$ but with $\tau$ replaced by $\delta$. From now on we abbreviate the Beta function into $\mathrm{B}[\substack{a\\b}]\equiv\frac{\Gamma[a]\Gamma[b-a]}{\Gamma[b]}$. The notation is different from the standard one in literature, but better suited to our purpose.

We then move on to the propagator $\Xi_{(1,1)}$, $\Xi_{(1,2)}$ and so on, gradually increasing the relative distance to $A_0$. It is straightforward to observe that for each propagator the analysis can be done exactly in the same way as above. As a result, given any tree-level diagram, after choosing the Mandelstam variables in $M_0$ to be indepedent, the remaining $\tau$ integrations can be perfomred without modifying $M_0$, so that we can express the new Mellin pre-amplitude as an integration over the Mandelstam variables only, against a new kernel $K$
\begin{equation}\label{formula:vertexinsertiontreesimple}
\checked{}
M[s]=\frac{1}{(-2)^{\sharp[\Xi]}}\int[\mathrm{d}\Xi]\,M_0[\Xi]\,K[\Xi,s],
\end{equation}
where $\sharp[\Xi]$ counts the total number of $\Xi$ variables, and
\begin{equation}\label{eq:treekernelgeneral}
\checked{}
K=\underbrace{\frac{\Gamma[\frac{\Delta_{A_0}-h\pm c_{(1)}}{2}]\Gamma[\frac{h+c_{(1)}-s_{(1)}}{2}]}{\Gamma[\frac{\Delta_{A_0}-s_{(1)}}{2}]}}_{K_2}
\underbrace{\prod_{a\in\{\text{vertices}\}}\!\!\!\!
\bfn{\frac{\Delta_{A_a}-\Xi_{(a)}+\sum_m \Xi_{(a,m)}}{2}}{\frac{\Delta_{A_a}-s_{(a)}+\sum_{m}s_{(a,m)}}{2}}}_{K_1}.
\end{equation}
In particular, $K_1$ is merely a product of Beta functions, arising from the set of original bulk vertices. There are several special cases that one has to pay slight attention:
\begin{itemize}
\item In the case when $A_{a}=\varnothing$, i.e., when $A_{a}$ does not contain any boundary points, the complementary set $T_{(a)}^{\rm c}$ is still non-empty, and so the corresponding contribution to the kernel remains the same apart from that $\Delta_{A_a}=0$.
\item If $A_b$ is an end leaf, for which no $A_{b,m}$ exists, then the constraints associate to it is
\begin{equation}
\checked{}
T_{(b)}=\frac{\Delta_{(b)}-\Xi_{(b)}}{2}\equiv\frac{\Delta_{A_b}-\Xi_{(b)}}{2},
\end{equation}
and no further decomposition is needed for $T_{(b)}$. Therefore its contribution reduces to
\begin{equation}\label{eq:kernelendleaf}
\checked{}
\bfn{\frac{\Delta_{A_b}-\Xi_{(b)}}{2}}{\frac{\Delta_{A_b}-s_{(b)}}{2}}
\end{equation}
\end{itemize}


\subsection{The kernel as a local operator (at any loops)}\label{sec:localityofinsertion}

As promised before, we now return to the question of why $M$ only depends on the Mandelstam variables in correspondence to the cuts of the tree. This is in fact tied to the more general feature that the operation of vertex insertion acts only \emph{locally}. While this is intuitively true from the diagram, and is in some sense a necessary condition in order that (tree) Witten diagrams are consistent with bulk unitarity, this is not at first sight obvious as an operation in the Mellin space \eqref{formula:vertexinsertion} nor \eqref{formula:vertexinsertiontreesimple} (since we are integrating away all the original Mellin variables). Nevertheless, as an important consequence the pre-amplitude $M$ of an arbitraray scalar Witten diagram (at any loop order) decomposes into factors each of which locally associates to a 1PI part of the diagram. When restricting to tree diagrams this immediately dictates the precise dependence on the Mandelstam variables, and further implies the existence of a set of diagrammatic rules for $M$.

Let us first understand precisely why the vertex insertion is a ``local'' operation. Here we do not restrict to tree diagrams.

\begin{figure}[ht]
\captionsetup{margin=2em}
\begin{center}
\begin{tikzpicture}
\begin{scope}[xshift=-3.2cm]
\node [anchor=center] at (0,-1.5) {(A)};
\draw [black,thick] (0,0) -- (0,4);
\draw [black,fill=black!15!white] (0,0) circle [radius=.6];
\draw [black,fill=black!15!white] (0,2) circle [radius=.6];
\draw [black,fill=black!15!white] ($(0,4)+(-.8,-.6)$) rectangle ($(0,4)+(.8,.6)$);
\node [anchor=center] at (0,0) {$A_{-1}$};
\node [anchor=center] at (0,2) {$A_{0}$};
\node [anchor=center] at (0,4) {$A_{(1)}$};
\node [anchor=west] at (0,1) {\scriptsize $(0)$};
\node [anchor=west] at (0,3) {\scriptsize $(1)$};
\end{scope}
\begin{scope}[xshift=3.2cm]
\node [anchor=center] at (0,-1.5) {(B)};
\draw [black,thick] (0,0) -- (0,2);
\draw [black,thick] (0,2) -- +(135:2.4);
\draw [black,thick] (0,2) -- +(45:2.4);
\draw [black,fill=black!15!white] (0,0) circle [radius=.6];
\draw [black,fill=black!15!white] (0,2) circle [radius=.6];
\draw [black,fill=black!15!white] ($(0,2)+(135:2.4)+(-.8,-.6)$) rectangle ($(0,2)+(135:2.4)+(.8,.6)$);
\draw [black,fill=black!15!white] ($(0,2)+(45:2.4)+(-.8,-.6)$) rectangle ($(0,2)+(45:2.4)+(.8,.6)$);
\node [anchor=center] at (0,0) {$A_{-1}$};
\node [anchor=center] at (0,2) {$A_{0}$};
\node [anchor=center] at ($(0,2)+(135:2.4)$) {$A_{(1)}$};
\node [anchor=center] at (0,3) {$\cdots$};
\node [anchor=center] at ($(0,2)+(45:2.4)$) {$A_{(r)}$};
\node [anchor=west] at (0,1) {\scriptsize $(0)$};
\node [anchor=north east] at ($(0,2)+(135:1)$) {\scriptsize $(1)$};
\node [anchor=north west] at ($(0,2)+(45:1)$) {\scriptsize $(r)$};
\end{scope}
\end{tikzpicture}
\end{center}
\vspace{-1.5em}\caption{Two consecutive vertex insertions. (A) the first insertion expands a connected diagram by a vertex $A_0$; (B) the first insertion glues a disconnected diagram into a connected one by a vertex $A_0$. Each grey box denotes a connected (component of the) diagram.}
\label{fig:recursiontermination}
\end{figure}
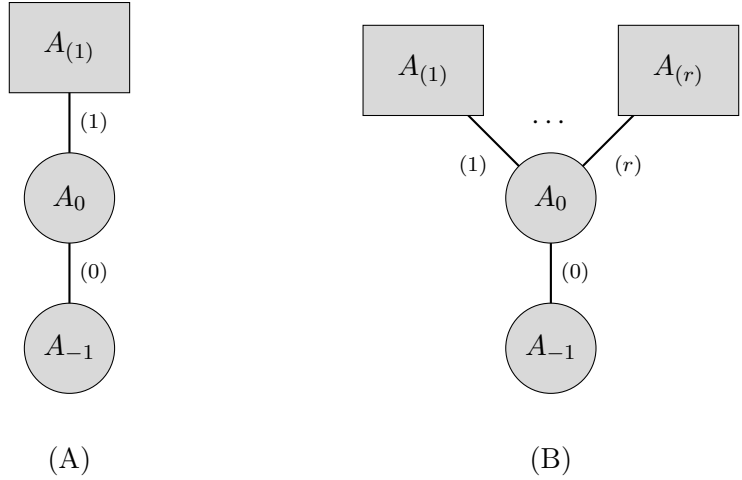

The basic idea is to consider $M_0$ itself as being obtained from another pre-amplitude $M'_0$ by inserting a vertex, as illustrated in Figure \ref{fig:recursiontermination} (A) (hence we start at least from an exchange diagram). In other words, we now study two consecutive insertions of vertices
\begin{equation}
\checked{}
M'_0[\mu]\xrightarrow{\text{insert }A_0}
M_0[\tau]\xrightarrow{\text{insert }A_{-1}}
M[\delta].
\end{equation}
Notation for the Mellin variables at the three stages are indicated above, and correspondingly we denote the Mandelstam variables as $u,t,s$ respectively (not to be confused with the three conventional Mandelstam variables for a 4-point scattering). We first insert a vertex $A_0$ to the diagram $M'_0$, and then further insert another vertex $A_{-1}$ on top of it.

It suffices to return to the formula \eqref{formula:vertexinsertion}. In the second step, we would like to separate the Mellin variables into two groups, depending on whether both labels already appear in $M'_0$ or not, i.e., $T_{(0)}=T_{(1)}\sqcup T_{(0)}^{\rm c}$. Note the overall constraint \eqref{eq:overallconstraint}
\begin{equation}
\checked{}
\sum_{i<j\in A_{(0)}}\tau_{i\,j}=\frac{\Delta_{(0)}-h+c_{(0)}}{2}
\end{equation}
implies
\begin{equation}
\checked{}
T_{(0)}^{\rm c}=\frac{\Delta_{A_0}-h+c_{(0)}+t_{(1)}}{2}.
\end{equation}
If we choose $t_{(1)}$ to be an independent integration variable, then this is the only constraint imposed on $T_{(0)}^{\rm c}$. So following the same analysis as in the previous subsection, we can already perform this part of the $\tau$ integrations such that $K_1$ in the second step (denote it as $K_1^{A_{-1}}[\tau,\delta]$) is replaced by
\begin{equation}
\checked{}
K_1^{A_{-1}}[\tau,\delta]\longrightarrow\bfn{\frac{\Delta_{A_0}-h+c_{(0)}+t_{(1)}}{2}}{\frac{\Delta_{A_0}-s_{(0)}+s_{(1)}}{2}}\,
\prod_{i<j\in A_{(1)}}\frac{\Gamma[\tau_{i\,j}]\Gamma[\delta_{i\,j}-\tau_{i\,j}]}{\Gamma[\delta_{i\,j}]}.
\end{equation}
With this result we can express $M$ in terms of $M'_0$ as
\begin{equation}
\begin{split}
\checked{}
M[s]=&\frac{1}{2}\int[\mathrm{d}\mu]\mathrm{d}t_{(1)}\,M'_0[\mu]\,K_2^{A_{0}}[t_{(1)}]\,K_2^{A_{-1}}[s_{(0)}]\\
&\times\underbrace{\bfn{\frac{\Delta_{A_0}-h+c_{(0)}+t_{(1)}}{2}}{\frac{\Delta_{A_0}-s_{(0)}+s_{(1)}}{2}}\int[\mathrm{d}\tau]_{\rm rest}\,
\prod_{i<j\in A_{(1)}}\!\!\!\frac{\Gamma[\delta_{i\,j}-\tau_{i\,j}]\Gamma[\tau_{i\,j}-\mu_{i\,j}]\Gamma[\mu_{i\,j}]}{\Gamma[\delta_{i\,j}]}}_{K_1^{A_{0}}K_1^{A_{-1}}}.
\end{split}
\end{equation}
The remaining $\tau$ variables do not enter $M'_0$ and again undergo only a single constraint $T_{(1)}=\frac{\Delta_{(1)}-t_{(1)}}{2}$. Hence iterating Barnes' first lemma for these integrals yields
\begin{equation}\label{eq:localactionontree}
\begin{split}
\checked{}
M[s]=&\underbrace{\int[\mathrm{d}\mu]\,M'_0[\mu]\,K_2^{A_0}[s_{(1)}]\prod_{i<j\in A_(1)}\!\!\!\frac{\Gamma[\mu_{i\,j}]\Gamma[\delta_{i\,j}-\mu_{i\,j}]}{\Gamma[\delta_{i\,j}]}}_{M_0[s]}\\
&\times\frac{1}{2}\int\mathrm{d}t_{(1)}\,\frac{K_2^{A_0}[t_{(1)}]}{K_2^{A_0}[s_{(1)}]}\,K_2^{A_{-1}}[s_{(0)}]\,\bfn{\frac{\Delta_{A_0}-h+c_{(0)}+t_{(1)}}{2}}{\frac{\Delta_{A_0}-s_{(0)}+s_{(1)}}{2}}\,\bfn{\frac{h-c_{(1)}-t_{(1)}}{2}}{\frac{h-c_{(1)}-s_{(1)}}{2}}.
\end{split}
\end{equation}
\temp{I have double-checked this result.} This result indicates that, if we take the diagram associated to $M_0$ and insert another vertex $A_{-1}$ to the vertex $A_0$ therein, then the new amplitude $M$ is as simple as taking $M_0$ itself (but with the Mandelstam variables $t$ substituted by the corresponding $s$) and multiplying by an extra factor as defined in the second line above. Furthermore, this extra factor does not at all depends on details in the diagram $M_0$ except for those locally attached to $A_0$ (i.e., $\{\Delta_{A_0},c_{(1)},s_{(1)}\}$). This confirms that vertex insertion acts only locally on the Mellin pre-amplitude, as it does to the diagram. Note in particular this does not depend on what $M'_0$ specifically is.

Of course in the above we were only looking at a special case because we implicitly assumed that the vertex $A_0$ is an end vertex in the original diagram (with only one bulk-to-bulk propagator attached to it). For completeness we also need to consider two consecutive vertex insertions, but generalize the first insertion to the situation that $M'_0$ consists of $r$ connected components (for some integer $r>1$) and that we add vertex $A_0$ to glue them into a connected diagram $M_0$. This is depicted in Figure \ref{fig:recursiontermination} (B). Details of this generalized vertex insertion is discussed in Appendix \ref{app:sec:forest2tree}. Here we only need one simple consequence of this generalization \eqref{eq:forest2treespecialmellingeneral}. Let us denote the pre-amplitudes of the components in $M'_0$ as $\{M_1,M_2,\ldots\}$, then the new pre-amplitude $M_0$ formed by $A_0$ insertion can be again written in terms of integrals
\begin{equation}
M_0[\tau]=\int\,K'[\tau]\,\prod_{a=1}^r\left([\mathrm{d}\mu]_aM_a[\mu]\prod_{i<j\in A_{(a)}}\frac{\Gamma[\mu_{i\,j}]\Gamma[\tau_{i\,j}-\mu_{i\,j}]}{\Gamma[\tau_{i\,j}]}\right).
\end{equation}
Note in particular that all the $\mu_{i\,j}$ with $i,j\in A_{(m)}$ for a specific $m$ only enter in $M_m$ and the product of Beta functions after that in the above formula. For each original component this is exactly the same as the formula \eqref{formula:vertexinsertion}. Correspondingly, for the second step in inserting $A_{-1}$ we group the $\tau$ variables into $T_{(0)}=T_{(1)}\sqcup\cdots\sqcup T_{(r)}\sqcup T_{(0)}^{\rm c}$, and by the identity
\begin{equation}
\checked{}
T_{(0)}^{\rm c}=\frac{\Delta_{A_0}-h+c_{(0)}+\sum_{a=1}^rt_{(a)}}{2},
\end{equation}
we can again directly integrate away the $\tau$ variables in $T_{(0)}^{\rm c}$, leading to the substitution
\begin{equation}
\checked{}
K_2^{A_{-1}}[\tau,\delta]\longrightarrow\bfn{\frac{\Delta_{A_0}-h+c_{(0)}+\sum_{a=1}^r t_{(a)}}{2}}{\frac{\Delta_{A_0}-s_{(0)}+\sum_{a=1}^rs_{(a)}}{2}}\,\prod_{a=1}^r\prod_{i<j\in A_{(a)}}\frac{\Gamma[\tau_{i\,j}]\Gamma[\delta_{i\,j}-\tau_{i\,j}]}{\Gamma[\delta_{i\,j}]}.
\end{equation}
Applying the same logic for each value of $a$, we thus conclude that in the most general setup, the pre-amplitude $M_0$ for the original diagram is not modified except for the local contribution from the vertex $A_0$, to each the new vertex $A_{-1}$ is inserted.

Before we move on to the next subsection for the diagrammatic rules specific to trees, let us emphasize again here that the arguments in this individual subsection does not at all depend on whether each connected component of $M'_0$ is a tree or not. Hence the same conclusion will be reached even if they are at generic loops. As a consequence, the pre-amplitude for an arbitrary diagram should admit of a decomposition into contributions that associates to each one-particle irreducible part of the diagram as well as the contributions from the tree propagators connecting them. Especially, these tree propagators manifestly depend on the Mandelstam variables in their corresponding channels, and in tree digrams these are the only ones that appear in the Mellin (pre-)amplitudes.

\subsection{Tree diagrammatic rules}\label{sec:treediagrammaticrules}

When we restrict the result from the previous subsection to tree diagrams, since each 1PI part is merely a bulk vertex, this should imply simple diagrammatic rules for the pre-amplitude $M$. Let us now work out these rules explicitly.

Already from the explicit result for the generic exchange diagram \eqref{eq:Mexchange} it is tempting to summarize the following rules for part of the diagrams
\begin{enumerate}
\item[1a.] For each end vertex $A_a$, i.e., vertex that has only one bulk-to-bulk propagator $(a)$ attached to it, we write down
\begin{equation}
\checked{}
\frac{\Gamma[\frac{\Delta_{A_a}-h\pm c_{(a)}}{2}]}{\Gamma[\frac{\Delta_{A_a}-s_{(a)}}{2}]}.
\end{equation}
\item[2.] For each propagator $(a)$, we write down
\begin{equation}
\checked{}
\Gamma[{\textstyle\frac{h\pm c_{(a)}-s_{(a)}}{2}}].
\end{equation}
\end{enumerate}
Since the above two points are associated to the addition of each extra vertex itself, to further confirm their validity it suffices to focus on the procedure as depicted in Figure \ref{fig:recursiontermination} (A), and directly grab the result \eqref{eq:localactionontree}
\begin{equation}\label{eq:vertexruleonM0}
\begin{split}
\checked{}
M[s]\!=&\frac{M_0[s]\,\Gamma[\frac{\Delta_{A_0}+h-c_{(0)}-s_{(1)}}{2}]}{\Gamma[\frac{\Delta_{A_0}-c_{(0)}\pm c_{(1)}}{2}]}\!
\left(\Gamma[{\textstyle\frac{\Delta_{A_0}\pm c_{(1)}\pm c_{(0)}}{2}}]\,C_{A_0}\right)\!\Gamma[{\textstyle\frac{h\pm c_{(0)}-s_{(0)}}{2}}]\frac{\Gamma[\frac{\Delta_{A_{-1}}-h\pm c_{(0)}}{2}]}{\Gamma[\frac{\Delta_{A_{-1}}-s_{(0)}}{2}]}.
\end{split}
\end{equation}
Here we re-arranged the expression such that $C_{A_0}$ is a function
\begin{equation}\label{eq:r2correction}
\checked{}
C_{A_0}\!=\!\frac{1}{2}\!\int
\frac{\mathrm{d}t_{(1)}\,\Gamma[\frac{h\pm c_{(1)}-t_{(1)}}{2}]\Gamma[\frac{t_{(1)}-s_{(1)}}{2}]\Gamma[\frac{\Delta_{A_{0}}-h+c_{(0)}+t_{(1)}}{2}]\Gamma[\frac{h-c_{(0)}-t_{(1)}-s_{(0)}+s_{(1)}}{2}]}{\Gamma[\frac{h-c_{(0)}-s_{(0)}}{2}]\Gamma[\frac{h\pm c_{(1)}-s_{(1)}}{2}]\Gamma[\frac{\Delta_{A_0}+c_{(0)}\pm c_{(1)}}{2}]\Gamma[\frac{\Delta_{A_{0}}-s_{(0)}+s_{(1)}}{2}]\Gamma[\frac{\Delta_{A_0}+h-c_{(0)}-t_{(1)}}{2}]}.
\end{equation}
\temp{I have double-checked this expression.} It can be explicitly checked that $C_{A_0}$ has no poles, and so all the poles are manifestly shown in \eqref{eq:vertexruleonM0} (and one can treat $C_{A_0}$ as merely providing corrections to the residues as compared to those from simple Gamma functions). There the combination in the first parentheses is the $M_0[s]$ excluding the contribution from the original end vertex $A_0$ (before insertion), and the last two factors are contributions from the propagator $(0)$ and the new end vertex $A_{-1}$, as according to the rules we proposed above. 

The combination in the second parentheses can then be interpreted as contribution from the new intermediate vertex $A_0$ of valency 2 (i.e., vertex with two bulk-to-bulk propagators attached). This is justified because this combination only depends on $\Delta_{A_0}$ of the vertex and $\{c_{(0)},c_{(1)},s_{(0)},s_{(1)}\}$ of the bulk-to-bulk propagators adjacent to it. In other words, all the data entering are local to this intermediate vertex itself. Furthermore, although not manifest, it can be verified that $C_{A_0}$ is invariant under $c_{(0)}\to-c_{(0)}$, and $c_{(1)}\to-c_{(1)}$, as well as under exchanging $\{c_{(0)},s_{(0)}\}$ with $\{c_{(1)},s_{(1)}\}$, thus enjoying the symmetry of this scalar vertex.

The above arguments provide a derivation for the two rules we proposed before, as well as the rule for any intermediate vertices of valency $2$. By applying the generalized vertex insertion, it is not hard to observe that universal local rules for intermediate vertices of higher valency also exists. A natural proposal is
\begin{enumerate}
\item[1b.] For any intermediate vertex $A_a$ of valency $r>1$, assuming the bulk-to-bulk propagators attached to it are labeled as $\{(1),(2),\ldots,(r)\}$, we write down
\begin{equation}\label{eq:generictreevertexrule}
\checked{}
\Gamma[{\textstyle\frac{\Delta_{A_a}+(r-2)h\pm c_{(1)}\pm\cdots\pm c_{(r)}}{2}}]\;C_{A_a}\!\!\left[\Delta_{A_a};\substack{c_{(1)},\ldots,c_{(r)}\\s_{(1)},\ldots,s_{(r)}}\right],
\end{equation}
\end{enumerate}
where $C_{A_a}$ is a universal function that only depends on the data local to $A_a$ and has no poles. In the case of $r=2$ this obviously reduces to the one we derived above. Furthermore, note also that the first rule we summarized for the end vertices also fit into this general one as a special case $r=1$, as long as we identify $C_{A_a}=1/\Gamma[\frac{\Delta_{A_a}-s_{(a)}}{2}]$ therein. Moreover, this of course implies the $M_{\rm contact}$ for a contact diagram ($r=0$), as we set $C_{A_a}=1$. Hence at tree level we have a universal rule for the bulk vertices and another for the bulk-to-bulk propagators.

\begin{figure}[ht]
\captionsetup{margin=2em}
\begin{center}
\begin{tikzpicture}
\draw [black,thick] (0,0) -- (0:2);
\draw [black,thick] (0,0) -- (45:2);
\draw [black,thick] (0,0) -- (135:2);
\draw [black,thick] (0,0) -- (180:2);
\draw [black,fill=black!15!white] (0,0) circle [radius=.6]; 
\draw [black,fill=black!15!white] (0:2) circle [radius=.6]; 
\draw [black,fill=black!15!white] (45:2) circle [radius=.6]; 
\draw [black,fill=black!15!white] (135:2) circle [radius=.6]; 
\draw [black,fill=black!15!white] (180:2) circle [radius=.6]; 
\node [anchor=center] at (0,0) {$A_0$};
\node [anchor=center] at (0:2) {$A_1$};
\node [anchor=center] at (45:2) {$A_2$};
\node [anchor=center] at (90:2) {$\cdots$};
\node [anchor=center] at (135:2) {$A_{r-1}$};
\node [anchor=center] at (180:2) {$A_r$};
\node [anchor=south] at (0:1) {$1$};
\node [anchor=south east] at (45:1) {$2$};
\node [anchor=south] at (180:1) {$r$};
\end{tikzpicture}
\end{center}
\caption{Satellite diagrams}
\label{fig:satellitediagrams}
\end{figure}
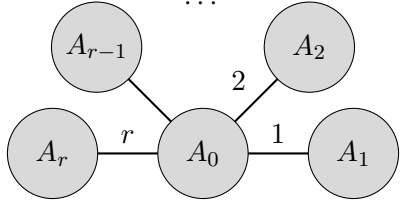
The remaining question is to work out the explicit formula for $C_{A_a}$ in the case of $r>2$. It suffices to focus on the special family of satellite diagrams, as depicted in Figure \ref{fig:satellitediagrams}: we just need to compute this diagram and chop off the contributions from all the end vertices and bulk-to-bulk propagators, and the remaining piece has to be exactly the contribution from the vertex in the center. We leave the detailed derivation to Appendix \ref{app:sec:forest2tree}, but only quote the result \eqref{app:eq:vertexcontributionspecialmellin} here
\begin{equation}\label{eq:vertexcorrectionspecial}
\begin{split}
\checked{}
C_{A_a}\!\!\left[\Delta_{A_a};\substack{c_{(1)},\ldots,c_{(r)}\\s_{(1)},\ldots,s_{(r)}}\right]\!=\!\!\int[\mathrm{d}w]&\,\frac{\Gamma[\frac{\Delta_{A_0}+(r-2)h+\sum_{b=1}^rc_{(b)}}{2}+\sum_{b=1}^rw_b]}{\Gamma[\frac{\Delta_{A_0}+(r-2)h\pm c_1\pm c_2\pm\cdots\pm c_r}{2}]}\\
&\times\prod_{b=1}^r\frac{\Gamma[-w_b]\Gamma[-c_{(b)}-w_b]\Gamma[\frac{h+c_{(b)}-s_{(b)}}{2}+w_b]}{\Gamma[\frac{h\pm c_{(b)}-s_{(b)}}{2}]},
\end{split}
\end{equation}
where as before $\Delta_{A_0}$ denotes the total conformal dimension of the boundary points attached to $A_0$. In the special case of $r=0$ this expression obviously reduces to 1, and for $r=1$ it can be easily verified that this reduces to $1/\Gamma[\frac{\Delta_{A_a}-s_{(1)}}{2}]$ as we desired. In the case of $r=2$, though not obvious, this is equivalent to \eqref{eq:r2correction}. In consequence, with the \eqref{eq:vertexcorrectionspecial} our \eqref{eq:generictreevertexrule} can be regarded as dictating the contribution from any scalar bulk vertex in a tree diagram.

\newpage

\section{One-loop Diagrams}\label{sec:oneloopdiagrams}

\temp{Apart from the marked places, no major revisions further needed in this section.}

As we analyzed the consequence of vertex insertions at tree level in Section \ref{sec:kernelgenerictrees}, we observed that with a proper choice of independent variables parametrizing the Mellin space, i.e., the Mandelstam variables that naturally appear in the original Mellin (pre-)amplitude, the corresponding integral recursion can be greatly simplified.

In this section we develop this intuition further and apply it to the construction of pre-amplitudes for generic one-loop scalar diagrams. While at tree level this technique seems a bit redundant since the Mellin pre-amplitudes already follow simple diagrammatic rules, at loop level this can bring about great convenience for precise understanding of analytic properties of the diagrams, as will be further examined in Section \ref{sec:polestructure} and \ref{sec:residuecomputation}.

This whole section is based on the formula for loop formation \eqref{formula:loopformation}
\begin{equation}
\checked{}
M[\delta]=\int[\mathrm{d}\tau]'_0\,M_0[\tau]\,K_{\rm pre}[\tau,\delta],
\end{equation}
with its kernel
\begin{equation}
\checked{}
K_{\rm pre}[\tau,\delta]=
\underbrace{\prod_{k=1}^n\frac{\Gamma[\tau_{0\,k}]\Gamma[\tau_{k\,n+1}]}{\Gamma[\tau_{0\,k}+\tau_{k\,n+1}]}}_{K_2}
\underbrace{\prod_{1\leq i<j\leq n}\!\!\frac{\Gamma[\tau_{i\,j}]\Gamma[\delta_{i\,j}-\tau_{i\,j}]}{\Gamma[\delta_{i\,j}]}}_{K_1}.
\end{equation}
The $\tau$ integrations here come with the consrtaints
\begin{align}
\checked{}
\Delta_k&=\sum_{j=0,\neq k}^{n+1}\tau_{j\,k},\quad k=1,\ldots,n,\\
2c&=\sum_{k=1}^n(\tau_{k\,n+1}-\tau_{0\,k}),\\
\label{eq:looptauijconstraint}\frac{\Sigma\Delta}{2}-h&=\sum_{1\leq i<j\leq n}\tau_{i\,j}.
\end{align}
Once more we divided the kernel into two parts, and the $K_1$ part shares the same structure as that in the kernel for vertex insertion \eqref{eq:vertexinsertion2}. Note that $\tau$'s with one label in $\{0,n+1\}$ only enter $K_2$ while all the other variables only enter $K_1$, and that $\tau_{0\,n+1}=0$.

The above formula again does not draw any assumption on $M_0[\tau]$. But as we solely focus on the contruction of one-loop diagrams in this section, here we assume that $M_0$ associates to a tree diagram and hence can be directly written down following the diagrammatic rules in Section \ref{sec:treediagrammaticrules}. In particular, as we showed before, $M_0$ only depends on the set of Mandelstam variables associated to the tree propagators in the diagram.

\subsection{Tadpoles}

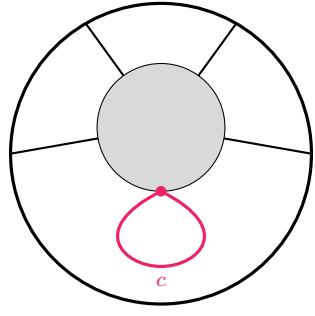
\begin{figure}[ht]
\captionsetup{margin=2em}
\begin{center}
\begin{tikzpicture}
\draw [black,thick] (0,.35) -- (0:2);
\draw [black,thick] (0,.35) -- (60:2);
\draw [black,thick] (0,.35) -- (120:2);
\draw [black,thick] (0,.35) -- (180:2);
\draw [WildStrawberry,very thick] (-90:.5) .. controls (-1,-1) and (-.5,-1.5) .. (-90:1.5) .. controls (.5,-1.5) and (1,-1) .. (-90:.5);
\draw [black,fill=black!15!white] (0,.35) circle [radius=.85];
\draw [black,very thick] (0,0) circle [radius=2];
\fill [WildStrawberry] (-90:.5) circle [radius=2pt];
\node [anchor=north] at (-90:1.5) {\color{WildStrawberry}\scriptsize $c$};
\end{tikzpicture}
\end{center}
\caption{Diagram with a tadpole. There is one bulk-to-bulk propagator whose two ends are anchored at the same bulk point. The remaining part of the diagram can be anything.}
\label{fig:tadpole}
\end{figure}

The simplest situation of forming a loop is of course the case of tadpole, as shown in Figure \ref{fig:tadpole}, where the loop is attached to a single vertex. In this case one can even directly compute the contribution of the loop, which merely gives rise to an extra factor as compared to the corresponding diagram without the tadpole
\begin{equation}
\checked{}
M=\frac{\Gamma[h]\Gamma[h\pm c]}{\Gamma[2h]}\,M_{\text{no tadpole}}.
\end{equation}
(Here the extra factor $\Gamma[h\pm c]$ is due to our specific convention on the normalization of $M$ in \eqref{eq:Mellinnormalization}) For consistency this same factor has to be reproduced by the kernel.

Here let us focus on the simplest case where the tadpole is attached to the unique vertex in a contact diagram, for which we should have
\begin{equation}\label{eq:tadpoleresult}
\checked{}
M=\frac{\Gamma[h]\Gamma[h\pm c]}{\Gamma[2h]}\Gamma[{\textstyle\frac{\Sigma\Delta}{2}-h}],\qquad
\Sigma\Delta=\sum_{i=1}^n\Delta_i.
\end{equation}

In this particular case the only constraints are listed above. The system only has $\frac{(n+2)(n-1)}{2}-1$ independent variables. Similar to what happens in the operation of vertex insertion, the presence of the last constraint \eqref{eq:looptauijconstraint} implies the impossibility of picking all $\tau_{i\,j}$ ($1\leq i<j\leq n$) to be independent. Nevertheless we can choose all of them to be independent except for an arbitrary one. This then means that we still need to choose $n-1$ extra independent variables from the set $\{\tau_{0\,k},\tau_{k\,n+1}\}$. For convenience we can choose these to be $\tau_{k\,n+1}$ ($k=2,\ldots,n$).


In this way it is easy to see that $K_2$ can be explicitly expressed so that $\tau_{0\,k}$'s and $\tau_{1\,n+1}$ are absent
\begin{equation}
\begin{split}
\checked{}
K_2=&\frac{\Gamma[\Delta_1-h-c-\sum_{k=2}^n(\tau_{1\,k}-\tau_{k\,n+1})]\Gamma[h+c-\sum_{k=2}^n\tau_{k\,n+1}]}{\Gamma[\Delta_1-\sum_{k=2}^n\tau_{1\,k}]}
\\
&\times\prod_{k=2}^n\frac{\Gamma[\Delta_k-\tau_{k\,n+1}-\sum_{j=1,\neq k}^n\tau_{j\,k}]\Gamma[\tau_{k\,n+1}]}{\Gamma[\Delta_k-\sum_{j=1,\neq k}^n\tau_{j\,k}]}.
\end{split}
\end{equation}
In this form it is clear that the $\tau_{k\,n+1}$ integration can be performed in the same way as the case of a vertex inserted on a contact diagram, whose effect is merely a modification of the Beta function in front of the product. As a result, these integrations leads to the replacement
\begin{equation}
\checked{}
K_2\longrightarrow\frac{\Gamma[h+c]\Gamma[\Sigma\Delta-h-c-2\sum_{1\leq i<j\leq n}\tau_{i\,j}]}{\Gamma[\Sigma\Delta-2\sum_{1\leq i<j\leq n}\tau_{i\,j}]}\equiv\frac{\Gamma[h\pm c]}{\Gamma[2h]}.
\end{equation}
Very nicely the dependence on $\tau_{i\,j}$ ($1\leq i<j\leq n$) is completely gone after these integrations. This allows us to directly borrow the result from the tree level study to deal with $K_1$, which results in (recall these variables are constrained by \eqref{eq:looptauijconstraint} only)
\begin{equation}
\checked{}
K_1\longrightarrow\frac{\Gamma[\frac{\Sigma\Delta}{2}-h]\Gamma[\sum_{1\leq i<j\leq n}\delta_{i\,j}-\frac{\Sigma\Delta}{2}+h]}{\Gamma[\sum_{1\leq i<j\leq n}\delta_{i\,j}]}\equiv\frac{\Gamma[\frac{\Sigma\Delta}{2}-h]\Gamma[h]}{\Gamma[\frac{\Sigma\Delta}{2}]}.
\end{equation}
Since $M_0=\Gamma[\frac{\Sigma\Delta}{2}]$, we see this explicitly recovers \eqref{eq:tadpoleresult}.

\subsection{Necklace diagrams}\label{sec:necklace}

\begin{figure}[ht]
\captionsetup{margin=2em}
\begin{center}
\begin{tikzpicture}
\draw [black,thick] (-75:3) arc [start angle=-75,end angle=0,radius=3];
\draw [black,thick,dashed] (0:3) arc [start angle=0,end angle=45,radius=3];
\draw [black,thick] (45:3) arc [start angle=45,end angle=135,radius=3];
\draw [black,thick,dashed] (135:3) arc [start angle=135,end angle=180,radius=3];
\draw [black,thick] (180:3) arc [start angle=180,end angle=255,radius=3];
\draw [black,thick,dotted] (255:3) arc [start angle=255,end angle=285,radius=3];
\draw [black,fill=black!15!white] (-45:3) circle [radius=.6];
\draw [black,fill=black!15!white] (0:3) circle [radius=.6];
\draw [black,fill=black!15!white] (45:3) circle [radius=.6];
\draw [black,fill=black!15!white] (90:3) circle [radius=.6];
\draw [black,fill=black!15!white] (135:3) circle [radius=.6];
\draw [black,fill=black!15!white] (180:3) circle [radius=.6];
\draw [black,fill=black!15!white] (225:3) circle [radius=.6];
\node [anchor=center] at (225:3) {$A_1$};
\node [anchor=center] at (180:3) {$A_2$};
\node [anchor=center] at (135:3) {$A_{a-1}$};
\node [anchor=center] at (90:3) {$A_a$};
\node [anchor=center] at (45:3) {$A_{a+1}$};
\node [anchor=center] at (0:3) {$A_{r-1}$};
\node [anchor=center] at (-45:3) {$A_r$};
\node [anchor=north] at (255:3) {\scriptsize $0$};
\node [anchor=north] at (285:3) {\scriptsize $n+1$};
\draw [black,fill=white] (255:3) circle [radius=2pt];
\draw [black,fill=white] (285:3) circle [radius=2pt];
\draw [red,very thick,dashed] (112.5:4) -- (112.5:3) .. controls (112.5:1) and (265:1) .. (265:3) -- (265:4);
\node [anchor=north] at (265:4) {$\Xi_a$};
\draw [red,very thick,dashed] (67.5:4) -- (67.5:3) .. controls (67.5:1) and (275:1) .. (275:3) -- (275:4);
\node [anchor=north] at (275:4) {$\Xi_{a+1}$};
\draw [orange,very thick,dashed] (119.5:4) -- (119.5:3) .. controls (119.5:1) and (241:1) .. (241:3) -- (241:4);
\node [anchor=east] at (119.5:4) {$\xi_{1,a-1}$};
\draw [orange,very thick,dashed] (74.5:4) -- (74.5:3) .. controls (74.5:1) and (248:1) .. (248:3) -- (248:4);
\node [anchor=south] at (74.5:4) {$\xi_{1,a}$};
\draw [orange,very thick,dashed] (105.5:4) -- (105.5:3) .. controls (105.5:1) and (292:1) .. (292:3) -- (292:4);
\node [anchor=south] at (105.5:4) {$\xi_{a,r}$};
\draw [orange,very thick,dashed] (60.5:4) -- (60.5:3) .. controls (60.5:1) and (299:1) .. (299:3) -- (299:4);
\node [anchor=west] at (60.5:4) {$\xi_{a+1,r}$};
\node [anchor=east] at (202.5:3) {$\Xi_2$};
\node [anchor=west] at (-22.5:3) {$\Xi_r$};
\end{tikzpicture}
\end{center}
\caption{Necklace diagram. Red lines indicate the OPE channels in the original chain, with the corresponding Mandelstam variables $\Xi$. Orange lines indicate the OPE channels of the newly formed necklace diagram; after replacing $\tau$ by $\delta$, $\xi_{1,a-1}$ and $\xi_{a,r}$ becomes the same Mandelstam variable for the channel (similarly $\xi_{1,a}$ and $\xi_{a+1,r}$), but in terms of $\tau$ they are different.}
\label{fig:polygons}
\end{figure}
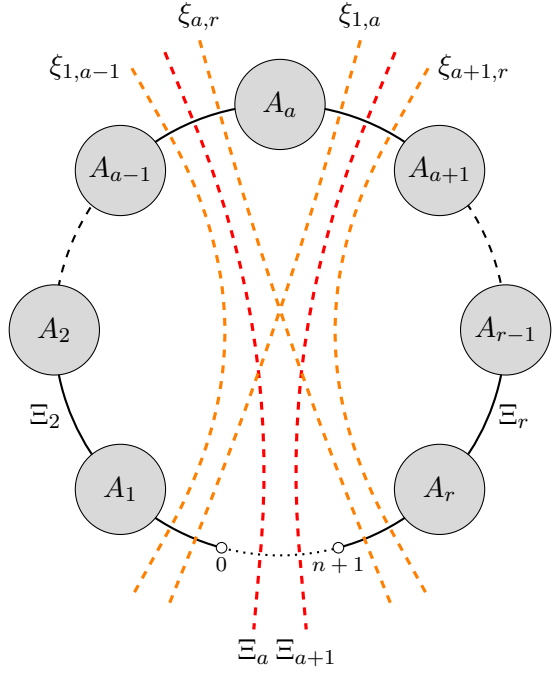

Let us now examine the possible simplification in the construction of a special type of one-loop diagrams, the necklace diagrams as shown in Figure \ref{fig:polygons}, where all bulk vertices sit sequentially on the loop. Such diagram can thus be formed from a chain, and we assume there are $r$ vertices. We use $\Delta_{A_1}$ and $\Delta_{A_r}$ for the total conformal dimensions of boundary points attached to $A_0$ and $A_r$ respectively, excluding $P_0$ and $P_{n+1}$.

We first work on $K_2$ and then $K_1$, because the $\tau$ variables entering the latter may potentially enter the former as well due to the linear constraints among the $\tau$'s.


Similar to what we did in the tadpole, we are going to take the $n-1$ variables $\tau_{k\,n+1}$ ($k=2,\ldots,n$) to be independent and solve the $\tau_{0\,k}$'s and $\tau_{1\,n+1}$ in terms of them, if no further constraints are imposed. However, these independent variables cannot all be trivially integrated because $M_0$ depends on the Mandelstam variables (or OPE channels) for the tree propagators of the chain, which are particular combination of them. Let us denote these Mandelstam variables as $\{\Xi_2,\Xi_2,\ldots,\Xi_r\}$. The idea is to trade $r-1$ of the independent $\tau_{k\,n+1}$'s for these $\Xi$. For this purpose, recall they are related by
\begin{equation}\label{eq:M0tconstraints}
\begin{split}\checked{}
\Xi_2&=h+c+\Delta_{A_2}+\cdots+\Delta_{A_r}-2\sum_{A_2\sqcup\cdots\sqcup A_r}\tau_{i\,j}-2\sum_{A_2\sqcup\cdots\sqcup A_r}\tau_{k\,n+1},\\
\Xi_3&=h+c+\Delta_{A_3}+\cdots+\Delta_{A_r}-2\sum_{A_3\sqcup\cdots\sqcup A_r}\tau_{i\,j}-2\sum_{A_3\sqcup\cdots\sqcup A_r}\tau_{k\,n+1},\\
&\cdots\\
\Xi_r&=h+c+\Delta_{A_r}-2\sum_{A_r}\tau_{i\,j}-2\sum_{A_r}\tau_{k\,n+1}.
\end{split}
\end{equation}
Let us also introduce additional Mandelstam variables for OPE channels that do not separate point $P_0$ from $P_{n+1}$ but are consistent with the natural planar ordering induced by the loop to be formed. For example,
\begin{align}
\checked{}
\xi_{a,r}&=\Delta_{A_a}+\cdots+\Delta_{A_r}-2\sum_{A_a\sqcup\cdots \sqcup A_r}\tau_{i\,j},\\
\xi_{2,3}&=\Delta_{A_2}+\Delta_{A_3}-2\sum_{A_2\sqcup A_3}\tau_{i\,j},\\
\xi_2\equiv\xi_{2,2}&=\Delta_{A_2}-2\sum_{A_2}\tau_{i\,j},
\end{align}
and so on (recall that $n+1\notin A_r$ and $1\notin A_1$). We then reorganize the constraint \eqref{eq:M0tconstraints} into
\begin{equation}\label{eq:Xixitau}
\checked{}
\Xi_a-\Xi_{a+1}=\xi_{a,r}-\xi_{a+1,r}-2\sum_{k\in A_a}\tau_{k\,n+1},\quad a=2,\ldots,r,
\end{equation}
where $\Xi_{r+1}=h+c$ and $\xi_{r+1,r}=0$. The set of $\tau$ variables appearing in each equation do not have any overlap with others, and none of them appear in $K_1$. It is then ideal to pick up one such $\tau_{k\,n+1}$ by choosing one boundary point $k$ from each $A_a$ ($a=2,\ldots,r$) and solve them as well in terms of the $\Xi$'s and $\xi$'s (note that by thinking about $\xi$ as independent variables will impose extra constraints on $\tau_{i\,j}$ with $1\leq i<j\leq n$ but not the $\tau_{k\,n+1}$'s, and so there is no extra $\tau_{k\,n+1}$ to be solved). Furthermore, the factor $K_2$ can be organized into
\begin{equation}
\begin{split}
\checked{}
K_2=&
\frac{\Gamma[\Delta_1-\frac{h+c+\Xi_2-\xi_{2,r}}{2}-\sum_{k=2}^n\tau_{1\,k}+\sum_{A_1\backslash\{1\}}\tau_{k\,n+1}]\Gamma[\frac{h+c+\Xi_2-\xi_{2,r}}{2}-\sum_{A_1\backslash\{1\}}\tau_{k\,n+1}]}{\Gamma[\Delta_1-\sum_{k=2}^n\tau_{1\,k}]}
\\
&\times\prod_{a=1}^r\prod_{k\in A_a}\frac{\Gamma[\Delta_k-\tau_{k\,n+1}-\sum_{j=1,\neq k}^n\tau_{j\,k}]\Gamma[\tau_{k\,n+1}]}{\Gamma[\Delta_k-\sum_{j=1,\neq k}^n\tau_{j\,k}]}.
\end{split}
\end{equation}
In the above, for $a=1$ the product over $k\in A_a$ excludes $k=1$. At this stage we are able to perform the remaining $n-r$ $\tau_{k\,n+1}$ integrations using Barnes' first lemma again, and by using \eqref{eq:Xixitau} this leads to the substitution
\begin{equation}\label{eq:K2oneloopreduced}
\begin{split}
\checked{}
K_2\longrightarrow&
\frac{\Gamma[\frac{h+c+\Xi_2-\xi_{2,r}}{2}]\Gamma[\frac{h-c-\Xi_2+\xi_{1}}{2}]}{\Gamma[h+\frac{\xi_{1}-\xi_{2,r}}{2}]}
\frac{\Gamma[\frac{h+c+\xi_{r}-\Xi_r}{2}]\Gamma[\frac{h-c-\xi_{1,r-1}+\Xi_r}{2}]}{\Gamma[h+\frac{\xi_{r}-\xi_{1,r-1}}{2}]}\\
&\times\prod_{a=2}^{r-1}\frac{\Gamma[\frac{(\xi_{a,r}-\xi_{a+1,r})-(\Xi_a-\Xi_{a+1})}{2}]\Gamma[\frac{(\Xi_a-\Xi_{a+1})-(\xi_{1,a-1}-\xi_{1,a})}{2}]}{\Gamma[\frac{(\xi_{a,r}-\xi_{a+1,r})-(\xi_{1,a-1}-\xi_{1,a})}{2}]}\\
&\equiv\prod_{a=1}^{r}\frac{\Gamma[\frac{(\xi_{a,r}-\xi_{a+1,r})-(\Xi_a-\Xi_{a+1})}{2}]\Gamma[\frac{(\Xi_a-\Xi_{a+1})-(\xi_{1,a-1}-\xi_{1,a})}{2}]}{\Gamma[\frac{(\xi_{a,r}-\xi_{a+1,r})-(\xi_{1,a-1}-\xi_{1,a})}{2}]},
\end{split}
\end{equation}
which is again manifestly a product of Beta functions, one for each bulk vertex, and here we naturally identify
\begin{equation}
\checked{}
\Xi_1=h-c,\quad \Xi_{r+1}=h+c,\quad\xi_{1,r}=2h,\quad\xi_{a,b}=0\;\;(\forall a>b).
\end{equation}
To better illustrate the structure in this result, for a fixed $a$ we explicitly draw out in Figure \ref{fig:polygons} the corresponding OPE channels associated to the Mandelstam variables that appear in the Beta factor labeled by $a$. 

Very nicely, apart from $\xi$'s, the result above does not depend on any of $\tau_{i\,j}$ with $1\leq i<j\leq n$. In order to deal with $K_1$, it is now clear that we need to take all the $\xi$'s as independent variables as well (there is no relation among them). This imposes constraints
\begin{align}
\label{eq:Aaconstraint}
\checked{}\sum_{i<j\in A_a}\tau_{i\,j}&=\frac{\Delta_{A_a}-\xi_{a}}{2},\\
\label{eq:Aaconstraint2}
\checked{}\sum_{i\in A_a,j\in A_{b}}\tau_{i\,j}&=\frac{\xi_{a,b-1}+\xi_{a+1,b}-\xi_{a+1,b-1}-\xi_{a,b}}{2},\quad b-a>1.
\end{align}
Note again that, similar to the constraints \eqref{eq:Xixitau}, there is no overlap between the sets of $\tau$ variables entering each equation. This suggests us to divide $K_1$ into groups of factors according to the pattern of their $\tau_{i\,j}$ labels, one group in correspondence to each equation above. Within each group we solve one variable using the corresponding equation, which then allows us to integrate away the rest by Barnes' first lemma. This in the end leads to another replacement
\begin{equation}
\begin{split}\label{eq:K1oneloopreduced}
\checked{}
K_1\longrightarrow&
\prod_{a=1}^r\bfn{\frac{\Delta_{A_a}-\xi_{a}}{2}}{\frac{\Delta_{A_a}-s_{a}}{2}}\,
\prod_{1\leq a<b\leq r}\bfn{\frac{\xi_{a,b-1}+\xi_{a+1,b}-\xi_{a+1,b-1}-\xi_{a,b}}{2}}{\frac{s_{a,b-1}+s_{a+1,b}-s_{a+1,b-1}-s_{a,b}}{2}}.
\end{split}
\end{equation}
Here the notation $s$ for the Mandelstam variables of the new Mellin pre-amplitude are defined in the same way as the $\xi$ variables, but with $\tau$'s substituted by $\delta$'s.

Additional simplification of $K_1$ occurs when some vertex, say $A_a$, is connected to only one boundary point. In this case $\xi_a$ merely reduces to the conformal dimension of that point. Furthermore, since there is no $\tau_{i\,j}$ belonging to the set, the constraint \eqref{eq:Aaconstraint} associated to $A_a$ is absent, as well as the corresponding contribution to $K_1$, i.e., $\bfn{\frac{\Delta_{A_a}-\xi_{a}}{2}}{\frac{\Delta_{A_a}-s_{a}}{2}}$.

Taking into consideration the Jacobian from the transformation of variables in \eqref{eq:Xixitau}, \eqref{eq:Aaconstraint} and \eqref{eq:Aaconstraint2}, we thus obtain a simplified formula
\begin{equation}\label{eq:oneloopformulareduced}
\checked{}
M[s]=\frac{1}{(-2)^{\sharp[\Xi,\xi]}}\int[\mathrm{d}\Xi][\mathrm{d}\xi]\,M_0[\Xi,\xi]\,K[\Xi,\xi,s],
\end{equation}
with $\sharp[\Xi,\xi]$ counting the total number of remaining integrations, and
\begin{equation}\label{eq:Koneloopreduced}
\begin{split}\checked{}
K=&\underbrace{\prod_{a=1}^{r}\frac{\Gamma[\frac{(\xi_{a,r}-\xi_{a+1,r})-(\Xi_a-\Xi_{a+1})}{2}]\Gamma[\frac{(\Xi_a-\Xi_{a+1})-(\xi_{1,a-1}-\xi_{1,a})}{2}]}{\Gamma[\frac{(\xi_{a,r}-\xi_{a+1,r})-(\xi_{1,a-1}-\xi_{1,a})}{2}]}}_{K_2}\\
&\times\underbrace{\prod_{a=1}^r\bfn{\frac{\Delta_{A_a}-\xi_{a}}{2}}{\frac{\Delta_{A_a}-s_{a}}{2}}\,
\prod_{1\leq a<b\leq r}\bfn{\frac{\xi_{a,b-1}+\xi_{a+1,b}-\xi_{a+1,b-1}-\xi_{a,b}}{2}}{\frac{s_{a,b-1}+s_{a+1,b}-s_{a+1,b-1}-s_{a,b}}{2}}}_{K_1}.
\end{split}
\end{equation}

\subsection{Examples}

We now provide several explicit examples to illustrate our discussions so far.

\subsubsection{Bubble diagrams}

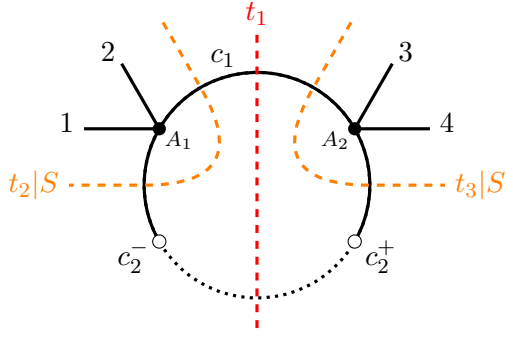
\begin{figure}[h]
\captionsetup{margin=2em}
\begin{center}
\begin{tikzpicture}
\draw [black,very thick,dotted] (0,0) circle [radius=1.5];
\draw [black,very thick] (-30:1.5) arc [start angle=-30,end angle=210,radius=1.5];
\coordinate (p1) at ($(150:1.5)+(180:1)$);
\node [anchor=center] at ($(0,0)!1.1!(p1)$) {$1$};
\coordinate (p2) at ($(150:1.5)+(120:1)$);
\node [anchor=center] at ($(0,0)!1.1!(p2)$) {$2$};
\draw [black,very thick] (p1) -- (150:1.5) -- (p2);
\node [anchor=center] at (150:1.2) {\scriptsize $A_1$};
\coordinate (p3) at ($(30:1.5)+(60:1)$);
\node [anchor=center] at ($(0,0)!1.1!(p3)$) {$3$};
\coordinate (p4) at ($(30:1.5)+(0:1)$);
\node [anchor=center] at ($(0,0)!1.1!(p4)$) {$4$};
\draw [black,very thick] (p3) -- (30:1.5) -- (p4);
\node [anchor=center] at (30:1.2) {\scriptsize $A_2$};
\node [anchor=center] at (210:1.9) {$c_2^-$};
\node [anchor=center] at (-30:1.9) {$c_2^+$};
\node [anchor=center] at (105:1.75) {$c_1$};
\fill [black] (150:1.5) circle [radius=2.5pt];
\fill [black] (30:1.5) circle [radius=2.5pt];
\draw [black,fill=white] (-30:1.5) circle [radius=2.5pt];
\draw [black,fill=white] (-150:1.5) circle [radius=2.5pt];
\draw [red,very thick,dashed] (90:2) -- (270:2);
\node [anchor=south] at (90:2) {\color{red}$t_1$};
\draw [orange,very thick,dashed] (120:2.5) -- (120:1.5) .. controls (120:.5) and (180:.5) .. (180:1.5) -- (180:2.5);
\node [anchor=east] at (180:2.5) {\color{orange}$t_2|S$};
\draw [orange,very thick,dashed] (60:2.5) -- (60:1.5) .. controls (60:.5) and (0:.5) .. (0:1.5) -- (0:2.5);
\node [anchor=west] at (0:2.5) {\color{orange}$t_3|S$};
\end{tikzpicture}
\end{center}
\caption{Construting the 4-point bubble. In this and latter diagrams we skip drawing the AdS boundary. Solid lines and vertices represent the original diagram, with the open legs being the bulk-to-boundary propagators. The two white nodes are the boundary points to be identified and integrated (indicated by the dotted line). The red and orange dashed lines denote OPE channels in correspondence to the $\Xi$ and $\xi$ variables in $K$. In the notation of the form $x|y$ for the orange lines, $x$ refers to the Mandelstam in the original Mellin variables $\tau$, while $y$ refers to that in the new Mellin variables $\delta$. For the red lines there is no counterpart in the new diagram, and so the notation only indicates the original Mellin variable.}
\label{fig:bubble4pt}
\end{figure}

$r=2$ corresponds to diagrams with a bubble. Here we look at a 4-point diagram with a bubble as constructed in Figure \ref{fig:bubble4pt}, such that $A_1=\{1,2\}$, $A_2=\{3,4\}$. 
The original diagram is thus a 6-point exchange diagram, whose pre-amplitude is
\begin{equation}\label{eq:M0bubble4pt}
\checked{}
M_0=\frac{\Gamma[\frac{\Delta_{12}-c_2\pm c_1}{2}]}{\Gamma[\frac{\Delta_{12}+h-c_2-t_1}{2}]}\,
\Gamma[{\textstyle\frac{h\pm c_1-t_1}{2}}]\,
\frac{\Gamma[\frac{\Delta_{34}+c_2\pm c_1}{2}]}{\Gamma[\frac{\Delta_{34}+h+c_2-t_1}{2}]},
\end{equation}
where $\Delta_{12}\equiv\Delta_1+\Delta_2$, $\Delta_{34}\equiv\Delta_3+\Delta_4$. Applying \eqref{eq:oneloopformulareduced}, we have the following specifications
\begin{equation}
\begin{split}\checked{}
&\Xi_1=h-c_2,\quad
\Xi_2=t_1,\quad
\Xi_3=h+c_2,\quad
\xi_{1,1}=t_2,\quad
\xi_{1,2}=2h,\quad
\xi_{2,2}=t_3,\\
&\Delta_{A_1}=\Delta_{12},\quad
\Delta_{A_2}=\Delta_{34},\quad
s_{1,1}=s_{2,2}=S,\quad
s_{1,2}=0,
\end{split}
\end{equation}
where $S$ denotes the $S$-channel Mandelstam for the new diagram. So the kernel reads
\begin{equation}\label{eq:Kbubble4pt}\checked{}
K=\underbrace{\bfn{\frac{h+c_2+t_1-t_3}{2}}{\frac{2h+t_2-t_3}{2}}\bfn{\frac{h+c_2-t_1+t_3}{2}}{\frac{2h-t_2+t_3}{2}}}_{K_2}\,
\underbrace{\bfn{\frac{\Delta_{12}-t_2}{2}}{\frac{\Delta_{12}-S}{2}}
\bfn{\frac{\Delta_{34}-t_3}{2}}{\frac{\Delta_{34}-S}{2}}
\bfn{\frac{t_2+t_3-2h}{2}}{S}}_{K_1}.
\end{equation}

\temp{I haven't yet double-checked the derivation below.}
There is an even simpler case that we can study, which is the 3-point bubbble related to Figure \ref{fig:bubble4pt} by deleting the leg 4. For this diagram $t_3=\Delta_3$, and so there are two integrations instead. The corresponding kernel is then
\begin{equation}
\checked{}
K=\underbrace{
\bfn{\frac{h+c_2+t_1-\Delta_3}{2}}{\frac{2h+t_2-\Delta_3}{2}}\,
\bfn{\frac{h+c_2-t_1+\Delta_3}{2}}{\frac{2h-t_2+\Delta_3}{2}}
}_{K_2}\,
\underbrace{
\bfn{\frac{\Delta_{12}-t_2}{2}}{\frac{\Delta_{12}-\Delta_3}{2}}\,
\bfn{\frac{t_2+\Delta_3-2h}{2}}{\Delta_3}
}_{K_1},
\end{equation}
and the original amplitude $M_0$ is the same as \eqref{eq:M0bubble4pt} but with $\Delta_{34}$ replaced by $\Delta_3$. In this simple case it is relatively easy that we directly perform the remaining integrations, by first observing that the integrand contains the following combination of $\Gamma$ functions, which we can replace by an integral using Barnes' first lemma
\begin{equation}
\frac{\Gamma[\frac{h+c_1-t_1}{2}]\Gamma[\frac{h-c_2-t_1+t_2}{2}]\Gamma[\frac{\Delta_{12}-t_2}{2}]}{\Gamma[\frac{h-c_2-t_1+\Delta_{12}}{2}]}=\frac{1}{2}\int\frac{\mathrm{d}t_3}{2\pi i}\frac{\Gamma[\frac{h-c_2-t_1+t_2-t_3}{2}]\Gamma[\frac{\Delta_{12}-c_1-c_2-t_3}{2}]\Gamma[\frac{t_3}{2}]\Gamma[\frac{c_1+c_2-t_2+t_3}{2}]}{\Gamma[\frac{\Delta_{12}-c_1-c_2}{2}]}.
\end{equation}
Although we temporarily lift the expression up to a 3-fold integral, this helps eliminate the $t_1$ dependence in the denominator of the integrand. Then it turns out that we are able to apply Barnes' first lemma for the $t_1$ integral, followed by Barnes' second lemma for the $t_2$ integral, and in the end Barnes' first lemma again for the $t_3$ integral. This yields the final result
\begin{equation}\label{eq:bubble3ptresult}
M=2\,\frac{\Gamma[\frac{\Delta_3\pm c_1\pm c_2}{2}]\,\Gamma[\frac{2h\pm c_1\pm c_2-\Delta_3}{2}]\,\Gamma[\frac{\Delta_{123}}{2}-h]}{\Gamma[h]\,\Gamma[2h-\Delta_3]\,\Gamma[\Delta_3]}.
\end{equation}
This same 3-point diagram can also be computed by straightforward spacetime integrations and then translate into Mellin space. One can explicitly verify that the two results are identical.

\subsubsection{Triangle diagrams}

\begin{figure}[h]
\captionsetup{margin=2em}
\begin{center}
\begin{tikzpicture}
\begin{scope}[xshift=-4cm]
\node [anchor=center] at (0,-3) {(A)};
\draw [black,very thick,dotted] (0,0) circle [radius=1.5];
\draw [black,very thick] (-60:1.5) arc [start angle=-60,end angle=240,radius=1.5];
\draw [black,very thick] (180:1.5) -- (180:2.5);
\node [anchor=center] at (180:2.7) {$1$};
\draw [black,very thick] (90:1.5) -- (90:2.5);
\node [anchor=center] at (90:2.75) {$2$};
\draw [black,very thick] (0:1.5) -- (0:2.5);
\node [anchor=center] at (0:2.7) {$3$};
\node [anchor=center] at (240:1.9) {$c_2^-$};
\node [anchor=center] at (-60:1.9) {$c_2^+$};
\node [anchor=center] at (135:1.75) {$c_3$};
\node [anchor=center] at (45:1.75) {$c_1$};
\fill [black] (180:1.5) circle [radius=2.5pt];
\fill [black] (90:1.5) circle [radius=2.5pt];
\fill [black] (0:1.5) circle [radius=2.5pt];
\node [anchor=center] at (180:1.2) {\scriptsize $A_1$};
\node [anchor=center] at (90:1.2) {\scriptsize $A_2$};
\node [anchor=center] at (0:1.2) {\scriptsize $A_3$};
\draw [black,fill=white] (-60:1.5) circle [radius=2.5pt];
\draw [black,fill=white] (-120:1.5) circle [radius=2.5pt];
\draw [red,very thick,dashed] (150:2.5) -- (150:1.5) .. controls (150:.5) and (-100:.5) .. (-100:1.5) -- (-100:2);
\node [anchor=center] at ($(0,0)!1.15!(150:2.5)$) {\color{red}$t_3$};
\draw [red,very thick,dashed] (30:2.5) -- (30:1.5) .. controls (30:.5) and (-80:.5) .. (-80:1.5) -- (-80:2);
\node [anchor=center] at ($(0,0)!1.15!(30:2.5)$) {\color{red}$t_1$};
\draw [orange,very thick,dashed] (120:2.5) -- (120:1.5) .. controls (120:.5) and (-30:.5) .. (-30:1.5) -- (-30:2.5);
\node [anchor=center] at ($(0,0)!1.2!(-30:2.5)$) {\color{orange}$t_4|\Delta_1$};
\draw [orange,very thick,dashed] (60:2.5) -- (60:1.5) .. controls (60:.5) and (210:.5) .. (210:1.5) -- (210:2.5);
\node [anchor=center] at ($(0,0)!1.2!(210:2.5)$) {\color{orange}$t_2|\Delta_3$};
\end{scope}
\begin{scope}[xshift=4cm]
\node [anchor=center] at (0,-3) {(B)};
\draw [black,very thick,dotted] (0,0) circle [radius=1.5];
\draw [black,very thick] (-60:1.5) arc [start angle=-60,end angle=240,radius=1.5];
\draw [black,very thick] (180:1.5) -- (180:2.5);
\node [anchor=center] at (180:2.7) {$4$};
\coordinate (p1) at ($(90:1.5)+(120:1)$);
\node [anchor=center] at ($(0,0)!1.1!(p1)$) {$1$};
\coordinate (p2) at ($(90:1.5)+(60:1)$);
\node [anchor=center] at ($(0,0)!1.1!(p2)$) {$2$};
\draw [black,very thick] (p1) -- (90:1.5) -- (p2);
\draw [black,very thick] (0:1.5) -- (0:2.5);
\node [anchor=center] at (0:2.7) {$3$};
\node [anchor=center] at (240:1.9) {$c_3^-$};
\node [anchor=center] at (-60:1.9) {$c_3^+$};
\node [anchor=center] at (146:1.75) {$c_1$};
\node [anchor=center] at (34:1.75) {$c_2$};
\fill [black] (180:1.5) circle [radius=2.5pt];
\fill [black] (90:1.5) circle [radius=2.5pt];
\fill [black] (0:1.5) circle [radius=2.5pt];
\node [anchor=center] at (180:1.2) {\scriptsize $A_1$};
\node [anchor=center] at (90:1.2) {\scriptsize $A_2$};
\node [anchor=center] at (0:1.2) {\scriptsize $A_3$};
\draw [black,fill=white] (-60:1.5) circle [radius=2.5pt];
\draw [black,fill=white] (-120:1.5) circle [radius=2.5pt];
\draw [red,very thick,dashed] (157.5:2.5) -- (157.5:1.5) .. controls (157.5:.5) and (-100:.5) .. (-100:1.5) -- (-100:2);
\node [anchor=center] at ($(0,0)!1.15!(157.5:2.5)$) {\color{red}$t_1$};
\draw [red,very thick,dashed] (22.5:2.5) -- (22.5:1.5) .. controls (22.5:.5) and (-80:.5) .. (-80:1.5) -- (-80:2);
\node [anchor=center] at ($(0,0)!1.15!(22.5:2.5)$) {\color{red}$t_2$};
\draw [orange,very thick,dashed] (135:2.5) -- (135:1.5) .. controls (135:.5) and (-30:.5) .. (-30:1.5) -- (-30:2.5);
\node [anchor=center] at ($(0,0)!1.2!(-30:2.5)$) {\color{orange}$t_5|\Delta_4$};
\draw [orange,very thick,dashed] (45:2.5) -- (45:1.5) .. controls (45:.5) and (210:.5) .. (210:1.5) -- (210:2.5);
\node [anchor=center] at ($(0,0)!1.2!(210:2.5)$) {\color{orange}$t_3|\Delta_3$};
\draw [orange,very thick,dashed] (112.5:2.5) -- (112.5:1.5) .. controls (112.5:.5) and (67.5:.5) .. (67.5:1.5) -- (67.5:2.5);
\node [anchor=center] at ($(0,0)!1.15!(67.5:2.5)$) {\color{orange}$t_4|S$};
\end{scope}
\end{tikzpicture}
\end{center}
\caption{Constructing the triangle diagrams: (A) 3-point; (B) 4-point.}
\label{fig:triangle}
\end{figure}
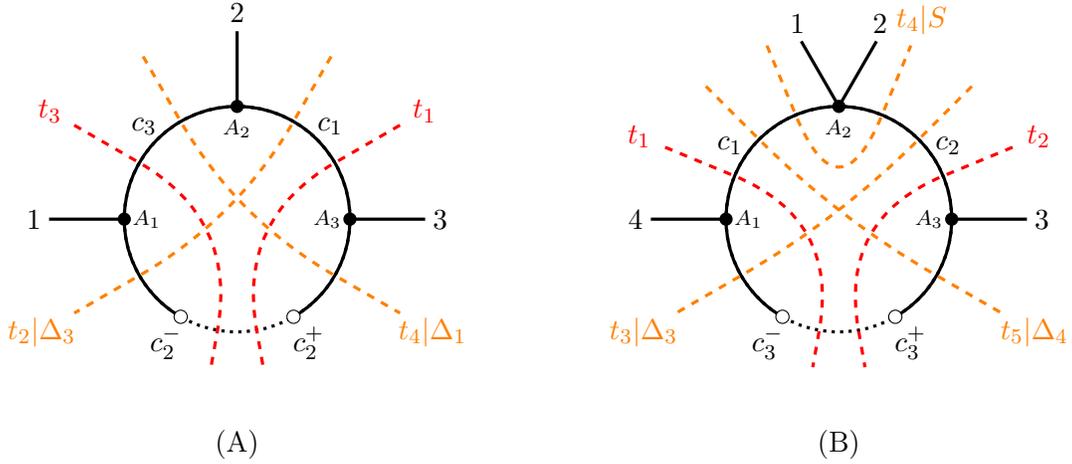

$r=3$ corresponds to triangle diagrams. We look at two cases, the 3-point diagram and a 4-point diagram, as shown in Figure \ref{fig:triangle} (A) (B) respectively.

For diagram (A), $A_1=\{1\}$, $A_2=\{2\}$ and $A_3=\{3\}$. The original pre-amplitude is
\begin{equation}\label{eq:M0triangle3pt}
\checked{}
\frac{M_0}{C_{A_2}}=\frac{\Gamma[\frac{\Delta_1-c_2\pm c_3}{2}]}{\Gamma[\frac{\Delta_1+h-c_2-t_3}{2}]}\,\Gamma[{\textstyle\frac{h\pm c_3-t_3}{2}}]\,\Gamma[{\textstyle\frac{\Delta_2\pm c_3\pm c_1}{2}}]\,\Gamma[{\textstyle\frac{h\pm c_1-t_1}{2}}]\,\frac{\Gamma[\frac{\Delta_3+c_2\pm c_1}{2}]}{\Gamma[\frac{\Delta_3+h+c_2-t_1}{2}]}.
\end{equation}
The substitution rules are
\begin{equation}
\begin{split}\checked{}
&\Xi_1=h-c_2,\quad
\Xi_2=t_3,\quad
\Xi_3=t_1,\quad
\Xi_4=h+c_2,\\
&\xi_{1,1}=\Delta_1,\quad
\xi_{1,2}=t_2,\quad
\xi_{1,3}=2h,\quad
\xi_{2,2}=\Delta_2,\quad
\xi_{2,3}=t_4,\quad
\xi_{3,3}=\Delta_3,\\
&s_{1,1}=s_{2,3}=\Delta_1,\quad
s_{2,2}=\Delta_2,\quad
s_{1,2}=s_{3,3}=\Delta_3,\quad
s_{1,3}=0.
\end{split}
\end{equation}
So we have
\begin{equation}\label{eq:Ktriangle3pt}
\begin{split}\checked{}
K=&\underbrace{
\bfn{\frac{h+c_2+t_3-t_4}{2}}{\frac{2h-t_4+\Delta_1}{2}}\,
\bfn{\frac{t_1-t_3+t_4-\Delta_3}{2}}{\frac{t_2+t_4-\Delta_1-\Delta_3}{2}}\,
\bfn{\frac{h+c_2-t_1+\Delta_3}{2}}{\frac{2h-t_2+\Delta_3}{2}}
}_{K_2}\\
&\times\underbrace{
\bfn{\frac{-t_2+\Delta_1+\Delta_2}{2}}{\frac{\Delta_1+\Delta_2-\Delta_3}{2}}\,
\bfn{\frac{-2h+t_2+t_4-\Delta_2}{2}}{\frac{\Delta_1-\Delta_2+\Delta_3}{2}}\,
\bfn{\frac{-t_4+\Delta_2+\Delta_3}{2}}{\frac{-\Delta_1+\Delta_2+\Delta_3}{2}}
}_{K_1}.
\end{split}
\end{equation}

For diagram (B), $A_1=\{4\}$, $A_2=\{1,2\}$ and $A_3=\{3\}$. The original pre-amplitude is
\begin{equation}\label{eq:M0triangle4ptquartic}
\checked{}
\frac{M_0}{C_{A_2}}=\frac{\Gamma[\frac{\Delta_4-c_3\pm c_1}{2}]}{\Gamma[\frac{\Delta_4+h-c_3-t_1}{2}]}\,\Gamma[{\textstyle\frac{h\pm c_1-t_1}{2}}]\,\Gamma[{\textstyle\frac{\Delta_{12}\pm c_1\pm c_2}{2}}]\,\Gamma[{\textstyle\frac{h\pm c_2-t_2}{2}}]\,\frac{\Gamma[\frac{\Delta_3+c_3\pm c_2}{2}]}{\Gamma[\frac{\Delta_3+h+c_2-t_2}{2}]}.
\end{equation}
The substitution rules are
\begin{equation}
\begin{split}\checked{}
&\Xi_1=h-c_3,\quad
\Xi_2=t_1,\quad
\Xi_3=t_2,\quad
\Xi_4=h+c_3,\\
&\xi_{1,1}=\Delta_4,\quad
\xi_{1,2}=t_3,\quad
\xi_{1,3}=2h,\quad
\xi_{2,2}=t_4,\quad
\xi_{2,3}=t_5,\quad
\xi_{3,3}=\Delta_3,\\
&s_{1,1}=s_{2,3}=\Delta_4,\quad
s_{1,2}=s_{3,3}=\Delta_3,\quad
s_{1,3}=0,\quad
s_{2,2}=S.
\end{split}
\end{equation}
So we have
\begin{equation}\label{eq:Ktriangle4ptquartic}
\begin{split}\checked{}
K=&\underbrace{
\bfn{\frac{h+c_3+t_1-t_5}{2}}{\frac{2h-t_5+\Delta_4}{2}}\,
\bfn{\frac{-t_1+t_2+t_5-\Delta_3}{2}}{\frac{t_3+t_5-\Delta_3-\Delta_4}{2}}\,
\bfn{\frac{h+c_3-t_2+\Delta_3}{2}}{\frac{2h-t_3+\Delta_3}{2}}
}_{K_2}\\
&\times\underbrace{
\bfn{\frac{\Delta_{12}-t_4}{2}}{\frac{\Delta_{12}-S}{2}}\,
\bfn{\frac{-t_3+t_4+\Delta_4}{2}}{\frac{S-\Delta_3+\Delta_4}{2}}\,
\bfn{\frac{-2h+t_3-t_4+t_5}{2}}{\frac{-S+\Delta_3+\Delta_4}{2}}\,
\bfn{\frac{t_4-t_5+\Delta_3}{2}}{\frac{S+\Delta_3-\Delta_4}{2}}
}_{K_1}.
\end{split}
\end{equation}

\subsubsection{Box diagrams}

\begin{figure}[h]
\captionsetup{margin=2em}
\begin{center}
\begin{tikzpicture}
\draw [black,very thick,dotted] (0,0) circle [radius=1.5];
\draw [black,very thick] (-66:1.5) arc [start angle=-66,end angle=246,radius=1.5];
\draw [black,very thick] (198:1.5) -- (198:2.5);
\node [anchor=center] at (198:2.7) {$1$};
\draw [black,very thick] (126:1.5) -- (126:2.5);
\node [anchor=center] at (126:2.75) {$2$};
\draw [black,very thick] (54:1.5) -- (54:2.5);
\node [anchor=center] at (54:2.7) {$3$};
\draw [black,very thick] (-18:1.5) -- (-18:2.5);
\node [anchor=center] at (-18:2.7) {$4$};
\node [anchor=center] at (246:1.9) {$c_1^-$};
\node [anchor=center] at (-66:1.9) {$c_1^+$};
\node [anchor=center] at (171:1.75) {$c_2$};
\node [anchor=center] at (98:1.75) {$c_3$};
\node [anchor=center] at (9:1.75) {$c_4$};
\fill [black] (198:1.5) circle [radius=2.5pt];
\fill [black] (126:1.5) circle [radius=2.5pt];
\fill [black] (54:1.5) circle [radius=2.5pt];
\fill [black] (-18:1.5) circle [radius=2.5pt];
\node [anchor=center] at (198:1.2) {\scriptsize $A_1$};
\node [anchor=center] at (126:1.2) {\scriptsize $A_2$};
\node [anchor=center] at (54:1.2) {\scriptsize $A_3$};
\node [anchor=center] at (-18:1.2) {\scriptsize $A_4$};
\draw [black,fill=white] (-66:1.5) circle [radius=2.5pt];
\draw [black,fill=white] (246:1.5) circle [radius=2.5pt];
\draw [red,very thick,dashed] (180:2.5) -- (180:1.5) .. controls (180:.5) and (258:.5) .. (258:1.5) -- (258:2);
\node [anchor=center] at ($(0,0)!1.15!(180:2.5)$) {\color{red}$t_2$};
\draw [red,very thick,dashed] (90:2) -- (270:2);
\node [anchor=center] at ($(0,0)!1.15!(90:2)$) {\color{red}$t_3$};
\draw [red,very thick,dashed] (0:2.5) -- (0:1.5) .. controls (0:.5) and (-78:.5) .. (-78:1.5) -- (-78:2);
\node [anchor=center] at ($(0,0)!1.15!(0:2.5)$) {\color{red}$t_4$};
\draw [orange,very thick,dashed] (214:2.5) -- (214:1.5) .. controls (214:.5) and (108:.5) .. (108:1.5) -- (108:2.5);
\node [anchor=center] at ($(0,0)!1.2!(214:2.5)$) {\color{orange}$t_1|S$};
\draw [orange,very thick,dashed] (230:2.5) -- (230:1.5) .. controls (230:.5) and (18:.5) .. (18:1.5) -- (18:2.5);
\node [anchor=center] at ($(0,0)!1.2!(18:2.5)$) {\color{orange}$t_5|\Delta_4$};
\draw [orange,very thick,dashed] (144:2.5) -- (144:1.5) .. controls (144:.5) and (36:.5) .. (36:1.5) -- (36:2.5);
\node [anchor=center] at ($(0,0)!1.15!(144:2.5)$) {\color{orange}$t_6|T$};
\draw [orange,very thick,dashed] (162:2.5) -- (162:1.5) .. controls (162:.5) and (-50:.5) .. (-50:1.5) -- (-50:2.5);
\node [anchor=center] at ($(0,0)!1.2!(162:2.5)$) {\color{orange}$t_7|\Delta_1$};
\draw [orange,very thick,dashed] (72:2.5) -- (72:1.5) .. controls (72:.5) and (-34:.5) .. (-34:1.5) -- (-34:2.5);
\node [anchor=center] at ($(0,0)!1.15!(-34:2.5)$) {\color{orange}$t_8|S$};
\end{tikzpicture}
\end{center}
\caption{Constructing the 4-point box.}
\label{fig:box4pt}
\end{figure}
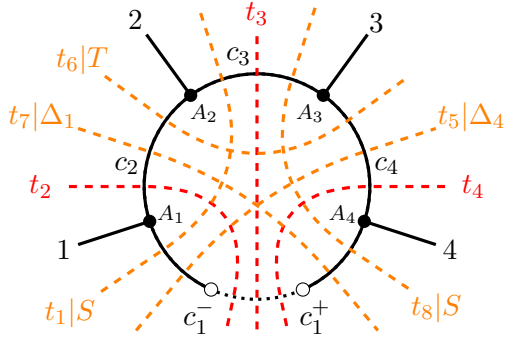

$r=4$ corresponds to box diagrams. We study a 4-point diagram in this case, as shown in Figure \ref{fig:box4pt}, where $A_1=\{1\}$, $A_2=\{2\}$, $A_3=\{3\}$ and $A_4=\{4\}$.
The original pre-amplitude is
\begin{equation}\label{eq:M0box4pt}
\begin{split}\checked{}
\frac{M_0}{C_{A_2}C_{A_3}}=&
\frac{\Gamma[\frac{\Delta_1-c_1\pm c_2}{2}]}{\Gamma[\frac{\Delta_1+h-c_1-t_2}{2}]}\,\Gamma[{\textstyle\frac{\Delta_2\pm c_2\pm c_3}{2}}]\,\Gamma[{\textstyle\frac{\Delta_3\pm c_3\pm c_4}{2}}]\,\frac{\Gamma[\frac{\Delta_4+c_1\pm c_3}{2}]}{\Gamma[\frac{\Delta_4+h+c_1-t_4}{2}]}\,\prod_{a=2}^4\Gamma[{\textstyle\frac{h\pm c_a-t_a}{2}}].
\end{split}
\end{equation}
The substitution rules are
\begin{equation}
\begin{split}\checked{}
&\Xi_1=h-c_1,\quad
\Xi_2=t_2,\quad
\Xi_3=t_3,\quad
\Xi_4=t_4,\quad
\Xi_5=h+c_1,\\
&\xi_{1,1}=s_{1,1}=s_{2,4}=\Delta_1,\quad
\xi_{1,2}=t_1,\quad
\xi_{1,3}=t_5,\quad
\xi_{1,4}=2h,\quad
\xi_{2,2}=s_{2,2}=\Delta_2,\\
&\xi_{2,3}=t_6,\quad
\xi_{2,4}=t_7,\quad
\xi_{3,3}=s_{3,3}=\Delta_3,\quad
\xi_{3,4}=t_8,\quad
\xi_{4,4}=s_{1,3}=s_{4,4}=\Delta_4,\\
&s_{1,2}=s_{3,4}=S,\quad
s_{1,4}=0,\quad
s_{2,3}=T.
\end{split}
\end{equation}
So we have
\begin{equation}\label{eq:Kbox4pt}
\begin{split}\checked{}
K=&\underbrace{
\bfn{\frac{h+c_1+t_2-t_7}{2}}{\frac{2h-t_7+\Delta_1}{2}}\,
\bfn{\frac{-t_2+t_3+t_7-t_8}{2}}{\frac{t_1+t_7-t_8-\Delta_1}{2}}\,
\bfn{\frac{-t_3+t_4+t_8-\Delta_4}{2}}{\frac{-t_1+t_5+t_8-\Delta_4}{2}}\,
\bfn{\frac{h+c_1-t_4+\Delta_4}{2}}{\frac{2h-t_5+\Delta_4}{2}}
}_{K_2}\\
&\times\underbrace{
\bfn{\frac{\Delta_{12}-t_1}{2}}{\frac{\Delta_{12}-S}{2}}\,
\bfn{\frac{t_1-t_5+t_6-\Delta_2}{2}}{\frac{S+T-\Delta_{24}}{2}}\,
\bfn{\frac{-2h+t_5-t_6+t_7}{2}}{\frac{-T+\Delta_{14}}{2}}\,
\bfn{\frac{\Delta_{23}-t_6}{2}}{\frac{\Delta_{23}-T}{2}}\,
\bfn{\frac{t_6-t_7+t_8-\Delta_3}{2}}{\frac{S+T-\Delta_{13}}{2}}\,
\bfn{\frac{\Delta_{34}-t_8}{2}}{\frac{\Delta_{34}-S}{2}}
}_{K_1}.
\end{split}
\end{equation}

\subsection{General one-loop diagrams}\label{sec:generaloneloop}

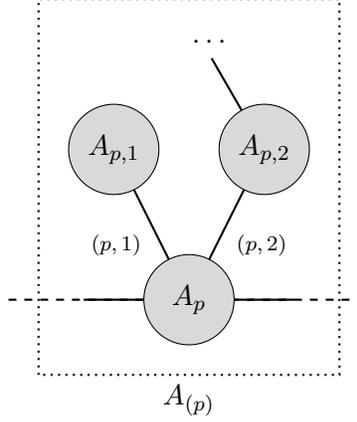
\begin{figure}[ht]
\captionsetup{margin=2em}
\begin{center}
\begin{tikzpicture}
\draw [black,thick,dashed] (-2.4,0) -- (2.4,0);
\draw [black,thick] (-1.4,0) -- (1.4,0);
\draw [black,thick] (-1,2) -- (0,0) -- (1,2) -- +(120:1.4);
\draw [black,fill=black!15!white] (0,0) circle [radius=.6];
\draw [black,fill=black!15!white] (-1,2) circle [radius=.6];
\draw [black,fill=black!15!white] (1,2) circle [radius=.6];
\node [anchor=center] at (0,0) {$A_p$};
\node [anchor=center] at (-1,2) {$A_{p,1}$};
\node [anchor=center] at (1,2) {$A_{p,2}$};
\node [anchor=south] at ($(1,2)+(120:1.4)$) {$\cdots$};
\node [anchor=north east] at (-.5,1) {\scriptsize $(p,1)$};
\node [anchor=north west] at (.5,1) {\scriptsize $(p,2)$};
\draw [black,thick,dotted] (-2,-1) rectangle (2,4);
\node [anchor=north] at (0,-1) {$A_{(p)}$};
\end{tikzpicture}
\end{center}
\caption{Trees upon one loop.}
\label{fig:treesonloop}
\end{figure}

Given the detailed knowledge about the construction of the necklace diagrams, the kernel for generic one-loop diagrams only requires a slight modification. It suffices to focus on one specific vertex on the loop, say $A_p$, where we have additional tree-level structures attached to it, as exemplified in Figure \ref{fig:treesonloop}.
We denote the tree propagators in the same way as in the discussion of generic trees in Section \ref{sec:kernelgenerictrees}, and correspondingly we have the identification $\Xi_{(p)}\equiv\xi_p\equiv\xi_{p,p}$.

First of all, as is obvious in \eqref{eq:K2oneloopreduced}, since $K_2$ only depends on Mandelstam variables essentially related to the loop, it does not receive any modification at all. So we just need to focus on $K_1$.

Following the same logic as in necklace diagrams, ideally here we would like to treat $A_p$ together with all its tree branches as a generalized vertex $A_{(p)}$ on the loop and directly apply the formula \eqref{eq:K1oneloopreduced} we had for the necklace diagrams. While this applies to the crossing terms (i.e., for $\tau_{i\,j}$, $i\in A_{(p)}$ and $j\in A_{(q)}$ with $p\neq q$), terms that fall into the same $A_{(p)}$
\begin{equation}
\checked{}
\prod_{i,j\in A_{(p)}}\frac{\Gamma[\tau_{i\,j}]\Gamma[\delta_{i\,j}-\tau_{i\,j}]}{\Gamma[\delta_{i\,j}]}
\end{equation}
needs extra care. However, this structure is no different from that for the original diagram in the tree-level vertex insertion, thus allowing us to directly borrow the formula \eqref{eq:treekernelgeneral} (excluding the factor $K_2$ therein). It is also easy to see that in the situation when there is no extra tree structure the corresponding contribution reduces to that appearing in the formula for necklace diagrams.

As a result, the $K_1$ part of the kernel for a generic one-loop diagram can be simplified to
\begin{equation}
\begin{split}\label{eq:K1oneloopgeneric}
\checked{}
K_1\longrightarrow&
\prod_{1\leq a<b\leq r}\!\!\bfn{\frac{\xi_{a,b-1}+\xi_{a+1,b}-\xi_{a+1,b-1}-\xi_{a,b}}{2}}{\frac{s_{a,b-1}+s_{a+1,b}-s_{a+1,b-1}-s_{a,b}}{2}}\,
\prod_{p=1}^r\prod_{a\in\{A_{(p)}\text{ vertices}\}}\!\!\!\bfn{\frac{\Delta_{A_a}-\Xi_{(a)}+\sum_m \Xi_{(a,m)}}{2}}{\frac{\Delta_{A_a}-s_{(a)}+\sum_{m}s_{(a,m)}}{2}}.
\end{split}
\end{equation}
In the last product (for every $p$) we scan over all vertices inside the gerenalize vertex $A_{(p)}$, i.e., all vertices in the dotted box in Figure \ref{fig:treesonloop}, including the one on the loop. However, when the vertex $A_p$ itself is trivalent (apart from the propagators in the loop it contains only one other propagator, either bulk-to-bulk or bulk-to-boundary), the contribution from vertex $A_p$ is absent from this product.

\subsubsection{4-point triangle with only cubic vertices}

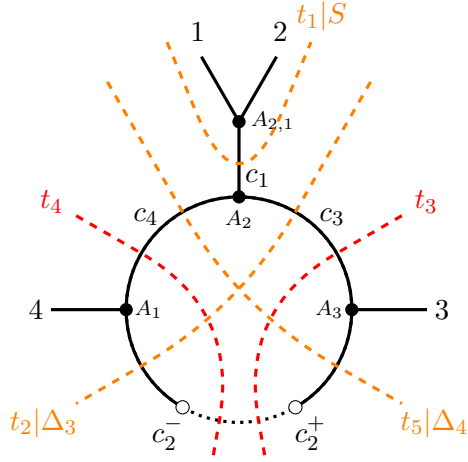
\begin{figure}[h]
\captionsetup{margin=2em}
\begin{center}
\begin{tikzpicture}
\draw [black,very thick,dotted] (0,0) circle [radius=1.5];
\draw [black,very thick] (-60:1.5) arc [start angle=-60,end angle=240,radius=1.5];
\draw [black,very thick] (180:1.5) -- (180:2.5);
\node [anchor=center] at (180:2.7) {$4$};
\draw [black,very thick] (90:1.5) -- (90:2.5);
\coordinate (p1) at ($(90:2.5)+(120:1)$);
\node [anchor=center] at ($(0,0)!1.1!(p1)$) {$1$};
\coordinate (p2) at ($(90:2.5)+(60:1)$);
\node [anchor=center] at ($(0,0)!1.1!(p2)$) {$2$};
\draw [black,very thick] (p1) -- (90:2.5) -- (p2);
\draw [black,very thick] (0:1.5) -- (0:2.5);
\node [anchor=center] at (0:2.7) {$3$};
\node [anchor=center] at (240:1.9) {$c_2^-$};
\node [anchor=center] at (-60:1.9) {$c_2^+$};
\node [anchor=center] at (135:1.75) {$c_4$};
\node [anchor=center] at (45:1.75) {$c_3$};
\node [anchor=center] at (.25,1.75) {$c_1$};
\fill [black] (180:1.5) circle [radius=2.5pt];
\fill [black] (90:1.5) circle [radius=2.5pt];
\fill [black] (90:2.5) circle [radius=2.5pt];
\fill [black] (0:1.5) circle [radius=2.5pt];
\node [anchor=center] at (180:1.2) {\scriptsize $A_1$};
\node [anchor=center] at (90:1.2) {\scriptsize $A_2$};
\node [anchor=center] at (.45,2.5) {\scriptsize $A_{2,1}$};
\node [anchor=center] at (0:1.2) {\scriptsize $A_3$};
\draw [black,fill=white] (-60:1.5) circle [radius=2.5pt];
\draw [black,fill=white] (-120:1.5) circle [radius=2.5pt];
\draw [red,very thick,dashed] (150:2.5) -- (150:1.5) .. controls (150:.5) and (-100:.5) .. (-100:1.5) -- (-100:2);
\node [anchor=center] at ($(0,0)!1.15!(150:2.5)$) {\color{red}$t_4$};
\draw [red,very thick,dashed] (30:2.5) -- (30:1.5) .. controls (30:.5) and (-80:.5) .. (-80:1.5) -- (-80:2);
\node [anchor=center] at ($(0,0)!1.15!(30:2.5)$) {\color{red}$t_3$};
\draw [orange,very thick,dashed] (120:3.5) -- (120:1.5) .. controls (120:.5) and (-30:.5) .. (-30:1.5) -- (-30:2.5);
\node [anchor=center] at ($(0,0)!1.2!(-30:2.5)$) {\color{orange}$t_5|\Delta_4$};
\draw [orange,very thick,dashed] (60:3.5) -- (60:1.5) .. controls (60:.5) and (210:.5) .. (210:1.5) -- (210:2.5);
\node [anchor=center] at ($(0,0)!1.2!(210:2.5)$) {\color{orange}$t_2|\Delta_3$};
\begin{scope}[yshift=1.25cm]
\draw [orange,very thick,dashed] (112.5:2.5) -- (112.5:1.5) .. controls (112.5:.5) and (67.5:.5) .. (67.5:1.5) -- (67.5:2.5);
\node [anchor=center] at ($(0,0)!1.15!(67.5:2.5)$) {\color{orange}$t_1|S$};
\end{scope}
\end{tikzpicture}
\end{center}
\caption{Constructing the 4-point triangle with only cubic vertices.}
\label{fig:triangle4ptC}
\end{figure}

Let us illustrate the above discussion with a 4-point triangle diagram with cubic vertices, as constructed in Figure \ref{fig:triangle4ptC}, which is only slightly different from Figure \ref{fig:triangle} (B).

According to the diagram we have $A_1=\{4\}$, $A_2=\varnothing$, $A_{2,1}=\{1,2\}$ and $A_3=\{3\}$. The original amplitude reads
\begin{equation}\label{eq:M0triangle4ptcubic}
\checked{}
\frac{M_0}{C_{A_2}}=\frac{\Gamma[\frac{\Delta_4-c_2\pm c_4}{2}]}{\Gamma[\frac{\Delta_4+h-c_2-t_4}{2}]}\,\frac{\Gamma[\frac{\Delta_3+c_2\pm c_3}{2}]}{\Gamma[\frac{\Delta_3+h+c_2-t_3}{2}]}\,\frac{\Gamma[\frac{\Delta_{12}-h\pm c_1}{2}]}{\Gamma[\frac{\Delta_{12}-t_1}{2}]}\,\Gamma[{\textstyle\frac{h\pm c_1\pm c_3\pm c_4}{2}}]\,\prod_{a=\{1,3,4\}}\!\!\Gamma[{\textstyle\frac{h\pm c_a-t_a}{2}}].
\end{equation}
The substitution rules are
\begin{equation}
\begin{split}
\checked{}
&\Xi_1=h-c_2,\quad
\Xi_2=t_4,\quad
\Xi_3=t_3,\quad
\Xi_4=h+c_2,\quad
\Xi_{(2,1)}=\xi_{2,2}=t_1,\\
&\xi_{1,1}=\Delta_4,\quad
\xi_{1,2}=t_2,\quad
\xi_{1,3}=2h,\quad
\xi_{2,3}=t_5,\quad
\xi_{3,3}=\Delta_3,\\
&s_{1,1}=s_{2,3}=\Delta_4,\quad
s_{1,2}=s_{3,3}=\Delta_3,\quad
s_{2,2}=s_{(2,1)}=S,\quad
s_{1,3}=0.
\end{split}
\end{equation}
So we have
\begin{equation}\label{eq:Ktriangle4ptcubic}
\begin{split}\checked{}
K=&\underbrace{
\bfn{\frac{h+c_2+t_4-t_5}{2}}{\frac{2h-t_5+\Delta_4}{2}}\,
\bfn{\frac{t_3-t_4+t_5-\Delta_3}{2}}{\frac{t_2+t_5-\Delta_3-\Delta_4}{2}}\,
\bfn{\frac{h+c_2-t_3+\Delta_3}{2}}{\frac{2h-t_2+\Delta_3}{2}}
}_{K_2}\\
&\times\underbrace{
\bfn{\frac{t_1-t_2+\Delta_4}{2}}{\frac{S-\Delta_3+\Delta_4}{2}}\,
\bfn{\frac{-2h-t_1+t_2+t_5}{2}}{\frac{-S+\Delta_3+\Delta_4}{2}}\,
\bfn{\frac{t_1-t_5+\Delta_3}{2}}{\frac{S+\Delta_3-\Delta_4}{2}}\,
\bfn{\frac{\Delta_{12}-t_1}{2}}{\frac{\Delta_{12}-S}{2}}
}_{K_1}.
\end{split}
\end{equation}
Although we use different labeling for the vertices and propagators here, attentive readers may have perceived that the integration kernel here shares the same structure as that of the 4-point triangle with a quartic vertex in Figure \ref{fig:triangle} (B).

\newpage

\section{Structure of Poles}\label{sec:polestructure}

\temp{Formulas are double-checked. No major revisions further needed except for the marked places.}

In the previous section we worked out kernels that builds up one-loop Mellin pre-amplitudes from those at tree level. Now we use these representations to examine analytic properties of pre-amplitudes for one-loop diagrams, prior to the spectrum integrations. Since these are in general delicate and unknown functions, the main aim is to identify its singularities as well as to directly analyze local properties around each singularity, instead of seeking for expansion of $M$ onto more familiar functions. 

This particular section is devoted to the analysis on the pole structure, for which we summarize a set of rules that we conjecture to hold to all loop levels.

Singularity of an integral can always be understood as arising from the inability of the integration contour to deform off the site of pinching by singularities of the integrand from two sides (see, e.g., Section 2.1 of \cite{Eden:1966dnq}). The construction prescribed in this paper always yields representations of pre-amplitudes in terms of (a special class of) the standard Mellin integrals. Especially, the singularities of the integrand always come as poles in the $\Gamma$ functions therein, by which one can quickly conclude that the singularities of the integral are always poles and that they always group into similar semi-infinite families (see Appendix \ref{app:sec:poleemergence} for details). In special cases they can again coincide with poles of some new $\Gamma$ function, e.g.,
\begin{equation}\label{eq:emergentpolessimpleeg}
\checked{}
\Gamma[a+z]\Gamma[b-z]\xrightarrow{\int\mathrm{d}z}\Gamma[a+b],\qquad
\Gamma[a+z]\Gamma[b-2z]\xrightarrow{\int\mathrm{d}z}\Gamma[2a+b],
\end{equation}
and so on. In principle, iterating this contour analysis provides a way to detect the poles of arbitrary multi-dimensional Mellin integrals. However, in practice the poles worked out with this na\"ive method are mostly fake, and so an improved method is in need for our purpose.

The discussions in this section and the next one are fully based on several techniques specialized to multi-dimensional Mellin integrals, on which we provide a careful account in Appendix \ref{app:sec:mellinintegrals}.  But in the following we begin by summarizing a resulting algorithm (Algorithm \ref{app:sec:polealgoritm}) without full explanation, which is what we mainly utilize in the rest of this section. Readers who are interested in understanding the technical details are highly suggested to read that appendix.

\subsection{An algorithm for the detection of poles}\label{sec:polealgorithm}

In analyzing the poles that emerge from integrations, it is important to keep track of their origin in the integrand. For this purpose it is useful to start by labeling all those $\Gamma$ factors in the integrand that depend on the integration variables. With this notation we are able to label each family of emergent poles by the collection of labels in the original $\Gamma$'s that are responsible for them. For instance \eqref{eq:emergentpolessimpleeg} can be re-written into
\begin{equation}\label{eq:emergentpolessimpleeglabeled}
\checked{}
\Gamma_1[a+z]\Gamma_2[b-z]\xrightarrow{\int\mathrm{d}z}\Gamma_{\{1,2\}}[a+b],\qquad
\Gamma_1[a+z]\Gamma_2[b-2z]\xrightarrow{\int\mathrm{d}z}\Gamma_{\{1,2\}}[2a+b],
\end{equation}
meaning that poles in $\Gamma_{\{1,2\}}$ arise in the $z$ integral as its contour is pinched by poles in $\Gamma_1$ and those in $\Gamma_2$. More generally we may not hope that each family of poles be evenly-spaced, e.g., those from $\int\frac{\mathrm{d}z}{2\pi i}\Gamma_1[a+2z]\Gamma_2[b-3z]$, which produces poles at $\frac{a}{2}+\frac{b}{3}+\frac{m_1}{2}+\frac{m_2}{3}=0$ for every $m_1,m_2\in\mathbb{N}$,
but we can still abstractly label the whole family by $\gamma_{\{1,2\}}$. Here we use $\gamma$ to denote a generic family regardless of whether the poles are evenly spaced or not (even when, say, the family contains only one pole, as we will encounter later on in the spectrum integrals).

In fact, at the preliminary stage we can ignore the detailed contents of the poles but just inquire about their existence. Correspondingly, for each factor in the original integrand generically of the form $\Gamma_i[a_0+\sum_{j=1}^ra_jz_j]$ (where all $z$'s are integration variables), we can denote its corresponding family of poles as $\gamma_i[\sum_{j=1}^ra_jz_j]$, where we set $a_0\mapsto0$ and understand the argument as being equivalent up to an arbitrary \emph{positive} overall rescaling
\begin{equation}
\checked{}
\gamma_i[{\textstyle\sum_{j=1}^ra_jz_j}]\sim\gamma_i[{\textstyle c\sum_{j=1}^ra_jz_j}],\quad c>0.
\end{equation}
Hence, e.g., both cases in \eqref{eq:emergentpolessimpleeglabeled} are encoded in the same expression $\gamma_1[z]\gamma_2[-z]\xrightarrow{\int\mathrm{d}z}\gamma_{\{1,2\}}[0]$. Most generally, in every step of integration, say $z_1$, for every pair of $\gamma$'s where this variables appear with coefficients of opposite signs, there should be a new $\gamma$ emerging from them, whose argument is obtained by eliminating $z_1$, i.e.,
\begin{equation}\label{eq:gammaemergence}
\checked{}
\gamma_{A}\!\left[\sum_{i=1}^ra_iz_i\right]\,\gamma_{B}\!\left[\sum_{i=1}^rb_iz_i\right]\xrightarrow{\int\mathrm{d}z_1}\gamma_{A\cup B}\!\left[\frac{1}{|a_1|}\sum_{i=2}^ra_iz_i+\frac{1}{|b_1|}\sum_{i=2}^rb_iz_i\right],
\end{equation}
where we assume $a_1$ and $b_1$ have opposite signs. Here $A$ and $B$ each can be some set of original labels and we label the emergent $\gamma$ by their \emph{union}. It is then apparent that after all the integrations are performed the argument in each $\gamma$ family is zero; these are potentially the poles of the whole integral. With this tool, we have the following algorithm in identifying all (families of) poles for a generic multi-dimensional Mellin integral.

\textbf{Algorithm \ref{app:sec:polealgoritm}:}
\begin{enumerate}[noitemsep,nolistsep]
\item[(a).] Turn every $\Gamma$ function in the (numerator of the) original integrand into its corresponding $\gamma$ notation and label them. (We temporarily put aside the remaining part of the integrand.)
\item[(b).] Pick up an arbitrary integration variable $z$, and replace the $\gamma$'s containing $z$ by all $\gamma$'s that can arise from the $z$ integration following \eqref{eq:gammaemergence}.
\item[(c).] In the resulting set of $\gamma$'s, collect the subscripts of those in the form $\gamma[0]$.
\item[(d).] Inspect each $\gamma$: if its corresponding subscript contains any subscript collected in the previous step as a proper subset, we delete this $\gamma$ (these are the so-called ``composite poles'', following Proposition \ref{app:prop:disconnectedpoles} in Appendix \ref{app:sec:poleglobalview}).
\item[(e).] Collecting the remaining $\gamma$'s. If there are any other integration variables, pick up an arbitrary one, and repeat the analysis from (b) to (d). After all the integration variables are analyzed, the resulting $\gamma$'s are all in the form $\gamma_A[0]$ for some set of labels $A$. This terminates the algorithm.
\end{enumerate}
The notion of the composite poles in step (d) above is the essential tool that helps get rid of the fake poles that arise in a na\"ive contour analysis. This algorithm returns a list of $\gamma$'s recording the families of poles that may exist (apart from those that directly descend from the integrand). 

For simplicity let us assume that the $\Gamma$ functions in the original integrand are all distinct, so that all poles are simple poles. Then generically poles in every $\gamma$ family are simple as well. For a family $\gamma_A[0]$, when we hit the pole in it the contour is pinched by $|A|$ sigularity hyperplanes $E_i+m_i=0$ (for some $m_i\in\mathbb{N}$), from the $\Gamma$ functions $\Gamma_i[E_i]$ ($i\in A$). Without loss of generality let us denote this set as $A=\{0,1,\ldots,r\}$ for some $r$. Then $r$ of the integrations are essentially pinched; let us denote the corresponding variables as $\{z_1,\ldots,z_r\}$. In this case we can solve these variables by the equations $E_i+m_i=0$ ($i=1,\ldots,r$), and then further evaluate $E_0+m_0=0$ on the unique solution, yielding an expression
\begin{equation}\label{eq:gammapolelocations}
\checked{}
E+\sum_{i=0}^rp_im_i=0,\quad m_0,\ldots,m_r\in\mathbb{N},
\end{equation}
for some $E$. Enumerating all possible $m$'s generates the locations for all poles in the family $\gamma_A[0]$. Assuming we have already performed an overall rescaling such that $\min(p_0,\ldots,p_r)=1$, if further all $p$'s are integers, then this whole family is identical to the poles in $\Gamma[E]$.

For each $\gamma_A$ family, we call the codim-$r$ hyperplane defined by the solution for the integration variables worked above as its corresponding \emph{pinching plane} $P_A$ \footnote{Hopefully this will not be confused with the notation for the boundary points of a Witten diagram.}, which is where the contour has to be constrained. It may happen that poles in different $\gamma$ families coincide, e.g., $\Gamma_A[E]\Gamma_B[E]$. In this case we should further inspect their pinching planes (it sufficies to focus on the leading pole $m_0=\cdots=m_r=0$). (1) If $P_A\cap P_B=\varnothing$, this just means that the same pole receives distinct contributions (and so when computing residues we need to sum them up) but the pole still remains simple. (2) If $P_A\cap P_B\neq\varnothing$, this is then an indication that the pole under study is at higher order, because the contour has to be pinched by even more singularities of the integrand. Details of these are discussed in Appendix \ref{app:sec:higherpoles}.

Similarly we also need to worry about possible zeros arising from the rest of the integrand as the integration contour is pinched by some of the singularities. This may potentially kill some of the $\gamma$ families from our algorithm. This will be further explained as we work with specific examples later on.


\subsection{A convenient strategy specific to the Witten diagrams}\label{sec:strategy}

As mentioned previously, the representation we have for Mellin (pre-)amplitudes are purely in terms of multi-dimensional Mellin integrals. Hence in principle for any specific diagram under study we should be able to analyze all the integrals together using the method described in the previous subsection.  For our applications to Witten diagrams, however, it is in practice more convenient to divide the analysis into several stages by separating the integrals into groups. On the one hand, this helps save unnecessary labors in the detection of poles, because it allows us to get rid of poles that are absent due to extra zeros from the integrand, so that they do not further affect later integrals. On the other hand, some of the intermediate results may either be of interest to us as well or possess elegant structures on their own.

To be specific, there are three types of separation of integrals that we carry out throughout this paper:
\begin{enumerate}[noitemsep,nolistsep]
\item As we have insisted on from the beginning of the paper, we postpone the spectrum integrations for all the bulk-to-bulk propagators to the end, and first focus on the Mellin pre-amplitudes $M$ instead of directly analyzing the Mellin amplitudes $\mathcal{M}$. It turns out that $M$ itself is an elegant object such that its entire pole structure follows simple diagrammatic rules, as well be presented in Section \ref{sec:polediagrammaticrules} to \ref{sec:enhancedrules}. Furthermore, while the analysis on the integrals encountered in $M$ is relatively simple, the $c$ integrals that bridge $M$ and $\mathcal{M}$ are in fact involved and call extra care. This latter point will be clarified as we move on to the next section.
\item We recruit knowledge on lower-loop diagrams in the construction of diagrams at higher loop. In particular, in analyzing the pole structure of a higher-loop diagram, we are going to directly work with the the assumed pole structure of the corresponding lower-loop diagram in its construction, instead of working with some integral representation for the lower-loop diagram. This will be illustrated further in Section \ref{sec:arbitraryloops}.
\item Even in the study of one-loop diagrams we directly apply our complete knowledge on the poles of pre-amplitudes of tree-diagrams. In other words, we intentionally divide the contribution from each bulk-vertex into the product of $\Gamma$ functions encoding the poles and an extra correction function which is free of poles (except for the vertices of valency 1, whose correction is as simple as a $1/\Gamma$). The vertex corrections will be relevant only when we actually confirm the absence/existence of a few certain $\gamma$ families at the end, or in the computation of residues later on.
\end{enumerate}

In the next several subsections let us apply the discussions so far to explicit examples to illustrate the method.

\subsection{Bubble diagrams}

A nice representation for the pre-amplitude of bubble diagrams has already been worked out in \cite{Penedones:2010ue}, in terms of just a single Mellin integral that mimics the spectrum integral associated to some single tree exchange merged from the original pair of bulk-to-bulk propagators, which is also reviewed in Appendix \ref{app:sec:bbpropagators}. The representation we obtained in \eqref{eq:M0bubble4pt} and \eqref{eq:Kbubble4pt} from our recursive construction thus does not have any particular advantage. Nevertheless, the construction described here can be more naturally generalized to generic Witten diagrams, and we use this simple example to illustrate the algorithm stated above in full detail, as well as to serve as some consistency check.

Taking the diagram constructed in Figure \ref{fig:bubble4pt} as an example, the integrand is simply the product $M_0\,K$. First of all we extract the $\Gamma$ functions in the numerator that depends on the $t$ variables and label them
\begin{equation}\label{eq:bubblelabeledintegrand}
\begin{split}\checked{}
&\Gamma_1[{\textstyle\frac{h-c_1-t_1}{2}}]\,
\Gamma_2[{\textstyle\frac{h+c_1-t_1}{2}}]\,
\Gamma_3[{\textstyle\frac{h-c_2+t_1-t_2}{2}}]\,
\Gamma_4[{\textstyle\frac{t_2-S}{2}}]\,
\Gamma_5[{\textstyle\frac{h-c_2-t_1+t_2}{2}}]\,
\Gamma_6[{\textstyle\frac{h+c_2+t_1-t_3}{2}}]\\
&
\Gamma_7[{\textstyle\frac{2h+2S-t_2-t_3}{2}}]\,
\Gamma_8[{\textstyle\frac{t_3-S}{2}}]\,
\Gamma_9[{\textstyle\frac{h+c_2-t_1+t_3}{2}}]\,
\Gamma_{10}[{\textstyle\frac{-2h+t_2+t_3}{2}}]\,
\Gamma_{11}[{\textstyle\frac{\Delta_{12}-t_2}{2}}]\,
\Gamma_{12}[{\textstyle\frac{\Delta_{34}-t_3}{2}}].
\end{split}
\end{equation}
Following our abstract $\gamma$ notation their corresponding poles are
\begin{equation}\label{eq:bubblelabeledintegrandpoles}
\begin{split}\checked{}
&\gamma_1[{\textstyle\frac{-t_1}{2}}]\,
\gamma_2[{\textstyle\frac{-t_1}{2}}]\,
\gamma_3[{\textstyle\frac{t_1-t_2}{2}}]\,
\gamma_4[{\textstyle\frac{t_2}{2}}]\,
\gamma_5[{\textstyle\frac{-t_1+t_2}{2}}]\,
\gamma_6[{\textstyle\frac{t_1-t_3}{2}}]\,
\gamma_7[{\textstyle\frac{-t_2-t_3}{2}}]\,
\gamma_8[{\textstyle\frac{t_3}{2}}]\\
&
\gamma_9[{\textstyle\frac{-t_1+t_3}{2}}]\,
\gamma_{10}[{\textstyle\frac{t_2+t_3}{2}}]\,
\gamma_{11}[{\textstyle\frac{-t_2}{2}}]\,
\gamma_{12}[{\textstyle\frac{-t_3}{2}}].
\end{split}
\end{equation}
Apart from these, there are also the remaining integrand
\begin{equation}\label{eq:bubble4ptirrelevantfactors}
\checked{}
\frac{\Gamma[{\textstyle\frac{\Delta_{12}\pm c_1-c_2}{2}}]\,
\Gamma[{\textstyle\frac{\Delta_{34}\pm c_1+c_2}{2}}]}{\Gamma[S]\,
\Gamma[{\textstyle\frac{2h+t_2-t_3}{2}}]\,
\Gamma[{\textstyle\frac{2h-t_2+t_3}{2}}]\,
\Gamma[{\textstyle\frac{\Delta_{12}-S}{2}}]\,\Gamma[{\textstyle\frac{\Delta_{12}+h-c_2-t_1}{2}}]\,
\Gamma[{\textstyle\frac{\Delta_{34}-S}{2}}]\,
\Gamma[{\textstyle\frac{\Delta_{34}+h+c_2-t_1}{2}}]}.
\end{equation}

To perform the three $t$ integrations we choose a specific sequence, say $\{t_1,t_2,t_3\}$.

For the $t_1$ integration, the relevant factors are $\gamma_3\gamma_6$ (left) and $\gamma_1\gamma_2\gamma_5\gamma_9$ (right). The na\"ive contour analysis suggests that in the result the poles are encoded in
\begin{equation}\label{eq:bubble4ptaftert1}
\checked{}
\gamma_4\,\gamma_7\,\gamma_8\,\gamma_{10}\,\gamma_{11}\,\gamma_{12}
\times\gamma_{\{1,3\}}\,\gamma_{\{1,6\}}\,\gamma_{\{2,3\}}\,\gamma_{\{2,6\}}\,\underline{\gamma_{\{3,5\}}}\,\gamma_{\{3,9\}}\,\gamma_{\{5,6\}}\,\underline{\gamma_{\{6,9\}}}.
\end{equation}
Here we underlined $\gamma$'s whose arguments are already zero. Then look for $\gamma$'s in the above result whose subscript properly contains either $\{3,5\}$, $\{6,9\}$. Obviously after the first integrand there should be none, and so we take this expression and directly move to the next integration.

For the $t_2$ integration, the na\"ive contour analysis amounts to the replacement
\begin{equation}
\begin{split}\checked{}
&\gamma_4\,\gamma_7\,\gamma_{10}\,\gamma_{11}\,\gamma_{\{1,3\}}\,\gamma_{\{2,3\}}\,\gamma_{\{3,9\}}\,\gamma_{\{5,6\}}\longrightarrow\\
&\gamma_{\{4,7\}}\,\underline{\gamma_{\{4,11\}}}\,\underline{\gamma_{\{7,10\}}}\,\gamma_{\{10,11\}}\,\underline{\gamma_{\{1,3,4\}}}\,\gamma_{\{1,3,10\}}\,\underline{\gamma_{\{2,3,4\}}}\,\gamma_{\{2,3,10\}}\,\gamma_{\{3,4,9\}}\,\gamma_{\{3,9,10\}}\\
&\times\gamma_{\{5,6,7\}}\,\gamma_{\{5,6,11\}}\,{\color{red}\cancel{\gamma}_{\{1,3,5,6\}}\,\cancel{\gamma}_{\{2,3,5,6\}}\,\underline{\cancel{\gamma}_{\{3,5,6,9\}}}}.
\end{split}
\end{equation}
This yields five additional $\gamma[0]$'s, as indicated by underlines again. Each of the last three families is composite, as its subscript properly contains $\{3,5\}$, and so is ruled out. Combining with \eqref{eq:bubble4ptaftert1} we thus result in
\begin{equation}
\begin{split}\checked{}
&\gamma_8\,\gamma_{12}
\times\gamma_{\{1,6\}}\,\gamma_{\{2,6\}}\,\underline{\gamma_{\{3,5\}}}\,\gamma_{\{4,7\}}\,\underline{\gamma_{\{4,11\}}}\,\underline{\gamma_{\{6,9\}}}\,\underline{\gamma_{\{7,10\}}}\,\gamma_{\{10,11\}}\\
&\times\underline{\gamma_{\{1,3,4\}}}\,\gamma_{\{1,3,10\}}\,\underline{\gamma_{\{2,3,4\}}}\,\gamma_{\{2,3,10\}}\,\gamma_{\{3,4,9\}}\,\gamma_{\{3,9,10\}}\,\gamma_{\{5,6,7\}}\,\gamma_{\{5,6,11\}}.
\end{split}
\end{equation}

Taking the above result, the $t_3$ integration produces a lot of composite poles, which we do not bother to explicitly write out here. After chopping off these poles we land on the result
\begin{equation}
\begin{split}\checked{}
&\gamma_{\{3,5\}}\,
{\color{ProcessBlue}\cancel{\gamma}_{\{4,11\}}}\,
\gamma_{\{6,9\}}\,
{\color{ProcessBlue}\cancel{\gamma}_{\{7,10\}}}\,
{\color{ProcessBlue}\cancel{\gamma}_{\{8,12\}}}\\&
\gamma_{\{1,3,4\}}\,
\gamma_{\{1,6,8\}}\,
\gamma_{\{2,3,4\}}\,
\gamma_{\{2,6,8\}}\,
{\color{ProcessBlue}\cancel{\gamma}_{\{4,7,8\}}}\,
\gamma_{\{10,11,12\}}\\&
\gamma_{\{1,3,6,10\}}\,
\gamma_{\{1,3,10,12\}}\,
\gamma_{\{1,6,10,11\}}\,
\gamma_{\{2,3,6,10\}}\,
\gamma_{\{2,3,10,12\}}\,
\gamma_{\{2,6,10,11\}}\\&
{\color{ProcessBlue}\cancel{\gamma}_{\{3,4,7,9\}}\,
\cancel{\gamma}_{\{3,4,9,12\}}\,
\cancel{\gamma}_{\{3,9,10,12\}}\,
\cancel{\gamma}_{\{5,6,7,8\}}\,
\cancel{\gamma}_{\{5,6,8,11\}}\,
\cancel{\gamma}_{\{5,6,10,11\}}}.
\end{split}
\end{equation}
Since the arguments in all families are zero, we do not further highlight them by underlines. It is easy to work out the explicit locations of the poles in each $\gamma$ family following the discussion around \eqref{eq:gammapolelocations}. To name one slightly non-trivial example, let us look at $\gamma_{\{1,3,6,10\}}$, for which we may conclude that the poles locates at $\frac{h-c_1}{2}+m_1+\frac{m_3+m_6+m_{10}}{2}$, hence multiplying by $2$ we note these coincide with poles of $\Gamma[h-c_1]$. The others are computed similarly.

Almost all the poles find out above are obviously simple poles. For the simple pole families marked blue, as we hit them it turns out there are extra zeros from the divergence of $\Gamma$ factors in the denominator of the remaining integrand \eqref{eq:bubble4ptirrelevantfactors}. This indicates that the contour is in fact not pinched in such configurations, and so the corresponding $\gamma$ families are ruled out. For instance, the poles $\Gamma_{\{5,6,8,11\}}[\frac{2h-S+\Delta_{12}}{2}]$ come from $\Gamma_5[\frac{h-c_2-t_1+t_2}{2}]$, $\Gamma_6[\frac{h+c_2+t_1-t_3}{2}]$, $\Gamma_8[\frac{t_3-S}{2}]$, $\Gamma_{11}[\frac{\Delta_{12}-t_2}{2}]$. At the pole $\frac{2h-S+\Delta_{12}}{2}+m=0$ the variables are localized to $t_1=h-c_2+\Delta_{12}+2(m_5+m_{11})$, $t_2=\Delta_{12}+2m_{11}$ and $t_3=2h+\Delta_{12}+2(m_5+m_6+m_{11})$ for some $m_5+m_6+m_8+m_{11}=m$. This forces $\Gamma[\frac{2h-t_2+t_3}{2}]=\Gamma[-m_5-m_6]$ and $\Gamma[\frac{h-c_2-t_1+\Delta_{12}}{2}]=\Gamma[-m_5-m_{11}]$, and so \eqref{eq:bubble4ptirrelevantfactors} produces a double zero.

The only exception to the above discussion is $\Gamma_{\{4,7,8\}}[h]\,\Gamma_{\{3,4,7,9\}}[2h]\,\Gamma_{\{5,6,7,8\}}[2h]$. Firstly $\gamma_{\{4,7,8\}}$ localizes $t_2=S-2m'$ and $t_3=S-2m''$ for some integers $m',m''$. $\gamma_{\{3,4,7,9\}}$ further imposes $t_1=\frac{2S+2c_2+m'''}{2}$ while $\gamma_{\{5,6,7,8\}}$ further impoese $t_1=\frac{2S-2c_2+m''''}{2}$. Hence the pinching planes $P_{\{3,4,7,9\}}$ and $P_{\{5,6,7,8\}}$ parallel each other. This means the poles at $h+m=0$ are double poles for integral $m$ and simple poles for half-odd-integral $m$. However, it can be verified that the denominator always leads to a double zero at the pinching, and so in effect there is no pole at these locations at all, and we also rule them out.

We now collect the remaining $\gamma$ families, which all turn out to be integrally-spaced.  Including the additional $\Gamma$'s in the numeartor of \eqref{eq:bubble4ptirrelevantfactors} that we put aside at the beginning, these make up the complete pole structure of the bubble diagram
\begin{equation}\label{eq:bubble4ptpoles}
\begin{split}\checked{}
&\underset{\{3,5\}}{\Gamma[h-c_2]}\,
\underset{\{6,9\}}{\Gamma[h+c_2]}\,\underset{\{1,3,6,10\}}{\Gamma[h-c_1]}\,
\underset{\{2,3,6,10\}}{\Gamma[h+c_1]}\,
\underset{\{10,11,12\}}{\Gamma[{\textstyle\frac{\Sigma\Delta}{2}-h}]}\\&
\underset{\{1,3,4\}}{\Gamma[{\textstyle\frac{2h-S-c_1-c_2}{2}}]}\,
\underset{\{1,6,8\}}{\Gamma[{\textstyle\frac{2h-S-c_1+c_2}{2}}]}\,
\underset{\{2,3,4\}}{\Gamma[{\textstyle\frac{2h-S+c_1-c_2}{2}}]}\,
\underset{\{2,6,8\}}{\Gamma[{\textstyle\frac{2h-S+c_1+c_2}{2}}]}\\&
\underset{\{1,3,10,12\}}{\Gamma[{\textstyle\frac{\Delta_{34}-c_1-c_2}{2}}]}\,
\underset{\{1,6,10,11\}}{\Gamma[{\textstyle\frac{\Delta_{12}-c_1+c_2}{2}}]}\,
\underset{\{2,3,10,12\}}{\Gamma[{\textstyle\frac{\Delta_{34}+c_1-c_2}{2}}]}\,
\underset{\{2,6,10,11\}}{\Gamma[{\textstyle\frac{\Delta_{12}+c_1+c_2}{2}}]}\\&
\Gamma[{\textstyle\frac{\Delta_{12}\pm c_1-c_2}{2}}]\,
\Gamma[{\textstyle\frac{\Delta_{34}\pm c_1+c_2}{2}}].
\end{split}
\end{equation}
(From now on we switch the notation to underscript whenever it makes the structure clearer.) Note that while we obtain this result from a specific ordering for the integrations, it is in fact independent of the ordering, which is guaranteed by the algorithm (for a detailed explanation, see Appendix \ref{app:sec:poleglobalview}).

From the above result, we see that the poles in the unique channel $S$ are encoded in the four families
\begin{equation}\label{eq:bubble4ptMpoleS}
\checked{}
\Gamma[{\textstyle\frac{2h\pm c_1\pm c_2-S}{2}}],
\end{equation}
which are obviously consistent with the symmetry of the diagram under the exchange of the two bulk-to-bulk propagators as well as the flipping of signs of the spectrum variables $c_1,c_2$. Apart from these, there are also poles in the conformal dimensions of the boundary points as well as the spectrum variables, which can be grouped into
\begin{equation}\label{eq:bubble4ptMpolerest}
\checked{}
\Gamma[h\pm c_1]\,
\Gamma[h\pm c_2]\,
\Gamma[{\textstyle\frac{\Delta_{12}\pm c_1\pm c_2}{2}}]\,\Gamma[{\textstyle\frac{\Delta_{34}\pm c_1\pm c_2}{2}}]\,\Gamma[{\textstyle\frac{\Sigma\Delta}{2}-h}].
\end{equation}


\subsection{Triangle diagrams}

For the triangle diagrams we provide three case studies: the 3-point triangle (Figure \ref{fig:triangle} (A)), the 4-point triangle with a quartic vertex (\ref{fig:triangle} (B)), and the 4-point triangle with cubic vertices only (Figure \ref{fig:triangle4ptC}). 

\subsubsection{The 3-point triangle}

The formulas are shown in \eqref{eq:M0triangle3pt} and \eqref{eq:Ktriangle3pt}. In the integrand we label the $\Gamma$ functions that depend on the $t$ variables as
\begin{equation}\label{eq:labeledpolesintegrandtriangle3pt}
\begin{split}\checked{}
&\Gamma_1[{\textstyle\frac{h-c_1-t_1}{2}}]\,
\Gamma_2[{\textstyle\frac{h+c_1-t_1}{2}}]\,
\Gamma_3[{\textstyle\frac{h-c_2+t_1-t_2}{2}}]\,
\Gamma_4[{\textstyle\frac{h-c_3-t_3}{2}}]\,
\Gamma_5[{\textstyle\frac{h+c_3-t_3}{2}}]\,
\Gamma_6[{\textstyle\frac{h+c_2+t_3-t_4}{2}}]\\&
\Gamma_7[{\textstyle\frac{-\Delta_1-t_1+t_2+t_3}{2}}]\,
\Gamma_8[{\textstyle\frac{t_4-\Delta_1}{2}}]\,
\Gamma_9[{\textstyle\frac{-2h+t_2+t_4-\Delta_2}{2}}]\,
\Gamma_{10}[{\textstyle\frac{\Delta_{12}-t_2}{2}}]\,
\Gamma_{11}[{\textstyle\frac{t_2-\Delta_3}{2}}]\\&
\Gamma_{12}[{\textstyle\frac{t_1-t_3+t_4-\Delta_3}{2}}]\,
\Gamma_{13}[{\textstyle\frac{2h-t_2-t_4+\Delta_{31}}{2}}]\,
\Gamma_{14}[{\textstyle\frac{\Delta_{23}-t_4}{2}}].
\end{split}
\end{equation}
Apart from these the remaining integrand is
\begin{equation}\label{eq:3pttriangleremainingintegrand}
\checked{}
\frac{\Gamma[\frac{\Delta_1-c_2\pm c_3}{2}]\,\Gamma[\frac{\Delta_2\pm c_3\pm c_1}{2}]\,\Gamma[\frac{\Delta_3\pm c_1+c_2}{2}]}{\Gamma[\frac{2h+\Delta_1-t_4}{2}]\,\Gamma[\frac{t_2+t_4-\Delta_{31}}{2}]\,\Gamma[\frac{2h+\Delta_3-t_2}{2}]\,\Gamma[\frac{\Delta_{12}-\Delta_3}{2}]\,\Gamma[\frac{\Delta_{23}-\Delta_1}{2}]\,\Gamma[\frac{\Delta_{31}-\Delta_2}{2}]}\,C_{A_2},
\end{equation}
where $C_{A_2}$ is the correction function associated to vertex $A_2$. Since $C_{A_2}$ is free of poles, all what we may care regarding this factor is whether it can develop zeros.

The formula involves four Mellin integrals. In order to illustrate intermediate details we still write out the result (after deleting composite poles) from each integration. $\gamma[0]$'s in the intermediate steps are again underlined.
\begin{enumerate}[noitemsep,nolistsep]
\item After $t_1$ integration:
\begin{equation}\checked{}
\begin{split}
\gamma_4\,\gamma_5\,\gamma_6\,\gamma_8\,\gamma_9\,\gamma_{10}\,\gamma_{11}\,\gamma_{13}\,\gamma_{14}\,\gamma_{\{1,3\}}\,\gamma_{\{1,12\}}\,\gamma_{\{2,3\}}\,\gamma_{\{2,12\}}\,\gamma_{\{3,7\}}\,\gamma_{\{7,12\}}.
\end{split}
\end{equation}
\item After $t_2$ integration:
\begin{equation}\checked{}
\begin{split}
&\gamma_4\,\gamma_5\,\gamma_6\,\gamma_8\,\gamma_{14}\,\gamma_{\{1,12\}}\,\gamma_{\{2,12\}}\,\gamma_{\{3,7\}}\,\gamma_{\{9,10\}}\,\underline{\gamma_{\{9,13\}}}\,\underline{\gamma_{\{10,11\}}}\,\gamma_{\{11,13\}}\\
&\gamma_{\{1,3,9\}}\,\underline{\gamma_{\{1,3,11\}}}\,\gamma_{\{2,3,9\}}\,\underline{\gamma_{\{2,3,11\}}}\,\gamma_{\{7,10,12\}}\,\underline{\gamma_{\{7,12,13\}}}\,\gamma_{\{1,3,7,12\}}\,\gamma_{\{2,3,7,12\}}
\end{split}
\end{equation}
\item After $t_3$ integration:
\begin{equation}\checked{}
\begin{split}
&\gamma_8\,\gamma_{14}\,\gamma_{\{4,6\}}\,\gamma_{\{5,6\}}\,\gamma_{\{9,10\}}\,\underline{\gamma_{\{9,13\}}}\,\underline{\gamma_{\{10,11\}}}\,\gamma_{\{11,13\}}\,\gamma_{\{1,3,9\}}\,\underline{\gamma_{\{1,3,11\}}}\,\underline{\gamma_{\{1,6,12\}}}\,\gamma_{\{2,3,9\}}\\
&\underline{\gamma_{\{2,3,11\}}}\,\underline{\gamma_{\{2,6,12\}}}\,\underline{\gamma_{\{3,4,7\}}}\,\underline{\gamma_{\{3,5,7\}}}\,\gamma_{\{7,10,12\}}\,\underline{\gamma_{\{7,12,13\}}}\,\gamma_{\{1,3,7,12\}}\,\gamma_{\{2,3,7,12\}}.
\end{split}
\end{equation}
\item After $t_4$ integration:
\begin{equation}\checked{}
\begin{split}
&{\color{ProcessBlue}\cancel{\gamma}_{\{8,14\}}\,
\cancel{\gamma}_{\{9,13\}}\,
\cancel{\gamma}_{\{10,11\}}}\,
\gamma_{\{1,3,11\}}\,
\gamma_{\{1,6,12\}}\,
\gamma_{\{2,3,11\}}\,
\gamma_{\{2,6,12\}}\,
\gamma_{\{3,4,7\}}\,
\gamma_{\{3,5,7\}}\,
\gamma_{\{4,6,8\}}\\
&\gamma_{\{5,6,8\}}\,
{\color{ProcessBlue}\cancel{\gamma}_{\{7,12,13\}}\,
\cancel{\gamma}_{\{8,11,13\}}}\,
\gamma_{\{9,10,14\}}\,
\gamma_{\{1,3,9,14\}}\,
\gamma_{\{2,3,9,14\}}\,
\gamma_{\{4,6,9,10\}}\,
\gamma_{\{5,6,9,10\}}\\
&{\color{ProcessBlue}\cancel{\gamma}_{\{7,10,12,14\}}}\,
\gamma_{\{1,3,4,6,9\}}\,
\gamma_{\{1,3,5,6,9\}}\,
{\color{ProcessBlue}\cancel{\gamma}_{\{1,3,7,12,14\}}}\,
\gamma_{\{2,3,4,6,9\}}\,
\gamma_{\{2,3,5,6,9\}}\,
{\color{ProcessBlue}\cancel{\gamma}_{\{2,3,7,12,14\}}}\\
&{\color{ProcessBlue}\cancel{\gamma}_{\{4,6,7,10,12\}}\,
\cancel{\gamma}_{\{5,6,7,10,12\}}}.
\end{split}
\end{equation}
\end{enumerate}
$\gamma$ families marked blue are killed by extra zeros from the remaining integrand \eqref{eq:3pttriangleremainingintegrand} (To be more precise, these zeros arise from the divergence of the $\Gamma$ factors explicitly shown in the denominator of \eqref{eq:3pttriangleremainingintegrand} when the contour is pinched. Hence for these $\gamma$'s there is even no need to worry about $C_{A_2}$. In order to confirm the existence of the remaining families, however, we need to verify that $C_{A_2}\neq0$). Working out the locations for the remaining families, the final result reads
\begin{equation}\label{eq:triangle3ptMpolesemerge}
\begin{split}\checked{}
&\underset{\{1,3,11\}}{\Gamma[{\textstyle\frac{2h-c_1-c_2-\Delta_3}{2}}]}\,
\underset{\{1,6,12\}}{\Gamma[{\textstyle\frac{2h-c_1+c_2-\Delta_3}{2}}]}\,
\underset{\{2,3,11\}}{\Gamma_[{\textstyle\frac{2h+c_1-c_2-\Delta_3}{2}}]}\,
\underset{\{2,6,12\}}{\Gamma[{\textstyle\frac{2h+c_1+c_2-\Delta_3}{2}}]}\\&
\underset{\{3,4,7\}}{\Gamma[{\textstyle\frac{2h-c_2-c_3-\Delta_1}{2}}]}\,
\underset{\{3,5,7\}}{\Gamma[{\textstyle\frac{2h-c_2+c_3-\Delta_1}{2}}]}\,
\underset{\{4,6,8\}}{\Gamma[{\textstyle\frac{2h+c_2-c_3-\Delta_1}{2}}]}\,
\underset{\{5,6,8\}}{\Gamma[{\textstyle\frac{2h+c_2+c_3-\Delta_1}{2}}]}\\&
\underset{\{1,3,9,14\}}{\Gamma[{\textstyle\frac{\Delta_3-c_1-c_2}{2}}]}\,
\underset{\{2,3,9,14\}}{\Gamma[{\textstyle\frac{\Delta_3+c_1-c_2}{2}}]}\,
\underset{\{4,6,9,10\}}{\Gamma[{\textstyle\frac{\Delta_1+c_2-c_3}{2}}]}\,
\underset{\{5,6,9,10\}}{\Gamma[{\textstyle\frac{\Delta_1+c_2+c_3}{2}}]}\\&
\underset{\{1,3,4,6,9\}}{\Gamma[{\textstyle\frac{2h-c_1-c_3-\Delta_2}{2}}]}\,
\underset{\{1,3,5,6,9\}}{\Gamma[{\textstyle\frac{2h-c_1+c_3-\Delta_2}{2}}]}\,
\underset{\{2,3,4,6,9\}}{\Gamma[{\textstyle\frac{2h+c_1-c_3-\Delta_2}{2}}]}\,
\underset{\{2,3,5,6,9\}}{\Gamma[{\textstyle\frac{2h+c_1+c_3-\Delta_2}{2}}]}\\&
\underset{\{9,10,14\}}{\Gamma[{\textstyle\frac{\Sigma\Delta}{2}-h}]}\,
\Gamma[{\textstyle\frac{\Delta_1-c_2\pm c_3}{2}}]\,
\Gamma[{\textstyle\frac{\Delta_2\pm c_1\pm c_3}{2}}]\,
\Gamma[{\textstyle\frac{\Delta_3\pm c_1+c_2}{2}}].
\end{split}
\end{equation}
Again  the factors without labels collect existing poles in \eqref{eq:3pttriangleremainingintegrand}. At this point, it is clear that the above poles can be grouped into
\begin{equation}\label{eq:triangle3ptpolesall}
\begin{split}\checked{}
&\Gamma[{\textstyle{\frac{\Delta_1\pm c_2\pm c_3}{2}}}]\,
\Gamma[{\textstyle{\frac{\Delta_2\pm c_3\pm c_1}{2}}}]\,
\Gamma[{\textstyle{\frac{\Delta_3\pm c_1\pm c_2}{2}}}]\,
\Gamma[{\textstyle\frac{\Sigma\Delta}{2}-h}]\\&
\Gamma[{\textstyle{\frac{2h\pm c_2\pm c_3-\Delta_1}{2}}}]\,
\Gamma[{\textstyle{\frac{2h\pm c_3\pm c_1-\Delta_2}{2}}}]\,
\Gamma[{\textstyle{\frac{2h\pm c_1\pm c_2-\Delta_3}{2}}}],
\end{split}
\end{equation}
which fully enjoys the permutation symmetry of the 3-point triangle diagram.

\subsubsection{4-point triangle with a quartic vertex}

The formula is shown in \eqref{eq:M0triangle4ptquartic} and \eqref{eq:Ktriangle4ptquartic}. We label the $\Gamma$ functions that depend on the $t$ variables as
\begin{equation}\label{eq:labeledpolesintegrandtriangle4pt}\checked{}
\begin{split}
&\Gamma_1[{\textstyle\frac{h-c_1-t_1}{2}}]\,
\Gamma_2[{\textstyle\frac{h+c_1-t_1}{2}}]\,
\Gamma_3[{\textstyle\frac{h-c_2-t_2}{2}}]\,
\Gamma_4[{\textstyle\frac{h+c_2-t_2}{2}}]\,
\Gamma_5[{\textstyle\frac{h-c_3+t_2-t_3}{2}}]\,
\Gamma_6[{\textstyle\frac{t_4-S}{2}}]\,
\Gamma_7[{\textstyle\frac{h+c_3+t_1-t_5}{2}}]\\&
\Gamma_8[{\textstyle\frac{-2h+t_3-t_4+t_5}{2}}]\,
\Gamma_9[{\textstyle\frac{S+t_3-t_4-\Delta_3}{2}}]\,
\Gamma_{10}[{\textstyle\frac{-t_1+t_2+t_5-\Delta_3}{2}}]\,
\Gamma_{11}[{\textstyle\frac{\Delta_3+t_4-t_5}{2}}]\,
\Gamma_{12}[{\textstyle\frac{t_1-t_2+t_3-\Delta_4}{2}}]\\&
\Gamma_{13}[{\textstyle\frac{S-t_4+t_5-\Delta_4}{2}}]\,
\Gamma_{14}[{\textstyle\frac{\Delta_4-t_3+t_4}{2}}]\,
\Gamma_{15}[{\textstyle\frac{2h-S-t_3+t_4-t_5+\Delta_{34}}{2}}]\,
\Gamma_{16}[{\scriptstyle\frac{\Delta_{12}-t_4}{2}}].
\end{split}
\end{equation}
The remaining integrand is ($C_{A_2}$ being the vertex correction)
\begin{equation}\checked{}
\frac{\Gamma[\frac{\Delta_3\pm c_2+c_3}{2}]\,\Gamma[\frac{\Delta_{12}\pm c_1\pm c_2}{2}]\,\Gamma[\frac{\Delta_4-c_3\pm c_1}{2}]}{\Gamma[\frac{2h-t_3+\Delta_3}{2}]\,\Gamma[\frac{t_3+t_5-\Delta_{34}}{2}]\,\Gamma[\frac{S+\Delta_3-\Delta_4}{2}]\,\Gamma[\frac{2h-t_5+\Delta_4}{2}]\,\Gamma[\frac{\Delta_4\pm(\Delta_3-S)}{2}]\,\Gamma[\frac{\Delta_{12}-S}{2}]}\,C_{A_2}.
\end{equation}

For this and later examples we skip all the intermediate computations and directly present the final result from Algorithm \ref{app:sec:polealgoritm} and also after chopping of poles that are killed by zeros from the $\Gamma$ factors in the denominator of the integrand (the vertex correction $C_{A_2}$ again plays no role in this particular case). Here the final poles selected by the algorithm are
\begin{equation}\label{eq:preMpolestriangle4pt}
\begin{split}\checked{}
&\underset{\{1,5,12\}}{\Gamma[{\textstyle\frac{2h-c_1-c_3-\Delta_4}{2}}]}\,
\underset{\{2,5,12\}}{\Gamma[{\textstyle\frac{2h+c_1-c_3-\Delta_4}{2}}]}\,
\underset{\{3,7,10\}}{\Gamma[{\textstyle\frac{2h-c_2+c_3-\Delta_3}{2}}]}\,
\underset{\{4,7,10\}}{\Gamma[{\textstyle\frac{2h+c_2+c_3-\Delta_3}{2}}]}\\&
\underset{\{1,6,7,13\}}{\Gamma[{\textstyle\frac{2h-c_1+c_3-\Delta_4}{2}}]}\,
\underset{\{1,7,8,14\}}{\Gamma[{\textstyle\frac{\Delta_4-c_1+c_3}{2}}]}\,
\underset{\{2,6,7,13\}}{\Gamma[{\textstyle\frac{2h+c_1+c_3-\Delta_4}{2}}]}\,
\underset{\{2,7,8,14\}}{\Gamma[{\textstyle\frac{\Delta_4+c_1+c_3}{2}}]}\\&
\underset{\{3,5,6,9\}}{\Gamma[{\textstyle\frac{2h-c_2-c_3-\Delta_3}{2}}]}\,
\underset{\{3,5,8,11\}}{\Gamma[{\textstyle\frac{\Delta_3-c_2-c_3}{2}}]}\,
\underset{\{4,5,6,9\}}{\Gamma[{\textstyle\frac{2h+c_2-c_3-\Delta_3}{2}}]}\,
\underset{\{4,5,8,11\}}{\Gamma[{\textstyle\frac{\Delta_3+c_2-c_3}{2}}]}\\&
\underset{\{1,3,5,6,7,8\}}{\Gamma[{\textstyle\frac{2h-S-c_1-c_2}{2}}]}\,
\underset{\{1,4,5,6,7,8\}}{\Gamma[{\textstyle\frac{2h-S-c_1+c_2}{2}}]}\,
\underset{\{2,3,5,6,7,8\}}{\Gamma[{\textstyle\frac{2h-S+c_1-c_2}{2}}]}\,
\underset{\{2,4,5,6,7,8\}}{\Gamma[{\textstyle\frac{2h-S+c_1+c_2}{2}}]}\\&
\underset{\{8,11,14,16\}}{\Gamma[{\textstyle\frac{\Sigma\Delta}{2}-h}]}\,
\Gamma[{\textstyle\frac{\Delta_3\pm c_2+c_3}{2}}]\,
\Gamma[{\textstyle\frac{\Delta_4\pm c_1-c_3}{2}}]\,
\Gamma[{\textstyle\frac{\Delta_{12}\pm c_1\pm c_2}{2}}].
\end{split}
\end{equation}
This can be grouped into
\begin{equation}\label{eq:preMtriangle4ptpolesall}
\begin{split}\checked{}
&\Gamma[{\textstyle\frac{\Delta_{12}\pm c_1\pm c_2}{2}}]\,
\Gamma[{\textstyle\frac{\Delta_3\pm c_2\pm c_3}{2}}]\,
\Gamma[{\textstyle\frac{\Delta_4\pm c_1\pm c_3}{2}}]\,
\Gamma[{\textstyle\frac{\Sigma\Delta}{2}-h}]\\&
\Gamma[{\textstyle\frac{2h\pm c_1\pm c_2-S}{2}}]\,
\Gamma[{\textstyle\frac{2h\pm c_2\pm c_3-\Delta_3}{2}}]\,
\Gamma[{\textstyle\frac{2h\pm c_1\pm c_3-\Delta_4}{2}}].
\end{split}
\end{equation}
This shares the same structure as the 3-pt triangle diagram.

\subsubsection{4-point triangle with only cubic vertices}

The formula is shown in \eqref{eq:M0triangle4ptcubic} and \eqref{eq:Ktriangle4ptcubic}. $\Gamma$ factors in the integrand are labeled as
\begin{equation}\label{eq:labeledpolesintegrandtriangle4ptC}\checked{}
\begin{split}
&\Gamma_1[{\textstyle\frac{h-c_1-t_1}{2}}]\,
\Gamma_2[{\textstyle\frac{h+c_1-t_1}{2}}]\,
\Gamma_3[{\textstyle\frac{t_1-S}{2}}]\,
\Gamma_4[{\textstyle\frac{h-c_3-t_3}{2}}]\,
\Gamma_5[{\textstyle\frac{h+c_3-t_3}{2}}]\,
\Gamma_6[{\textstyle\frac{h-c_2-t_2+t_3}{2}}]\,
\Gamma_7[{\textstyle\frac{h-c_4-t_4}{2}}]\\&
\Gamma_8[{\textstyle\frac{h+c_4-t_4}{2}}]\,
\Gamma_9[{\textstyle\frac{h+c_2+t_4-t_5}{2}}]\,
\Gamma_{10}[{\textstyle\frac{-2h-t_1+t_2+t_5}{2}}]\,
\Gamma_{11}[{\textstyle\frac{S-t_1+t_2-\Delta_3}{2}}]\,
\Gamma_{12}[{\textstyle\frac{t_3-t_4+t_5-\Delta_3}{2}}]\\&
\Gamma_{13}[{\textstyle\frac{t_1-t_5+\Delta_3}{2}}]\,
\Gamma_{14}[{\textstyle\frac{t_2-t_3+t_4-\Delta_4}{2}}]\,
\Gamma_{15}[{\textstyle\frac{S-t_1+t_5-\Delta_4}{2}}]\,
\Gamma_{16}[{\textstyle\frac{t_1-t_2+\Delta_4}{2}}]\,
\Gamma_{17}[{\textstyle\frac{2h-S+t_1-t_2-t_5+\Delta_{34}}{2}}].
\end{split}
\end{equation}
The remaining integrand is
\begin{equation}\label{eq:remainingintegrand3ptcubic}
\checked{}
\frac{\Gamma[\frac{\Delta_3+c_2\pm c_3}{2}]\,\Gamma[\frac{h\pm c_1\pm c_3\pm c_4}{2}]\,\Gamma[\frac{\Delta_4-c_2\pm c_4}{2}]\,\Gamma[\frac{\Delta_{12}-h\pm c_1}{2}]}{\Gamma[\frac{2h-t_2+\Delta_3}{2}]\,\Gamma[\frac{t_2+t_5-\Delta_{34}}{2}]\,\Gamma[\frac{S\pm(\Delta_3-\Delta_4)}{2}]\,\Gamma[\frac{2h-t_5+\Delta_4}{2}]\,\Gamma[\frac{\Delta_{34}-S}{2}]\,\Gamma[\frac{\Delta_{12}-S}{2}]}\,C_{A_2}.
\end{equation}

The final result for this diagram is
\begin{equation}\label{eq:poleresult4pttrianglecubic}
\begin{split}\checked{}
&\underset{\{1,3\}}{\Gamma[{\textstyle\frac{h-S-c_1}{2}}]}\,
\underset{\{2,3\}}{\Gamma[{\textstyle\frac{h-S+c_1}{2}}]}\,
\underset{\{1,10,13,16\}}{\Gamma[{\textstyle\frac{\Delta_{34}-h-c_1}{2}}]}\,
\underset{\{2,10,13,16\}}{\Gamma[{\textstyle\frac{\Delta_{34}-h+c_1}{2}}]}\\&
\underset{\{4,9,12\}}{\Gamma[{\textstyle\frac{2h+c_2-c_3-\Delta_3}{2}}]}\,
\underset{\{5,9,12\}}{\Gamma[{\textstyle\frac{2h+c_2+c_3-\Delta_3}{2}}]}\,
\underset{\{3,4,6,11\}}{\Gamma[{\textstyle\frac{2h-c_2-c_3-\Delta_3}{2}}]}\,
\underset{\{3,5,6,11\}}{\Gamma[{\textstyle\frac{2h-c_2+c_3-\Delta_3}{2}}]}\\&
\underset{\{6,7,14\}}{\Gamma[{\textstyle\frac{2h-c_2-c_4-\Delta_4}{2}}]}\,
\underset{\{6,8,14\}}{\Gamma[{\textstyle\frac{2h-c_2+c_4-\Delta_4}{2}}]}\,
\underset{\{3,7,9,15\}}{\Gamma[{\textstyle\frac{2h+c_2-c_4-\Delta_4}{2}}]}\,
\underset{\{3,8,9,15\}}{\Gamma[{\textstyle\frac{2h+c_2+c_4-\Delta_4}{2}}]}\\&
\underset{\{4,6,10,13\}}{\Gamma[{\textstyle\frac{\Delta_3-c_2-c_3}{2}}]}\,
\underset{\{5,6,10,13\}}{\Gamma[{\textstyle\frac{\Delta_3-c_2+c_3}{2}}]}\,
\underset{\{7,9,10,16\}}{\Gamma[{\textstyle\frac{\Delta_4+c_2-c_4}{2}}]}\,
\underset{\{8,9,10,16\}}{\Gamma[{\textstyle\frac{\Delta_4+c_2+c_4}{2}}]}\\&
{\color{ForestGreen}
\underset{\{3,4,6,7,9,10\}}{\cancel{\Gamma}[{\textstyle\frac{2h-S-c_3-c_4}{2}}]}\,
\underset{\{3,4,6,8,9,10\}}{\cancel{\Gamma}[{\textstyle\frac{2h-S-c_3+c_4}{2}}]}\,
\underset{\{3,5,6,7,9,10\}}{\cancel{\Gamma}[{\textstyle\frac{2h-S+c_3-c_4}{2}}]}\,
\underset{\{3,5,6,8,9,10\}}{\cancel{\Gamma}[{\textstyle\frac{2h-S+c_3+c_4}{2}}]}}\\&
\Gamma[{\textstyle\frac{h\pm c_1\pm c_3\pm c_4}{2}}]\,
\Gamma[{\textstyle\frac{\Delta_3+c_2\pm c_3}{2}}]\,
\Gamma[{\textstyle\frac{\Delta_4-c_2\pm c_4}{2}}]\,
\Gamma[{\textstyle\frac{\Delta_{12}-h\pm c_1}{2}}].
\end{split}
\end{equation}
Here we highlighted the four families marked in green, which, different from the previous examples, are absent due to the vanishing of the vertex correction $C_{A_2}$. This will be verified in detail in Appendix \ref{app:sec:checkfakepoletriangle}. The remaining pole families can then be grouped into
\begin{equation}\label{eq:poleresult4pttrianglecubicshort}
\begin{split}\checked{}
&\Gamma[{\textstyle\frac{\Delta_{12}-h\pm c_1}{2}}]\,
\Gamma[{\textstyle\frac{h\pm c_1\pm c_3\pm c_4}{2}}]\,
\Gamma[{\textstyle\frac{\Delta_3\pm c_2\pm c_3}{2}}]\,
\Gamma[{\textstyle\frac{\Delta_4\pm c_2\pm c_4}{2}}]\,
\Gamma[{\textstyle\frac{\Delta_{34}-h\pm c_1}{2}}]\\
&
\Gamma[{\textstyle\frac{h\pm c_1-S}{2}}]\,
\Gamma[{\textstyle\frac{2h\pm c_2\pm c_3-\Delta_3}{2}}]\,
\Gamma[{\textstyle\frac{2h\pm c_2\pm c_4-\Delta_4}{2}}].
\end{split}
\end{equation}
Note in particular, the only families of poles of the Mandelstam variable $S$ are tied to the unique tree propagator in the diagram, which is essentially different from the previous example.

\subsection{Box diagrams}\label{sec:boxpoles}

The formula is shown in \eqref{eq:M0box4pt} and \eqref{eq:Kbox4pt}. We label the $\Gamma$ functions in the integrand as
\begin{equation}\label{eq:labeledpolesintgrandbox}\checked{}
\begin{split}
&\Gamma_1[{\textstyle\frac{t_1-S}{2}}]\,
\Gamma_2[{\textstyle\frac{h-c_2-t_2}{2}}]\,
\Gamma_3[{\textstyle\frac{h+c_2-t_2}{2}}]\,
\Gamma_4[{\textstyle\frac{h-c_3-t_3}{2}}]\,
\Gamma_5[{\textstyle\frac{h+c_3-t_3}{2}}]\,
\Gamma_6[{\textstyle\frac{h-c_4-t_4}{2}}]\,
\Gamma_7[{\textstyle\frac{h+c_4-t_4}{2}}]\\&
\Gamma_8[{\textstyle\frac{h-c_1+t_4-t_5}{2}}]\,
\Gamma_9[{\textstyle\frac{-t_1+t_3-t_4+t_5}{2}}]\,
\Gamma_{10}[{\textstyle\frac{t_6-T}{2}}]\,
\Gamma_{11}[{\textstyle\frac{h+c_1+t_2-t_7}{2}}]\,
\Gamma_{12}[{\textstyle\frac{-2h+t_5-t_6+t_7}{2}}]\\&
\Gamma_{13}[{\textstyle\frac{-t_2+t_3+t_7-t_8}{2}}]\,
\Gamma_{14}[{\textstyle\frac{t_8-S}{2}}]\,
\Gamma_{15}[{\textstyle\frac{t_1+t_2-t_3-\Delta_1}{2}}]\,
\Gamma_{16}[{\textstyle\frac{S+T-t_6+t_7-t_8-\Delta_1}{2}}]\\&
\Gamma_{17}[{\textstyle\frac{t_1-t_5+t_6-\Delta_2}{2}}]\,
\Gamma_{18}[{\textstyle\frac{\Delta_{12}-t_1}{2}}]\,
\Gamma_{19}[{\textstyle\frac{t_6-t_7+t_8-\Delta_3}{2}}]\,
\Gamma_{20}[{\textstyle\frac{\Delta_{23}-t_6}{2}}]\,
\Gamma_{21}[{\textstyle\frac{S+T-t_1+t_5-t_6-\Delta_4}{2}}]\\&
\Gamma_{22}[{\textstyle\frac{-t_3+t_4+t_8-\Delta_4}{2}}]\,
\Gamma_{23}[{\textstyle\frac{2h-T-t_5+t_6-t_7+\Delta_{41}}{2}}]\,
\Gamma_{24}[{\textstyle\frac{\Delta_{34}-t_8}{2}}].
\end{split}
\end{equation}
The remaining integrand is
\begin{equation}
\begin{split}\checked{}
&C_{A_2}\,C_{A_3}\,\Gamma[{\textstyle\frac{\Delta_1-c_1\pm c_2}{2}}]\,\Gamma[{\textstyle\frac{\Delta_2\pm c_2\pm c_3}{2}}]\,\Gamma[{\textstyle\frac{\Delta_3\pm c_3\pm c_4}{2}}]\,\Gamma[{\textstyle\frac{\Delta_4\pm c_4+c_1}{2}}]/\big(\Gamma[{\textstyle\frac{t_1+t_7-t_8-\Delta_1}{2}}]\\&\times\Gamma[{\textstyle\frac{2h-t_7+\Delta_1}{2}}]\,\Gamma[{\textstyle\frac{\Delta_{12}-S}{2}}]\,\Gamma[{\textstyle\frac{S+T-\Delta_{13}}{2}}]\,\Gamma[{\textstyle\frac{\Delta_{23}-T}{2}}]\,\Gamma[{\textstyle\frac{-t_1+t_5+t_8-\Delta_4}{2}}]\,\Gamma[{\textstyle\frac{S+T-\Delta_{24}}{2}}]\\&\times\Gamma[{\textstyle\frac{2h-t_5+\Delta_4}{2}}]\,\Gamma[{\textstyle\frac{\Delta_{41}-T}{2}}]\,\Gamma[{\textstyle\frac{\Delta_{34}-S}{2}}]\big).
\end{split}
\end{equation}

The final result is
\begin{equation}\label{eq:resultpolesM4pt}
\begin{split}\checked{}
&\underset{\{1,4,8,9\}}{\Gamma[{\textstyle\frac{2h-S-c_1-c_3}{2}}]}\,
\underset{\{1,5,8,9\}}{\Gamma[{\textstyle\frac{2h-S-c_1+c_3}{2}}]}\,
\underset{\{4,11,13,14\}}{\Gamma[{\textstyle\frac{2h-S+c_1-c_3}{2}}]}\,
\underset{\{5,11,13,14\}}{\Gamma[{\textstyle\frac{2h-S+c_1+c_3}{2}}]}\\&
\underset{\{2,8,9,15\}}{\Gamma[{\textstyle\frac{2h-\Delta_1-c_1-c_2}{2}}]}\,
\underset{\{3,8,9,15\}}{\Gamma[{\textstyle\frac{2h-\Delta_1-c_1+c_2}{2}}]}\,
\underset{\{2,10,11,14,16\}}{\Gamma[{\textstyle\frac{2h-\Delta_1+c_1-c_2}{2}}]}\,
\underset{\{3,10,11,14,16\}}{\Gamma[{\textstyle\frac{2h-\Delta_1+c_1+c_2}{2}}]}\\&
\underset{\{6,11,13,22\}}{\Gamma[{\textstyle\frac{2h-\Delta_4+c_1-c_4}{2}}]}\,
\underset{\{7,11,13,22\}}{\Gamma[{\textstyle\frac{2h-\Delta_4+c_1+c_4}{2}}]}\,
\underset{\{1,6,8,10,21\}}{\Gamma[{\textstyle\frac{2h-\Delta_4-c_1-c_4}{2}}]}\,
\underset{\{1,7,8,10,21\}}{\Gamma[{\textstyle\frac{2h-\Delta_4-c_1+c_4}{2}}]}\\&
\underset{\{2,11,12,17,18\}}{\Gamma[{\textstyle\frac{\Delta_1+c_1-c_2}{2}}]}\,
\underset{\{3,11,12,17,18\}}{\Gamma[{\textstyle\frac{\Delta_1+c_1+c_2}{2}}]}\,
\underset{\{6,8,12,19,24\}}{\Gamma[{\textstyle\frac{\Delta_4-c_1-c_4}{2}}]}\,
\underset{\{7,8,12,19,24\}}{\Gamma[{\textstyle\frac{\Delta_4-c_1+c_4}{2}}]}\\&
\underset{\{2,6,8,10,11,12\}}{\Gamma[{\textstyle\frac{2h-T-c_2-c_4}{2}}]}\,
\underset{\{2,7,8,10,11,12\}}{\Gamma[{\textstyle\frac{2h-T-c_2+c_4}{2}}]}\,
\underset{\{3,6,8,10,11,12\}}{\Gamma[{\textstyle\frac{2h-T+c_2-c_4}{2}}]}\,
\underset{\{3,7,8,10,11,12\}}{\Gamma[{\textstyle\frac{2h-T+c_2+c_4}{2}}]}\\&
\underset{\{2,4,8,9,11,12,17\}}{\Gamma[{\textstyle\frac{2h-\Delta_2-c_2-c_3}{2}}]}\,
\underset{\{2,5,8,9,11,12,17\}}{\Gamma[{\textstyle\frac{2h-\Delta_2-c_2+c_3}{2}}]}\,
\underset{\{3,4,8,9,11,12,17\}}{\Gamma[{\textstyle\frac{2h-\Delta_2+c_2-c_3}{2}}]}\,
\underset{\{3,5,8,9,11,12,17\}}{\Gamma[{\textstyle\frac{2h-\Delta_2+c_2+c_3}{2}}]}\\&
\underset{\{4,6,8,11,12,13,19\}}{\Gamma[{\textstyle\frac{2h-\Delta_3-c_3-c_4}{2}}]}\,
\underset{\{4,7,8,11,12,13,19\}}{\Gamma[{\textstyle\frac{2h-\Delta_3-c_3+c_4}{2}}]}\,
\underset{\{5,6,8,11,12,13,19\}}{\Gamma[{\textstyle\frac{2h-\Delta_3+c_3-c_4}{2}}]}\,
\underset{\{5,7,8,11,12,13,19\}}{\Gamma[{\textstyle\frac{2h-\Delta_3+c_3+c_4}{2}}]}\\&
\underset{\{12,17,18,19,20,24\}}{\Gamma[{\textstyle\frac{\Sigma\Delta}{2}-h}]}\,
{\color{ForestGreen}\underset{\{4,8,9,11,12,13,17,19,20\}}{\cancel{\Gamma}[{\textstyle h-c_3}]}}\,
{\color{ForestGreen}\underset{\{5,8,9,11,12,13,17,19,20\}}{\cancel{\Gamma}[{\textstyle h+c_3}]}}\\&
{\color{ForestGreen}\underset{\{4,8,9,12,17,19,20,24\}}{\cancel{\Gamma}[{\textstyle\frac{\Delta_{34}-c_1-c_3}{2}}]}}\,
{\color{ForestGreen}\underset{\{4,11,12,13,17,18,19,20\}}{\cancel{\Gamma}[{\textstyle\frac{\Delta_{12}+c_1-c_3}{2}}]}}\,
{\color{ForestGreen}\underset{\{5,8,9,12,17,19,20,24\}}{\cancel{\Gamma}[{\textstyle\frac{\Delta_{34}-c_1+c_3}{2}}]}}\,
{\color{ForestGreen}\underset{\{5,11,12,13,17,18,19,20\}}{\cancel{\Gamma}[{\textstyle\frac{\Delta_{12}+c_1+c_3}{2}}]}}\\&
\Gamma[{\textstyle\frac{\Delta_1-c_1\pm c_2}{2}}]\,
\Gamma[{\textstyle\frac{\Delta_2\pm c_2\pm c_3}{2}}]\,
\Gamma[{\textstyle\frac{\Delta_3\pm c_3\pm c_4}{2}}]\,
\Gamma[{\textstyle\frac{\Delta_4+c_1\pm c_4}{2}}].
\end{split}
\end{equation}
Similar to the previous example, the families marked green turn out to be absent due to the vanishing of the vertex corrections $C_{A_2}$ or/and $C_{A_3}$ at the corresponding pinching. In fact, in this example even without these checks we are already able to rule these poles out simply by taking the symmetry of the box diagram into consideration. Nevertheless, we collect the detailed checks in Appendix \ref{app:sec:checkfakepolesbox} in order to verify the consistency of the method.

In consequence, genuine poles of the pre-amplitude of the 4-point box are encoded in
\begin{equation}\label{eq:box4ptMpolesshort}
\begin{split}\checked{}
&\Gamma[{\textstyle\frac{\Delta_1\pm c_1\pm c_2}{2}}]\,
\Gamma[{\textstyle\frac{\Delta_2\pm c_2\pm c_3}{2}}]\,
\Gamma[{\textstyle\frac{\Delta_3\pm c_3\pm c_4}{2}}]\,
\Gamma[{\textstyle\frac{\Delta_4\pm c_4\pm c_1}{2}}]\,
\Gamma[{\textstyle\frac{\Sigma\Delta}{2}-h}]\\&
\Gamma[{\textstyle\frac{2h\pm c_1\pm c_2-\Delta_1}{2}}]\,
\Gamma[{\textstyle\frac{2h\pm c_2\pm c_3-\Delta_2}{2}}]\,
\Gamma[{\textstyle\frac{2h\pm c_3\pm c_4-\Delta_3}{2}}]\,
\Gamma[{\textstyle\frac{2h\pm c_4\pm c_1-\Delta_4}{2}}]\\&
\Gamma[{\textstyle\frac{2h\pm c_1\pm c_3-S}{2}}]\,
\Gamma[{\textstyle\frac{2h\pm c_2\pm c_4-T}{2}}].
\end{split}
\end{equation}

\subsection{Universality of the pole structure of pre-amplitudes}\label{sec:polediagrammaticrules}

It is good to pause at this stage and seek for general patterns in the pole structure that we have worked out in the various examples so far. Again, let us emphasize that the object in our current focus is the Mellin pre-amplitude $M$, prior to all the spectrum integrals, rather than the more commonly studied Mellin amplitude $\mathcal{M}$.

\subsubsection{Preliminary observations}\label{sec:preliminaryobservations}

Recall that at tree level we have established diagrammatic rules for the pre-amplitude, according to which the poles of $M$ can be classified into two types
\begin{align}
\checked{}\label{eq:treevertexpoles}\text{bulk vertex }A_a\text{ of valency }r:&\qquad\Gamma[{\textstyle\frac{\Delta_{A_a}+(r-2)h\pm c_{(1)}\pm\cdots\pm c_{(r)}}{2}}],\\
\checked{}\label{eq:treepropagatorpoles}\text{bulk-to-bulk propagator }(a):&\qquad\Gamma[\textstyle\frac{h\pm c_{(a)}-s_{(a)}}{2}].
\end{align}
We naturally expect a set of universal rules (at least for the poles) for arbitrary diagrams that reduce to the above when restricting to tree level.  With this regard, let us draw some observations at one loop.

\paragraph*{Vertex contributions.}

Contribution from each vertex follows the same rule as \eqref{eq:treevertexpoles}. Evidence from our examples includes $r=2$ for all the vertices on the loop, as well as the $r=1$ case $\Gamma[{\textstyle\frac{\Delta_{12}-h\pm c_1}{2}}]$ and the $r=3$ case $\Gamma[{\textstyle\frac{h\pm c_1\pm c_3\pm c_4}{2}}]$ in the 4-point triangle diagram with only cubic vertices, etc.

\paragraph*{Channel contributions.}

While we see the same rule \eqref{eq:treepropagatorpoles} holds for the tree propagator in a one-loop diagram, e.g., the factor $\Gamma[{\textstyle\frac{h\pm c_1-S}{2}}]$ in the 4-point triangle diagram with only cubic vertices, the notion of contribution from bulk-to-bulk propagators necessarily has to be extended to contribution from each OPE channel tied to the diagram. Here we include trivial channels where we separate just one boundary point from the rest.

In the most general situation, let us consider an OPE channel that separates the boundary points into two groups, whose corresponding Mandelstam variable we denote as $S$. For a given diagram this can be visualized as drawing a surface across the diagram so as to cut it into two \emph{connected} pieces, with the two groups of points on each side, as was discussed in Figure \ref{fig:bubblegeneralization} of Section \ref{sec:bubblereview}. For a specific choice of drawing let us assume this surface runs across $r$ bulk-to-bulk propagators, denoted as $\{(1),(2),\ldots,(r)\}$. Then each such drawing corresponds to poles of $M$ encoded in
\begin{equation}
\checked{}
\Gamma[{\textstyle\frac{r\,h\pm c_{(1)}\pm\cdots\pm c_{(r)}-S}{2}}].
\end{equation}
Pay attention that there can be different cuts associated to the same channel, and we need to enumerate all possible choices.

This rule obviously reduces to \eqref{eq:treepropagatorpoles} for tree propagators. In our existing one-loop examples it also passes all checks (except for a subtlety in the 4-pt triangle with only cubic vertices, as will be addressed later). For example, in the 4-point triangle diagram with a quartic vertex the $S$ channel poles take the form $\Gamma[\frac{2h\pm c_1\pm c_2-S}{2}]$. The same diagram also contains two other trivial channels, corresponding to cutting the triangle in other ways, e.g., the one separating point 3 from the rest. In this case the Mandelstam variable just reduces to the conformal dimension $\Delta_3$, and correspondingly we have poles of the form $\Gamma[\frac{2h\pm c_2\pm c_3-\Delta_3}{2}]$. The same holds for other examples as well.

\paragraph*{Loop contraction.}

An essentially new feature arising in loop-level diagrams is extra families of poles associated to new diagrams induced from the diagram under study by contracting loops. An obvious evidence is the 4-point triangle diagram with only cubic vertices, where we observe poles corresponding to $\Gamma[\frac{\Delta_{34}-h\pm c_1}{2}]$. This is not associated to any existing vertices nor propagators in the diagram. But if we contract the triangle to a bulk point such that the boundary points 3 and 4 and the propagator $(1)$ are attached to it, then these poles can be considered as arising from this emergent bulk vertex.

Intuitively one can think about the necessity of considering such loop-contracted diagrams as due to the fact that in the integration of the bulk interaction vertices over the AdS bulk there exists region where the three vertices on the triangle are kept coincident. This is naturally expected to hold for other one-loop diagrams as well as at higher loops.

As a further evidence, note that once we land on a tree diagram no further contractions are allowed. While the necklace diagrams contract to a contact diagram, the 4-point triangle with only cubic vertices has to end up at an exchange diagram. Correspondingly, poles associated to $\Gamma[\frac{\Sigma\Delta}{2}-h]$ exist always in the former but are absent in the latter.

Consequently, we summarize the following loop contraction rule for arbitrary diagrams: for a given diagram we consider all possible ways of contracting the loops in it, and for every choice of contraction the newly emergent vertex leads to poles of the diagram that follows the same rule \eqref{eq:treevertexpoles} in the vertex contributions.

\vspace{.5em}

From the above discussions we can draw an empirical guess for the structure of the genuine poles of scalar Witten diagrams at arbitrary loop level:
\begin{conjecture}{Guess}
For an arbitrary scalar Witten diagram, we work out all the poles following the above three rules: (1) vertex rule, (2) channel rule, (3) loop contraction rule. Then these contain all the genuine poles of its corresponding Mellin pre-amplitude $M$.
\end{conjecture}

While this guess applies nicely to almost all the poles in the examples we have studied so far, one may already notice some subtle issues. These are listed below:
\begin{itemize}[noitemsep,nolistsep]
\item In the 4-point triangle diagram with only cubic vertices (Figure \ref{fig:triangle4ptC}), from cutting the propagators $3$ and $4$ in the $S$ channel we should read out four families of poles $\Gamma[\frac{2h\pm c_3\pm c_4-S}{2}]$. But these are absent from the final result \eqref{eq:poleresult4pttrianglecubicshort}. Although this does not contradict the above guess, this definitely indicates that it is not constraining enough.
\item In the 4-point bubble diagram (Figure \ref{fig:bubble4pt}) we observed four additional families of poles $\Gamma[h\pm c_1]\Gamma[h\pm c_2]$, which do not follow the rules provided above. 
\item In \eqref{eq:bubble3ptresult} we also worked out the explicit result for a 3-point bubble diagram. When comparing the poles therein with the above rules, we note that the families $\Gamma[\frac{\Delta_{12}\pm c_1\pm c_2}{2}]$ (induced by the quartic vertex) are missing. In addition, while this diagram is structurally similar to the 4-point bubble, it however does not possess poles $\Gamma[h\pm c_1]\Gamma[h\pm c_2]$.
\end{itemize}
These are the only facts in our existing examples that do not quite fit the above guess, and so we need to further modify and refine it.

\subsubsection{Pre-amplitudes for the generalized bubbles}\label{sec:Mgbubbles}

To gain a better understanding of the subtleties in the pole structure of pre-amplitudes, let us return to the generalized bubble diagrams, because some of the qualitative features should already appear in diagrams belonging to that special class.

This requires us to quantify our discussion in Section \ref{sec:bubblereview}. Let us assume two effective bulk-to-bulk propagators $G_1$ and $G_2$, defined by
\begin{equation}
\checked{}
G_a[X,Y]=\int\frac{\mathrm{d}c_a}{2\pi i}\,g_a[c_a]\,\Omega_{c_a}[X,Y],\quad
\Omega_c[X,Y]=\!\!\underset{\partial\text{AdS}}{\int}\!\!\mathrm{d}P\,\frac{\mathcal{N}_c}{(-2P\cdot X)^{h+c}(-2P\cdot Y)^{h-c}},
\end{equation}
for some spectral function $g_a[c_a]$ (in the special case of actual bulk-to-bulk propagators we have the identification $g_a[c_a]=1/((\underline\Delta_a-h)^2-c_a^2)$). We study their parallel and series products
\begin{align}
\checked{}G_1[X,Y]\,G_2[X,Y]&\equiv\int\frac{\mathrm{d}c}{2\pi i}\,g_{\{1,2\}}[c]\,\Omega_a[X,Y],\\
\checked{}\underset{\rm AdS}{\int}\mathrm{d}Y\,G_1[X,Y]\,G_2[Y,Z]&\equiv\int\frac{\mathrm{d}c}{2\pi i}\,g^{\{1,2\}}[c]\,\Omega_c[X,Z].
\end{align}
These products can again be treated as some effective propagator and be expanded on $\Omega_c$, and their corresponding spectral functions are related to $g_1$ and $g_2$ as
\begin{align}
\checked{}\label{eq:parallelprod}g_{\{1,2\}}[c]&=\frac{1}{8\pi^h}\int\frac{\mathrm{d}c_1\mathrm{d}c_2}{(2\pi i)^2}\,g_1[c_1]\,g_2[c_2]\,\frac{\Gamma[\frac{h\pm c\pm c_1\pm c_2}{2}]}{\Gamma[h]\Gamma[h\pm c]\Gamma[\pm c_1]\Gamma[\pm c_2]},\\
\checked{}\label{eq:seriesprod}g^{\{1,2\}}[c]&=g_1[c]\,g_2[c].
\end{align}
Details are reviewed in Appendix \ref{app:sec:bbpropagators}.

For concreteness, let us return to the explicit 3-loop example discussed before, in Figure \ref{fig:bubblegeneralization}. By  iterating \eqref{eq:parallelprod} and \eqref{eq:seriesprod}, it is straightforward to work out the spectral function for the effective propagator induced by all the six bulk-to-bulk propagators together
\begin{equation}
\checked{}
g_{\rm eff}[c]\equiv\frac{1}{(8\pi^h)^3}\int\prod_{a=1}^6\frac{[\mathrm{d}c_a]_{\underline\Delta_a}}{\Gamma[\pm c_a]}\,\tilde{g}_{\rm eff}[c],
\end{equation}
with
\begin{equation}\label{eq:geffexample}
\begin{split}\checked{}
\tilde{g}_{\rm eff}[c]
=&\frac{1}{\Gamma[h]^3}\frac{\Gamma[\frac{h\pm c_4\pm c_5\pm c_6}{2}]}{\Gamma[h\pm c_4]}\,
\frac{\Gamma[\frac{h\pm c_2\pm c_3\pm c_4}{2}]}{\Gamma[h\pm c_2]}\,
\frac{\Gamma_1[\frac{h\pm c\pm c_1\pm c_2}{2}]}{\Gamma[h\pm c]}.
\end{split}
\end{equation}
Taking into consideration the Mellin pre-amplitude of a single exchange \eqref{eq:Mexchange} (assuming the total conformal dimension of boundary operators on the two sides to be $\Delta_{A_0}$ and $\Delta_{A_1}$ again)
\begin{equation}\label{eq:Mexchange2}
\begin{split}\checked{}
M_{\rm exchange}[c]&=\frac{\Gamma_2[\frac{\Delta_{A_0}-h\pm c}{2}]}{\Gamma[\frac{\Delta_{A_0}-S}{2}]}
\Gamma_3[{\textstyle\frac{h\pm c-S}{2}}]
\frac{\Gamma_4[\frac{\Delta_{A_1}-h\pm c}{2}]}{\Gamma[\frac{\Delta_{A_1}-S}{2}]},
\end{split}
\end{equation}
the pre-amplitude of the generalized bubble shown in Figure \ref{fig:bubblegeneralization} is then
\begin{equation}\label{eq:Mfromeffectivespectrum}
\checked{}
M=\int\frac{\mathrm{d}c}{2\pi i}\,\frac{\tilde{g}_{\rm eff}[c]}{\Gamma[\pm c]}\,M_{\rm exchange}[c].
\end{equation}
So even though we are studying a 3-loop diagram there is only one remaining Mellin integral in the above representation. While it is not hard to further move on to its Mellin amplitude and expand it in terms of a summation over physical poles in $S$, here we are more interested in the pole structure of $M$. Following the analysis in Section \ref{sec:polealgorithm}, in \eqref{eq:geffexample} and \eqref{eq:Mexchange2} we labeled the $\Gamma$ functions that depend on $c$ (here it suffices to use a single label for a bunch of $\Gamma$'s that differ only by signs of the $c$'s, because poles from the $c$ integral cannot emerge from poles with the same label due to the presence of $\Gamma[\pm c]$ in \eqref{eq:Mfromeffectivespectrum}). Then we can easily conclude with the following poles after the $c$ integration
\begin{equation}\label{eq:generalizedbubblepoleseg}
\begin{split}\checked{}
&\underset{\{1,2\}}{\Gamma[{\textstyle\frac{\Delta_{A_0}\pm c_1\pm c_2}{2}}]}\,
\underset{\{1,3\}}{\Gamma[{\textstyle\frac{2h\pm c_1\pm c_2-S}{2}}]}\,
\underset{\{1,4\}}{\Gamma[{\textstyle\frac{\Delta_{A_1}\pm c_1\pm c_2}{2}}]}\,
\underset{\{2,4\}}{\Gamma[{\textstyle\frac{\Delta_{A_0}+\Delta_{A_1}}{2}-h}]}\\
&\times\Gamma[{\textstyle\frac{h\pm c_4\pm c_5\pm c_6}{2}}]\,
\Gamma[{\textstyle\frac{h\pm c_2\pm c_3\pm c_4}{2}}].
\end{split}
\end{equation}
Na\"ively there are also poles $\Gamma_{\{2,3\}}[\frac{\Delta_{A_0}-S}{2}]$ and $\Gamma_{\{3,4\}}[\frac{\Delta_{A_1}-S}{2}]$, but obviously they are killed by the denominator in \eqref{eq:Mexchange2}.

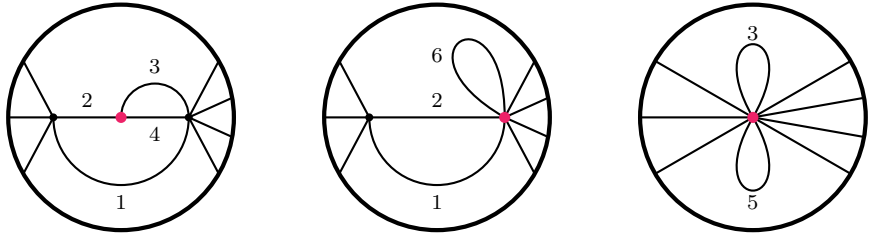
\begin{figure}[ht]
\captionsetup{margin=2em}
\begin{center}
\begin{tikzpicture}
\begin{scope}[scale=.6]
\begin{scope}[xshift=-7cm]
\draw [black,ultra thick] (0,0) circle [radius=2.5];
\draw [black,thick] (-1.5,0) -- (150:2.5);
\draw [black,thick] (-1.5,0) -- (180:2.5);
\draw [black,thick] (-1.5,0) -- (210:2.5);
\draw [black,thick] (1.5,0) -- (30:2.5);
\draw [black,thick] (1.5,0) -- (10:2.5);
\draw [black,thick] (1.5,0) -- (-10:2.5);
\draw [black,thick] (1.5,0) -- (-30:2.5);
\draw [black,thick] (-1.5,0) -- (1.5,0);
\draw [black,thick] (180:1.5) arc [start angle=180,end angle=360,radius=1.5] arc [start angle=0,end angle=180,radius=.75];
\fill [black] (-1.5,0) circle [radius=2.5pt];
\fill [WildStrawberry] (0,0) circle [radius=3.5pt];
\fill [black] (1.5,0) circle [radius=2.5pt];
\node [anchor=north] at (0,-1.5) {\scriptsize $1$};
\node [anchor=south] at (-.75,0) {\scriptsize $2$};
\node [anchor=south] at (.75,.75) {\scriptsize $3$};
\node [anchor=north] at (.75,0) {\scriptsize $4$};
\end{scope}
\begin{scope}
\draw [black,ultra thick] (0,0) circle [radius=2.5];
\draw [black,thick] (-1.5,0) -- (150:2.5);
\draw [black,thick] (-1.5,0) -- (180:2.5);
\draw [black,thick] (-1.5,0) -- (210:2.5);
\draw [black,thick] (1.5,0) -- (30:2.5);
\draw [black,thick] (1.5,0) -- (10:2.5);
\draw [black,thick] (1.5,0) -- (-10:2.5);
\draw [black,thick] (1.5,0) -- (-30:2.5);
\draw [black,thick] (-1.5,0) -- (1.5,0);
\draw [black,thick] (180:1.5) arc [start angle=180,end angle=360,radius=1.5] .. controls (1.5,3) and ($(1.5,0)+(150:3)$) .. (1.5,0);
\fill [black] (-1.5,0) circle [radius=2.5pt];
\fill [WildStrawberry] (1.5,0) circle [radius=3.5pt];
\node [anchor=north] at (0,-1.5) {\scriptsize $1$};
\node [anchor=south] at (0,0) {\scriptsize $2$};
\node [anchor=south] at (0,1) {\scriptsize $6$};
\end{scope}
\begin{scope}[xshift=7cm]
\draw [black,ultra thick] (0,0) circle [radius=2.5];
\draw [black,thick] (0,0) -- (150:2.5);
\draw [black,thick] (0,0) -- (180:2.5);
\draw [black,thick] (0,0) -- (210:2.5);
\draw [black,thick] (0,0) -- (30:2.5);
\draw [black,thick] (0,0) -- (10:2.5);
\draw [black,thick] (0,0) -- (-10:2.5);
\draw [black,thick] (0,0) -- (-30:2.5);
\draw [black,thick] (0,0) .. controls (60:2.5) and (120:2.5) .. (0,0) .. controls (-60:2.5) and (-120:2.5) .. (0,0);
\fill [WildStrawberry] (0,0) circle [radius=3.5pt];
\node [anchor=south] at (0,1.5) {\scriptsize $3$};
\node [anchor=north] at (0,-1.5) {\scriptsize $5$};
\end{scope}
\end{scope}
\end{tikzpicture}
\end{center}
\vspace{-.5em}\caption{Contraction of loops in the generalized bubble of Figure \ref{fig:bubblegeneralization}.}
\label{fig:bubblecontraction}
\end{figure}

Let us look into the poles worked out in \eqref{eq:generalizedbubblepoleseg} more closely, especially in comparison to what the empirical guess in Section \ref{sec:preliminaryobservations} suggests. It is obvious in Figure \ref{fig:bubblegeneralization} that this diagram contains four vertices and three possible cuts in its unique $S$ channel. Then the vertex rule and the channel rule in Section \ref{sec:preliminaryobservations} indicate the following possible poles
\begin{align}
\checked{}\label{eq:verticespoleseg}\text{vertices:}&\quad\Gamma[{\textstyle\frac{\Delta_{A_0}\pm c_1\pm c_2}{2}}]\,
{\color{ForestGreen}\Gamma[{\textstyle\frac{2h\pm c_2\pm c_3\pm c_5\pm c_6}{2}}]}\,
\Gamma[{\textstyle\frac{h\pm c_4\pm c_5\pm c_6}{2}}]\,
{\color{ForestGreen}\Gamma[{\textstyle\frac{\Delta_{A_1}\pm c_1\pm c_3\pm c_4}{2}}]};\\
\checked{}\label{eq:cutspoleseg}\text{cuts:}&\quad
\Gamma[{\textstyle\frac{2h\pm c_1\pm c_2-S}{2}}]\,
{\color{ForestGreen}\Gamma[{\textstyle\frac{4h\pm c_1\pm c_3\pm c_5\pm c_6-S}{2}}]}\,
{\color{ForestGreen}\Gamma[{\textstyle\frac{3h\pm c_1\pm c_3\pm c_4-S}{2}}]}.
\end{align}
Among these the poles marked green are however absent in \eqref{eq:generalizedbubblepoleseg}. Let us next check various contractions of the existing loops into vertices. The three possible contraction of single loops are shown in Figure \ref{fig:bubblecontraction} (marked red).
The three emergent vertices from these contractions indicate the poles
\begin{equation}
\text{loop contractions:}\quad
\Gamma[{\textstyle\frac{h\pm c_2\pm c_3\pm c_4}{2}}]\,
\Gamma[{\textstyle\frac{\Delta_{A_1}\pm c_1\pm c_2}{2}}]\,
\Gamma[{\textstyle\frac{\Delta_{A_0}+\Delta_{A_1}}{2}-h}],
\end{equation}
all of which appear in the result \eqref{eq:generalizedbubblepoleseg}. Furthermore, all the poles in \eqref{eq:generalizedbubblepoleseg} are captured from the above rules, thus verifying our empirical guess in this specific case. Here notice that, e.g., in the middle diagram although there is a tadpole attached to the emergent vertex it does not contribute to the corresponding $\Gamma$ function. In principle we should consider contractions of more loops, but apparently they do not induce new emergent vertices.

Already in this case of generalized bubble diagram we observe similar phenomena of absent poles as pointed out in the previous subsubsection, which is stronger than what the empirical guess suggests. There are both poles associated to the cuts in \eqref{eq:cutspoleseg} and those associated to the vertices in \eqref{eq:verticespoleseg} that are absent.

It is in fact simple to observe the qualitative difference between these absent poles and the poles that genuinely exist: they can further emerge from the genuine poles by performing Mellin integrations on some other variables, while the genuine poles do not. In other words, these poles are ``composite''\footnote{This notion is well-defined, because the argument in the $\Gamma$ function for the emergent poles are purely summation of arguments in the original $\Gamma$'s in the integrand, and no subtraction ever occur.}. In this particular example they arise from the spectrum integrations. Explicitly, the four families of poles marked green in \eqref{eq:verticespoleseg} and \eqref{eq:cutspoleseg} emerge by
\begin{align}
\checked{}
\Gamma[{\textstyle\frac{h\pm c_2\pm c_3\pm c_4}{2}}]\,\Gamma[{\textstyle\frac{h\pm c_4\pm c_5\pm c_6}{2}}]&\xrightarrow{\int\mathrm{d}c_4}\Gamma[{\textstyle\frac{2h\pm c_2\pm c_3\pm c_5\pm c_6}{2}}],\\
\Gamma[{\textstyle\frac{\Delta_{A_1}\pm c_1\pm c_2}{2}}]\,\Gamma[{\textstyle\frac{h\pm c_2\pm c_3\pm c_4}{2}}]&\xrightarrow{\int\mathrm{d}c_2}\Gamma[{\textstyle\frac{\Delta_{A_1}\pm c_1\pm c_3\pm c_4}{2}}],\\
\Gamma[{\textstyle\frac{2h\pm c_1\pm c_2-S}{2}}]\,\Gamma[{\textstyle\frac{2h\pm c_2\pm c_3\pm c_5\pm c_6}{2}}]&\xrightarrow{\int\mathrm{d}c_2}\Gamma[{\textstyle\frac{4h\pm c_1\pm c_3\pm c_5\pm c_6-S}{2}}],\\
\Gamma[{\textstyle\frac{2h\pm c_1\pm c_2-S}{2}}]\,\Gamma[{\textstyle\frac{h\pm c_2\pm c_3\pm c_4}{2}}]&\xrightarrow{\int\mathrm{d}c_2}\Gamma[{\textstyle\frac{3h\pm c_1\pm c_3\pm c_4-S}{2}}].
\end{align}
Note that in the poles in the third line above actually emerge from two integrals, because one of the factors on LHS comes from the first line.

This helps solve part of the issues we had before. Going back to the example of 4-point triangle diagram with only cubic vertices, we see that the poles associated to the double cut can also be related to other existing poles via a spectrum integration
\begin{equation}
\checked{}
\Gamma[{\textstyle\frac{h\pm c_1-S}{2}}]\,\Gamma[{\textstyle\frac{h\pm c_1\pm c_3\pm c_4}{2}}]\xrightarrow{\int\mathrm{d}c_1}\Gamma[{\textstyle\frac{2h\pm c_3\pm c_4-S}{2}}].
\end{equation}
Hence they turn out to be absent. Furthermore, in the example of 3-point bubble \eqref{eq:bubble3ptresult} the poles read off from the quartic vertex are also absent due to the same reason, but these time by pretending that the conformal dimension $\Delta_3$ is integrated
\begin{equation}
\checked{}
\Gamma[{\textstyle\frac{2h\pm c_1\pm c_2-\Delta_3}{2}}]\,\Gamma[{\textstyle\frac{\Sigma\Delta}{2}-h}]\xrightarrow{\int\mathrm{d}\Delta_3}\Gamma[{\textstyle\frac{\Delta_{12}\pm c_1\pm c_2}{2}}].
\end{equation}

Now the only remaining issue is the presence of poles $\Gamma[h\pm c_1]\Gamma[h\pm c_2]$ in the 4-point bubble diagram as constructed in Figure \ref{fig:bubble4pt}, which do not arise from any rules proposed in Section \ref{sec:preliminaryobservations}. These seems to be some special feature tied to the generalized bubbles. In order to resolve this, we propose the \textbf{generalized bubble rule}: for any diagram or subdiagram that is a generalized bubble (i.e., can be reduced to an effective bulk-to-bulk propagator following \eqref{eq:parallelprod} and \eqref{eq:seriesprod}), we associated poles $\Gamma[h\pm c_a]$ to each propagator $a$ in it.

Of course the above simple proposal trivially address our previous issue with the 4-point bubble diagram. But in order that it is consistent with the rules we summarized before, we should also understand why the same poles are however absent in the 3-point bubble \eqref{eq:bubble3ptresult}. Very nicely, these are absent due to their compositeness, e.g.,
\begin{equation}
\checked{}
\Gamma[{\textstyle\frac{\Delta_3\pm c_1\pm c_2}{2}}]\,\Gamma[{\textstyle\frac{2h\pm c_1\pm c_2-\Delta_3}{2}}]\xrightarrow{\int\mathrm{d}c_2}\Gamma[h\pm c_1].
\end{equation}

\subsubsection{A universal conjecture}\label{sec:enhancedrules}

From the evidence and intuitions collected from the previous discussions in this subsection, at this point we are able to draw a precise conjecture on the complete pole structure of the Mellin pre-amplitude of a given diagram. 

Let us first emphasize again that, when we list out a set of $\Gamma$ functions that encode the poles of some function, they may include $\Gamma$ functions which can be considered as emerging from the remaining $\Gamma$ functions in the set via Mellin integrating some of the variables (which can be the spectrum variables, as well as the conformal dimensions of boundary points), and to distinguish such $\Gamma$ functions we call them \emph{composite}. With this notion, we have the following conjecture:
\begin{conjecture}{Conjecture on $M$}
For an arbitrary scalar Witten diagram, we work out all the poles following the four rules in Section \ref{sec:preliminaryobservations} and \ref{sec:Mgbubbles}: (1) vertex rule, (2) channel rule, (3) loop contraction rule, (4) generalized bubble rule. Among the resulting families of poles, after eliminating the composite ones, the remaining families exactly constitute all the genuine poles of the pre-amplitude of the given diagram.
\end{conjecture}

Note this conjecture is still purely empirical, and we do not seek for a general proof for it in this paper. However, to further support this conjecture we performed explicit computations on several more non-trivial examples of generalized bubble diagrams (using the method of effective propagators in Appendix \ref{app:sec:bbpropagators}), including the graph topologies
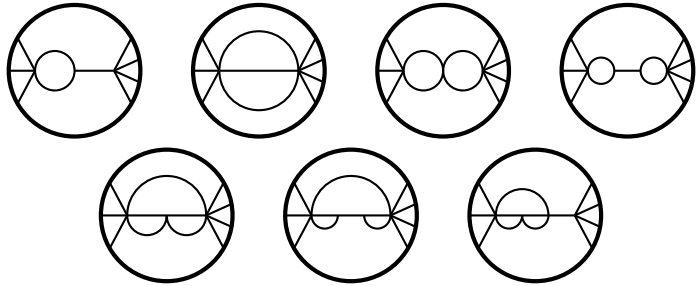
\begin{center}
\begin{tikzpicture}
\begin{scope}[scale=.35]
\begin{scope}
\draw [black,ultra thick] (0,0) circle [radius=2.5];
\draw [black,thick] (-1.5,0) -- (150:2.5);
\draw [black,thick] (-1.5,0) -- (180:2.5);
\draw [black,thick] (-1.5,0) -- (210:2.5);
\draw [black,thick] (1.5,0) -- (30:2.5);
\draw [black,thick] (1.5,0) -- (10:2.5);
\draw [black,thick] (1.5,0) -- (-10:2.5);
\draw [black,thick] (1.5,0) -- (-30:2.5);
\draw [black,thick] (-.75,0) circle [radius=.75];
\draw [black,thick] (0,0) -- (1.5,0);
\end{scope}
\begin{scope}[xshift=7cm]
\draw [black,ultra thick] (0,0) circle [radius=2.5];
\draw [black,thick] (-1.5,0) -- (150:2.5);
\draw [black,thick] (-1.5,0) -- (180:2.5);
\draw [black,thick] (-1.5,0) -- (210:2.5);
\draw [black,thick] (1.5,0) -- (30:2.5);
\draw [black,thick] (1.5,0) -- (10:2.5);
\draw [black,thick] (1.5,0) -- (-10:2.5);
\draw [black,thick] (1.5,0) -- (-30:2.5);
\draw [black,thick] (0,0) circle [radius=1.5];
\draw [black,thick] (-1.5,0) -- (1.5,0);
\end{scope}
\begin{scope}[xshift=14cm]
\draw [black,ultra thick] (0,0) circle [radius=2.5];
\draw [black,thick] (-1.5,0) -- (150:2.5);
\draw [black,thick] (-1.5,0) -- (180:2.5);
\draw [black,thick] (-1.5,0) -- (210:2.5);
\draw [black,thick] (1.5,0) -- (30:2.5);
\draw [black,thick] (1.5,0) -- (10:2.5);
\draw [black,thick] (1.5,0) -- (-10:2.5);
\draw [black,thick] (1.5,0) -- (-30:2.5);
\draw [black,thick] (-.75,0) circle [radius=.75];
\draw [black,thick] (.75,0) circle [radius=.75];
\end{scope}
\begin{scope}[xshift=21cm]
\draw [black,ultra thick] (0,0) circle [radius=2.5];
\draw [black,thick] (-1.5,0) -- (150:2.5);
\draw [black,thick] (-1.5,0) -- (180:2.5);
\draw [black,thick] (-1.5,0) -- (210:2.5);
\draw [black,thick] (1.5,0) -- (30:2.5);
\draw [black,thick] (1.5,0) -- (10:2.5);
\draw [black,thick] (1.5,0) -- (-10:2.5);
\draw [black,thick] (1.5,0) -- (-30:2.5);
\draw [black,thick] (-1,0) circle [radius=.5];
\draw [black,thick] (1,0) circle [radius=.5];
\draw [black,thick] (-.5,0) -- (.5,0);
\end{scope}
\begin{scope}[xshift=3.5cm,yshift=-5.5cm]
\draw [black,ultra thick] (0,0) circle [radius=2.5];
\draw [black,thick] (-1.5,0) -- (150:2.5);
\draw [black,thick] (-1.5,0) -- (180:2.5);
\draw [black,thick] (-1.5,0) -- (210:2.5);
\draw [black,thick] (1.5,0) -- (30:2.5);
\draw [black,thick] (1.5,0) -- (10:2.5);
\draw [black,thick] (1.5,0) -- (-10:2.5);
\draw [black,thick] (1.5,0) -- (-30:2.5);
\draw [black,thick] (-1.5,0) -- (1.5,0);
\draw [black,thick] (1.5,0) arc [start angle=0,end angle=180,radius=1.5] arc [start angle=180,end angle=360,radius=.75] arc [start angle=180,end angle=360,radius=.75];
\end{scope}
\begin{scope}[xshift=10.5cm,yshift=-5.5cm]
\draw [black,ultra thick] (0,0) circle [radius=2.5];
\draw [black,thick] (-1.5,0) -- (150:2.5);
\draw [black,thick] (-1.5,0) -- (180:2.5);
\draw [black,thick] (-1.5,0) -- (210:2.5);
\draw [black,thick] (1.5,0) -- (30:2.5);
\draw [black,thick] (1.5,0) -- (10:2.5);
\draw [black,thick] (1.5,0) -- (-10:2.5);
\draw [black,thick] (1.5,0) -- (-30:2.5);
\draw [black,thick] (-1.5,0) -- (1.5,0);
\draw [black,thick] (.5,0) arc [start angle=-180,end angle=0,radius=.5] arc [start angle=0,end angle=180,radius=1.5] arc [start angle=180,end angle=360,radius=.5];
\end{scope}
\begin{scope}[xshift=17.5cm,yshift=-5.5cm]
\draw [black,ultra thick] (0,0) circle [radius=2.5];
\draw [black,thick] (-1.5,0) -- (150:2.5);
\draw [black,thick] (-1.5,0) -- (180:2.5);
\draw [black,thick] (-1.5,0) -- (210:2.5);
\draw [black,thick] (1.5,0) -- (30:2.5);
\draw [black,thick] (1.5,0) -- (10:2.5);
\draw [black,thick] (1.5,0) -- (-10:2.5);
\draw [black,thick] (1.5,0) -- (-30:2.5);
\draw [black,thick] (-1.5,0) -- (1.5,0);
\draw [black,thick] (-.5,0) arc [start angle=-180,end angle=0,radius=.5] arc [start angle=0,end angle=180,radius=1] arc [start angle=180,end angle=360,radius=.5];
\end{scope}
\end{scope}
\end{tikzpicture}
\end{center}
The detailed results are summarized in Appendix \ref{app:sec:polegbubbles}, and they are match this conjecture perfectly.  Evidence from non-trivial diagrams at higher loops beyond generalized bubbles will be provided in Section \ref{sec:arbitraryloops}.

\newpage

\section{Residues and the Spectrum}\label{sec:residuecomputation}

\temp{No major revisions further needed except for marked places.}

In the previous section we examined the pole structures of Mellin pre-amplitudes, where a set of diagrammatic patterns for $M$ at arbitrary loops naturally emerge, and a universal conjecture was proposed at the end.

In this section we push the analysis forward and take very preliminary steps in studying the analytic structure of $\mathcal{M}$.

The very first question regarding $\mathcal{M}$ is of course again about its pole structure.  In principle this follows the same line of analysis as we have performed on $M$ in the previous section, essentially because the spectrum integrals are of the Mellin type as well. However, the level of complexity in the spectrum integrals is in a sense higher than the ones we encountered during the recursive construction of $M$, as will be discussed in detail in Section \ref{sec:MtoM}.  This is one of the reasons why we carry out the strategy mentioned in Section \ref{sec:strategy} and postpone all the spectrum integrals until we obtain a good knowledge about the structure of $M$ itself.

Nevertheless, it is amusing that the emergence of poles of $\mathcal{M}$ in the physics channels, i.e., those in the Mandelstam variables of the diagram, appear to follow simple patterns despite of the complication in other poles.  This allows us to still straightforwardly analyze the residue at these poles.  In this paper we only aim at displaying how the analysis is carried out in general,  and draw some preliminary investigations of their structure and the consistency with unitarity. So we will mostly restrict our scope to the leading poles of the Mandelstam variables in $\mathcal{M}$.  By the qualitative estimation in Section \ref{sec:bubblereview} these should correspond to the primary double-trace operators composed of exchanging modes that appear in the OPE expansion in the given channel.


\subsection{From $M$ to $\mathcal{M}$}\label{sec:MtoM}

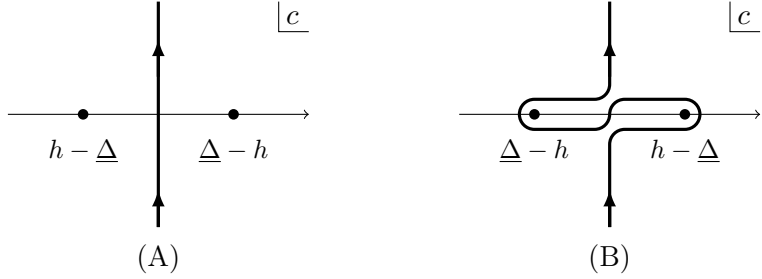
\begin{figure}[ht]
\captionsetup{margin=2em}
\begin{center}
\begin{tikzpicture}
\begin{scope}[xshift=-3cm]
\node [anchor=north] at (0,-1.6) {(A)};
\draw [black,->] (-2,0) -- (2,0);
\fill [black] (-1,0) circle [radius=2pt];
\fill [black] (1,0) circle [radius=2pt];
\node [anchor=north] at (-1,-.2) {\small $h-\underline\Delta$};
\node [anchor=north] at (1,-.2) {\small $\underline\Delta-h$};
\draw [black,very thick] (0,-1.5) -- (0,1.5);
\draw [black,very thick,-latex] (0,-1.5) -- +(0,.5);
\draw [black,very thick,-latex] (0,.5) -- +(0,.5);
\node [anchor=north east] at (2,1.5) {$c$};
\draw [black] (2,1.1) -- ++(-.4,0) -- ++(0,.4);
\end{scope}
\begin{scope}[xshift=3cm]
\node [anchor=north] at (0,-1.6) {(B)};
\draw [black,->] (-2,0) -- (2,0);
\fill [black] (-1,0) circle [radius=2pt];
\fill [black] (1,0) circle [radius=2pt];
\node [anchor=north] at (-1,-.2) {\small $\underline\Delta-h$};
\node [anchor=north] at (1,-.2) {\small $h-\underline\Delta$};
\draw [black,very thick] (0,-1.5) -- (0,-.4) arc [start angle=180,end angle=90,radius=.2] -- (1,-.2) arc [start angle=-90,end angle=90,radius=.2] -- (.2,.2) arc [start angle=90,end angle=180,radius=.2] arc [start angle=0,end angle=-90,radius=.2] -- (-1,-.2) arc [start angle=-90,end angle=-270,radius=.2] -- (-.2,.2) arc [start angle=-90,end angle=0,radius=.2] -- (0,1.5);
\draw [black,very thick,-latex] (0,-1.5) -- +(0,.5);
\draw [black,very thick,-latex] (0,.5) -- +(0,.5);
\node [anchor=north east] at (2,1.5) {$c$};
\draw [black] (2,1.1) -- ++(-.4,0) -- ++(0,.4);
\end{scope}
\end{tikzpicture}
\end{center}
\vspace{-1.5em}\caption{Contour for the spectrum integral: (A) $\underline\Delta-h>0$; (B) $\underline\Delta-h<0$.}
\label{fig:ccontour}
\end{figure}

Let us begin by commenting on some general features about the relation between $M$ and $\mathcal{M}$, which are bridged by the spectrum integrals \eqref{eq:Mellinnormalization}
\begin{equation}
\checked{}
\mathcal{M}=\int\mathcal{N}\,M,\qquad
\mathcal{N}=\frac{\pi^{(1-L)h}}{2^{2V+L-1}}
\prod_{i=1}^n\frac{\mathcal{C}_{\Delta_i}}{\Gamma[\Delta_i]}\prod_{a}\frac{[\mathrm{d}c_a]_{\underline\Delta_a}}{\Gamma[\pm c_a]}.
\end{equation}
In order to extract information about the poles in $\mathcal{M}$, note that the measure $\mathcal{N}$ introduces additional poles in the spectrum variables
\begin{equation}\label{eq:extracpoles}
\checked{}
\prod_{a}\frac{1}{(\underline\Delta_a-h)^2-c_a^2}\equiv\prod_a\frac{\Gamma[(\underline\Delta_a-h)\pm c_a]}{\Gamma[(\underline\Delta_a-h)\pm c_a+1]},
\end{equation}
as well as additional zeros $\prod_a\frac{1}{\Gamma[\pm c_a]}$. The re-writing on RHS above is merely a reminder of how the $c_a$ contour should be properly specified as a Mellin integral. In other words, as illustrated in Figure \ref{fig:ccontour}, here we should regard the pole $\frac{1}{(\underline\Delta_a-h)+c_a}$ as always sitting to the left of the $c_a$ contour while the pole $\frac{1}{(\underline\Delta_a-h)-c_a}$ as always to the right, regardless of the sign of $\underline\Delta_a-h$. This is equivalent to picking up the proper monodromy in the shadow formalism for the contribution of a specific operator to the OPE \cite{SimmonsDuffin:2012uy}.
In practice, as we apply Algorithm \ref{app:sec:polealgoritm} to these spectrum integrals, for each $a$ it suffices to book-keep the corresponding two poles as forming two $\gamma$ families on their own $\gamma[(\underline\Delta_a-h)+c_a]\gamma[(\underline\Delta_a-h)-c_a]$, and we just need to keep in mind that despite of the notation each family in fact consists of only a unique pole. 

Since from the previous section we have gained a detailed understanding of the entire pole structure of the pre-amplitude $M$, when analyzing the poles of $\mathcal{M}$ from the $c$ integrals it is most convenient to treat $M$ as being literally the product of $\Gamma$ function associated to its poles, multiplying an extra pole-free ``correction function'' that compensates whatever difference between the actual $M$ and this na\"ive product.  This helps quickly list out the set of $\gamma$ families after the $c$ integrals. Of course, as before, afterwards we should inspect the possible absence of some of the $\gamma$ families due to extra zeros from the rest of the integrand at the corresponding pinching.

While in principle this analysis fully parallels that of $M$, it is in practice more involved.  Attentive readers may have already noticed that the pre-amplitudes enjoy fair amount of simplicity, because the poles in different $\gamma$ families almost never overlap each other (except for those resulting in $\Gamma[h]$). This guarantees that each family of poles are always simple poles, and they are eliminated once we find the rest of the integrand vanishes at the pinching.  

In comparison, however, from the $c$ integrals we will observe that the same family of poles frequently arise from a bunch of different $\gamma$ families, i.e., from the shrinking of different simplices formed by the singularities of the integrand (following the discussion in Appendix \ref{app:sec:pinchinggeometry}).  In this situation, we need to further inspect the geometry of the pinching planes associated to the entire set of these $\gamma$ families.  If it so happens that the pinching planes have empty intersection with each other, then the poles in the family stays to be simple poles and they merely receive additive contribution from each pinching configurations (i.e., they arise because the contour is pinched at several \emph{distinct} sites).  If, on the other hand, there turn out to be some non-trivial intersections, then the poles are necessarily at some higher order determined by the detailed geometry.   Of course, in the end we again need to check possible zeros from the rest of the integrand at the pinching, but here in general the order of zero will be relevant.  The necessary technical details are discussed in Appendix \ref{app:sec:residues}.

In this section we do not dive into the details about this analysis in actual applications.  Instead, we summarize the final results on the pole structures of $\mathcal{M}$ for the examples we discussed so far in Appendix \ref{app:sec:resultpolesofamplitudes}. There in particular, below each family of poles we also list out the set of distinct pinching configurations that contribute to it.

\subsection{Poles in the Mandelstam variables}\label{sec:polesinMandelstam}

Although the way that poles of $\mathcal{M}$ emerges from $M$ via the spectrum integrals is in general involved, the good news is that the emergence of poles of the Mandelstam variables in $\mathcal{M}$ stays relatively simple and even turns out to follow some generic pattern, which we describe in this subsection.

Here our discussion is based on the assumption that Conjecture \ref{sec:enhancedrules} is valid. We do not seek for a proof in this paper, but merely summarize empirical observations. While detailed verification following Appendix \ref{app:sec:residues} is needed in order for complete rigor, in many explicit examples it is relatively easy to see that they hold.

For our purpose it suffices to focus on a specific OPE channel. Let us again call its corresponding Mandelstam variable $S$. As we separate a given diagram in this channel we may usually encounter different choices of combinations of bulk-to-bulk propagators bridging the two parts, as can be represented by the graphical cut discussed in Section \ref{sec:bubblereview}, and each combination indicates a multi-trace operator that in principle may contribute to the OPE expansion of the correlator. It can be verified that the poles in the Mandelstam variables are always at locations in correspondence to the twist of these multi-trace operators, i.e.
\begin{equation}\label{eq:cutpolesinM}
\checked{}
\Gamma[{\textstyle\frac{\underline\Delta_1+\cdots+\underline\Delta_r-S}{2}}]
\end{equation}
for some positive integer $r$.

The cuts under consideration also have their counterparts at the level of the pre-amplitude 
\begin{equation}\label{eq:cutpolesinpreM}
\checked{}
\Gamma[{\textstyle\frac{rh\pm c_1\pm\cdots\pm c_{r}-S}{2}}],
\end{equation}
as explicitly dictated by the channel rule. This gives rise to a potential correspondence between poles in $M$ and poles in $\mathcal{M}$ for any cut. However, some of these poles in $M$ may turn out to be absent due to compositeness. Hence in discussing the poles of $\mathcal{M}$ in the Mandelstam variables we need to distinguish between two relatively different situations in general. Let us call the poles \emph{minimal} if the correspondence between \eqref{eq:cutpolesinM} and \eqref{eq:cutpolesinpreM} indeed exists (and we also call their corresponding cuts \emph{minimal}), and otherwise \emph{non-minimal}.

\subsubsection{Minimal poles}\label{sec:minimalpoles}

For minimal poles, their corresponding poles \eqref{eq:cutpolesinpreM} in $M$ exist. In this case the poles \eqref{eq:cutpolesinM} of $\mathcal{M}$ arise from the spectrum integrals $\int\mathrm{d}c_1\cdots\mathrm{d}c_{r}$ by colliding the poles in the above with either of the two poles of $\frac{1}{(\underline\Delta_a-h)^2-c_a^2}$ for each $c_a$. So these poles always emerge after and only after $r$ spectrum integrations. 

Here note that the same physical pole always arises in several different ways of contractions. For example, in the simplest case when we have just a single propagator, then the poles of $\Gamma[\frac{\underline\Delta_1-S}{2}]$ emerge from both
\begin{equation}
\begin{split}
\checked{}&\Gamma[{\textstyle\frac{h+c_1-S}{2}}]\,\frac{1}{(\underline\Delta_1-h)-c_1}\xrightarrow{\int\mathrm{d}c_1}\Gamma[{\textstyle\frac{\underline\Delta_1-S}{2}}],\\
\checked{}&\Gamma[{\textstyle\frac{h-c_1-S}{2}}]\,\frac{1}{(\underline\Delta_1-h)+c_1}\xrightarrow{\int\mathrm{d}c_1}\Gamma[{\textstyle\frac{\underline\Delta_1-S}{2}}].
\end{split}
\end{equation}
Obviously, in the generic case of $r_{\rm min}$ propagators the physical poles always emerge from $2^{r_{\rm min}}$ ways of contractions. As is obvious from the localized values of the relevant $c$ variables at the pinching, the pinching planes do not have any non-empty intersection with each other at all. Hence the poles are simple poles, and as we compute the residues at these physical poles, we should always sum up contributions over all of them.  

In practice, due to the symmetry under $c\leftrightarrow-c$ it is easy to observe that the contribution to the residue from each pinching have to be all identical, and so it suffices to compute only one of them as long as we also take care of the multiplicity $2^{r_{\rm min}}$.

\subsubsection{Non-minimal poles}\label{sec:nonminimalpoles}

For non-minimal poles their corresponding poles in $M$ according to the channel rule are absent. At first sight this seems to cause some puzzle regarding how these may possibly emerge from the spectrum integrals. To answer this, note that the poles of $M$ also follow diagrammatic rules associated to each bulk vertex, as well as to each emergent vertex from contracting loops, as described in Section \ref{sec:polediagrammaticrules}. It is these poles associated to the vertices that provides a bridge.

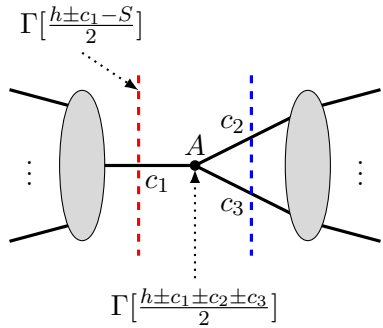
\begin{figure}[ht]
\captionsetup{margin=2em}
\begin{center}
\begin{tikzpicture}
\draw [black,very thick] (0,0) -- (1.5,0);
\draw [black,very thick] (3,-.75) -- (1.5,0) -- (3,.75);
\draw [black,very thick] (0,.75) -- +(165:1);
\draw [black,very thick] (0,-.75) -- +(-165:1);
\draw [black,very thick] (3,.75) -- +(15:1);
\draw [black,very thick] (3,-.75) -- +(-15:1);
\node [anchor=center] at (-.7,0) {$\vdots$};
\node [anchor=center] at (3.7,0) {$\vdots$};
\fill [black] (1.5,0) circle [radius=2pt];
\draw [black,fill=black!15!white] (0,1) .. controls (.2,1) and (.3,.5) .. (.3,0) .. controls (.3,-.5) and (.2,-1) .. (0,-1) .. controls (-.2,-1) and (-.3,-.5) .. (-.3,0) .. controls (-.3,.5) and (-.2,1) .. cycle;
\draw [black,fill=black!15!white] (3,1) .. controls (3.2,1) and (3.3,.5) .. (3.3,0) .. controls (3.3,-.5) and (3.2,-1) .. (3,-1) .. controls (2.8,-1) and (2.7,-.5) .. (2.7,0) .. controls (2.7,.5) and (2.8,1) .. cycle;
\draw [red,very thick,dashed] (.75,1.2) -- (.75,-1.2);
\draw [blue,very thick,dashed] (2.25,1.2) -- (2.25,-1.2); 
\node [anchor=north] at (1,0) {$c_1$};
\node [anchor=south] at (1.5,0) {$A$};
\node [anchor=south] at (2,.3) {$c_2$};
\node [anchor=north] at (2,-.3) {$c_3$};
\node [anchor=south] (c1) at (0,1.5) {$\Gamma[\frac{h\pm c_1-S}{2}]$};
\node [anchor=north] (A) at (1.5,-1.5) {$\Gamma[\frac{h\pm c_1\pm c_2\pm c_3}{2}]$};
\draw [black,thick,dotted,-latex] (c1.south) -- (.75,1);
\draw [black,thick,dotted,-latex] (A.north) -- (1.5,-.05);
\end{tikzpicture}
\end{center}
\vspace{-1.5em}\caption{Origin of non-minimal poles. The red dashed line denotes the cut corresponding to a minimal pole, and the blue one denotes that to a non-minimal pole.}
\label{fig:highertracepole}
\end{figure}

To see this more explicitly, let us consider the simplest case where the given channel allows a cut across a single propagator 1, as well as a cut across a pair of propagators 2 and 3, such that all these propagators are adjacent to the same vertex $A$, as shown in Figure \ref{fig:highertracepole}. Here in $M$ the single propagator cut gives rise to poles $\Gamma[\frac{h\pm c_1-S}{2}]$, and the vertex $A$ gives rise to poles $\Gamma[\frac{h\pm c_1\pm c_2\pm c_3}{2}]$. Although $M$ does not contain poles encoded in $\Gamma[\frac{2h\pm c_2\pm c_3-S}{2}]$ that are supposed to associate to the double propagator cut, they can precisely arise from the $c_1$ integration as the contraction of the former two. After this, the emergence of the physical poles $\Gamma[\frac{\underline\Delta_2+\underline\Delta_3-S}{2}]$ follow in exactly the same way as the minimal ones.

At this point we see the emergence of non-minimal poles necessarily involves more integrations, and are ultimately connected to the counterpart in $M$ of the minimal poles in the same channel.

\begin{figure}[ht]
\captionsetup{margin=2em}
\begin{center}
\begin{tikzpicture}
\begin{scope}
\node [anchor=north] at (1.2,-1.6) {(A)};
\draw [black,very thick] (0,1) -- (2.4,1);
\draw [black,very thick] (1.2,1) -- (2.4,0);
\draw [black,very thick] (0,-1) -- (2.4,-1);
\draw [black,very thick] (0,0) -- (1.2,-1);
\draw [red,very thick,dashed] (.2,1.4)  -- (.2,1) .. controls (.2,.5) and (2.2,-.5) .. (2.2,-1) -- (2.2,-1.4);
\draw [blue,very thick,dashed] (2.2,1.4) -- (2.2,1) .. controls (2.2,.5) and (.2,-.5) .. (.2,-1) -- (.2,-1.4);
\fill [black] (1.2,1) circle [radius=2pt];
\fill [black] (1.2,-1) circle [radius=2pt];
\node [anchor=south] at (.7,1) {$1$};
\node [anchor=south] at (1.7,1) {$3$};
\node [anchor=north] at (2,.3) {$4$};
\node [anchor=south] at (.4,-.3) {$5$};
\node [anchor=north] at (.7,-1) {$6$};
\node [anchor=north] at (1.7,-1) {$2$};
\end{scope}
\begin{scope}[xshift=4cm]
\node [anchor=north] at (1.5,-1.6) {(B)};
\draw [black,very thick] (0,0) -- (1,0) -- (3,1);
\draw [black,very thick] (1,0) -- (3,-1);
\draw [black,very thick] (2,.5) -- (2,-.5);
\fill [black] (1,0) circle [radius=2pt];
\fill [black] (2,.5) circle [radius=2pt];
\fill [black] (2,-.5) circle [radius=2pt];
\draw [black,very thick,dotted] (1.6,0) circle [radius=.8];
\draw [red,very thick,dashed] (.2,1.4) -- (.2,-1.4);
\draw [blue,very thick,dashed] (2.8,1.4) -- (2.8,-1.4);
\node [anchor=south] at (.5,0) {$1$};
\node [anchor=south] at (2.4,.7) {$2$};
\node [anchor=north] at (2.4,-.7) {$3$};
\node [anchor=south] at (1.5,.2) {$4$};
\node [anchor=north] at (1.5,-.2) {$5$};
\node [anchor=west] at (1.9,0) {$6$};
\node [anchor=center] (emerge) at (1.6,2) {$\Gamma[\frac{h\pm c_1\pm c_2\pm c_3}{2}]$};
\draw [black,very thick,dotted,-latex] (emerge.south) -- (1.6,.8);
\end{scope}
\end{tikzpicture}
\end{center}
\vspace{-1.5em}\caption{More examples of connections between minimal and non-minimal cuts (the diagrams here only indicates some sub-diagram inside some other diagrams). (A) structure between the two cut is disconnected; (B) structure between the two cuts involves loops.}
\label{fig:topologyinbetweencuts}
\end{figure}
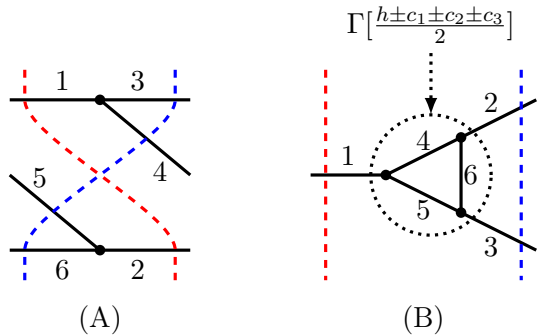

Generalizing the above discussion, it is obvious that the propagators involved in a non-minimal cut are always connected to the propagators involved in a minimal cut via a collection of bulk vertices. In the most generic situation we can talk about the sub-diagram structure sandwiched between a minimal cut and a non-minimal cut in the same channel, i.e., the part of the diagram that links the two corresponding cut surfaces. This structure can be either connected (as in Figure \ref{fig:highertracepole}) or disconnected (as in Figure \ref{fig:topologyinbetweencuts} (A)), and can even involve loops (as in Figure \ref{fig:topologyinbetweencuts} (B)).  Obviously, in the case of disconnected sub-diagram structure we basically follow the same treatment as before for every its connected component.

The case involving loops deserves some more comments. Here each individual vertices and propagators on the loop are not effective. Instead, we need to first contract any loop that appears into a single vertex. Recall our discussion in Section \ref{sec:polediagrammaticrules} that each such emergent vertex also indicates one family of genuine poles in $M$. The fully contracted structure is effectively tree-like, and we perform the same analysis using the $\Gamma$ functions associated to the emergent vertices. For instance, in Figure \ref{fig:topologyinbetweencuts} (B) we first contract the triangle loop into a vertex, for which there is a factor $\Gamma[\frac{h\pm c_1\pm c_2\pm c_3}{2}]$. Then this case exactly reduces to that of Figure \ref{fig:highertracepole}.

Before we end this subsection and move on to actual applications, let us further point out that in general in a given channel there can be more than one choice of minimal cuts. When this happens, in considering the emergence of non-minimal poles, we should take into consideration the connection between their corresponding cuts with every possible minimal cut (as long as the connection exists). Differing from the case of a unique minimal cut, here in general we should further inspect whether the pinching planes for different $\gamma$ families may have non-empty intersection or not, and perform the analysis accordingly following Appendix \ref{app:sec:residues}. An explicit example of such situation will be discussed in Section \ref{sec:doubletriangleresidueS}.

\subsection{Bubble diagrams}

Let us start with the bubble diagram constructed in Figure \ref{fig:bubble4pt}, which we can use for a consistency check, since the result is known in literature.

We are interested in computing the residue at the leading pole of the Mandelstam $S=\underline{\Delta}_{12}$. This has to be detected by pinching the $c_1$ and $c_2$ contours. To begin with, let us collect the factors from the result of the other integrations in \eqref{eq:bubble4ptpoles} that are relevant
\begin{equation}\label{eq:bubble4ptpolesduplicated}
\begin{split}\checked{}
\underset{\{1,3,4\}}{\Gamma[{\textstyle\frac{2h-S-c_1-c_2}{2}}]}\,
\underset{\{1,6,8\}}{\Gamma[{\textstyle\frac{2h-S-c_1+c_2}{2}}]}\,
\underset{\{2,3,4\}}{\Gamma[{\textstyle\frac{2h-S+c_1-c_2}{2}}]}\,
\underset{\{2,6,8\}}{\Gamma[{\textstyle\frac{2h-S+c_1+c_2}{2}}]}.
\end{split}
\end{equation}
In addition there are also poles in $c$'s from the normalization factor $\mathcal{N}$, which are (we label them as well)
\begin{equation}\label{eq:bubblenormpoles}
\checked{}
\gamma_{17}[(\underline{\Delta}_1-h)+c_1]\,
\gamma_{18}[(\underline{\Delta}_1-h)-c_1]\,
\gamma_{19}[(\underline{\Delta}_2-h)+c_2]\,
\gamma_{20}[(\underline{\Delta}_2-h)-c_2],
\end{equation}
Note these are the only source where the actual conformal dimension of each bulk-to-bulk propagator enters.

From the description in Section \ref{sec:minimalpoles} (and can also be checked explicitly, as shown in Appendix \ref{app:sec:resultpolesofamplitudes}) this same physical pole $S=\underline{\Delta}_{12}$ only arises from the four emergent $\gamma$ families
\begin{equation}
\checked{}
\gamma_{\{1,3,4,17,19\}}\,
\gamma_{\{1,6,8,17,20\}}\,
\gamma_{\{2,3,4,18,19\}}\,
\gamma_{\{2,6,8,18,20\}}.
\end{equation} 
So the residue is a summation of those computed by each of the above pinching families. It suffices to focus on the first one, $\gamma_{\{1,3,4,17,19\}}$. To do the computation we need to further reveal the previous integrals that lead to $\gamma_{\{1,3,4\}}$, which is indicated in the subscripts. So the set of factors in the original integrand responsible for this contribution to the pole $S=\underline{\Delta}_{12}$ are
\begin{equation}
\checked{}
\Gamma_1[{\textstyle\frac{h-c_1-t_1}{2}}],\quad
\Gamma_3[{\textstyle\frac{h-c_2+t_1-t_2}{2}}],\quad
\Gamma_4[{\textstyle\frac{t_2-S}{2}}],\quad
\frac{1}{(\underline{\Delta}_1-h)+c_1},\quad
\frac{1}{(\underline{\Delta}_2-h)+c_2}.
\end{equation}
As this $\gamma$ arises from a shrinking $4$-simplex, we just need to localize the relevant integrals to the five (leading) boundaries indicated above. While in general we can first transform it to a canonical simplex and then perform the computation as prescribed in Appendix \ref{app:sec:genericresiduecontour}, here it is also simple to straightforwardly write out the corresponding residue contour, which yields
\begin{equation}\label{eq:bubble4ptleadingresidue}
\begin{split}\checked{}
&(-1)^2\int\frac{\mathrm{d}t_3}{2\pi i}\,
\residue{S=\underline{\Delta}_{12}}\;
\residue{c_1=h-\underline{\Delta}_1}\;
\residue{c_2=h-\underline{\Delta_2}}\;
\residue{t_2=2h-c_1-c_2}\;
\residue{t_1=h-c_1}\;\mathcal{N}\,\frac{M_0\,K}{(-2)^3}\\
&=\frac{\pi^{2h}}{8}\prod_{i=1}^4\frac{\mathcal{C}_{\Delta_i}}{\Gamma[\Delta_i]}\,\frac{\mathcal{C}_{\underline\Delta_1}\mathcal{C}_{\underline\Delta_2}}{\Gamma[\underline\Delta_{12}]}\times
\underbrace{\Gamma[{\textstyle\frac{\underline\Delta_{12}+\Delta_{12}}{2}-h}]}_{\mathcal{O}_1\mathcal{O}_2[\mathcal{O}_{\underline1}\mathcal{O}_{\underline2}]}
\times\underbrace{\Gamma[{\textstyle\frac{\underline\Delta_{12}+\Delta_{34}}{2}-h}]}_{[\mathcal{O}_{\underline1}\mathcal{O}_{\underline2}]\mathcal{O}_3\mathcal{O}_4}.
\end{split}
\end{equation}
The other three $\gamma$ pinching all give rise to identical contributions, and so the actual residue $\text{Res}_{S=\underline{\Delta}_{12}}\mathcal{M}$ is four times the above result.  Here apart from a normalization factor we see that the residue factorizes into two amplitudes, each of which associates to a 3-point contact diagram where the new boundary point is a double-trace primary $[\mathcal{O}_{\underline1}\mathcal{O}_{\underline2}]_{0,0}$.


\subsection{4-point triangle with a quartic vertex}

We now study the 4-point triangle with a quartic vertex, where we are interested again in the pole $S=\underline\Delta_{12}$. Similar to the analysis for the bubble diagram, this pole arises only from the following poles in the pre-amplitude \eqref{eq:preMpolestriangle4pt}
\begin{equation}\label{eq:triangle4ptrelevant}
\checked{}
\underset{\{1,3,5,6,7,8\}}{\Gamma[{\textstyle\frac{2h-S-c_1-c_2}{2}}]}\,
\underset{\{1,4,5,6,7,8\}}{\Gamma[{\textstyle\frac{2h-S-c_1+c_2}{2}}]}\,
\underset{\{2,3,5,6,7,8\}}{\Gamma[{\textstyle\frac{2h-S+c_1-c_2}{2}}]}\,
\underset{\{2,4,5,6,7,8\}}{\Gamma[{\textstyle\frac{2h-S+c_1+c_2}{2}}]},
\end{equation}
by the collision of their poles to the extra poles 
\begin{equation}
\checked{}
\gamma_{25}[(\underline{\Delta}_1-h)+c_1]\,
\gamma_{26}[(\underline{\Delta}_1-h)-c_1]\,
\gamma_{27}[(\underline{\Delta}_2-h)+c_2]\,
\gamma_{28}[(\underline{\Delta}_2-h)-c_2],
\end{equation}
from the normalization $\mathcal{N}$. Again following Section \ref{sec:nonminimalpoles} they leads to four $\gamma$ families for the same physical pole
\begin{equation}
\checked{}
\gamma_{\{1, 3, 5, 6, 7, 8, 25, 27\}}\,
\gamma_{\{1, 4, 5, 6, 7, 8, 25, 28\}}\,
\gamma_{\{2, 3, 5, 6, 7, 8, 26, 27\}}\,
\gamma_{\{2, 4, 5, 6, 7, 8, 26, 28\}}.
\end{equation}
Take $\gamma_{\{1, 3, 5, 6, 7, 8, 25, 27\}}$ as an example; the relevant factors in the original integrand are
\begin{equation}
\begin{split}\checked{}
&\Gamma_1[{\textstyle\frac{h-c_1-t_1}{2}}],\quad
\Gamma_3[{\textstyle\frac{h-c_2-t_2}{2}}],\quad
\Gamma_5[{\textstyle\frac{h-c_3+t_2-t_3}{2}}],\quad
\Gamma_6[{\textstyle\frac{t_4-S}{2}}],\quad
\Gamma_7[{\textstyle\frac{h+c_3+t_1-t_5}{2}}],\\
&\Gamma_8[{\textstyle-2h+t_3-t_4+t_5}],\quad
\frac{1}{(\underline\Delta_1-h)+c_1},\quad
\frac{1}{(\underline\Delta_2-h)+c_2}.
\end{split}
\end{equation}
Note the $c_3$ integration remains unmodified in this residue computation. From the above list it is obvious the corresponding contribution to the residue is
\begin{equation}
\begin{split}\checked{}
&(-1)^5\,\residue{S=\underline\Delta_{12}}\;
\residue{c_2=h-\underline\Delta_2}\;
\residue{c_1=h-\underline\Delta_1}\;
\residue{t_5=2h-c_1+c_3}\;
\residue{t_4=S}\;
\residue{t_3=2h-c_2-c_3}\;
\residue{t_2=h-c_2}\;
\residue{t_1=h-c_1}\;\mathcal{N}\,\frac{M_0\,K}{(-2)^5}\\
&=\prod_{i=1}^4\frac{\mathcal{C}_{\Delta_i}}{\Gamma[\Delta_i]}\times\frac{\int[\mathrm{d}c_3]_{\underline\Delta_3}}{\Gamma[\pm c_3]}\,\frac{\Gamma[\frac{\underline\Delta_1\pm(h-\Delta_4)\pm c_3}{2}]\Gamma[\frac{\underline\Delta_2\pm(h-\Delta_3)\pm c_3}{2}]\Gamma[\frac{\Delta_{12}\pm(\underline\Delta_1-\underline\Delta_2)}{2}]\Gamma[\frac{\Delta_{12}\pm(2h-\underline\Delta_{12})}{2}]}{-2\,\Gamma[\underline\Delta_1-h+1]\Gamma[\underline\Delta_2-h+1]\Gamma[h+\frac{\underline\Delta_{12}-\Delta_{34}}{2}]\Gamma[\frac{\underline\Delta_{12}\pm(\Delta_3-\Delta_4)}{2}]}C_{A_2}.
\end{split}
\end{equation}
Here we have not yet explicitly computed the vertex correction $C_{A_2}$. Since $C_{A_2}$ has no poles, it suffices to directly evaluate it at the values of the variables determined above.

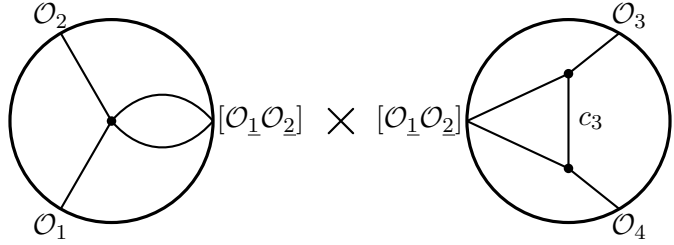
\begin{figure}[ht]
\captionsetup{margin=2em}
\begin{center}
\begin{tikzpicture}
\begin{scope}[scale=.9]
\draw [black,very thick] (0,0) circle [radius=1.5];
\draw [black,thick] (120:1.5) -- (0,0) -- (-120:1.5);
\begin{scope}[xshift=.75cm]
\draw [black,thick] (0:.75) .. controls (60:.6) and (120:.6) .. (180:.75) .. controls (-120:.6) and (-60:.6) .. (0:.75);
\end{scope}
\fill [black] (0,0) circle [radius=2pt];
\node [anchor=center] at (120:1.8) {$\mathcal{O}_2$};
\node [anchor=center] at (-120:1.8) {$\mathcal{O}_1$};
\node [anchor=center] at (0:2.2) {$[\mathcal{O}_{\underline1}\mathcal{O}_{\underline2}]$};
\end{scope}
\node [anchor=center] at (3.05,0) {\huge $\times$};
\begin{scope}[scale=.9,xshift=6.75cm]
\draw [black,very thick] (0,0) circle [radius=1.5];
\draw [black,thick] (180:1.5) -- (90:.7) -- (60:1.5);
\draw [black,thick] (180:1.5) -- (-90:.7) -- (-60:1.5);
\draw [black,thick] (-90:.7) -- (90:.7);
\fill [black] (90:.7) circle [radius=2pt];
\fill [black] (-90:.7) circle [radius=2pt];
\node [anchor=center] at (180:2.2) {$[\mathcal{O}_{\underline1}\mathcal{O}_{\underline2}]$};
\node [anchor=center] at (60:1.8) {$\mathcal{O}_3$};
\node [anchor=center] at (-60:1.8) {$\mathcal{O}_4$};
\node [anchor=west] at (0,0) {$c_3$};
\end{scope}
\end{tikzpicture}
\end{center}
\caption{Leading residue of the 4-point triangle with a quartic vertex.}
\label{fig:quartictriangleleadingresidue}
\end{figure}

The contributions from the other three $\gamma$ families are of exactly the same form, but with the $C_{A_2}$ evaluated at different configurations. We can easily work out
\begin{equation}
\checked{}
\sum_{\substack{c_1=\pm(\underline\Delta_1-h)\\c_2=\pm(\underline\Delta_2-h)}}C_{A_2}\big|_{\substack{t_1=\underline\Delta_1\\t_2=\underline\Delta_2}}=\frac{1}{\Gamma[h+\frac{\Delta_{12}-\underline\Delta_{12}}{2}]\Gamma[\frac{\Delta_{12}\pm(\underline\Delta_1-\underline\Delta_2)}{2}]},
\end{equation}
which stays the same for all the four families. Hence in total we obtain
\begin{equation}\label{eq:quartictriangleleadingresidue}
\begin{split}
\residue{S=\underline\Delta_{12}}\mathcal{M}=&
-8\pi^{2h}\prod_{i=1}^4\frac{\mathcal{C}_{\Delta_i}}{\Gamma[\Delta_i]}\prod_{a=1}^2\frac{\mathcal{C}_{\underline\Delta_a}}{\Gamma[\underline\Delta_a]}\\
&\times\underbrace{\Gamma[{\textstyle\frac{\Delta_{12}+\underline\Delta_{12}}{2}-h}]}_{\mathcal{O}_1\mathcal{O}_2[\mathcal{O}_{\underline1}\mathcal{O}_{\underline2}]}\times
\underbrace{\int\frac{[\mathrm{d}c_3]_{\underline\Delta_3}}{\Gamma[\pm c_3]}\frac{\Gamma[\frac{\underline\Delta_1\pm(h-\Delta_4)\pm c_3}{2}]\Gamma[\frac{\underline\Delta_2\pm(h-\Delta_3)\pm c_3}{2}]}{\Gamma[h+\frac{\underline\Delta_{12}-\Delta_{34}}{2}]\Gamma[\frac{\underline\Delta_{12}\pm(\Delta_3-\Delta_4)}{2}]}}_{[\mathcal{O}_{\underline1}\mathcal{O}_{\underline2}]\mathcal{O}_3\mathcal{O}_4}.
\end{split}
\end{equation}
Apart from a normalization factor this naturally factorizes into two parts, each of which is the Mellin amplitude of a 3-point diagram that invovles a double-trace primary $[\mathcal{O}_{\underline1}\mathcal{O}_{\underline2}]_{0,0}$ on the boundary, as shown in Figure \ref{fig:quartictriangleleadingresidue}. The amplitude of the right diagram can be computed quite straightforwardly.

\subsection{4-point triangle with only cubic vertices}

For the 4-point triangle diagram with only cubic vertices, we are interested in two possible families of poles in $S$. One is related to cutting the single propagator 1, and the other to cutting the pair of propagators 3 and 4.

\subsubsection*{Leading minimal pole}

Let us first look at cutting the propagator 1. This is much like the residue of a tree diagram. The relevant factors in $M$ are \eqref{eq:poleresult4pttrianglecubic}
\begin{equation}
\checked{}
\underset{\{1,3\}}{\Gamma[{\textstyle\frac{h-S-c_1}{2}}]}\,
\underset{\{2,3\}}{\Gamma[{\textstyle\frac{h-S+c_1}{2}}]},
\end{equation}
which together with
\begin{equation}
\checked{}
\gamma_{32}[(\underline\Delta_1-h)+c_1]\,
\gamma_{33}[(\underline\Delta_1-h)-c_1]
\end{equation}
create poles at $\Gamma[\frac{\underline\Delta_1-S}{2}]$ from the $c_1$ integration. These involves two $\gamma$ families 
\begin{equation}
\checked{}
\gamma_{\{1,3,32\}}\,
\gamma_{\{2,3,33\}}.
\end{equation}
We focus on the contribution from $\gamma_{\{1,3,32\}}$ to the residue at the leading pole $S=\underline\Delta_1$, corresponding to which the set of original poles are
\begin{equation}
\checked{}
\Gamma_1[{\textstyle\frac{h-c_1-t_1}{2}}],\quad
\Gamma_3[{\textstyle\frac{t_1-S}{2}}],\quad
\frac{1}{(\underline\Delta_1-h)+c_1}.
\end{equation}
It can be directly verified that this contributes to the residue by
\begin{equation}\label{eq:triangle4ptCleadingminimalres}\checked{}
\begin{split}
&\residue{S=\underline\Delta_1}\int[\mathrm{d}c]_{\rm rest}[\mathrm{d}t]_{\rm rest}\,\residue{c_1=h-\underline\Delta_1}\,\residue{t_1=S}\,\mathcal{N}\,\frac{M_0\,K}{(-2)^5}\\
&=-\frac{\pi^h}{8}\prod_{i=1}^4\frac{\mathcal{C}_{\Delta_i}}{\Gamma[\Delta_i]}\frac{\mathcal{C}_{\underline\Delta_1}}{\Gamma[\underline\Delta_1]}\times\underbrace{\Gamma[{\textstyle\frac{\underline\Delta_1+\Delta_{12}}{2}-h}]}_{\mathcal{O}_1\mathcal{O}_2\mathcal{O}_{\underline1}}\times \int\prod_{a=2}^4\frac{[\mathrm{d}c_a]_{\underline\Delta_a}}{\Gamma[\pm c_a]}\,M^{(1)}_{\mathcal{O}_3\mathcal{O}_4\mathcal{O}_{\underline1}},
\end{split}
\end{equation}
where $M^{(1)}_{\mathcal{O}_3\mathcal{O}_4\mathcal{O}_{\underline1}}$ refers to the Mellin pre-amplitude of the 3-point triangle diagram of three single-trace scalar primaries with conformal dimensions $\{\Delta_3,\Delta_4,\underline\Delta_1\}$ and the propagators on the triangle the same as those of the original triangle. Here it is easy to observe this remaining triangle because there are four $t$ integrals left over, and the resulting expression exactly matches the representation for the 3-point triangle diagram in terms of 4-fold Mellin integrals that we obtained in \eqref{eq:Ktriangle4ptquartic} (after a proper re-labeling).

The other contribution, from $\gamma_{\{2,3,33\}}$, leads to exactly the same result, and so the actual residue at $S=\underline\Delta_1$ doubles the above expression.

\subsubsection*{Leading non-minimal pole}

Let us then move on to the leading pole associated to cutting the propagators $\underline3$ and $\underline4$, which is expected to correspond to a double-trace primary operator $[\mathcal{O}_{\underline3}\mathcal{O}_{\underline4}]_{0,0}$. As we confirmed in Appendix \ref{app:sec:checkfakepoletriangle} the poles $\Gamma[\frac{2h\pm c_3\pm c_4-S}{2}]$ are all absent in $M$, and so the poles $\Gamma[\frac{\underline\Delta_3+\underline\Delta_4-S}{2}]$ are non-minimal and have to arise in a more interesting way.

Following the discussions in Section \ref{sec:nonminimalpoles}, the relevant poles in $M$ are
\begin{equation}\checked{}
\begin{split}
&\underset{\{1,3\}}{\Gamma[{\textstyle\frac{h-S-c_1}{2}}]}\,
\underset{\{2,3\}}{\Gamma[{\textstyle\frac{h-S+c_1}{2}}]}\,
\Gamma_{18}[{\textstyle\frac{h- c_1- c_3- c_4}{2}}]\,
\Gamma_{19}[{\textstyle\frac{h+ c_1- c_3- c_4}{2}}]\,
\Gamma_{20}[{\textstyle\frac{h- c_1+ c_3- c_4}{2}}]\\
&\Gamma_{21}[{\textstyle\frac{h+ c_1+ c_3- c_4}{2}}]\,
\Gamma_{22}[{\textstyle\frac{h- c_1- c_3+ c_4}{2}}]\,
\Gamma_{23}[{\textstyle\frac{h+ c_1- c_3+ c_4}{2}}]\,
\Gamma_{24}[{\textstyle\frac{h- c_1+ c_3+ c_4}{2}}]\,
\Gamma_{25}[{\textstyle\frac{h+ c_1+ c_3+ c_4}{2}}].
\end{split}
\end{equation}
Note that the poles $\Gamma[{\textstyle\frac{h\pm c_1\pm c_3\pm c_4}{2}}]$ are from the original tree amplitude. Together with the additional poles from the normalization
\begin{equation}
\begin{split}\checked{}
&\gamma_{32}[(\underline\Delta_1-h)+c_1]\,
\gamma_{33}[(\underline\Delta_1-h)-c_1]\,
\gamma_{36}[(\underline\Delta_3-h)+c_3]\,
\gamma_{37}[(\underline\Delta_3-h)-c_3]\\
&\gamma_{38}[(\underline\Delta_4-h)+c_4]\,
\gamma_{39}[(\underline\Delta_4-h)-c_4]
\end{split}
\end{equation}
these creates altogether 8 $\gamma$ families contributing to the same physical poles
\begin{equation}
\begin{split}\checked{}
&\gamma_{\{1, 3, 19, 36, 38\}}\,
\gamma_{\{1, 3, 21, 37, 38\}}\,
\gamma_{\{1, 3, 23, 36, 39\}}\,
\gamma_{\{1, 3, 25, 37, 39\}}\\
&\gamma_{\{2, 3, 18, 36, 38\}}\,
\gamma_{\{2, 3, 20, 37, 38\}}\,
\gamma_{\{2, 3, 22, 36, 39\}}\,
\gamma_{\{2, 3, 24, 37, 39\}}.
\end{split}
\end{equation}
Their pinching planes do not intersect each other, as can be seen from the localized values for the spectrum variables. Without loss of generality let us focus on $\gamma_{\{1, 3, 19, 36, 38\}}$, which emerges from the original poles
\begin{equation}
\checked{}
\Gamma_1[{\textstyle\frac{h-c_1-t_1}{2}}],\quad
\Gamma_3[{\textstyle\frac{t_1-S}{2}}],\quad
\Gamma_{19}[{\textstyle\frac{h+c_1-c_3-c_4}{2}}],\quad
\frac{1}{(\underline\Delta_3-h)+c_3},\quad
\frac{1}{(\underline\Delta_4-h)+c_4}.
\end{equation}
Its corresponding pinching plane is defined by
\begin{equation}\label{eq:triangle3ptCdoublepinchingeg}
\checked{}
t_1=\underline\Delta_{34},\quad
c_1=h-\underline\Delta_{34},\quad
c_3=h-\underline\Delta_3,\quad
c_4=h-\underline\Delta_4,
\end{equation}
under the condition $S=\underline\Delta_{34}$. It is easy to see that apart from the vertex correction $C_{A_2}$ the remaining integrand stays finite at this pinching.  As for $C_{A_2}$, note that the pole under study originates in part from the pole $\Gamma_{19}[{\textstyle\frac{h+c_1-c_3-c_4}{2}}]$, which is tied to vertex $A_2$. By the formula for $C_{A_2}$ \eqref{eq:Corr3} we immediately see that the three integrals therein are localized at the above values. Specifically
\begin{equation}
\begin{split}\checked{}
C_{A_2}=&-\frac{1}{2}\,\residue{c_1=-h+c_3+c_4}\;\residue{w_4=-c_4}\;\residue{w_3=-c_3}\;\residue{w_1=0}\left(\Gamma[{\textstyle\frac{h+c_1-c_3-c_4}{2}}]\,c_{A_2}\right)\\
=&\frac{1}{\Gamma[h]\Gamma[h-c_3]\Gamma[h-c_4]\Gamma[c_3+c_4]\Gamma[\frac{2h-c_3-c_4-t_1}{2}]\Gamma[\frac{h+c_3-t_3}{2}]\Gamma[\frac{h+c_4-t_4}{2}]},
\end{split}
\end{equation}
which, however, vanishes at the pinching \eqref{eq:triangle3ptCdoublepinchingeg}. 

It can be verified similarly that the correction function $C_{A_2}$ vanishes as well at the pinching for every other seven $\gamma$ families. This indicates that the contour is not pinched at all, and so the triangle diagram under study is in fact free of this double-trace operator pole.

\subsection{Box diagrams}\label{eq:boxresidues}

Finally let us study the 4-point box constructed in Figure \ref{fig:box4pt}. This example involves some new feature as it contains poles in both $S$ and $T$ channels, and our construction does not make the symmetry between these two channels manifest.

\subsubsection*{A consistency check}

Before we explore the physical poles, let us inspect the residue of $M$ at the leading pole of $\Gamma_{\{12,17,18,19,20,24\}}[\frac{\Delta_{1234}}{2}-h]$. Since this family emerges from the contraction of the loop and thus respects the rotation and reflection symmetry of the diagram, its residue should in principle respects the same set of symmetries, which is non-trivial due to the asymmetric nature of our recursive construction.

As indicated in the subscript, the original $\Gamma$ functions that are relevant for this pole are
\begin{equation}
\begin{split}\checked{}
&\Gamma_{12}[{\textstyle\frac{-2h+t_5-t_6+t_7}{2}}]\,
\Gamma_{17}[{\textstyle\frac{t_1-t_5+t_6-\Delta_2}{2}}]\,
\Gamma_{18}[{\textstyle\frac{\Delta_{12}-t_1}{2}}]\,
\Gamma_{19}[{\textstyle\frac{t_6-t_7+t_8-\Delta_3}{2}}]\\
&\Gamma_{20}[{\textstyle\frac{\Delta_{23}-t_6}{2}}]\,
\Gamma_{24}[{\textstyle\frac{\Delta_{34}-t_8}{2}}].
\end{split}
\end{equation}
Hence the residue at $h=\frac{\Delta_{1234}}{2}$ can be computed by imposing the contour
\begin{equation}\label{eq:boxtotalres}
\checked{}
\residue{h=\frac{\Delta_{1234}}{2}}M=(-1)^5\int[\mathrm{d}t]_{\rm rest}\,\residue{h=\frac{\Delta_{1234}}{2}}\;
\residue{t_8=\Delta_{34}}\;
\residue{t_7=\Delta_2+t_8}\;
\residue{t_6=\Delta_{23}}\;
\residue{t_5=\Delta_1+t_6}\;
\residue{t_1=\Delta_{12}}\,\frac{M_0\,K}{(-2)^8}.
\end{equation}
Note that in this case the integrals inside the two correction functions $C_{A_2}$ and $C_{A_3}$ are still non-trivial. However, putting every together, it turns out that all the remaining integration (after the six residue contour integrals) can be performed one by one using Barnes' lemmas, resulting in purely a product of $\Gamma$ functions
\begin{equation}\label{eq:box4ptconsistencycheck}
\checked{}
\residue{h=\frac{\Delta_{1234}}{2}}M=
\frac{\Gamma[\frac{\Delta_1\pm c_1\pm c_2}{2}]\Gamma[\frac{\Delta_2\pm c_2\pm c_3}{2}]\Gamma[\frac{\Delta_3\pm c_3\pm c_4}{2}]\Gamma[\frac{\Delta_4\pm c_4\pm c_1}{2}]}{\Gamma[\frac{\Delta_{1234}}{2}]\prod_{i=1}^4\Gamma[\Delta_i]}.
\end{equation}
This fully meets our expectation regarding the symmetries. More interestingly, this residue is completely free of the Mandelstam variables $S$ and $T$.

\subsubsection*{$S$ channel}

Now let us look at the leading pole in the $S$ channel, i.e., at $S=\underline\Delta_{12}$. The relavent poles in $M$ are encoded in \eqref{eq:resultpolesM4pt}
\begin{equation}
\checked{}
\underset{\{1,4,8,9\}}{\Gamma[{\textstyle\frac{2h-S-c_1-c_3}{2}}]}\,
\underset{\{1,5,8,9\}}{\Gamma[{\textstyle\frac{2h-S-c_1+c_3}{2}}]}\,
\underset{\{4,11,13,14\}}{\Gamma[{\textstyle\frac{2h-S+c_1-c_3}{2}}]}\,
\underset{\{5,11,13,14\}}{\Gamma[{\textstyle\frac{2h-S+c_1+c_3}{2}}]}.
\end{equation}
Together with the poles from the normalization
\begin{equation}
\checked{}
\gamma_{37}[(\underline\Delta_1-h)+c_1]\,
\gamma_{38}[(\underline\Delta_1-h)-c_1]\,
\gamma_{41}[(\underline\Delta_3-h)+c_3]\,
\gamma_{42}[(\underline\Delta_3-h)-c_3],
\end{equation}
they yields four $\gamma$ families for the $S$ channel pole under study
\begin{equation}
\checked{}
\gamma_{\{1, 4, 8, 9, 37, 41\}}\,
\gamma_{\{1, 5, 8, 9, 37, 42\}}\,
\gamma_{\{4, 11, 13, 14, 38, 41\}}\,
\gamma_{\{5, 11, 13, 14, 38, 42\}}.
\end{equation}

Considering the contribution from the family $\gamma_{\{1, 4, 8, 9, 37, 41\}}$, the relevant original poles are
\begin{equation}
\begin{split}\checked{}
&\Gamma_1[{\textstyle\frac{t_1-S}{2}}],\quad
\Gamma_4[{\textstyle\frac{h-c_3-t_3}{2}}],\quad
\Gamma_8[{\textstyle\frac{h-c_1+t_4-t_5}{2}}],\quad
\Gamma_9[{\textstyle\frac{-t_1+t_3-t_4+t_5}{2}}],\\
&\frac{1}{(\underline\Delta_1-h)+c_1},\quad
\frac{1}{(\underline\Delta_3-h)+c_3}.
\end{split}
\end{equation}
which indicates that the residue contour to be used localizes $t_1$, $t_3$, $t_5$ (or $t_4$), $c_1$ and $c_3$. Hence the residue still have seven non-compact contour integrals to be performed (although one can verify that the two correction functions both reduce to simple products of $\Gamma$ functions). To analytically simplify the resulting expression is not an easy task, and so instead of calculating this case further we turn to the leading pole in the $T$ channel, which is equivalent to this case by the rotation symmetry of the diagram but turns out to be simpler to work with using the representation we obtained in \eqref{eq:Kbox4pt}.

\subsubsection*{$T$ channel}

For the leading pole in the $T$ channel, $T=\underline\Delta_{13}$, the relevant poles in $M$ are
\begin{equation}\label{eq:polesrelevantinT}
\begin{split}\checked{}
\underset{\{2,6,8,10,11,12\}}{\Gamma[{\textstyle\frac{2h-T-c_2-c_4}{2}}]}\,
\underset{\{2,7,8,10,11,12\}}{\Gamma[{\textstyle\frac{2h-T-c_2+c_4}{2}}]}\,
\underset{\{3,6,8,10,11,12\}}{\Gamma[{\textstyle\frac{2h-T+c_2-c_4}{2}}]}\,
\underset{\{3,7,8,10,11,12\}}{\Gamma[{\textstyle\frac{2h-T+c_2+c_4}{2}}]},
\end{split}
\end{equation}
which together with the additional ones from the normalization
\begin{equation}
\checked{}
\gamma_{39}[(\underline\Delta_2-h)+c_2]\,
\gamma_{40}[(\underline\Delta_2-h)-c_2]\,
\gamma_{43}[(\underline\Delta_4-h)+c_4]\,
\gamma_{44}[(\underline\Delta_4-h)-c_4]
\end{equation}
produce again four pinching families
\begin{equation}
\checked{}
\gamma_{\{2, 6, 8, 10, 11, 12, 39, 43\}}\,
\gamma_{\{2, 7, 8, 10, 11, 12, 39, 44\}}\,
\gamma_{\{3, 6, 8, 10, 11, 12, 40, 43\}}\,
\gamma_{\{3, 7, 8, 10, 11, 12, 40, 44\}}.
\end{equation}

Take the contribution from $\gamma_{\{2, 6, 8, 10, 11, 12, 39, 43\}}$ for example. The poles in the original integrand responsible for it are
\begin{equation}
\begin{split}\checked{}
&\Gamma_2[{\textstyle\frac{h-c_2-t_2}{2}}],\quad
\Gamma_6[{\textstyle\frac{h-c_4-t_4}{2}}],\quad
\Gamma_8[{\textstyle\frac{h-c_1+t_4-t_5}{2}}],\quad
\Gamma_{10}[{\textstyle\frac{t_6-T}{2}}],\quad
\Gamma_{11}[{\textstyle\frac{h+c_1+t_2-t_7}{2}}],\\
&\Gamma_{12}[{\textstyle\frac{-2h+t_5-t_6+t_7}{2}}],\quad
\frac{1}{(\underline\Delta_2-h)+c_2},\quad
\frac{1}{(\underline\Delta_4-h)+c_4}.
\end{split}
\end{equation}
These force the integration variables to localize at
\begin{equation}
\begin{split}\checked{}
&t_2=\underline\Delta_2,\quad
t_4=\underline\Delta_4,\quad
t_5=h-c_1+\underline\Delta_4,\quad
t_6=\underline\Delta_{24},\quad
t_7=h+c_1+\underline\Delta_2,\\
&c_2=h-\underline\Delta_2,\quad
c_4=h-\underline\Delta_4,
\end{split}
\end{equation}
under the condition $T=\underline\Delta_{24}$. The proper residue contour is specified as
\begin{equation}\label{eq:Tresiduepart1}
(-1)^4\int[\mathrm{d}t]_{\rm rest}\,\residue{T=\underline\Delta_{24}}\,
\residue{c_4=h-\underline\Delta_4}\,
\residue{c_2=h-\underline\Delta_2}\,
\residue{t_7=2h+c_1-c_2}\,
\residue{t_6=T}\,
\residue{t_5=2h-c_1-c_4}\,
\residue{t_4=h-c_4}\,
\residue{t_2=h-c_2}\,
\mathcal{N}\,\frac{M_0\,K}{(-2)^8}.
\end{equation}
The localization of $\{t_2,t_4,c_2,c_4\}$ also forces the two correction functions to simplify into
\begin{align}
\checked{}C_{A_2}&=\frac{1}{\Gamma[\frac{h\pm c_3+\Delta_2-\underline\Delta_2}{2}]\Gamma[\frac{\Delta_2+\underline\Delta_2-t_3}{2}]},\\
\checked{}C_{A_3}&=\frac{1}{\Gamma[\frac{h\pm c_3+\Delta_3-\underline\Delta_4}{2}]\Gamma[\frac{\Delta_3+\underline\Delta_4-t_3}{2}]}.
\end{align}
One can verify that the contribution from the other three families of poles of $M$ in \eqref{eq:polesrelevantinT} produces exactly the same result. Hence the total residue at the leading pole in $T$ is 4 times \eqref{eq:Tresiduepart1}, which explicitly is
\begin{equation}\label{eq:leadingTresidue}
\begin{split}\checked{}
\residue{T=\underline\Delta_{13}}\mathcal{M}=&\int[\mathrm{d}t]_{\rm rest}\,
\frac{\mathcal{N}'}{D}\,
\Gamma_{1}[{\textstyle\frac{t_1-S}{2}}]\,
\Gamma_{2}[{\textstyle\frac{h-c_3-t_3}{2}}]\,
\Gamma_{3}[{\textstyle\frac{h+c_3-t_3}{2}}]\,
\Gamma_{4}[{\textstyle\frac{h-c_1-t_1+t_3}{2}}]\\
&\times\Gamma_{5}[{\textstyle\frac{h+c_1+t_3-t_8}{2}}]\,
\Gamma_{6}[{\textstyle\frac{t_8-S}{2}}]\,
\Gamma_{7}[{\textstyle\frac{\Delta_{12}-t_1}{2}}]\,
\Gamma_{8}[{\textstyle\frac{\Delta_{34}-t_8}{2}}]\\
&\times\Gamma_{9}[{\textstyle\frac{t_1-t_3-\Delta_1+\underline\Delta_2}{2}}]\,
\Gamma_{10}[{\textstyle\frac{h+S+c_1-t_8-\Delta_1+\underline\Delta_2}{2}}]\,
\Gamma_{11}[{\textstyle\frac{-h+c_1+t_1-\Delta_2+\underline\Delta_2}{2}}]\\
&\times\Gamma_{12}[{\textstyle\frac{-h-c_1+t_8-\Delta_3+\underline\Delta_4}{2}}]\,
\Gamma_{13}[{\textstyle\frac{h+S-c_1-t_1-\Delta_4+\underline\Delta_4}{2}}]\,
\Gamma_{14}[{\textstyle\frac{-t_3+t_8-\Delta_4+\underline\Delta_4}{2}}]\\
&\times\Gamma[{\textstyle\frac{\Delta_1+\underline\Delta_2-h-c_1}{2}}]\,
\Gamma[{\textstyle\frac{\Delta_2+\underline\Delta_2-h\pm c_3}{2}}]\,
\Gamma[{\textstyle\frac{\Delta_3+\underline\Delta_4-h\pm c_3}{2}}]\,
\Gamma[{\textstyle\frac{\Delta_4+\underline\Delta_4-h+c_1}{2}}],
\end{split}
\end{equation}
where $\mathcal{N}'$ is the normalization factor proper for this residue
\begin{equation}
\checked{}
\mathcal{N}'=-\frac{1}{2^{10}}\prod_{i=1}^4\frac{\mathcal{C}_{\Delta_i}}{\Gamma[\Delta_i]}\prod_{a\in\{1,3\}}\frac{[\mathrm{d}c_a]_{\underline\Delta_a}}{\Gamma[\pm c_a]},
\end{equation}
and the denominator $D$ is
\begin{equation}\checked{}
\begin{split}
D=&\Gamma[{\textstyle\underline\Delta_2-h+1}]\,
\Gamma[{\textstyle\underline\Delta_4-h+1}]\,
\Gamma[{\textstyle\frac{\Delta_{12}-S}{2}}]\,
\Gamma[{\textstyle\frac{\Delta_{34}-S}{2}}]\,
\Gamma[{\textstyle\frac{h+c_1+t_1-t_8-\Delta_1+\underline\Delta_2}{2}}]\\
&\times\Gamma[{\textstyle\frac{\Delta_2+\underline\Delta_2-t_3}{2}}]\,
\Gamma[{\textstyle\frac{\Delta_3+\underline\Delta_4-t_3}{2}}]\,
\Gamma[{\textstyle\frac{h-c_1-t_1+t_8-\Delta_4+\underline\Delta_4}{2}}]\,
\Gamma[{\textstyle\frac{S-\Delta_{13}+\underline\Delta_{24}}{2}}]\,
\Gamma[{\textstyle\frac{S-\Delta_{24}+\underline\Delta_{24}}{2}}].
\end{split}
\end{equation}
In the above we have re-labeled the $\Gamma$ functions that depend on the remaining integration variables $\{t_1,t_3,t_8\}$.

In principle there is possibility that the result \eqref{eq:leadingTresidue} can be further simplified, but it is not very obvious how. Nevertheless it is still simple to learn about its pole structure. Applying the Algorithm \ref{app:sec:polealgoritm}, the three $t$ integrals yield the following emergent poles
\begin{equation}\checked{}
\begin{split}
&\underset{\{1,2,4\}}{\Gamma[{\textstyle\frac{2h-S-c_1-c_3}{2}}]}\,
\underset{\{2,5,6\}}{\Gamma[{\textstyle\frac{2h-S+c_1-c_3}{2}}]}\,
\underset{\{1,3,4\}}{\Gamma[{\textstyle\frac{2h-S-c_1+c_3}{2}}]}\,
\underset{\{3,5,6\}}{\Gamma[{\textstyle\frac{2h-S+c_1+c_3}{2}}]}\\
&\underset{\{4,9\}}{\Gamma[{\textstyle\frac{h-c_1-\Delta_1+\underline\Delta_2}{2}}]}\,
\underset{\{6,10\}}{\Gamma[{\textstyle\frac{h+c_1-\Delta_1+\underline\Delta_2}{2}}]}\,
\underset{\{7,11\}}{\Gamma[{\textstyle\frac{-h+c_1+\Delta_1+\underline\Delta_2}{2}}]}\,
\underset{\{2,4,11\}}{\Gamma[{\textstyle\frac{h-c_3-\Delta_2+\underline\Delta_2}{2}}]}\,
\underset{\{3,4,11\}}{\Gamma[{\textstyle\frac{h+c_3-\Delta_2+\underline\Delta_2}{2}}]}\\
&\underset{\{2,5,12\}}{\Gamma[{\textstyle\frac{h-c_3-\Delta_3+\underline\Delta_4}{2}}]}\,
\underset{\{3,5,12\}}{\Gamma[{\textstyle\frac{h+c_3-\Delta_3+\underline\Delta_4}{2}}]}\,
\underset{\{1,3\}}{\Gamma[{\textstyle\frac{h-c_1-\Delta_4+\underline\Delta_4}{2}}]}\,
\underset{\{5,14\}}{\Gamma[{\textstyle\frac{h+c_1-\Delta_4+\underline\Delta_4}{2}}]}\,
\underset{\{8,12\}}{\Gamma[{\textstyle\frac{-h-c_1+\Delta_4+\underline\Delta_4}{2}}]}.
\end{split}
\end{equation}
Including the additional poles from the original integrand in \eqref{eq:leadingTresidue} all these families can be grouped into
\begin{equation}
\checked{}
\Gamma[{\textstyle\frac{2h\pm c_1\pm c_3-S}{2}}]\times
\underbrace{\Gamma[{\textstyle\frac{\underline\Delta_4\pm(\Delta_4-h)\pm c_1}{2}}]\,
\Gamma[{\textstyle\frac{\underline\Delta_2\pm(\Delta_1-h)\pm c_1}{2}}]}_{\mathcal{O}_4\mathcal{O}_1[\mathcal{O}_{\underline2}\mathcal{O}_{\underline4}]}\times
\underbrace{\Gamma[{\textstyle\frac{\underline\Delta_2\pm(\Delta_2-h)\pm c_3}{2}}]\,
\Gamma[{\textstyle\frac{\underline\Delta_4\pm(\Delta_3-h)\pm c_3}{2}}]}_{\mathcal{O}_2\mathcal{O}_3[\mathcal{O}_{\underline2}\mathcal{O}_{\underline4}]}.
\end{equation}
This pole structure naturally factorizes into three parts. By comparing with \eqref{eq:quartictriangleleadingresidue}, we immediately observe that the second part are identical to the pole structure of the leading connected correlation among $\mathcal{O}_4$, $\mathcal{O}_1$ and the double-trace primary $[\mathcal{O}_{\underline2}\mathcal{O}_{\underline4}]_{0,0}$, and the third part among $\mathcal{O}_2$, $\mathcal{O}_3$ and $[\mathcal{O}_{\underline2}\mathcal{O}_{\underline4}]_{0,0}$. In addition, there are also poles in the first part that associates to the double-trace operators in the $S$ channel. This seems to suggest that the residue at the leading $T$ channel pole can again be interpreted in terms of factorized diagrams, though the detailed expression may not be a simple product.

Due to the presence of the $S$ channel poles in the residue at the leading $T$ channel pole, we can proceed to further compute the residue at, say the leading pole in $S$ channel. There are again four contributions corresponding to each choice of signs in $\Gamma[\frac{2h\pm c_1\pm c_3-S}{2}]$. The computations follow the same procedure as we described in previous examples and we do not duplicate here. Again these four contributions are identical to each other, and the total residue turns out to be another simple product of $\Gamma$ functions
\begin{equation}\label{eq:box4ptSTres}
\checked{}
\residue{S=\underline\Delta_{13}}\,\residue{T=\underline\Delta_{24}}\,\mathcal{M}=
4\pi^{4h}\prod_{i=1}^4\frac{\mathcal{C}_{\Delta_i}}{\Gamma[\Delta_i]}\prod_{a=1}^4\frac{\mathcal{C}_{\underline\Delta_a}}{\Gamma[\underline\Delta_a]}\times\frac{\Gamma[\frac{\underline\Delta_{12}-\Delta_1}{2}]\Gamma[\frac{\underline\Delta_{12}+\Delta_1}{2}-h]\times\text{cyclic}}{\Gamma[\frac{\underline\Delta_{1234}-\Delta_{13}}{2}]\Gamma[\frac{\underline\Delta_{1234}-\Delta_{24}}{2}]}.
\end{equation}
It is an interesting question how this residue should be properly interpreted. We leave this for the future.

Although we did not present before, the same factorized pole structure of the residue can also be easily concluded for the leading pole of $\mathcal{M}$ in the $S$ channel (despite of the more complicated expression), which in particular contains all the original poles in the $T$ channel. And as we further compute its residue at the leading $T$ channel pole we land on the same result as above.

\newpage

\section{Arbitrary Loops}\label{sec:arbitraryloops}

In the seek for a proper integration kernel for the construction of one-loop diagrams such that only non-trivial integrations of Mandelstam variables remain, we observed the necessity of Mandelstam variables extra to what are already in $M_0$. These extra variables $\xi$ can naturally be chosen in correspondence to the possible OPE channels newly emerged in the one-loop diagram to be constructed. This fact has the appealing implication that the structure of kernels in the higher loop constructions largely follows that of the one loop. This is because such $\xi$ variables may actually enter $M_0$ as well if the original diagram is a loop diagram. 

Another evidence for this possibility is that, when simplifying the $K_1$ part of the kernel in \eqref{eq:K1oneloopreduced} and also in \eqref{eq:K1oneloopgeneric} we grouped the original $\frac{\Gamma[\tau_{i\,j}]\Gamma[\delta_{i\,j}-\tau_{i\,j}]}{\Gamma[\delta_{i\,j}]}$ factors according to the vertices of the diagram, which is not tied to any notion of loops, and so can in principle be straightforwardly carried over to higher loops.

Therefore the essential idea for the derivation of a simplified integration kernel suitable for the construction of diagrams at arbitrary loop level is to make use of the analogy between the formation of the new loop at higher loops and that of the unique loop at one loop, such that we are able to directly borrow the previous results at one loop.

We first discuss in the next two subsections about two new features that arise at higher loops, which are important for the analogy to be drawn. We then present the general prescription for higher-loop construction in Section \ref{sec:generalkernel}.  As an application of this construction we analyze in detail the 4-point double-triangle diagram in Section \ref{sec:doubletrianglediagram}.  Based on this explicit result we also further propose a conjecture regarding the non-minimal poles that we expect to hold to all loops, which is stated in Section \ref{sec:conjectureonamplitude}.

\subsection{Emergence of the new loop}\label{sec:emergehigherloop}

In order to draw the analogy with the construction of a one-loop diagram, the first thing we need to address is the notion of a loop in a higher-loop diagram. An obvious notion is of course any closed curve formed by bulk-to-bulk propagators. Correspondingly in order to form such a loop we seek for a chain of bulk-to-bulk propagators in the original diagram with its to ends anchored at the two boundary points $0$ and $n+1$ that we are going to glue. At one loop this is what we exactly used to further seek for a simplified integration kernel, and there is a unique choice. While obviously at higher loops the choice is non-unique, this notion may even be insufficient.

The main reason for the inconvenience of this na\"ive notion is that individual bulk-to-bulk propagator loses physical significance at loop level, if it is not a tree propagator. Suppose we are given certain analytic result for a Mellin amplitude but without knowing the corresponding diagram(s), then it could be hard to figure out what the set of existing bulk-to-bulk propagators are. In fact this is not even necessary. In comparison, what is always manifest in a Mellin amplitude is the set of non-trivial OPE channels, for which poles are present in the amplitude. This is true regardless of specific loop level, and even regardless of whether we are looking at a single diagram or not. While it is also hard to tell at first sight whether a given pole arises from a tree effect or a loop effect, the good news is that one can basically treat any pole as coming from a ``tree'', even though the set of non-trivial OPE channels that appear are not mutually consistent in general (in the sense that we can sequetially do OPEs so as to explore all of them).

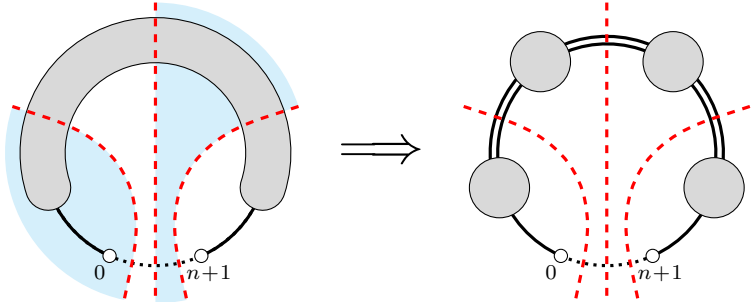
\begin{figure}[ht]
\captionsetup{margin=2em}
\begin{center}
\begin{tikzpicture}
\begin{scope}[xshift=-3cm]
\fill [ProcessBlue!15!white] (162:2) -- (162:1.5) .. controls (162:.5) and (258:.5) .. (258:1.5) -- (258:2) arc [start angle=258,end angle=162,radius=2];
\fill [ProcessBlue!15!white] (18:2) -- (18:1.5) .. controls (18:.5) and (-78:.5) .. (-78:1.5) -- (-78:2) arc [start angle=-78,end angle=-90,radius=2] -- (90:2) arc [start angle=90,end angle=18,radius=2];
\draw [black,very thick,dotted] (0,0) circle [radius=1.5];
\draw [black,very thick] (-66:1.5) arc [start angle=-66,end angle=246,radius=1.5];
\fill [black] (198:1.5) circle [radius=2.5pt];
\fill [black] (126:1.5) circle [radius=2.5pt];
\fill [black] (54:1.5) circle [radius=2.5pt];
\fill [black] (-18:1.5) circle [radius=2.5pt];
\draw [black,fill=white] (-66:1.5) circle [radius=2.5pt];
\draw [black,fill=white] (246:1.5) circle [radius=2.5pt];
\node [anchor=center] at (245:1.75) {\scriptsize $0$};
\node [anchor=center] at (-66:1.75) {\scriptsize $n\!+\!1$};
\draw [fill=black!15!white] (198:1.2) arc [start angle=198,end angle=-18,radius=1.2] arc [start angle=162,end angle=342,radius=.3] arc [start angle=-18,end angle=198,radius=1.8] arc [start angle=-162,end angle=18,radius=.3];
\draw [red,very thick,dashed] (162:2) -- (162:1.5) .. controls (162:.5) and (258:.5) .. (258:1.5) -- (258:2);
\draw [red,very thick,dashed] (90:2) -- (270:2);
\draw [red,very thick,dashed] (18:2) -- (18:1.5) .. controls (18:.5) and (-78:.5) .. (-78:1.5) -- (-78:2);
\end{scope}
\node [anchor=center] at (0,0) {\huge $\Longrightarrow$};
\begin{scope}[xshift=3cm]
\draw [black,very thick,dotted] (246:1.5) arc [start angle=246,end angle=294,radius=1.5];
\draw [black,very thick] (-66:1.5) arc [start angle=-66,end angle=-18,radius=1.5];
\draw [black,very thick] (-18:1.45) arc [start angle=-18,end angle=198,radius=1.45];
\draw [black,very thick] (-18:1.55) arc [start angle=-18,end angle=198,radius=1.55];
\draw [black,very thick] (198:1.5) arc [start angle=198,end angle=246,radius=1.5];
\node [anchor=center] at (245:1.75) {\scriptsize $0$};
\node [anchor=center] at (-66:1.75) {\scriptsize $n\!+\!1$};
\draw [fill=black!15!white] (198:1.5) circle [radius=.4];
\draw [fill=black!15!white] (126:1.5) circle [radius=.4];
\draw [fill=black!15!white] (54:1.5) circle [radius=.4];
\draw [fill=black!15!white] (-18:1.5) circle [radius=.4];
\draw [black,fill=white] (-66:1.5) circle [radius=2.5pt];
\draw [black,fill=white] (246:1.5) circle [radius=2.5pt];
\draw [red,very thick,dashed] (162:2) -- (162:1.5) .. controls (162:.5) and (258:.5) .. (258:1.5) -- (258:2);
\draw [red,very thick,dashed] (90:2) -- (270:2);
\draw [red,very thick,dashed] (18:2) -- (18:1.5) .. controls (18:.5) and (-78:.5) .. (-78:1.5) -- (-78:2);
\end{scope}
\end{tikzpicture}
\end{center}
\caption{A chain of three non-trivial OPE channels. We use blue shades to distinguish between neighboring regions separated by the channels. When the chain is maximal, each region can effectively be treated as a generalized bulk vertex in the sense of Section \ref{sec:generaloneloop}.}
\label{fig:OPEchain}
\end{figure}

From this intuition, a more proper way to view the formation of a new loop is as follows. We start by collecting the set of OPE channels captured by poles in $M_0$. Within this set, we identify a chain of OPE channels, such that the two boundary points $0$ and $n+1$ sit at the two ends in these OPEs. To be precise about the ``chain'' here, since each OPE channel considered here necessarily has to separate point $0$ and $n+1$ on its two sides, we can label it by the set of boundary ponits on the side that contains point $0$. Let us consider $r$ such channels, and hence $r$ label sets $S_1,S_2,\ldots,S_r$. Then in order that they form a chain of channels we require that they obey a natural ordering induced by proper inclusion
\begin{equation}\label{eq:OPEordering}
\checked{}
\{0\}\subset S_1\subset S_2\subset\cdots\subset S_r.
\end{equation}
When this holds, we are allowed to introduce the notion of regions between each pair of adjacent OPE channels, such that the region between $S_a$ and $S_{a+1}$ contains boundary points in the set $S_{a+1}\backslash S_a$. This is illustrated in Figure \ref{fig:OPEchain}.

In order to set up for the integration kernel later on, we further introduce the notion of a maximal chain, in the sense that no extra OPE channel can be added into the collection such that an ordering like \eqref{eq:OPEordering} still exists.

Note that even when we obtain a maximal chain of OPE channels for $M_0$, each region may still contain extra OPE channels that are consistent with the existing ones in the chain. However, obviously such extra channels can no longer separate point $0$ and $n+1$ on two sides, since otherwise it violates our definition of the maximal chain. In fact, it can at most only further separate a subset of boundary points within the region from all the rest.

A given maximal chain of OPE channels  induces an effective (tree) diagram for the original amplitude $M_0$. Each OPE channel in the chain is replaced by an explicit propagator in the diagram. Each region is replaced by a generalized vertex on the chain as exemplified in Figure \ref{fig:treesonloop}, and in the special case of no further consistent OPE channel within the region, it is merely replaced by a normal vertex. If we purely focus on 1PI diagram, then each region is strictly replaced by a single vertex.  For the time being let us just focus on the 1PI diagrams.

At this stage it is clear that, once a maximal chain of OPE channels are specified, then there is a unique notion of the newly formed loop after gluing, which is the one obtained from the explicit chain in the effective diagram.


So far our discussion does not rely on specific diagrams, but assumes the knowledge about the set of OPE channels observed by $M_0$. In actual practice we more often encounter specific diagrams at some loop level to begin with, but may not already know for sure its Mandelstam dependence. Although we obtained a conjecture about the pole structure of pre-amplitudes of scalar diagrams in general, a full proof or verification is yet needed. Here we merely assume its validity. In other words, $M_0$ depends on Mandelstam variables corresponding to non-trivial cuts that separate the diagram into two connected sub-diagrams.

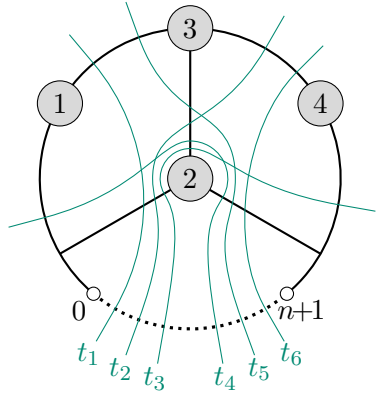
\begin{figure}[ht]
\captionsetup{margin=2em}
\begin{center}
\begin{tikzpicture}
\draw [black,very thick,dotted] (-130:2) arc [start angle=-130,end angle=-50,radius=2];
\draw [black,thick] (-50:2) arc [start angle=-50,end angle=230,radius=2];
\draw [black,thick] (0,0) -- (-30:2);
\draw [black,thick] (0,0) -- (90:2);
\draw [black,thick] (0,0) -- (210:2);
\draw [black,fill=white] (-50:2) circle [radius=2.5pt];
\draw [black,fill=white] (230:2) circle [radius=2.5pt];
\node [anchor=center] at (230:2.3) {$0$};
\node [anchor=center] at (-50:2.3) {$n\!\!+\!\!1$};
\draw [black,fill=black!15!white] (150:2) circle [radius=.3];
\node [anchor=center] at (150:2) {$1$};
\draw [black,fill=black!15!white] (0,0) circle [radius=.3];
\node [anchor=center] at (0,0) {$2$};
\draw [black,fill=black!15!white] (90:2) circle [radius=.3];
\node [anchor=center] at (90:2) {$3$};
\draw [black,fill=black!15!white] (30:2) circle [radius=.3];
\node [anchor=center] at (30:2) {$4$};
\draw [PineGreen] (240:2.5) -- (240:2) .. controls (240:.8) and (130:.8) .. (130:2) -- (130:2.5);
\node [anchor=center] at (240:2.7) {\color{PineGreen} $t_1$};
\draw [PineGreen] (-70:2.5) -- (-70:2) .. controls (-70:.7) and (.6,-.6) .. (.6,0) .. controls (.6,1) and (110:.7) .. (110:2) -- (110:2.5);
\node [anchor=center] at (-70:2.7) {\color{PineGreen} $t_5$};
\draw [PineGreen] (-60:2.5) -- (-60:2) .. controls (-60:1) and (45:1) .. (45:2) -- (45:2.5);
\node [anchor=center] at (-60:2.7) {\color{PineGreen} $t_6$};
\draw [PineGreen] (250:2.5) -- (250:2) .. controls (250:.7) and (-.5,-.4) .. (-.5,0) .. controls (-.5,.8) and (60:.7) .. (60:2) -- (60:2.5);
\node [anchor=center] at (250:2.7) {\color{PineGreen} $t_2$};
\draw [PineGreen] (260:2.5) -- (260:2) .. controls (260:.7) and ($(-150:.4)+(-60:.3)$) .. (-150:.4) arc [start angle=-150,end angle=-270,radius=.4] .. controls (.4,.4) and (-10:.7) .. (-10:2) -- (-10:2.5);
\node [anchor=center] at (260:2.7) {\color{PineGreen} $t_3$};
\draw [PineGreen] (-80:2.5) -- (-80:2) .. controls (-80:.7) and ($(-30:.5)+(-120:.3)$) .. (-30:.5) arc [start angle=-30,end angle=90,radius=.5] .. controls (-.4,.5) and (195:.7) .. (195:2) -- (195:2.5);
\node [anchor=center] at (-80:2.7) {\color{PineGreen} $t_4$};
\end{tikzpicture}
\end{center}
\caption{A three-loop case}
\label{fig:3loopexample}
\end{figure}
\begin{figure}[ht]
\captionsetup{margin=2em}
\begin{center}
\begin{tikzpicture}
\begin{scope}[xshift=-3.5cm]
\node [anchor=center] at (0,-3.2) {(A)};
\fill [ProcessBlue!15!white] (240:2.5) -- (240:2) .. controls (240:.8) and (130:.8) .. (130:2) -- (130:2.5) arc [start angle=130,end angle=240,radius=2.5];
\fill [ProcessBlue!15!white] (-70:2.5) -- (-70:2) .. controls (-70:.7) and (.6,-.6) .. (.6,0) .. controls (.6,1) and (110:.7) .. (110:2) -- (110:2.5) arc [start angle=110,end angle=45,radius=2.5] -- (45:2) .. controls (45:1) and (-60:1) .. (-60:2) -- (-60:2.5) arc [start angle=-60,end angle=-70,radius=2.5];
\draw [black,very thick,dotted] (-130:2) arc [start angle=-130,end angle=-50,radius=2];
\draw [black,thick] (-50:2) arc [start angle=-50,end angle=230,radius=2];
\draw [black,thick] (0,0) -- (-30:2);
\draw [black,thick] (0,0) -- (90:2);
\draw [black,thick] (0,0) -- (210:2);
\draw [black,fill=white] (-50:2) circle [radius=2.5pt];
\draw [black,fill=white] (230:2) circle [radius=2.5pt];
\node [anchor=center] at (230:2.3) {$0$};
\node [anchor=center] at (-50:2.3) {$n\!\!+\!\!1$};
\draw [black,fill=black!15!white] (150:2) circle [radius=.3];
\node [anchor=center] at (150:2) {$1$};
\draw [black,fill=black!15!white] (0,0) circle [radius=.3];
\node [anchor=center] at (0,0) {$2$};
\draw [black,fill=black!15!white] (90:2) circle [radius=.3];
\node [anchor=center] at (90:2) {$3$};
\draw [black,fill=black!15!white] (30:2) circle [radius=.3];
\node [anchor=center] at (30:2) {$4$};
\draw [red,thick,dashed] (240:2.5) -- (240:2) .. controls (240:.8) and (130:.8) .. (130:2) -- (130:2.5);
\node [anchor=center] at (240:2.7) {\color{red} $t_1$};
\draw [red,thick,dashed] (-70:2.5) -- (-70:2) .. controls (-70:.7) and (.6,-.6) .. (.6,0) .. controls (.6,1) and (110:.7) .. (110:2) -- (110:2.5);
\node [anchor=center] at (-70:2.7) {\color{red} $t_5$};
\draw [red,thick,dashed] (-60:2.5) -- (-60:2) .. controls (-60:1) and (45:1) .. (45:2) -- (45:2.5);
\node [anchor=center] at (-60:2.7) {\color{red} $t_6$};
\end{scope}
\begin{scope}[xshift=3.5cm]
\node [anchor=center] at (0,-3.2) {(B)};
\fill [ProcessBlue!15!white] (240:2.5) -- (240:2) .. controls (240:.8) and (130:.8) .. (130:2) -- (130:2.5) arc [start angle=130,end angle=240,radius=2.5];
\fill [ProcessBlue!15!white] (250:2.5) -- (250:2) .. controls (250:.7) and (-.5,-.4) .. (-.5,0) .. controls (-.5,.8) and (60:.7) .. (60:2) -- (60:2.5) arc [start angle=60,end angle=-10,radius=2.5] -- (-10:2) .. controls (-10:.7) and (.4,.4) .. (90:.4) arc [start angle=90,end angle=210,radius=.4] .. controls ($(-150:.4)+(-60:.3)$) and (260:.7) .. (260:2) -- (260:2.5) arc [start angle=260,end angle=250,radius=2.5];
\draw [black,very thick,dotted] (-130:2) arc [start angle=-130,end angle=-50,radius=2];
\draw [black,thick] (-50:2) arc [start angle=-50,end angle=230,radius=2];
\draw [black,thick] (0,0) -- (-30:2);
\draw [black,thick] (0,0) -- (90:2);
\draw [black,thick] (0,0) -- (210:2);
\draw [black,fill=white] (-50:2) circle [radius=2.5pt];
\draw [black,fill=white] (230:2) circle [radius=2.5pt];
\node [anchor=center] at (230:2.3) {$0$};
\node [anchor=center] at (-50:2.3) {$n\!\!+\!\!1$};
\draw [black,fill=black!15!white] (150:2) circle [radius=.3];
\node [anchor=center] at (150:2) {$1$};
\draw [black,fill=black!15!white] (0,0) circle [radius=.3];
\node [anchor=center] at (0,0) {$2$};
\draw [black,fill=black!15!white] (90:2) circle [radius=.3];
\node [anchor=center] at (90:2) {$3$};
\draw [black,fill=black!15!white] (30:2) circle [radius=.3];
\node [anchor=center] at (30:2) {$4$};
\draw [red,thick,dashed] (240:2.5) -- (240:2) .. controls (240:.8) and (130:.8) .. (130:2) -- (130:2.5);
\node [anchor=center] at (240:2.7) {\color{red} $t_1$};
\draw [red,thick,dashed] (250:2.5) -- (250:2) .. controls (250:.7) and (-.5,-.4) .. (-.5,0) .. controls (-.5,.8) and (60:.7) .. (60:2) -- (60:2.5);
\node [anchor=center] at (250:2.7) {\color{red} $t_2$};
\draw [red,thick,dashed] (260:2.5) -- (260:2) .. controls (260:.7) and ($(-150:.4)+(-60:.3)$) .. (-150:.4) arc [start angle=-150,end angle=-270,radius=.4] .. controls (.4,.4) and (-10:.7) .. (-10:2) -- (-10:2.5);
\node [anchor=center] at (260:2.7) {\color{red} $t_3$};
\end{scope}
\end{tikzpicture}
\end{center}
\vspace{-1.5em}\caption{Maximal chains and division into regions. (A) $\{t_1,t_5,t_6\}$. (B) $\{t_1,t_2,t_3\}$.}
\label{fig:divisions}
\end{figure}
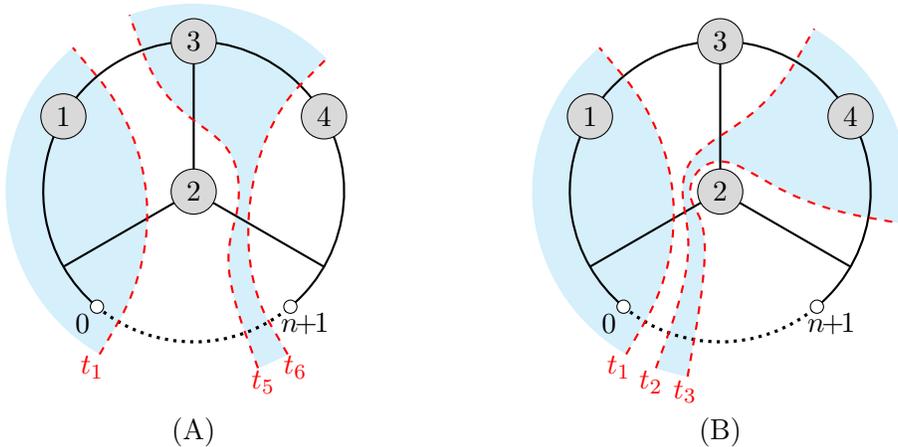

Consider the construction of the 3-loop diagram in Figure \ref{fig:3loopexample} for example, where the four nodes indicated by grey blobs each has some boundary points attached to it. The original 2-loop diagram has six non-trivial OPE channels that separates point $0$ and $n+1$, as indicated by the green curves together with their Mandelstam variables $\{t_1,\ldots,t_6\}$. These channels are not all consistent with each other, and a maximal chain contains three channels. The choice of the maximal chain is not unique though. For instance, we can choose it to be $\{t_1,t_5,t_6\}$, or $\{t_1,t_2,t_3\}$, etc. The first two choices are illustrated in Figure \ref{fig:divisions}, where we also indicate the corresponding division of the diagram into regions. Effective (chain) diagrams can easily be read off therein. 

From this example we see that for 1PI diagrams the number of regions is the same as the number of bulk vertices that have boundary points attached. Vertices that are purely internal, in the sense that they are only incident to bulk-to-bulk propagators, are trivial because the choice of maximal chain of OPE channels are completely blind to them. For instance, if vertex 3 is of this type, then $t_2$ and $t_3$ as well as $t_5$ and $t_6$ are identified, so that we have only three regions, and this vertex can be moved into any of these regions without changing the effective diagram.

Furthermore, here we observe an essential difference from the original notion of loop formation: the bulk vertices in the neighboring regions may not always be connected by bulk-to-bulk propagators. In Figure \ref{fig:divisions} (A) the vertex $1$ and $2$ are not directly connected to each other, even though they are separated by an actual OPE channel.

\subsection{Independent set of Mandelstam variables}\label{sec:Mandelstamindependence}

Since we have at hand an effective diagram from $M_0$ induced by a choice of maximal chain of OPE channels, which is always a tree, it is quite tempting to directly apply our result for the kernel \eqref{eq:K2oneloopreduced} and \eqref{eq:K1oneloopreduced} in the one-loop construction. 

More precisely, the Mandelstam variables associated to the maximal chain already make up the $\Xi$ variables for the effective diagram. We then collect the $\xi$ variables induced by the effective diagram (we label the effective vertices as $A_1,A_2,\ldots,A_r$ in sequence as before). These $\Xi$ and $\xi$ variables are the independent integration variables to be used in the same kernel. Here note that, as we expected at the beginning of this section, some of the $\xi$ variables indeed correspond to actual OPE channels of $M_0$. For instance in the construction in Figure \eqref{fig:divisions} (A), $\xi_{3,4}$ is a non-trivial OPE channel of the original 2-loop diagram, while $\xi_{1,2}$ is not.

However, there is one potential worry in doing this, because the original amplitude $M_0$ may also depend on other Mandelstam variables that do not fall into the set $\{\Xi,\xi\}$. From our experience with the derivation of the one-loop kernel, we cannot obtain a simplified kernel by totally ignoring extra variables that $M_0$ depends on.

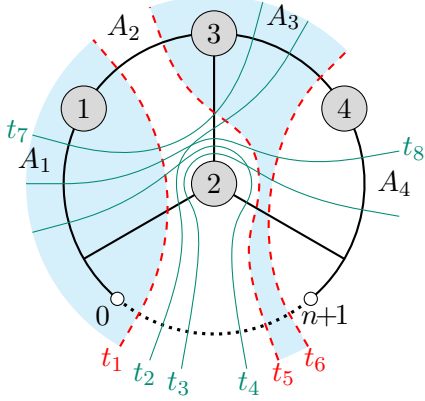
\begin{figure}[ht]
\captionsetup{margin=2em}
\begin{center}
\begin{tikzpicture}
\fill [ProcessBlue!15!white] (240:2.5) -- (240:2) .. controls (240:.8) and (130:.8) .. (130:2) -- (130:2.5) arc [start angle=130,end angle=240,radius=2.5];
\fill [ProcessBlue!15!white] (-70:2.5) -- (-70:2) .. controls (-70:.7) and (.6,-.6) .. (.6,0) .. controls (.6,1) and (110:.7) .. (110:2) -- (110:2.5) arc [start angle=110,end angle=45,radius=2.5] -- (45:2) .. controls (45:1) and (-60:1) .. (-60:2) -- (-60:2.5) arc [start angle=-60,end angle=-70,radius=2.5];
\draw [black,very thick,dotted] (-130:2) arc [start angle=-130,end angle=-50,radius=2];
\draw [black,thick] (-50:2) arc [start angle=-50,end angle=230,radius=2];
\draw [black,thick] (0,0) -- (-30:2);
\draw [black,thick] (0,0) -- (90:2);
\draw [black,thick] (0,0) -- (210:2);
\draw [black,fill=white] (-50:2) circle [radius=2.5pt];
\draw [black,fill=white] (230:2) circle [radius=2.5pt];
\node [anchor=center] at (230:2.3) {$0$};
\node [anchor=center] at (-50:2.3) {$n\!\!+\!\!1$};
\draw [black,fill=black!15!white] (150:2) circle [radius=.3];
\node [anchor=center] at (150:2) {$1$};
\node [anchor=center] at (172.5:2.4) {$A_1$};
\draw [black,fill=black!15!white] (0,0) circle [radius=.3];
\node [anchor=center] at (0,0) {$2$};
\node [anchor=center] at (120:2.4) {$A_2$};
\draw [black,fill=black!15!white] (90:2) circle [radius=.3];
\node [anchor=center] at (90:2) {$3$};
\node [anchor=center] at (67.5:2.4) {$A_3$};
\draw [black,fill=black!15!white] (30:2) circle [radius=.3];
\node [anchor=center] at (30:2) {$4$};
\node [anchor=center] at (0:2.4) {$A_4$};
\draw [red,thick,dashed] (240:2.5) -- (240:2) .. controls (240:.8) and (130:.8) .. (130:2) -- (130:2.5);
\node [anchor=center] at (240:2.7) {\color{red} $t_1$};
\draw [red,thick,dashed] (-70:2.5) -- (-70:2) .. controls (-70:.7) and (.6,-.6) .. (.6,0) .. controls (.6,1) and (110:.7) .. (110:2) -- (110:2.5);
\node [anchor=center] at (-70:2.7) {\color{red} $t_5$};
\draw [red,thick,dashed] (-60:2.5) -- (-60:2) .. controls (-60:1) and (45:1) .. (45:2) -- (45:2.5);
\node [anchor=center] at (-60:2.7) {\color{red} $t_6$};
\draw [PineGreen] (250:2.5) -- (250:2) .. controls (250:.7) and (-.5,-.4) .. (-.5,0) .. controls (-.5,.8) and (60:.7) .. (60:2) -- (60:2.5);
\node [anchor=center] at (250:2.7) {\color{PineGreen} $t_2$};
\draw [PineGreen] (260:2.5) -- (260:2) .. controls (260:.7) and ($(-150:.4)+(-60:.3)$) .. (-150:.4) arc [start angle=-150,end angle=-270,radius=.4] .. controls (.4,.4) and (-10:.7) .. (-10:2) -- (-10:2.5);
\node [anchor=center] at (260:2.7) {\color{PineGreen} $t_3$};
\draw [PineGreen] (-80:2.5) -- (-80:2) .. controls (-80:.7) and ($(-30:.5)+(-120:.3)$) .. (-30:.5) arc [start angle=-30,end angle=90,radius=.5] .. controls (-.4,.5) and (195:.7) .. (195:2) -- (195:2.5);
\node [anchor=center] at (-80:2.7) {\color{PineGreen} $t_4$};
\draw [PineGreen] (165:2.5) -- (165:2) .. controls (165:.7) and (75:.7) .. (75:2) -- (75:2.5);
\node [anchor=center] at (165:2.7) {\color{PineGreen} $t_7$};
\draw [PineGreen] (180:2.5) -- (180:2) .. controls (180:.8) and (-.4,.6) .. (0,.6) .. controls (.4,.6) and (10:.8) .. (10:2) -- (10:2.5);
\node [anchor=center] at (10:2.7) {\color{PineGreen} $t_8$};
\end{tikzpicture}
\end{center}
\caption{Extra variables in $M_0$}
\label{fig:extravariables}
\end{figure}

It turns out that this worry is absent, because any extra variables in $M_0$ are merely linear combinations of $\{\Xi,\xi\}$. 
As an explicit example, let us return to Figure \ref{fig:divisions} (A). In this case there are altogether five extra variables $\{t_2,t_3,t_4,t_7,t_8\}$ in $M_0$, as explicitly shown in Figure \ref{fig:extravariables}.
But it is simple to check that they can be expressed in terms of $\{\Xi,\xi\}$ as 
\begin{align}\checked{}
t_2&=\xi_{2,2}-\xi_{2,3}+\xi_{3,3}+\Xi_2-\Xi_3+\Xi_4,\\
t_3&=\Delta_{n+1}+\xi_{2,2}-\xi_{2,4}+\xi_{3,4}+\Xi_2-\Xi_3,\\
t_4&=\Delta_0+\xi_{1,1}-\xi_{1,2}+\xi_{2,2}-\Xi_2+\Xi_3,\\
t_7&=\xi_{1,1}-\xi_{1,2}+\xi_{1,3}+\xi_{2,2}-\xi_{2,3}+\xi_{3,3},\\
t_8&=2h+\xi_{1,1}-\xi_{1,2}+\xi_{2,2}-\xi_{2,4}+\xi_{3,4},
\end{align}
where we have the identification
\begin{equation}
\checked{}
\Xi_1=\Delta_0,\quad
\Xi_2=t_1,\quad
\Xi_3=t_5,\quad
\Xi_4=t_6,\quad
\Xi_5=\Delta_{n+1}.
\end{equation}

The linear dependence of the extra variables in $M_0$ on $\{\Xi,\xi\}$ is in fact necessary in order to be consistent with the non-uniqueness in the choice of the maximal chain of OPE channels. As we choose the maximal chain $\{t_1,t_2,t_3\}$ corresponding to Figure \ref{fig:divisions} (B), from the above relations it is obvious that now we can solve $t_5$ and $t_6$ in terms of $t_2$ and $t_3$.

With explicit diagrams it is simple to understand why the extra variables in $M_0$ has to linearly depend on $\{\Xi,\xi\}$. For simplicity let us just focus on 1PI diagrams. As discussed before, for these diagrams the regions induced by any choice of maximal chain of OPE channels each has to contain exactly one bulk vertex that is adjacent to some bulk-to-boundary propagators. We can regard the group of boundary points attached to the same vertex $A_a$ as a single ``off-shell particle'', in the sense that they have a fixed total conformal dimension $\Delta_{A_a}$ but the Mandelstam variable for the OPE channel separating them from the other boundary points $\xi_{a,a}$ does not necessarily equals it (in the case of a single boundary point we have the ``on-shell condition'' $\xi_{a,a}=\Delta_{A_a}$). Taking into consideration the two additional on-shell particles $0$ and $n+1$, the effective diagram induces a planar ordering for these particles. Then our choice of $\Xi$ and $\xi$ variables are nothing but the set of Mandelstam varaibles associated to all possible separations of these particles into two groups that are consistent with the planar ordering. It is well-known that such set of Mandelstam variables are independent and provide a basis for any other Mandelstam variables that can be constructed from the particles under consideration.  Note that according to our strong conjecture there is no OPE channel observed by the Mellin ampliutde that separates points attached to the same bulk vertex, which indicates that any Mandelstam variable appearing in $M$ has to belong to the set constructible from the effective particles discussed above. Therefore they has to linearly depend on $\{\Xi,\xi\}$.

\subsection{The generic prescription for the kernel}\label{sec:generalkernel}

Now we have enough ingredients to state the structure of the simplified kernel suitable for the construction at arbitrary loop level. In this section we again just focus on 1PI diagrams. As we discussed, the maximal chain of OPE channels induces an effective diagram. For the recursion formula we then directly borrow the result from the construction at one loop \eqref{eq:K2oneloopreduced} and \eqref{eq:K1oneloopreduced}, i.e.,
\begin{equation}
\checked{}
M[s]=\frac{1}{(-2)^{\sharp[\Xi,\xi]}}\int[\mathrm{d}\Xi][\mathrm{d}\xi]\,M_0[\Xi,\xi]\,K[\Xi,\xi,s],
\end{equation}
with $K\equiv K_1K_2$ such that
\begin{equation}
\begin{split}
\checked{}
K_1=&
\prod_{a=1}^r\bfn{\frac{\Delta_{A_a}-\xi_{a}}{2}}{\frac{\Delta_{A_a}-s_{a}}{2}}\,
\prod_{1\leq a<b\leq r}\bfn{\frac{\xi_{a,b-1}+\xi_{a+1,b}-\xi_{a+1,b-1}-\xi_{a,b}}{2}}{\frac{s_{a,b-1}+s_{a+1,b}-s_{a+1,b-1}-s_{a,b}}{2}}.
\end{split}
\end{equation}
and
\begin{equation}
\begin{split}
\checked{}
K_2&=\prod_{a=1}^{r}\frac{\Gamma[\frac{(\xi_{a,r}-\xi_{a+1,r})-(\Xi_a-\Xi_{a+1})}{2}]\Gamma[\frac{(\Xi_a-\Xi_{a+1})-(\xi_{1,a-1}-\xi_{1,a})}{2}]}{\Gamma[\frac{(\xi_{a,r}-\xi_{a+1,r})-(\xi_{1,a-1}-\xi_{1,a})}{2}]}.
\end{split}
\end{equation}

Recall that the counterpart of $\xi$ variables, $s_{a,b}$, all corresponds to actual OPE channels when the newly constructed diagram is at one loop. At higher loops there can be other Mandelstam variables associated to actual channels seen in the new amplitude but not directly showing up in the kernel. In this case they always come as linear combinations of $s_{a,b}$'s, and poles in these combination of variables should be correctly produced in the result. This is going to be illustrated in some non-planar diagrams in Appendix \ref{app:sec:nonplanar2loop}.

Beforing ending this subsection, let us comment that in the case when several different choices of constructions exist for the same diagram, while these different constructions should ultimately lead to the same result, their equivalence is very non-trivial. Hence the consistency between them in principle should impose strong constraints on any formula we obtain for the target Mellin (pre-)amplitude. It would be interesting to see whether this consistency condition can be utilized to generate useful information about the final result. We leave this for the future.

\begin{figure}[ht]
\captionsetup{margin=2em}
\begin{center}
\begin{tikzpicture}
\draw [black,very thick] (170:3) -- (0,-1) -- (10:3);
\draw [black,very thick] (-170:3) -- (0,1) -- (-10:3);
\draw [black,very thick] (0,-1) -- (0,1);
\fill [black] (0,-1) circle [radius=1.75pt];
\fill [black] (0,1) circle [radius=1.75pt];
\fill [black] (-1.95,0) circle [radius=1.75pt];
\fill [black] (1.95,0) circle [radius=1.75pt];
\coordinate [label=170:$2$] (p2) at (170:3);
\coordinate [label=-170:$1$] (p1) at (-170:3);
\coordinate [label=-10:$4$] (p4) at (-10:3);
\coordinate [label=10:$3$] (p3) at (10:3);
\node [anchor=center] at (150:1.4) {$c_2$};
\node [anchor=center] at (-150:1.4) {$c_1$};
\node [anchor=center] at (-30:1.4) {$c_4$};
\node [anchor=center] at (30:1.4) {$c_3$};
\node [anchor=center] at (.3,0) {$c_5$};
\end{tikzpicture}
\end{center}
\caption{Double triangle diagram.}
\label{fig:doubletriangle}
\end{figure}
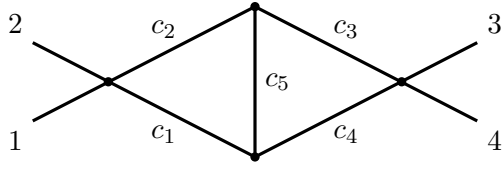

\subsection{The 4-point double triangle at two loops}\label{sec:doubletrianglediagram}

In this subsection we illustrate the general prescription discussed above by applying it to a 4-point two-loop double triangle diagram, as defined in Figure \ref{fig:doubletriangle}, where we have labeled the bulk-to-bulk propagators by their associated spectrum variables.
We begin by skipping the explicit constrution of the preliminary one-loop diagram and assuming it is known, while focusing on the construction of the two-loop diagram by form the propagator $c_1$.

\begin{figure}[ht]
\captionsetup{margin=2em}
\begin{center}
\begin{tikzpicture}
\begin{scope}[xshift=-4cm]
\node [anchor=center] at (0,-3) {(A)};
\fill [ProcessBlue!15!white] (90:2) arc [start angle=90,end angle=-90,radius=2] --cycle;
\draw [black,very thick,dotted] (0,0) circle [radius=1.5];
\draw [black,very thick] (-60:1.5) arc [start angle=-60,end angle=240,radius=1.5];
\coordinate (p1) at ($(180:1.5)+(210:1)$);
\coordinate (p2) at ($(180:1.5)+(150:1)$);
\draw [black,very thick] (p1) -- (180:1.5) -- (p2);
\node [anchor=center] at ($(0,0)!1.1!(p1)$) {$1$};
\node [anchor=center] at ($(0,0)!1.1!(p2)$) {$2$};
\coordinate (p3) at ($(45:1.5)+(75:1)$);
\coordinate (p4) at ($(45:1.5)+(15:1)$);
\draw [black,very thick] (p3) -- (45:1.5) -- (p4);
\node [anchor=center] at ($(0,0)!1.1!(p3)$) {$3$};
\node [anchor=center] at ($(0,0)!1.1!(p4)$) {$4$};
\draw [black,very thick] (90:1.5) arc [start angle=180,end angle=270,radius=1.5];
\node [anchor=center] at (240:1.9) {$c_1^-$};
\node [anchor=center] at (-60:1.9) {$c_1^+$};
\fill [black] (180:1.5) circle [radius=2.5pt];
\fill [black] (90:1.5) circle [radius=2.5pt];
\fill [black] (0:1.5) circle [radius=2.5pt];
\fill [black] (45:1.5) circle [radius=2.5pt];
\draw [black,fill=white] (-60:1.5) circle [radius=2.5pt];
\draw [black,fill=white] (-120:1.5) circle [radius=2.5pt];
\draw [red,very thick,dashed] (90:2) -- (270:2);
\node [anchor=center] at ($(0,0)!1.15!(90:2)$) {\color{red}$t_1$};
\draw [orange,very thick,dashed] (150:2.5) -- (150:1.5) .. controls (150:.5) and (210:.5) .. (210:1.5) -- (210:2.5);
\node [anchor=center] at ($(0,0)!1.2!(150:2.5)$) {\color{orange}$t_2|S$};
\draw [orange,very thick,dashed] (75:2.5) -- (75:1.5) .. controls (75:.75) and (15:.75) .. (15:1.5) -- (15:2.5);
\node [anchor=center] at ($(0,0)!1.2!(15:2.5)$) {\color{orange}$t_3|S$};
\end{scope}
\node [anchor=center] at (0,0) {\huge $\Rightarrow$};
\begin{scope}[xshift=4cm]
\node [anchor=center] at (0,-3) {(B)};
\draw [black,very thick,dotted] (0,0) circle [radius=1.5];
\draw [black,very thick] (-30:1.5) arc [start angle=-30,end angle=210,radius=1.5];
\coordinate (p1) at ($(150:1.5)+(180:1)$);
\node [anchor=center] at ($(0,0)!1.1!(p1)$) {$1$};
\coordinate (p2) at ($(150:1.5)+(120:1)$);
\node [anchor=center] at ($(0,0)!1.1!(p2)$) {$2$};
\draw [black,very thick] (p1) -- (150:1.5) -- (p2);
\coordinate (p3) at ($(30:1.5)+(60:1)$);
\node [anchor=center] at ($(0,0)!1.1!(p3)$) {$3$};
\coordinate (p4) at ($(30:1.5)+(0:1)$);
\node [anchor=center] at ($(0,0)!1.1!(p4)$) {$4$};
\draw [black,very thick] (p3) -- (30:1.5) -- (p4);
\node [anchor=center] at (210:1.9) {$c_1^-$};
\node [anchor=center] at (-30:1.9) {$c_1^+$};
\fill [black] (150:1.5) circle [radius=2.5pt];
\fill [black] (30:1.5) circle [radius=2.5pt];
\draw [black,fill=white] (-30:1.5) circle [radius=2.5pt];
\draw [black,fill=white] (-150:1.5) circle [radius=2.5pt];
\draw [red,very thick,dashed] (90:2) -- (270:2);
\node [anchor=south] at (90:2) {\color{red}$t_1$};
\draw [orange,very thick,dashed] (120:2.5) -- (120:1.5) .. controls (120:.5) and (180:.5) .. (180:1.5) -- (180:2.5);
\node [anchor=east] at (180:2.5) {\color{orange}$t_2|S$};
\draw [orange,very thick,dashed] (60:2.5) -- (60:1.5) .. controls (60:.5) and (0:.5) .. (0:1.5) -- (0:2.5);
\node [anchor=west] at (0:2.5) {\color{orange}$t_3|S$};
\end{scope}
\end{tikzpicture}
\end{center}
\caption{Double triangle from a triangle. (A) the original diagram; (B) the effective tree diagram induced by the maximal chain of channels $\{t_1\}$.}
\label{fig:triangle2doubletriangle}
\end{figure}
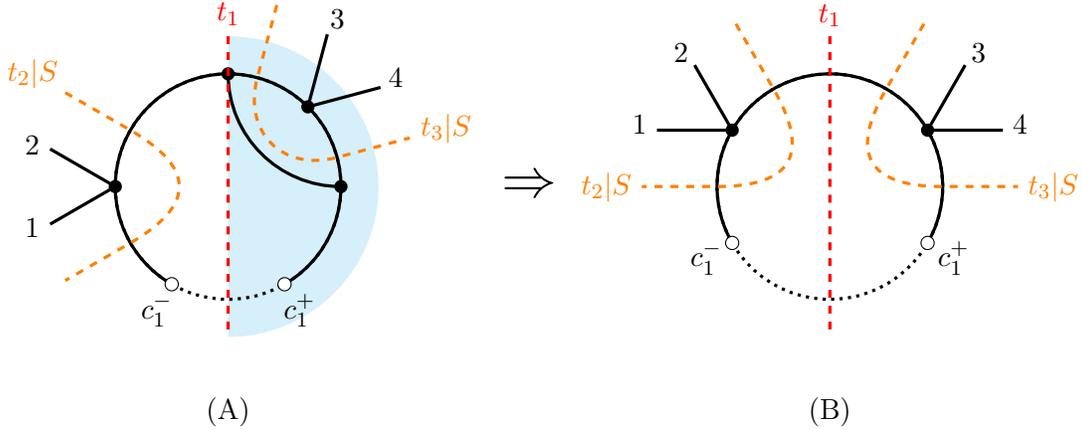

\subsubsection{Construction}

The construction we choose is shown in Figure \ref{fig:triangle2doubletriangle}. There (A) depicts the original one-loop triangle diagram as well as the two boundary points to be glued. The poles of the original pre-amplitude, expressed in terms of $\Gamma$ functions, are 
\begin{equation}\label{eq:oneloopM0poles}
\begin{split}\checked{}
M_0\big|_{\rm poles}
=&\Gamma[{\textstyle\frac{\Delta_{12}-c_1\pm c_2}{2}}]\,
\Gamma[{\textstyle\frac{h\pm c_2\pm c_3\pm c_5}{2}}]\,
\Gamma[{\textstyle\frac{\Delta_{34}\pm c_3\pm c_4}{2}}]\,
\Gamma[{\textstyle\frac{h\pm c_1\pm c_4\pm c_5}{2}}]\,
\Gamma[{\textstyle\frac{\Delta_{34}+c_1\pm c_2}{2}}]\\
&\times\Gamma[{\textstyle\frac{h\pm c_2-t_1}{2}}]\,
\Gamma[{\textstyle\frac{2h\pm c_3\pm c_4-t_3}{2}}].
\end{split}
\end{equation}
Note here that while the one-loop triangle diagram we start with slightly differs from the one constructed in Figure \ref{fig:triangle4ptC}, it can still be verified that the poles encoded in $\Gamma[{\textstyle\frac{2h\pm c_3\pm c_5-t_1}{2}}]$ are completely absent, by similar analysis as in Appendix \ref{app:sec:checkfakepoletriangle}, and so this is also consistent with Conjecture \ref{sec:enhancedrules}.

Since two of the bulk vertices here are not connected to any boundary points, the maximal chain in our choice consists of two regions, as is separated by the red dashed line. In each region we effectively have only a single vertex, and so this induces the effective diagram as shown in (B), which is qualitatively similar to that we encountered in the construction of one-loop bubble diagram in Figure \ref{fig:bubble4pt}.

In this case there are altogether three non-trivial OPE channels consistent with the planar ordering dictated by the chosen maximal chain. In terms of the effective diagram (B) we thus conclude the following substitution
\begin{equation}\label{eq:doubletrianglekernel}
\begin{split}\checked{}
&\Xi_1=h-c_1,\quad
\Xi_2=t_1,\quad
\Xi_3=h+c_1,\quad
\xi_{1,1}=t_2,\quad
\xi_{1,2}=2h,\quad
\xi_{2,2}=t_3,\\
&\Delta_{A_1}=\Delta_{12},\quad
\Delta_{A_2}=\Delta_{34},\quad
s_{1,1}=s_{2,2}=S,\quad
s_{1,2}=0.
\end{split}
\end{equation}
The kernel is thus that of the bubble together with the above substitutions.

\subsubsection{Pole structure}

We now use the above construction to determine the pole structure of this double triangle.

In the preliminary step of making a coarse estimation, it suffices to just know the poles of the one-loop diagram without further details about its analytic structure. We collect the $\Gamma$ functions for these poles both in $M_0$ and in the kernel that depend on the three integration variables $\{t_1,t_2,t_3\}$, and label them as
\begin{equation}\label{eq:relevantpolesinM0fortwoloop}\checked{}
\begin{split}
&\Gamma_{41}[{\textstyle\frac{h-c_2-t_1}{2}}]\,
\Gamma_{42}[{\textstyle\frac{h+c_2-t_1}{2}}]\,
\Gamma_{43}[{\textstyle\frac{2h-c_3-c_4-t_3}{2}}]\,
\Gamma_{44}[{\textstyle\frac{2h+c_3-c_4-t_3}{2}}]\,
\Gamma_{45}[{\textstyle\frac{2h-c_3+c_4-t_3}{2}}]\\
&\Gamma_{46}[{\textstyle\frac{2h+c_3+c_4-t_3}{2}}]\,
\Gamma_{51}[{\textstyle\frac{h-c_1+t_1-t_2}{2}}]\,
\Gamma_{52}[{\textstyle\frac{t_2-S}{2}}]\,
\Gamma_{53}[{\textstyle\frac{h-c_1-t_1+t_2}{2}}]\,
\Gamma_{54}[{\textstyle\frac{h+c_1+t_1-t_3}{2}}]\\
&\Gamma_{55}[{\textstyle\frac{2h+2S-t_2-t_3}{2}}]\,
\Gamma_{56}[{\textstyle\frac{t_3-S}{2}}]\,
\Gamma_{57}[{\textstyle\frac{h+c_1-t_1+t_3}{2}}]\,
\Gamma_{58}[{\textstyle\frac{-2h+t_2+t_3}{2}}]\,
\Gamma_{59}[{\textstyle\frac{\Delta_{12}-t_2}{2}}]\,
\Gamma_{60}[{\textstyle\frac{\Delta_{34}-t_3}{2}}].
\end{split}
\end{equation}
Here we employed a non-standard labeling for the $\Gamma$'s, for notational convenience in later analysis. Clearly $\Gamma_{41}\cdots\Gamma_{46}$ come from the one-loop pre-amplitude \eqref{eq:oneloopM0poles}, while $\Gamma_{51}\cdots\Gamma_{60}$ from from the kernel. Apart from these there are additional poles from $M_0$ that are free of the $t$ variables, as shown in the first line of \eqref{eq:oneloopM0poles}. 

Using Algorithm \ref{app:sec:polealgoritm} the new poles emerged from the three $t$ integrals are (we already eliminate the families that are ruled out by the zero from $K$ at the pinching, and all the remaining families are simple poles)
\begin{equation}\label{eq:resultdoubletrianglepolepreM}\checked{}
\begin{split}
&{\color{ForestGreen}\underset{\{51, 53\}}{\cancel{\Gamma}[{\textstyle h-c_1}]}}\,
{\color{ForestGreen}\underset{\{54, 57\}}{\cancel{\Gamma}[{\textstyle h+c_1}]}}\,
{\color{ForestGreen}\underset{\{41, 51, 54, 58\}}{\cancel{\Gamma}[{\textstyle h-c_2}]}}\,
{\color{ForestGreen}\underset{\{42, 51, 54, 58\}}{\cancel{\Gamma}[{\textstyle h+c_2}]}}\,
\underset{\{58, 59, 60\}}{\Gamma[{\textstyle\frac{\Delta_{1234}}{2}-h}]}\,
\underset{\{41, 51, 58, 60\}}{\Gamma[{\textstyle\frac{\Delta_{34}-c_1-c_2}{2}}]}\,
\underset{\{42, 51, 58, 60\}}{\Gamma[{\textstyle\frac{\Delta_{34}-c_1+c_2}{2}}]}\\
&\underset{\{41, 51, 52\}}{\Gamma[{\textstyle\frac{2h-S-c_1-c_2}{2}}]}\,
\underset{\{41, 54, 56\}}{\Gamma[{\textstyle\frac{2h-S+c_1-c_2}{2}}]}\,
\underset{\{42, 51, 52\}}{\Gamma[{\textstyle\frac{2h-S-c_1+c_2}{2}}]}\,
\underset{\{42, 54, 56\}}{\Gamma[{\textstyle\frac{2h-S+c_1+c_2}{2}}]}\\
&\underset{\{43, 56\}}{\Gamma[{\textstyle\frac{2h-S-c_3-c_4}{2}}]}\,
\underset{\{44, 56\}}{\Gamma[{\textstyle\frac{2h-S+c_3-c_4}{2}}]}\,
\underset{\{45, 56\}}{\Gamma[{\textstyle\frac{2h-S-c_3+c_4}{2}}]}\,
\underset{\{46, 56\}}{\Gamma[{\textstyle\frac{2h-S+c_3+c_4}{2}}]}\\
&{\color{ForestGreen}\underset{\{41, 43, 51, 58\}}{\cancel{\Gamma}[{\textstyle\frac{2h-c_1-c_2-c_3-c_4}{2}}]}}\,
{\color{ForestGreen}\underset{\{42, 43, 51, 58\}}{\cancel{\Gamma}[{\textstyle\frac{2h-c_1+c_2-c_3-c_4}{2}}]}}\,
{\color{ForestGreen}\underset{\{41, 44, 51, 58\}}{\cancel{\Gamma}[{\textstyle\frac{2h-c_1-c_2+c_3-c_4}{2}}]}}\,
{\color{ForestGreen}\underset{\{42, 44, 51, 58\}}{\cancel{\Gamma}[{\textstyle\frac{2h-c_1+c_2+c_3-c_4}{2}}]}}\\
&{\color{ForestGreen}\underset{\{41, 45, 51, 58\}}{\cancel{\Gamma}[{\textstyle\frac{2h-c_1-c_2-c_3+c_4}{2}}]}}\,
{\color{ForestGreen}\underset{\{42, 45, 51, 58\}}{\cancel{\Gamma}[{\textstyle\frac{2h-c_1+c_2-c_3+c_4}{2}}]}}\,
{\color{ForestGreen}\underset{\{41, 46, 51, 58\}}{\cancel{\Gamma}[{\textstyle\frac{2h-c_1-c_2+c_3+c_4}{2}}]}}\,
{\color{ForestGreen}\underset{\{42, 46, 51, 58\}}{\cancel{\Gamma}[{\textstyle\frac{2h-c_1+c_2+c_3+c_4}{2}}]}}\\
&\underset{\{41, 54, 58, 59\}}{\Gamma[{\textstyle\frac{\Delta_{12}+c_1-c_2}{2}}]}\,
\underset{\{42, 54, 58, 59\}}{\Gamma[{\textstyle\frac{\Delta_{12}+c_1+c_2}{2}}]}\,
\underset{\{43, 58, 59\}}{\Gamma[{\textstyle\frac{\Delta_{12}-c_3-c_4}{2}}]}\,
\underset{\{44, 58, 59\}}{\Gamma[{\textstyle\frac{\Delta_{12}+c_3-c_4}{2}}]}\,
\underset{\{45, 58, 59\}}{\Gamma[{\textstyle\frac{\Delta_{12}-c_3+c_4}{2}}]}\,
\underset{\{46, 58, 59\}}{\Gamma[{\textstyle\frac{\Delta_{12}+c_3+c_4}{2}}]}.
\end{split}
\end{equation}
In the above, the families of poles marked green are also absent, due to the obvious conflict with the reflection symmetries of the double triangle diagram as well as the symmetries in $c_a\mapsto-c_a$ ($\forall\text{}a$)\footnote{Though we do not present in this paper, we also checked the absence of these other poles by explicit residue computation. This check requires the knowledge about the one-loop diagram $M_0$, which itself can be contructed further from a tree, similar to our later discussion on the residues at the physical poles. But in practice the symmetry argument is strong enough to land on a quick conclusion.}.  Collecting the remaining poles above as well as those in the first line of \eqref{eq:oneloopM0poles}, we thus conclude that the poles of the 4-point double-triangle diagram in Figure \ref{fig:doubletriangle} are all encoded in
\begin{equation}\label{eq:doubletriangleMpolesshort}\checked{}
\begin{split}
&\Gamma[{\textstyle\frac{\Delta_{12}\pm c_1\pm c_2}{2}}]\,
\Gamma[{\textstyle\frac{h\pm c_1\pm c_4\pm c_5}{2}}]\,
\Gamma[{\textstyle\frac{h\pm c_2\pm c_3\pm c_5}{2}}]\,
\Gamma[{\textstyle\frac{\Delta_{34}\pm c_3\pm c_4}{2}}]\\
&\Gamma[{\textstyle\frac{\Delta_{12}\pm c_3\pm c_4}{2}}]\,
\Gamma[{\textstyle\frac{\Delta_{34}\pm c_1\pm c_2}{2}}]\,
\Gamma[{\textstyle\frac{\Delta_{1234}}{2}-h}]\\
&\Gamma[{\textstyle\frac{2h\pm c_1\pm c_2-S}{2}}]\,
\Gamma[{\textstyle\frac{2h\pm c_3\pm c_4-S}{2}}].
\end{split}
\end{equation}
The first line is consistent with the vertex rule, the second line with the loop contraction rule, and the third line with the channel rule. Since neither this diagram nor any of its sub-diagram belong to the generalized bubbles, we do not have any pole of the form $\Gamma[h\pm c]$. Furthermore, according to the channel rule there should also be poles at
\begin{equation}\label{eq:doubletriangleabsentSpoles}
\checked{}
\Gamma[{\textstyle\frac{3h\pm c_1\pm c_3\pm c_5-S}{2}}]\,
\Gamma[{\textstyle\frac{3h\pm c_2\pm c_4\pm c_5-S}{2}}],
\end{equation}
but these are composite, and indeed they did not show up in the result \eqref{eq:resultdoubletrianglepolepreM}.  Hence this two-loop double triangle also confirms the validity of Conjecture \ref{sec:enhancedrules}.

\subsubsection{Residues at poles in the $S$ channel}\label{sec:doubletriangleresidueS}

Now we move on to study properties of the amplitude $\mathcal{M}$.  We do not seek to determine the complete pole structure of this double triangle diagram.   Instead we will fully focus on the most interesting poles, the ones in the unique Mandelstam variable $S$ that the amplitude depends on.   As discussed in Section \ref{sec:polesinMandelstam} emerges in a relatively simple and organized way.  Due to the absence of the poles \eqref{eq:doubletriangleabsentSpoles} in $M$ we see that this diagram allows both minimal cuts and non-minimal cuts.   In the following, after preparing the one-loop data necessary for the anlaysis, we compute the residue at the leading pole for each of them separately.

\paragraph{The one-loop pre-amplitude $M_0$.}

In order for the residue computation we need a precise expression for $M_0$.  We choose to construct the one-loop triangle diagram $M_0$ used previously (where the propagator $c_1$ has not been created) by forming the propagator $c_4$.  Diagrammatically this is similar to Figure \ref{fig:triangle4ptC} and we do not further draw it explicitly.  The tree-level diagram entering this construction involves one vertex of valency 3 together with 3 end vertices, whose corresponding pre-amplitude we denote as $M'_0$, with
\begin{equation}\label{eq:doubletriangletreeM0}\checked{}
\begin{split}
\frac{M'_0}{C_{A_2}}=&
\frac{\Gamma[{\textstyle\frac{\Delta_{12}-c_1\pm c_2}{2}}]}{\Gamma[\frac{\Delta_{12}+h-c_1-u_2}{2}]}\,
\frac{\Gamma[{\textstyle\frac{\Delta_{34}\pm c_3+c_4}{2}}]}{\Gamma[\frac{\Delta_{34}+h+c_4-u_3}{2}]}\,
\frac{\Gamma[{\textstyle\frac{h+c_1-c_4\pm c_5}{2}}]\,
\Gamma[{\textstyle\frac{h\pm c_2\pm c_3\pm c_5}{2}}]}{\Gamma[\frac{2h+c_1-c_4-u_5}{2}]}\!\!\!
\prod_{a\in\{2,3,5\}}\!\!\!\!\Gamma[{\textstyle\frac{h\pm c_a-u_a}{2}}],
\end{split}
\end{equation}
where $\{u_2,u_3,u_5\}$ are the Mandelstam variables associated to the three tree propagators, and 
\begin{equation}
\checked{}
C_{A_2}\equiv C_{A_2}\left[0;\substack{c_2,c_3,c_5\\u_2,u_3,u_5}\right]
\end{equation}
is the correction function for the trivalent vertex, following the definition in \eqref{eq:vertexcorrectionspecial}. We denote the integration kernel from tree to one loop as $K'$, specified with the substitution
\begin{equation}\label{eq:doubletriangletreekernel}
\begin{split}
\checked{}
&\Xi_1=h-c_4,\quad
\Xi_2=u_5,\quad
\Xi_3=u_3,\quad
\Xi_4=h+c_4,\quad
\Xi_{(2,1)}=\xi_{2,2}=u_2,\\
&\Delta_{A_1}=h+c_1,\quad
\Delta_{A_2}=0,\quad
\Delta_{A_{2,1}}=\Delta_{12}+h-c_1,\quad
\Delta_{A_3}=\Delta_{34},\\
&\xi_{1,1}=h+c_1,\quad
\xi_{1,2}=u_1,\quad
\xi_{1,3}=2h,\quad
\xi_{2,3}=u_4,\quad
\xi_{3,3}=u_6,\\
&s_{1,1}=s_{2,3}=h+c_1,\quad
s_{1,2}=s_{3,3}=t_3,\quad
s_{2,2}=s_{(2,1)}=t_1,\quad
s_{1,3}=0.
\end{split}
\end{equation}

To set up further notations, we label the poles in $M'_0K'$ that are relevant for the newly emergent poles of $M_0$ (i.e., those depending on $u$'s) as
\begin{equation}\label{eq:polesinMp0Kprelevant}\checked{}
\begin{split}
&\Gamma_{1}[{\textstyle\frac{h-c_2-u_2}{2}}]\,
\Gamma_{2}[{\textstyle\frac{h+c_2-u_2}{2}}]\,
\Gamma_{3}[{\textstyle\frac{t_1-t_3+u_1-u_2}{2}}]\,
\Gamma_{4}[{\textstyle\frac{u_2-t_1}{2}}]\,
\Gamma_{5}[{\textstyle\frac{h+c_1-u_1+u_2}{2}}]\,
\Gamma_{6}[{\textstyle\frac{h-c_3-u_3}{2}}]\\
&\Gamma_{7}[{\textstyle\frac{h+c_3-u_3}{2}}]\,
\Gamma_{8}[{\textstyle\frac{h-c_4-u_1+u_3}{2}}]\,
\Gamma_{9}[{\textstyle\frac{3h+c_1-t_1+t_3-u_1+u_2-u_4}{2}}]\,
\Gamma_{10}[{\textstyle\frac{-2h+u_1-u_2+u_4}{2}}]\\
&\Gamma_{11}[{\textstyle\frac{h-c_5-u_5}{2}}]\,
\Gamma_{12}[{\textstyle\frac{h+c_5-u_5}{2}}]\,
\Gamma_{13}[{\textstyle\frac{-h-c_1+u_1-u_3+u_5}{2}}]\,
\Gamma_{14}[{\textstyle\frac{h+c_4-u_4+u_5}{2}}]\\
&\Gamma_{15}[{\textstyle\frac{-h-c_1+t_1+t_3-u_2+u_4-u_6}{2}}]\,
\Gamma_{16}[{\textstyle\frac{u_3+u_4-u_5-u_6}{2}}]\,
\Gamma_{17}[{\textstyle\frac{u_6-t_3}{2}}]\,
\Gamma_{18}[{\textstyle\frac{h+c_4-u_3+u_6}{2}}]\\
&\Gamma_{19}[{\textstyle\frac{u_2-u_4+u_6}{2}}]\,
\Gamma_{20}[{\textstyle\frac{\Delta_{34}-u_6}{2}}],
\end{split}
\end{equation}
and we further identify their emergent poles $\Gamma$ in the one-loop pre-amplitude \eqref{eq:oneloopM0poles} as
\begin{equation}\label{eq:polesoneloopemerge}\checked{}
\begin{split}
&\gamma_{\{8,11,13\}}\equiv\gamma_{35},\quad
\gamma_{\{4, 11, 14, 15, 17\}}\equiv\gamma_{36},\quad
\gamma_{\{5, 10, 11, 14\}}\equiv\gamma_{37},\quad
\gamma_{\{8, 12, 13\}}\equiv\gamma_{38},\\
&\gamma_{\{4, 12, 14, 15, 17\}}\equiv\gamma_{39},\quad
\gamma_{\{5, 10, 12, 14\}}\equiv\gamma_{40},\quad
\gamma_{\{1, 4\}}\equiv\gamma_{41},\quad
\gamma_{\{2, 4\}}\equiv\gamma_{42},\\
&\gamma_{\{3, 4, 6, 8\}}\equiv\gamma_{43},\quad
\gamma_{\{3, 4, 7, 8\}}\equiv\gamma_{44},\quad
\gamma_{\{6, 14, 16, 17\}}\equiv\gamma_{45},\quad
\gamma_{\{7, 14, 16, 17\}}\equiv\gamma_{46},\\
&\gamma_{\{1, 5, 10, 19, 20\}}\equiv\gamma_{47},\quad
\gamma_{\{2, 5, 10, 19, 20\}}\equiv\gamma_{48},\quad
\gamma_{\{6, 8, 10, 19, 20\}}\equiv\gamma_{49},\quad
\gamma_{\{7, 8, 10, 19, 20\}}\equiv\gamma_{50}.
\end{split}
\end{equation}
Note in particular that the $\gamma_{41}\cdots\gamma_{46}$ are exactly the $\Gamma_{41}\cdots\Gamma_{46}$ in \eqref{eq:relevantpolesinM0fortwoloop} that enter the two-loop construction. Apart from these the remaining poles directly entering $M_0$ as
\begin{equation}\label{eq:polesoneloopdirect}\checked{}
\begin{split}
&\Gamma_{21}[{\textstyle\frac{h-c_2-c_3-c_5}{2}}]\,
\Gamma_{22}[{\textstyle\frac{h+c_2-c_3-c_5}{2}}]\,
\Gamma_{23}[{\textstyle\frac{h-c_2+c_3-c_5}{2}}]\,
\Gamma_{24}[{\textstyle\frac{h+c_2+c_3-c_5}{2}}]\,
\Gamma_{25}[{\textstyle\frac{h+c_1-c_4-c_5}{2}}]\\
&\Gamma_{26}[{\textstyle\frac{h-c_2-c_3+c_5}{2}}]\,
\Gamma_{27}[{\textstyle\frac{h+c_2-c_3+c_5}{2}}]\,
\Gamma_{28}[{\textstyle\frac{h-c_2+c_3+c_5}{2}}]\,
\Gamma_{29}[{\textstyle\frac{h+c_2+c_3+c_5}{2}}]\,
\Gamma_{30}[{\textstyle\frac{h+c_1-c_4+c_5}{2}}]\\
&\Gamma_{31}[{\textstyle\frac{\Delta_{12}-c_1-c_2}{2}}]\,
\Gamma_{32}[{\textstyle\frac{\Delta_{12}-c_1+c_2}{2}}]\,
\Gamma_{33}[{\textstyle\frac{\Delta_{34}-c_3+c_4}{2}}]\,
\Gamma_{34}[{\textstyle\frac{\Delta_{34}+c_3+c_4}{2}}].
\end{split}
\end{equation}

Finally, to compute the amplitude in the end there are poles from the normalization $\mathcal{N}$, labeled as (again each family contains only one pole)
\begin{equation}\label{eq:polestwoloopnormalization}\checked{}
\begin{split}
&\gamma_{61}[(\underline\Delta_1-h)+c_1]\,
\gamma_{62}[(\underline\Delta_1-h)-c_1]\,
\gamma_{63}[(\underline\Delta_2-h)+c_2]\,
\gamma_{64}[(\underline\Delta_2-h)-c_2]\\
&\gamma_{65}[(\underline\Delta_3-h)+c_3]\,
\gamma_{66}[(\underline\Delta_3-h)-c_3]\,
\gamma_{67}[(\underline\Delta_4-h)+c_4]\,
\gamma_{68}[(\underline\Delta_4-h)-c_4]\\
&\gamma_{69}[(\underline\Delta_5-h)+c_5]\,
\gamma_{70}[(\underline\Delta_5-h)-c_5].
\end{split}
\end{equation}

In brief, among poles listed in \eqref{eq:relevantpolesinM0fortwoloop} and \eqref{eq:polesinMp0Kprelevant} to \eqref{eq:polestwoloopnormalization}, $\gamma_1\cdots\gamma_{34}\gamma_{51}\cdots\gamma_{70}$ are all the poles in $\mathcal{N}M'_0K'K$, from which the final amplitude $\mathcal{M}$ is obtained via the $u$, $t$ and $c$ integrals. $\gamma_{35}\cdots\gamma_{50}$ are the poles emerged in the intermediate one-loop per-amplitude $M_0$.

\paragraph{The minimal poles.}

The minimal poles consist of two families $\Gamma[\frac{\underline\Delta_{12}-S}{2}]$ and $\Gamma[\frac{\underline\Delta_{34}-S}{2}]$.  Due to the symmetry of the diagram we only focus on the leading pole of $\Gamma[\frac{\underline\Delta_{12}-S}{2}]$.

$\Gamma[\frac{\underline\Delta_{12}-S}{2}]$ emerges from the following corresponding poles in $M$ via the $c_1$ and $c_2$ integrals
\begin{equation}
\checked{}
\underset{\{41, 51, 52\}}{\Gamma[{\textstyle\frac{2h-S-c_1-c_2}{2}}]}\,
\underset{\{41, 54, 56\}}{\Gamma[{\textstyle\frac{2h-S+c_1-c_2}{2}}]}\,
\underset{\{42, 51, 52\}}{\Gamma[{\textstyle\frac{2h-S-c_1+c_2}{2}}]}\,
\underset{\{42, 54, 56\}}{\Gamma[{\textstyle\frac{2h-S+c_1+c_2}{2}}]}.
\end{equation}
Taking into all responsible poles in the integrand $\mathcal{N}M'_0K'K$ together this receive equal contribution from four $\gamma$ families (note here we also apply the identification \eqref{eq:polesoneloopemerge})
\begin{equation}
\checked{}
\gamma_{\{1, 4, 51, 52, 61, 63\}}\,
\gamma_{\{1, 4, 54, 56, 62, 63\}}\,
\gamma_{\{2, 4, 51, 52, 61, 64\}}\,
\gamma_{\{2, 4, 54, 56, 62, 64\}}.
\end{equation}
Their corresponding pinching planes do not have any intersection with each other, and so $\Gamma[\frac{\underline\Delta_{12}-S}{2}]$ are simple poles.

To compute the residue, let us focus on the family $\gamma_{\{1, 4, 51, 52, 61, 63\}}$, emerged from
\begin{equation}
\checked{}
\Gamma_{1}[{\textstyle\frac{h-c_2-u_2}{2}}]\,
\Gamma_{4}[{\textstyle\frac{u_2-t_1}{2}}]\,
\Gamma_{51}[{\textstyle\frac{h-c_1+t_1-t_2}{2}}]\,
\Gamma_{52}[{\textstyle\frac{t_2-S}{2}}]\,
\gamma_{61}[(\underline\Delta_1-h)+c_1]\,
\gamma_{63}[(\underline\Delta_3-h)+c_3].
\end{equation}
Hence the integrals $\{u_2,t_1,t_2,c_1,c_2\}$ are localized (and $\Gamma_{1}[{\textstyle\frac{h-c_2-u_2}{2}}]$ further forces the $w_2$ integral to be localized inside $C_{A_2}$), and the contribution to the residue from this pinching family is
\begin{equation}\label{eq:doubletriangleres12expr}\checked{}
\begin{split}
&(-1)^4\int[\mathrm{d}ctu]_{\rm rest}\,
\residue{S=\underline\Delta_{12}}\;
\residue{c_2=h-\underline\Delta_2}\;
\residue{c_1=h-\underline\Delta_1}\;
\residue{t_2=2h-c_1-c_2}\;
\residue{t_1=h-c_2}\;
\residue{u_2=h-c_2}\;
\residue{w_2=-c_2}\,
\frac{\mathcal{N}\,M'_0\,K'\,K}{(-2)^9}\\
&=\frac{1}{2^{14}\pi^h}\int\prod_{a=3}^5\frac{[\mathrm{d}c_a]_{\underline\Delta_a}}{\Gamma[\pm c_a]}\frac{\mathrm{d}u_1\mathrm{d}u_3\mathrm{d}u_4\mathrm{d}u_6\mathrm{d}w_5}{(2\pi i)^5}\frac{N'}{D'}\frac{\Gamma[\frac{\underline\Delta_{12}+\Delta_{12}}{2}-h]}{\Gamma[h]\Gamma[\underline\Delta_{12}]\Gamma[\frac{\Delta_{34}-\underline\Delta_{12}}{2}]\prod_{a=1}^2\Gamma[\underline\Delta_a-h+1]}.
\end{split}
\end{equation}
After the localized integrals it turns out that the integrals $\{w_3,t_3,u_5\}$ can be done by Barnes' lemmas, thus landing on the second line above, with
\begin{equation}\checked{}
\begin{split}
N'=&
\Gamma[{\textstyle\frac{h\pm c_3-u_3}{2}}]\,
\Gamma[{\textstyle\frac{h-c_4-u_1+u_3}{2}}]\,
\Gamma[{\textstyle\frac{h+c_4\pm(u_3-u_6)}{2}}]\,
\Gamma[-c_5-w_5]\,\Gamma[-w_5]\\
&\times\Gamma[{\textstyle\frac{2h+c_4+c_5-u_4+2w_5}{2}}]\,
\Gamma[{\textstyle\frac{\Delta_{34}\pm c_3+c_4}{2}}]\,
\Gamma[{\textstyle\frac{\Delta_{34}-u_6}{2}}]\,
\Gamma[{\textstyle\frac{2h-c_4\pm c_5-\underline\Delta_1}{2}}]\\
&\times\Gamma[{\textstyle\frac{h\pm(3h-u_1-u_4+u_6-\underline\Delta_1)}{2}}]\,
\Gamma[\underline\Delta_1]\,
\Gamma[{\textstyle\frac{-2h+u_4+\underline\Delta_1}{2}}]\,
\Gamma[{\textstyle\frac{-h+c_5+u_1-u_3+2w_5+\underline\Delta_1}{2}}]\\
&\times\Gamma[{\textstyle\frac{u_1-\underline\Delta_{12}}{2}}]\,
\Gamma[{\textstyle\frac{u_6-\underline\Delta_{12}}{2}}]\,
\Gamma[{\textstyle\frac{\underline\Delta_2-u_4+u_6}{2}}]\,
\Gamma[{\textstyle\frac{\underline\Delta_2\pm c_3+c_5+2w_5}{2}}]\,
\Gamma[{\textstyle\frac{2h-u_1-\underline\Delta_1+\underline\Delta_2}{2}}],
\end{split}
\end{equation}
and
\begin{equation}\checked{}
\begin{split}
D'=&\Gamma[{\textstyle\frac{2h-u_1+u_6}{2}}]\,
\Gamma[{\textstyle\frac{\Delta_{34}+h+c_4-u_3}{2}}]\,
\Gamma[{\textstyle\frac{\underline\Delta_1+c_4+c_5+u_1-u_6+2w_5}{2}}]\\
&\times\Gamma[{\textstyle\frac{4h-u_4+u_6-2\underline\Delta_1-\underline\Delta_2}{2}}]\,
\Gamma[{\textstyle\frac{\underline\Delta_2+h+c_5-u_3+2w_5}{2}}].
\end{split}
\end{equation}
Note that the data $\Delta_{12}$ does not enter $N'/D'$ at all, which is consistent with the expectation from factorization of the diagram. While the remaining integrals in \eqref{eq:doubletriangleres12expr} are still non-trivial, its pole structure (prior to the remaining $c$ integrals) can again be easily checked, which is
\begin{equation}
\Gamma[{\textstyle\frac{\underline\Delta_1\pm c_4\pm c_5}{2}}]\,
\Gamma[{\textstyle\frac{\underline\Delta_2\pm c_3\pm c_5}{2}}]\,
\Gamma[{\textstyle\frac{2h\pm c_3\pm c_4-\underline\Delta_{12}}{2}}]\,
\Gamma[{\textstyle\frac{\underline\Delta_{12}+\Delta_{34}}{2}-h}]\times
\Gamma[{\textstyle\frac{\underline\Delta_{12}+\Delta_{12}}{2}-h}].
\end{equation}

\temp{About this result, I might want to further commenting on the resulting pole structure.}

\paragraph{The non-minimal poles.}

The non-minimal poles, as seen from the double triangle diagram, are expected to belong to two families $\Gamma[\frac{\underline\Delta_{135}-S}{2}]$ and $\Gamma[\frac{\underline\Delta_{245}-S}{2}]$.  Again we only focus on the leading pole of $\Gamma[\frac{\underline\Delta_{245}-S}{2}]$, due to the symmetry of the diagram.

This case involves some new features as compared to the example we had at one loop, because the diagram allows for two minimal cuts, in correspondence to the poles
\begin{equation}\label{eq:doubletriangleintermediate12}\checked{}
\underset{\{41, 51, 52\}}{\Gamma[{\textstyle\frac{2h-S-c_1-c_2}{2}}]}\,
\underset{\{41, 54, 56\}}{\Gamma[{\textstyle\frac{2h-S+c_1-c_2}{2}}]}\,
\underset{\{42, 51, 52\}}{\Gamma[{\textstyle\frac{2h-S-c_1+c_2}{2}}]}\,
\underset{\{42, 54, 56\}}{\Gamma[{\textstyle\frac{2h-S+c_1+c_2}{2}}]}
\end{equation}
and
\begin{equation}\label{eq:doubletriangleintermediate34}\checked{}
\underset{\{43, 56\}}{\Gamma[{\textstyle\frac{2h-S-c_3-c_4}{2}}]}\,
\underset{\{44, 56\}}{\Gamma[{\textstyle\frac{2h-S+c_3-c_4}{2}}]}\,
\underset{\{45, 56\}}{\Gamma[{\textstyle\frac{2h-S-c_3+c_4}{2}}]}\,
\underset{\{46, 56\}}{\Gamma[{\textstyle\frac{2h-S+c_3+c_4}{2}}]}
\end{equation}
respectively in $M$ and the non-minimal poles under study can be connected to both of them. Together with other poles in $\mathcal{N}M'_0K'K$ that are responsible for the emergence of $\Gamma[\frac{\underline\Delta_{245}-S}{2}]$, following the discussions in Section \ref{sec:nonminimalpoles}, we see that these physical poles are contributed by $32\equiv16+16$ $\gamma$ families
\begin{equation}\label{eq:gammafromminimal12}\checked{}
\begin{split}
&\gamma_{\{1, 4, 25, 51, 52, 63, 67, 69\}}\,
\gamma_{\{1, 4, 30, 51, 52, 63, 67, 70\}}\,
\gamma_{\{2, 4, 25, 51, 52, 64, 67, 69\}}\,
\gamma_{\{2, 4, 30, 51, 52, 64, 67, 70\}}\\
&\gamma_{\{1, 4, 8, 11, 13, 54, 56, 63, 67, 69\}}\,
\gamma_{\{1, 4, 8, 12, 13, 54, 56, 63, 67, 70\}}\,
\gamma_{\{2, 4, 8, 11, 13, 54, 56, 64, 67, 69\}}\\
&\gamma_{\{2, 4, 8, 12, 13, 54, 56, 64, 67, 70\}}\,
\gamma_{\{1, 4, 5, 10, 11, 14, 51, 52, 63, 68, 69\}}\,
\gamma_{\{1, 4, 5, 10, 12, 14, 51, 52, 63, 68, 70\}}\\
&\gamma_{\{1, 4, 11, 14, 15, 17, 54, 56, 63, 68, 69\}}\,
\gamma_{\{1, 4, 12, 14, 15, 17, 54, 56, 63, 68, 70\}}\,
\gamma_{\{2, 4, 5, 10, 11, 14, 51, 52, 64, 68, 69\}}\\
&\gamma_{\{2, 4, 5, 10, 12, 14, 51, 52, 64, 68, 70\}}\,
\gamma_{\{2, 4, 11, 14, 15, 17, 54, 56, 64, 68, 69\}}\,
\gamma_{\{2, 4, 12, 14, 15, 17, 54, 56, 64, 68, 70\}}
\end{split}
\end{equation}
and
\begin{equation}\label{eq:gammafromminimal34}\checked{}
\begin{split}
&\gamma_{\{3, 4, 6, 8, 23, 56, 63, 67, 69\}}\,
\gamma_{\{3, 4, 6, 8, 24, 56, 64, 67, 69\}}\,
\gamma_{\{3, 4, 6, 8, 28, 56, 63, 67, 70\}}\,
\gamma_{\{3, 4, 6, 8, 29, 56, 64, 67, 70\}}\\
&\gamma_{\{3, 4, 7, 8, 21, 56, 63, 67, 69\}}\,
\gamma_{\{3, 4, 7, 8, 22, 56, 64, 67, 69\}}\,
\gamma_{\{3, 4, 7, 8, 26, 56, 63, 67, 70\}}\\
&\gamma_{\{3, 4, 7, 8, 27, 56, 64, 67, 70\}}\,
\gamma_{\{6, 14, 16, 17, 23, 56, 63, 68, 69\}}\,
\gamma_{\{6, 14, 16, 17, 24, 56, 64, 68, 69\}}\\
&\gamma_{\{6, 14, 16, 17, 28, 56, 63, 68, 70\}}\,
\gamma_{\{6, 14, 16, 17, 29, 56, 64, 68, 70\}}\,
\gamma_{\{7, 14, 16, 17, 21, 56, 63, 68, 69\}}\\
&\gamma_{\{7, 14, 16, 17, 22, 56, 64, 68, 69\}}\,
\gamma_{\{7, 14, 16, 17, 26, 56, 63, 68, 70\}}\,
\gamma_{\{7, 14, 16, 17, 27, 56, 64, 68, 70\}},
\end{split}
\end{equation}
where the former group is obtained via the intermediate poles $\Gamma[\frac{2h\pm c_1\pm c_2-S}{2}]$ in \eqref{eq:doubletriangleintermediate12} while the latter via $\Gamma[\frac{2h\pm c_3\pm c_4-S}{2}]$ in \eqref{eq:doubletriangleintermediate34}.  Each $\gamma$ family on their own are simple poles, but here in the situation of multiple minimal cuts we should take care to inspect whether the pinching planes associated to different $\gamma$ families have non-empty intersections or not.  If yes, then following Appendix \ref{app:sec:higherpoles} we have to work out all the greatest-codimension intersections such that they do not further intersect with any other $\gamma$ families in the list.

In this double triangle diagram, it turns out that the pinching planes for these $\gamma$ families intersect in pairs, and these intersection do not intersect any further. Let us denote these intersections again by $\gamma$, labeled by the union of the labels from both original $\gamma$'s yielding it. Altogether there are 16 families
\begin{equation}\label{eq:doubletriangleactualpinchinggamma}\checked{}
\begin{split}
&\gamma_{\{1, 3, 4, 6, 8, 11, 13, 23, 54, 56, 63, 67, 69\}}\,
\gamma_{\{1, 3, 4, 6, 8, 12, 13, 28, 54, 56, 63, 67, 70\}}\\
&\gamma_{\{1, 3, 4, 6, 8, 23, 25, 51, 52, 56, 63, 67, 69\}}\,
\gamma_{\{1, 3, 4, 6, 8, 28, 30, 51, 52, 56, 63, 67, 70\}}\\
&\gamma_{\{1, 3, 4, 7, 8, 11, 13, 21, 54, 56, 63, 67, 69\}}\,
\gamma_{\{1, 3, 4, 7, 8, 12, 13, 26, 54, 56, 63, 67, 70\}}\\
&\gamma_{\{1, 3, 4, 7, 8, 21, 25, 51, 52, 56, 63, 67, 69\}}\,
\gamma_{\{1, 3, 4, 7, 8, 26, 30, 51, 52, 56, 63, 67, 70\}}\\
&\gamma_{\{2, 3, 4, 6, 8, 11, 13, 24, 54, 56, 64, 67, 69\}}\,
\gamma_{\{2, 3, 4, 6, 8, 12, 13, 29, 54, 56, 64, 67, 70\}}\\
&\gamma_{\{2, 3, 4, 6, 8, 24, 25, 51, 52, 56, 64, 67, 69\}}\,
\gamma_{\{2, 3, 4, 6, 8, 29, 30, 51, 52, 56, 64, 67, 70\}}\\
&\gamma_{\{2, 3, 4, 7, 8, 11, 13, 22, 54, 56, 64, 67, 69\}}\,
\gamma_{\{2, 3, 4, 7, 8, 12, 13, 27, 54, 56, 64, 67, 70\}}\\
&\gamma_{\{2, 3, 4, 7, 8, 22, 25, 51, 52, 56, 64, 67, 69\}}\,
\gamma_{\{2, 3, 4, 7, 8, 27, 30, 51, 52, 56, 64, 67, 70\}}\\
&\gamma_{\{1, 4, 6, 11, 14, 15, 16, 17, 23, 54, 56, 63, 68, 69\}}\,
\gamma_{\{1, 4, 6, 12, 14, 15, 16, 17, 28, 54, 56, 63, 68, 70\}}\\
&\gamma_{\{1, 4, 7, 11, 14, 15, 16, 17, 21, 54, 56, 63, 68, 69\}}\,
\gamma_{\{1, 4, 7, 12, 14, 15, 16, 17, 26, 54, 56, 63, 68, 70\}}\\
&\gamma_{\{2, 4, 6, 11, 14, 15, 16, 17, 24, 54, 56, 64, 68, 69\}}\,
\gamma_{\{2, 4, 6, 12, 14, 15, 16, 17, 29, 54, 56, 64, 68, 70\}}\\
&\gamma_{\{2, 4, 7, 11, 14, 15, 16, 17, 22, 54, 56, 64, 68, 69\}}\,
\gamma_{\{2, 4, 7, 12, 14, 15, 16, 17, 27, 54, 56, 64, 68, 70\}}\\
&\gamma_{\{1, 4, 5, 6, 10, 11, 14, 16, 17, 23, 51, 52, 56, 63, 68, 69\}}\,
\gamma_{\{1, 4, 5, 6, 10, 12, 14, 16, 17, 28, 51, 52, 56, 63, 68, 70\}}\\
&\gamma_{\{1, 4, 5, 7, 10, 11, 14, 16, 17, 21, 51, 52, 56, 63, 68, 69\}}\,
\gamma_{\{1, 4, 5, 7, 10, 12, 14, 16, 17, 26, 51, 52, 56, 63, 68, 70\}}\\
&\gamma_{\{2, 4, 5, 6, 10, 11, 14, 16, 17, 24, 51, 52, 56, 64, 68, 69\}}\,
\gamma_{\{2, 4, 5, 6, 10, 12, 14, 16, 17, 29, 51, 52, 56, 64, 68, 70\}}\\
&\gamma_{\{2, 4, 5, 7, 10, 11, 14, 16, 17, 22, 51, 52, 56, 64, 68, 69\}}\,
\gamma_{\{2, 4, 5, 7, 10, 12, 14, 16, 17, 27, 51, 52, 56, 64, 68, 70\}}.
\end{split}
\end{equation} 
Amusingly, each (pinching plane) of these new $\gamma$ is the intersection of exactly one $\gamma$ in \eqref{eq:gammafromminimal12} and one $\gamma$ from \eqref{eq:gammafromminimal34}, e.g.,
\begin{equation}\label{eq:doubletrianglenonminimalpoleeg}
\checked{}
\gamma_{\{1, 3, 4, 6, 8, 11, 13, 23, 54, 56, 63, 67, 69\}}\equiv
\underbrace{\gamma_{\{1, 4, 8, 11, 13, 54, 56, 63, 67, 69\}}}_{\text{minimal cut }(12)}\cap
\underbrace{\gamma_{\{3, 4, 6, 8, 23, 56, 63, 67, 69\}}}_{\text{minimal cut }(34)}.
\end{equation}
As discussed in Appendix \ref{app:sec:higherpoles} each new $\gamma$ in \eqref{eq:doubletriangleactualpinchinggamma} indicates a polytope configuration of integrand singularities that pinch the integration contour simultaneously, and on their own the corresponding emergent $\gamma$ poles have to be \emph{double} poles in this case.  

We should care about possible additional zeros arising from the remaining part of the integrand that suppress the order of these poles or even kill them.  Take the family \eqref{eq:doubletrianglenonminimalpoleeg} for example. At its corresponding pinching the integration variables are localized to (upon $S=\underline\Delta_{245}$)
\begin{equation}\checked{}
\begin{split}
&u_1=\underline\Delta_{245},\quad
u_2=\underline\Delta_2,\quad
u_3=\underline\Delta_{23},\quad
u_5=\underline\Delta_5,\quad
t_1=\underline\Delta_2,\quad
t_3=\underline\Delta_{245},\\
&c_1=\underline\Delta_{45}-h,\quad
c_2=h-\underline\Delta_2,\quad
c_3=h-\underline\Delta_{25},\quad
c_4=h-\underline\Delta_4,\quad
c_5=h-\underline\Delta_5.
\end{split}
\end{equation}
It can be verified that at these values the remaining part of the integrand contains a simple zero (from the vertex correction $C_{A_2}$).  Hence, if the pole at $S=\underline\Delta_{245}$ receives contribution from only this pinching pattern, then it can at most be a simple pole\footnote{However, at this point it is not enough to determine that this is \emph{exactly} a simple pole, because depending on the specific source of the zero the order of the pole may drop by more than one.}.

To further determine the precise residue, a good practice is to slightly deform the integrand so as to split up this double pole family into a pair of simple pole families and perform the residue computation (as illustrated by the simple example \eqref{app:eq:doublepoleexample}). To preform this check, let us again focus on the family in \eqref{eq:doubletrianglenonminimalpoleeg}. Here for example we can first deform by an infinitesimal parameter $\epsilon$
\begin{equation}
\checked{}
\Gamma_3[{\textstyle\frac{t_1-t_3+u_1-u_2}{2}}]\longrightarrow
\Gamma_3[{\textstyle\frac{\epsilon+t_1-t_3+u_1-u_2}{2}}]
\end{equation}
Then the original double pole family \eqref{eq:doubletrianglenonminimalpoleeg} for $\Gamma[\frac{\underline\Delta_{245}-S}{2}]$ separates into
\begin{align}
\checked{}\gamma_{\{1, 4, 8, 11, 13, 54, 56, 63, 67, 69\}}\quad&\text{for}\quad\Gamma[{\textstyle\frac{\underline\Delta_{245}-S}{2}}],\\
\gamma_{\{3, 4, 6, 8, 23, 56, 63, 67, 69\}}\quad&\text{for}\quad\Gamma[{\textstyle\frac{\epsilon+\underline\Delta_{245}-S}{2}}],
\end{align}
both of which are simple poles at distinct locations. Hence we can compute the residues at their leading poles individually. For the family $\gamma_{\{1, 4, 8, 11, 13, 54, 56, 63, 67, 69\}}$ the relevant poles in the original integrand are
\begin{equation}\checked{}
\begin{split}
&\Gamma_{1}[{\textstyle\frac{h-c_2-u_2}{2}}]\,
\Gamma_{4}[{\textstyle\frac{u_2-t_1}{2}}]\,
\Gamma_{8}[{\textstyle\frac{h-c_4-u_1+u_3}{2}}]\,
\Gamma_{11}[{\textstyle\frac{h-c_5-u_5}{2}}]
\Gamma_{13}[{\textstyle\frac{-h-c_1+u_1-u_3+u_5}{2}}]\\
&\Gamma_{54}[{\textstyle\frac{h+c_1+t_1-t_3}{2}}]\,
\Gamma_{56}[{\textstyle\frac{t_3-S}{2}}]\,
\gamma_{63}[(\underline\Delta_2-h)+c_2]\,
\gamma_{67}[(\underline\Delta_4-h)+c_4]\,
\gamma_{69}[(\underline\Delta_5-h)+c_5].
\end{split}
\end{equation}
So its contribution to the residue is specified as
\begin{equation}\checked{}
\begin{split}
(-1)^6\int[\mathrm{d}ctu]_{\rm rest}&
\residue{S=\underline\Delta_{245}}\;
\residue{c_1=\underline\Delta_{45}-h}\;
\residue{c_5=h-\underline\Delta_5}\;
\residue{c_4=h-\underline\Delta_4}\;
\residue{c_2=h-\underline\Delta_2}\;
\residue{t_3=S}\;
\residue{t_1=h-c_2}\\
&\residue{u_3=-h+c_4+u_1}\;
\residue{u_5=h-c_5}\;
\residue{w_5=-c_5}\;
\residue{u_2=h-c_2}\;
\residue{w_2=-c_2}\,
\int\frac{\mathrm{d}w_3}{2\pi i}\frac{\mathcal{N}\,M'_0\,K'\,K}{(-2)^9}.
\end{split}
\end{equation}
It can be verified that even before the rest integrals are performed the integrand already vanishes after the above sequence of ordinary residue computations. On the other hand, for the family $\gamma_{\{3, 4, 6, 8, 23, 56, 63, 67, 69\}}$ the relevant poles in the original integrnd are
\begin{equation}\checked{}
\begin{split}
&\Gamma_{3}[{\textstyle\frac{\epsilon+t_1-t_3+u_1-u_2}{2}}]\,
\Gamma_{4}[{\textstyle\frac{u_2-t_1}{2}}]\,
\Gamma_{6}[{\textstyle\frac{h-c_3-u_3}{2}}]\,
\Gamma_{8}[{\textstyle\frac{h-c_4-u_1+u_3}{2}}]\,
\Gamma_{23}[{\textstyle\frac{h-c_2+c_3-c_5}{2}}]\\
&\Gamma_{56}[{\textstyle\frac{t_3-S}{2}}]\,
\gamma_{63}[(\underline\Delta_2-h)+c_2]\,
\gamma_{67}[(\underline\Delta_4-h)+c_4]\,
\gamma_{69}[(\underline\Delta_5-h)+c_5].
\end{split}
\end{equation}
For this family it is even more obvious why the residue has to vanish, because both poles $\Gamma_{6}[{\textstyle\frac{h-c_3-u_3}{2}}]$ and $\Gamma_{23}[{\textstyle\frac{h-c_2+c_3-c_5}{2}}]$ come from the vertex correction $C_{A_2}$ (ignoring the denominator therein), but the former requires the $w_3$ integral to be localized to $w_3=-c_3$ while the latter to $w_3=0$. Since $c_3$ is not localized to 0 at the pinching, there is a contradition.

Consequently, we observe that the leading non-minimal pole at $S=\underline\Delta_{245}$ in fact turns out to be absent!  By symmetry this also mean that the leading pole in the other family $\Gamma[\frac{\underline\Delta_{135}-S}{2}]$ is absent as well.  While the check at the subleading poles are more involved, it is in principle possible to verify the same conclusion for them, in similar ways as discussed in the verification of the absence of $\Gamma[\frac{2h\pm c_1\pm c_2-S}{2}]$ in the triangle diagram with only cubic vertices that we perform in Appendix \ref{app:sec:checkfakepoletriangle}.   In brief, the only poles of the Mandelstam variable $S$ in the Mellin amplitude for our double triangle diagram are the two minimal families.

\subsection{A conjecture on the non-minimal poles}\label{sec:conjectureonamplitude}

So far we have two explicit examples, the 4-point triangle diagram with only cubic vertices (Figure \ref{fig:triangle4ptC}) and the 4-point double triangle diagram (Figure \ref{fig:doubletriangle}), where we confirmed that in the Mellin amplitude $\mathcal{M}$ there are no poles present in correspondence to the non-minimal cuts.  In the former example there is a clear phsyical motivation, because the triangle therein only results in a one-loop renormalization of the 3-point coupling \cite{Aharony:2016dwx}.  But for the absence of the poles for the three-propagator cuts in the latter example we are not aware of a simple physical explanation. Nevertheless, the notion of (non-)minimal cuts/poles provides a clear criteria for what poles of the Mandelstam variables are actually present in the Mellin amplitude. Hence we propose the following conjecture:
\begin{conjecture}{Conjecture on $\mathcal{M}$}
For diagrams that do not contain any sub-diagram (including itself) as a generalized bubble, the only families of poles of the Mandelstam variables present in the Mellin amplitude are in one-to-one correspondence to its minimal cuts.
\end{conjecture}
Here we emphasize to exclude diagrams containing generalized bubbles, because there are clear evidence from diagrams of this type that their Mellin amplitudes indeed contain non-minimal poles, for which we provide several examples in Appendix \ref{app:sec:polegbubbles}.  It would be interesting to understand more clearly how the generalized bubble diagrams differ from generic ones in future. Nevertheless, for diagrams that are free of generalized bubbles, we expect the above conjecture to be valid to all loops. Additional evidence is provided by two non-planar diagrams at two loops in Appendix \ref{app:sec:nonplanar2loop}.

\newpage

\acknowledgments

EYY would like to thank Nima Afkhami-Jeddi, Nima Arkani-Hamed, Jacob Bourjaily, Carlos Cardona, Bartek Czech, Bo Feng, Song He, Johannes Henn, Yu-tin Huang, Mauricio Romo, Shu-heng Shao, Marcus Spradlin, Anastasia Volovich, Pedro Vieira, and in particular Shota Komatsu and Eric Perlmutter, for useful discussions, and would like to further thank Eric Perlmutter for many useful suggestions on the first draft of this paper.
EYY is supported by the DOE under grant DE-SC0009988 and by a Carl P.~Feinberg Founders Circle Membership.

\appendix

\newpage

\section{Bulk-to-Bulk Propagators}\label{app:sec:bbpropagators}

\temp{This appendix is already in a good shape.}

In this appendix we discuss in more detail the properties of bulk-to-bulk propagators mentioned in Section \ref{sec:bubblereview}. While this is mostly a review of existing results \cite{Penedones:2010ue,Hijano:2015zsa}, we present it in a way so that they naturally fit into our context of discussion.

In general we can define the so-called AdS harmonic functions
\begin{equation}
\checked{}
\Omega_c[X,Y]=\underset{\partial\text{AdS}}{\int}\mathrm{d}P\frac{\mathcal{N}_c}{(-2P\cdot X)^{h+c}(-2P\cdot Y)^{h-c}},
\end{equation}
where $\mathcal{N}_c=\frac{\Gamma[h\pm c]}{2\pi^{2h}\Gamma[\pm c]}$ is the same normalization factor as introduced in \eqref{eq:bbpropagatordef}. These functions nicely span an orthogonal basis, so that we can expand any functions $F[X,Y]$ as
\begin{equation}
\checked{}
F[X,Y]=\int\frac{\mathrm{d}c}{2\pi i}\,f[c]\,\Omega_c[X,Y],
\end{equation}
where
\begin{equation}
\checked{}
f[c]=\frac{1}{\Omega_c[Y,Y]}\underset{\partial\text{AdS}}{\int}\mathrm{d}X\,F[X,Y]\,\Omega_c[X,Y],\qquad
\Omega_c[Y,Y]=\frac{\pi^h\Gamma[h]}{\Gamma[2h]}\mathcal{N}_c.
\end{equation}
Obviously the spectrum function $f[c]$ for the bulk-to-bulk propagator $G_{\rm bb}^{\underline\Delta}[X,Y]$ is
\begin{equation}\label{app:eq:gDeltac}
\checked{}
g^{\underline\Delta}[c]\equiv\frac{1}{(\underline\Delta-h)^2-c^2}.
\end{equation}

With this tool we can work out the spectrum function for products of bulk-to-bulk propagators, which are discussed in the following.

\subsection{Parallel product}

Let us first look at the product of $r$ bulk-to-bulk propagators
\begin{equation}\label{app:eq:rbbpropagators}
\checked{}
G_{(r)}[X,Y]\equiv \prod_{a=1}^rG_{\rm bb}^{\underline\Delta_a}[X,Y]\equiv\int\frac{\mathrm{d}c}{2\pi i}\,g_{(r)}[c]\,\Omega_c[X,Y].
\end{equation}
$g_{(a)}$ can be determined recursively using the obvious relation $G_{(a)}=G_{(a-1)}G_{\rm bb}^{\underline\Delta_a}$
\begin{equation}\label{app:eq:garecursion}
\begin{split}
\checked{}
g_{(a)}[c]=&\int\frac{\mathrm{d}c_{a-1}\mathrm{d}c_a}{(2\pi i)^2}\underset{\text{AdS}}{\int}\mathrm{d}X\,g_{(a-1)}[c_{a-1}]\,g^{\underline\Delta_a}[c_a]\,\frac{\Omega_{c_{a-1}}[X,Y]\,\Omega_{c_a}[X,Y]\,\Omega_c[X,Y]}{\Omega_c[Y,Y]}\\
=&\frac{1}{8\pi^h}\int\frac{\mathrm{d}c_{a-1}\mathrm{d}c_a}{(2\pi i)^2}\,g_{(a-1)}[c_{a-1}]\,g^{\underline\Delta_a}[c_a]\,\frac{\Gamma[\frac{h\pm c\pm c_{a-1}\pm c_a}{2}]}{\Gamma[h]\Gamma[h\pm c]\Gamma[\pm c_{a-1}]\Gamma[\pm c_a]}.
\end{split}
\end{equation}
\temp{I have double-checked this result.}

In particular, $g_{(2)}[c]$ is obtained from $g^{\Delta_1}[c_1]$ and $g^{\Delta_2}[c_2]$ by integrating $c_1$ and $c_2$. It can be easily checked that any pole of $g_{(2)}$ in $c$ can arise only from pinching the $c_1$ and $c_2$ contours by one pole of $g^{\Delta_1}$, one pole of $g^{\Delta_2}$, and a third pole from $\Gamma[\frac{h\pm c\pm c_1\pm c_2}{2}]$ (any other pole that are suggested by the contour analysis are killed by the $\Gamma$ functions in the denominator). Thus this type of poles are all encoded in
\begin{equation}
\checked{}
\Gamma[{\textstyle\frac{\underline\Delta_{12}-h\pm c}{2}}].
\end{equation}
Note that if we treat this product of two propagators as a single effective propagator, and accordingly use $g_{(2)}$ instead of $g^{\underline\Delta}$ in the normalization for the Mellin amplitude $\mathcal{M}$, then the pre-amplitude $M$ is the same as that of an ordinary exchange diagram \eqref{eq:Mexchange}, whose only poles depending on the Mandelstam variable $S$ are in $\Gamma[\frac{h\pm c-S}{2}]$. Hence from the $c$ integration the only physical poles of a bubble diagram are from
\begin{equation}
\checked{}
\Gamma[{\textstyle\frac{\underline\Delta_{12}-S}{2}}],
\end{equation}
corresponding to the presence of the double-trace operators $[\underline{\mathcal{O}}_1\underline{\mathcal{O}}_2]_{k,\ell}$ at the twist $\underline{\Delta}_{12}+2k$ in the OPE expansion in this channel.

A large portion of our interest is in the pre-amplitude of the original diagram, hence in general we define an object $\tilde{g}_{(r)}$ by
\begin{equation}
\checked{}
g_{(r)}[c]=\int\prod_{a=1}^r\frac{[\mathrm{d}c_a]_{\underline\Delta_a}}{\Gamma[\pm c_a]}\,\tilde{g}_{(r)}[c_1,\ldots,c_a;c].
\end{equation}
We can write down an explicit expression for $\tilde{g}_{(r)}$ by iterating \eqref{app:eq:garecursion}, and applying similar logic to the above, it is not hard to observe that the only poles of $\tilde{g}_{(r)}$ are encoded in
\begin{equation}
\checked{}
\Gamma[{\textstyle\frac{(r-1)h\pm c\pm c_1\pm\cdots\pm c_r}{2}}],
\end{equation}
(to be brief, there are $r-2$ integrations involved, and the actual poles can arise only by pinching these contours using one pole from each $\Gamma[\frac{h\pm c_a\pm c_b\pm c_c}{2}]$ in the original integrand, because any other choice necessarily involves two or more poles from the same group of factors, which then localizes some of the integration variables to values that force the denominator to diverge). And so after taking into consideration the resulting effective  single exchange diagram as before, the poles of the pre-amplitude that depend on $S$ are
\begin{equation}
\checked{}
\Gamma[{\textstyle\frac{r\,h\pm c_1\pm\cdots\pm c_r-S}{2}}].
\end{equation}
Of course this immediately indicates that the physical pole coming from the product of $r$ bulk-to-bulk propagators \eqref{app:eq:rbbpropagators} are
\begin{equation}
\checked{}
\Gamma[{\textstyle\frac{\sum_{a=1}^r\underline\Delta_a-S}{2}}].
\end{equation}

\subsection{Series product}

Now let us instead look at another type of product. In the simplest cast we assume two bulk-to-bulk propagators
\begin{align}
\checked{}G_1[X,Y]&=\int\frac{\mathrm{d}c_1}{2\pi i}\,g_1[c_1]\,\Omega_{c_1}[X,Y],\\
\checked{}G_2[Y,Z]&=\int\frac{\mathrm{d}c_2}{2\pi i}\,g_2[c_2]\,\Omega_{c_2}[Y,Z],
\end{align}
which share only one common bulk point. Here we only assume $G_1$ and $G_2$ to be some effective propagators (for which $G_{(r)}$ introduced above is a prototype), i.e., we do not require $g_a$ to be exactly $g^{\underline\Delta_a}$ as defined in \eqref{app:eq:gDeltac}. We then study the product
\begin{equation}
\checked{}
G^{(2)}[X,Z]\equiv\underset{\text{AdS}}{\int}\mathrm{d}Y\,G_1[X,Y]\,G_2[Y,Z]\equiv\int\frac{\mathrm{d}c}{2\pi i}\,g^{(2)}[c]\,\Omega_c[X,Z].
\end{equation}
Inversing the transformation we can express $g^{(2)}$ in terms of $g_1$ and $g_2$ as
\begin{equation}
\begin{split}
\checked{}
g^{(2)}[c]=&\int\frac{\mathrm{d}c_1\mathrm{d}c_2}{(2\pi i)^2}\underset{\text{AdS}}{\int}\mathrm{d}X\mathrm{d}Y\,g_1[c_1]\,g_2[c_2]\,\frac{\Omega_c[X,Z]\,\Omega_{c_1}[X,Y]\,\Omega_{c_2}[Y,Z]}{\Omega_c[Z,Z]}.
\end{split}
\end{equation}
Here we will again encounter the same issue of regularization as discussed around \eqref{eq:I3regulated}, which forces the $c_1$ and $c_2$ integration to be localized, yielding
\begin{equation}
\checked{}
g^{(2)}[c]=g_1[c]\,g_2[c].
\end{equation}
This simple relation can as well be iterated as we consider product of effective bulk-to-bulk propagators of the same type.

In the special case when the two original propagators are ordinary bulk-to-bulk propagators, with conformal dimensions $\underline\Delta_1$ and $\underline\Delta_2$ respectively, such product reduces to
\begin{equation}
\begin{split}
\checked{}
G^{(2)}[X,Z]=&\int\frac{\mathrm{d}c}{2\pi i}\,\frac{1}{(\underline\Delta_1-h)^2-c^2}\,\frac{1}{(\underline\Delta_2-h)^2-c^2}\,\Omega_c[X,Z]\\
=&\frac{1}{m_1^2-m_2^2}\left(G_{\rm bb}^{\underline\Delta_2}[X,Z]-G_{\rm bb}^{\underline\Delta_1}[X,Z]\right),
\end{split}
\end{equation}
where $m_a^2=\underline\Delta_a(\underline\Delta_a-2h)$ follows the usual definition of the mass of AdS paricle associated to the operator $\underline{\mathcal{O}}_a$. The second line above arise from the first line just by a simple algebraic identity of the integrand, and exactly reproduces the result in (4.8) of \cite{Hijano:2015zsa}.

\newpage

\section{Gluing a Disconnected Diagram by a Single Vertex}\label{app:sec:forest2tree}

\temp{I have double-checked the expressions in this entire section except for the last subsection. No major revisions further needed.}

\begin{figure}[ht]
\captionsetup{margin=2em}
\begin{center}
\begin{tikzpicture}
\begin{scope}[xshift=-3.4cm]
\draw [WildStrawberry,very thick,dotted] (-90:1) -- (-150:2) -- (160:.75);
\draw [WildStrawberry,very thick,dotted] (-90:1) -- (-30:2) -- (20:.75);
\draw [black,thick] (-90:1) -- (-110:2);
\draw [black,thick] (-90:1) -- (-90:2);
\draw [black,thick] (-90:1) -- (-70:2);
\fill [black] (-90:1) circle [radius=3pt];
\draw [black,thick] (160:.75) -- (170:2);
\draw [black,thick] (160:.75) -- (150:2);
\draw [black,thick] (160:.75) -- (130:2);
\draw [black,thick] (20:.75) -- (10:2);
\draw [black,thick] (20:.75) -- (30:2);
\draw [black,thick] (20:.75) -- (50:2);
\draw [black,fill=black!15!white] (160:.75) circle [radius=.5];
\draw [black,fill=black!15!white] (20:.75) circle [radius=.5];
\draw [black,very thick] (0,0) circle [radius=2cm];
\fill [WildStrawberry] (-150:2) circle [radius=2pt];
\node [anchor=north east] at (-150:2) {$Q_1$};
\fill [WildStrawberry] (-30:2) circle [radius=2pt];
\node [anchor=north west] at (-30:2) {$Q_2$};
\end{scope}
\node [anchor=center] at (0,0) {\Huge\color{WildStrawberry} $\longrightarrow$};
\node [anchor=south] at (0,.1) {\scriptsize\color{WildStrawberry} $\int\prod_{a}\!\!\frac{[\mathrm{d}c_{(a)}]_{\underline\Delta_a}\mathcal{N}_{c_{(a)}}}{\mathcal{C}_{h\pm c_{(a)}}}$};
\node [anchor=north] at (0,-.1) {\scriptsize\color{WildStrawberry} $\int\prod_{a}\mathrm{d}^dQ_a$};
\begin{scope}[xshift=3.4cm]
\draw [WildStrawberry,very thick,dotted] (20:.75) -- (-90:1) -- (160:.75);
\draw [black,thick] (-90:1) -- (-110:2);
\draw [black,thick] (-90:1) -- (-90:2);
\draw [black,thick] (-90:1) -- (-70:2);
\fill [black] (-90:1) circle [radius=3pt];
\draw [black,thick] (160:.75) -- (170:2);
\draw [black,thick] (160:.75) -- (150:2);
\draw [black,thick] (160:.75) -- (130:2);
\draw [black,thick] (20:.75) -- (10:2);
\draw [black,thick] (20:.75) -- (30:2);
\draw [black,thick] (20:.75) -- (50:2);
\draw [black,fill=black!15!white] (160:.75) circle [radius=.5];
\draw [black,fill=black!15!white] (20:.75) circle [radius=.5];
\draw [black,very thick] (0,0) circle [radius=2cm];
\node [anchor=north east] at ($(160:.75)!.5!(-90:1)$) {\color{WildStrawberry}$\underline\Delta_1$};
\node [anchor=north west] at ($(20:.75)!.5!(-90:1)$) {\color{WildStrawberry}$\underline\Delta_2$};
\end{scope}
\end{tikzpicture}
\end{center}
\caption{Vertex insertion that glues a disconnected diagram.}
\label{app:fig:vertexinsertiontoforest}
\end{figure}
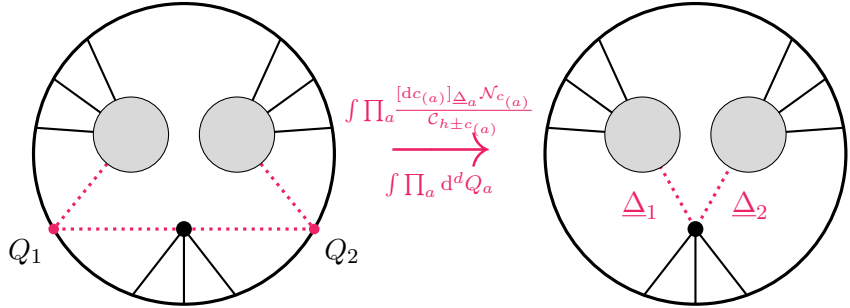

In this appendix we derive the kernel for the operation of gluing a disconnected diagram into a connected one by an additional bulk vertex, as illustrated in Figure \ref{app:fig:vertexinsertiontoforest}. This is a natural generalization for the operation of inserting a vertex to a connected diagram studied in Section \ref{sec:elementaryoperations}. This gives rise to a symmetric expression for the correction function of arbitrary bulk vertex, as will be worked out in Appendix \ref{app:sec:vertexcorrection}. As shown in Appendix \ref{app:sec:proofofequivalence} the result also makes the connection to previous literature manifest.

\subsection{Derivation of the kernel}

\subsubsection{New vertex that is internal}

\temp{I have checked the normalization constant in this subsection; in particular the resulting $\mathcal{N}$ is consistent with that in the introduction at tree level. The other expressions are also double-checked.}

To make the discussion simple, we first inspect the special case that the new bulk vertex we insert is purely internal, i.e., not incident to any bulk-to-boundary propagators.

Without loss of generlity let us assume the original diagram consists of $r$ connected components. We associate the Mellin pre-amplitude $M_a$ to the $a^{\rm th}$ component, with the Mellin variables $\{\tau_{i_a\,j_a}\}$. Correspondingly we consider an extra contact diagram $M_0$ with $r$ boundary points, which we label by $\{1,2,\ldots,r\}$, and we use this to bridge the $r$ components.

Recall the definition of the contact diagram
\begin{equation}
\checked{}
\underset{\rm AdS}{\int}\mathrm{d}X\prod_{a=1}^r\frac{\mathcal{C}_{h+c_{(a)}}}{(-2Q_a\cdot X)^{h+c_{(a)}}}.
\end{equation}
So we can directly write down the expression for the correlator resulting from the gluing
\begin{equation}
\begin{split}\checked{}
\mathcal{I}=\frac{2}{\pi^{(r+1)h}}\int\!\mathcal{N}&\int[\mathrm{d}\tau]\prod_{a=1}^r\left(\!M_a\prod_{i_a<j_a}\frac{\Gamma[\tau_{i_a\,j_a}]}{P_{i_a\,j_a}^{\tau_{i_a\,j_a}}}\!\right)
\\
&\times\underset{\rm AdS}{\int}\mathrm{d}X\prod_{a=1}^r\underset{\partial\text{AdS}}{\int}\!\mathrm{d}Q_a\frac{\Gamma[h+c_{(a)}]}{(-2Q_a\cdot X)^{h+c_{(a)}}}\prod_{k_a}\frac{\Gamma[\tau_{a_a\,k_a}]}{(-2Q_a\cdot P_{k_a})^{\tau_{a_a\,k_a}}},
\end{split}
\end{equation}
with the normalization $\mathcal{N}=\frac{\pi^{(r+1)h}}{2}\prod_{a=1}^r\frac{\mathcal{N}_a\,[\mathrm{d}c_{(a)}]_{\underline\Delta_a}\mathcal{N}_{c_{(a)}}}{\mathcal{C}_{h- c_{(a)}}\Gamma[h+c_{(a)}]}$. Here the sampling of label $i_a$ for each connected component $M_a$ is always implicitly assumed to be over all boundary points except for the one point, $a_a$, which we use to glue (with conformal dimension $h+c_a$). Hence in particular we have
\begin{equation}
\checked{}
\sum_{i_a}\tau_{a_a\,i_a}=h-c_{(a)},\qquad\forall a,
\end{equation}
which guarantees that the total weight of $Q_a$ in the integral is correctly $2h$. We thus integrate out every $Q_a$, 
and by apply the Mellin transformation $\frac{1}{(1+W_a)^{h}}=\int\frac{\mathrm{d}w_a}{2\pi i}\frac{\Gamma[h+w_a]\Gamma[-w_a]}{\Gamma[h]W_a^{h+w_a}}$ for each $a$ we have
\begin{equation}
\begin{split}\checked{}
\mathcal{I}=\frac{2}{\pi^h}\!\int\!\mathcal{N}&\int[\mathrm{d}\tau]\prod_{a=1}^r\left(M_a\prod_{i_a<j_a}\frac{\Gamma[\tau_{i_a\,j_a}]}{P_{i_a\,j_a}^{\tau_{i_a\,j_a}}}\right)
\\
&\times\int[\mathrm{d}\alpha]\int[\mathrm{d}w]\underset{\rm AdS}{\int}\mathrm{d}X\prod_{a=1}^r\frac{\Gamma[h+w_a]\Gamma[-w_a]\prod_{k_a}\alpha_{a_a\,k_a}^{\tau_{a_a\,k_a}-1}}{(-2X\cdot Y_a-Y_a^2)^{h+w_a}}.
\end{split}
\end{equation}
with $Y_a\equiv \sum_{k_a}\alpha_{a_a\,k_a}P_{k_a}$. This is further equivalent to
\begin{equation}
\begin{split}\checked{}
\mathcal{I}=&\frac{2^{r+1}}{\pi^h}\!\!\int\!\mathcal{N}\int[\mathrm{d}\tau]\prod_{a=1}^r\left(M_a\prod_{i_a<j_a}\frac{\Gamma[\tau_{i_a\,j_a}]}{P_{i_a\,j_a}^{\tau_{i_a\,j_a}}}\right)\text{exp}\!\!\left[-\sum_{a=1}^r\sum_{i_a<j_a}\alpha_{a_a\,i_a}\alpha_{a_a\,j_a}P_{i_a\,j_a}\right]\\
&\quad\times\int[\mathrm{d}\alpha]\int[\mathrm{d}w]\underset{\rm AdS}{\int}\mathrm{d}X\prod_{a=1}^r\frac{\Gamma[-w_a]\Gamma[h+c_{(a)}+2w_a]\prod_{k_a}\alpha_{a_a\,k_a}^{\tau_{a_a\,k_a}-1}}{(-2X\cdot \sum_{k_a}\alpha_{a\,k_a}P_{k_a})^{h+c_{(a)}+2w_a}},
\end{split}
\end{equation}
After integrating $X$, we can arrange the expression to
\begin{equation}
\begin{split}\checked{}
\mathcal{I}=&2^{r+1}\!\!\int\mathcal{N}\,\int[\mathrm{d}w]\Gamma[{\textstyle\frac{(r-2)h+\sum_{a=1}^rc_{(a)}}{2}+\sum_{a=1}^rw_a}]\int[\mathrm{d}\tau]\prod_{a=1}^r\left(M_a\prod_{i_a<j_a}\frac{\Gamma[\tau_{i_a\,j_a}]}{P_{i_a\,j_a}^{\tau_{i_a\,j_a}}}\right)\\
&\times\int[\mathrm{d}\alpha][\mathrm{d}\beta]\prod_{a=1}^r\left(\Gamma[-w_a]\beta_a^{h+c_{(a)}+2w_a-1}\prod_{k_a}\alpha_{a_a\,k_a}^{\tau_{a_a\,k_a}-1}\right)\\
&\times\exp\!\!\left[-\sum_{a=1}^r\sum_{i_a<j_a}(1+\beta_a^2)\alpha_{a_a\,i_a}\alpha_{a_a\,j_a}P_{i_a,j_a}-\!\!\!\!\sum_{1\leq a<b\leq r}\sum_{i_a,j_b}\beta_a\beta_b\alpha_{a_a\,j_a}\alpha_{b_b\,j_b}P_{i_a\,j_b}\right].
\end{split}
\end{equation}
This can then be used to extract the Mellin amplitude associated to the new diagram. 

At this stage we can choose to first integrate away $w$ variables, and with a further redefinition $\beta_a\mapsto\sqrt{\beta_a}$ we obtain
\begin{equation}\label{app:eq:satellitediagramspecial}
\begin{split}\checked{}
M[\delta]=&\int[\mathrm{d}\tau]\prod_{a=1}^r\left(M_a\prod_{i_a<j_a}\frac{\Gamma[\tau_{i_a\,j_a}]\Gamma[\delta_{i_a\,j_a}-\tau_{i_a\,j_a}]}{\Gamma[\delta_{i_a\,j_a}]}\right)\\
&\times\int[\mathrm{d}\beta]
\frac{\Gamma[\frac{(r-2)h+\sum_{a=1}^rc_{(a)}}{2}]}{(1+\sum_{a=1}^r\beta_a)^{\frac{(r-2)h+\sum_{a=1}^rc_{(a)}}{2}}}\prod_{a=1}^r\frac{\beta_a^{\frac{h+c_{(a)}-s_{(a)}}{2}-1}}{(1+\beta_a)^{\frac{h-c_{(a)}-s_{(a)}}{2}}}.
\end{split}
\end{equation}
Alternatively we can also choose to integrate out $\beta$ variables first, leading to
\begin{equation}\label{eq:forest2treespecialmellin}
\begin{split}\checked{}
M[\delta]=&\int[\mathrm{d}\tau][\mathrm{d}w]\prod_{a=1}^r\left(M_a\prod_{i_a<j_a}\frac{\Gamma[\tau_{i_a\,j_a}]\Gamma[\delta_{i_a\,j_a}-\tau_{i_a\,j_a}]}{\Gamma[\delta_{i_a\,j_a}]}\right)\\
&\times\,\Gamma[{\textstyle\frac{(r-2)h+\sum_{a=1}^rc_{(a)}}{2}}+\sum_{a=1}^rw_a]\prod_{a=1}^r\frac{\Gamma[-w_a]\Gamma[-c_{(a)}-w_a]\Gamma[\frac{h+c_{(a)}-s_{(a)}}{2}+w_a]}{\Gamma[\frac{h-c_{(a)}-s_{(a)}}{2}]}.
\end{split}
\end{equation}
Depending on specific purposes one of the two expressions can be more convenient than the other. The remaining $\beta$ or $w$ integrations are non-trivial and we leave them aside here.

\subsubsection{Generic new vertex}

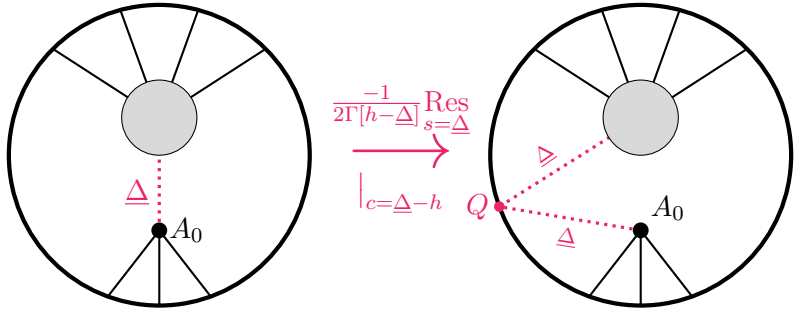
\begin{figure}[ht]
\captionsetup{margin=2em}
\begin{center}
\begin{tikzpicture}
\begin{scope}[xshift=3.2cm]
\draw [WildStrawberry,very thick,dotted] (-90:1) -- (-160:2) -- (90:.5);
\node [anchor=south west] at (-90:1) {$A_0$};
\node [anchor=south,rotate=35] at ($(-160:2)!.4!(90:.5)$) {\color{WildStrawberry}\scriptsize $\underline\Delta$};
\node [anchor=north,rotate=-9] at ($(-160:2)!.5!(-90:1)$) {\color{WildStrawberry}\scriptsize $\underline\Delta$};
\draw [black,thick] (-90:1) -- (-110:2);
\draw [black,thick] (-90:1) -- (-90:2);
\draw [black,thick] (-90:1) -- (-70:2);
\fill [black] (-90:1) circle [radius=3pt];
\draw [black,thick] (90:.5) -- (45:2);
\draw [black,thick] (90:.5) -- (75:2);
\draw [black,thick] (90:.5) -- (105:2);
\draw [black,thick] (90:.5) -- (135:2);
\draw [black,fill=black!15!white] (90:.5) circle [radius=.5];
\draw [black,ultra thick] (0,0) circle [radius=2cm];
\fill [WildStrawberry] (-160:2) circle [radius=2pt];
\node [anchor=east] at (-160:2) {\color{WildStrawberry}$Q$};
\end{scope}
\node [anchor=center] at (0,0) {\Huge\color{WildStrawberry} $\longrightarrow$};
\node [anchor=south] at (0,.1) {\color{WildStrawberry} $\frac{-1}{2\Gamma[h-\underline\Delta]}\residue{s=\underline\Delta}$};
\node [anchor=north] at (0,-.1) {\color{WildStrawberry} $\big|_{c=\underline\Delta-h}$};
\begin{scope}[xshift=-3.2cm]
\draw [WildStrawberry,very thick,dotted] (-90:1) -- (90:.5);
\node [anchor=east] at (-90:.5) {\color{WildStrawberry}$\underline\Delta$};
\node [anchor=west] at (-90:1) {$A_0$};
\draw [black,thick] (-90:1) -- (-110:2);
\draw [black,thick] (-90:1) -- (-90:2);
\draw [black,thick] (-90:1) -- (-70:2);
\fill [black] (-90:1) circle [radius=3pt];
\draw [black,thick] (90:.5) -- (45:2);
\draw [black,thick] (90:.5) -- (75:2);
\draw [black,thick] (90:.5) -- (105:2);
\draw [black,thick] (90:.5) -- (135:2);
\draw [black,fill=black!15!white] (90:.5) circle [radius=.5];
\draw [black,ultra thick] (0,0) circle [radius=2cm];
\end{scope}
\end{tikzpicture}
\end{center}
\caption{Cutting a tree propagator.}
\label{fig:cutpropagator}
\end{figure}

To obtain a useful formula for the insertion of a generic vertex (thus allowing boundary points to be attached), it is most convenient to start from the formula obtained above and ``chop off'' a subset of the components in the original diagram. As illustrated in Figure \ref{fig:cutpropagator}, this requires the operation of cutting up existing tree bulk-to-bulk propagators in the diagram, which is the inverse of the gluing we performed in Section \ref{sec:vertexinsertiongeneral}.
This is a permissible operation on the Mellin amplitude since it has to factorize when a tree propagator is ``on-shell''. Here we would like that the boundary point coming from this cut be a scalar with its conformal dimension the same as that of the original propagator, $\underline\Delta$. At the level of Mellin pre-amplitude, this amounts to identifying $c=\underline\Delta-h$ (or $c=h-\underline\Delta$, which leads to the same result since $M$ is invariant under changing the sign of $c$), and then taking the residue at $s=\underline\Delta$, and then further chop of a factor $-\frac{1}{2\Gamma[h-\underline\Delta]}$. Take the exchange diagram \eqref{eq:Mexchange} for instance. We thus have
\begin{equation}
\begin{split}
\checked{}
-\frac{1}{2\Gamma[h-\underline\Delta_1]}\residue{s_{(1)}=\underline\Delta_1}M_{\rm exchange}\Big|_{c_{(1)}=\underline\Delta_1-h}&=\Gamma[{\textstyle\frac{\Delta_{A_0}+\underline\Delta_1}{2}-h}]\,\Gamma[{\textstyle\frac{\Delta_{A_1}+\underline\Delta_1}{2}-h}],
\end{split}
\end{equation}
which exactly results in the product of the pre-amplitudes associated to the two individual bulk vertices, where the new boundary points from the cut has dimension $\underline\Delta_1$.

Let us now apply this procedure to the insertion of a generic vertex to a disconnected diagram. Without loss of generality let us assume there are $r$ components in the original diagram, and that the vertex itself is incident to $r+q$ propagators. To obtain a general formula for the insertion, we start by pretending that the original diagram has $q$ additional components, each being a contact diagram, hence \eqref{eq:forest2treespecialmellin} becomes available. We then cut the tree propagators associated to these additional components, following the discussion above (we denote their conformal dimensions as $\Delta_{a}$, $a=r+1,\ldots,r+q$), and further chopping off their pre-amplitudes. Note that after we evaluate $c_{(a)}=\Delta_a-h$ for $a>\tilde{r}$ the pole in $s_{(a)}$ only arises from the $w_a$ integration
\begin{equation}
\begin{split}\checked{}
&\int[\mathrm{d}\tau][\mathrm{d}w]\prod_{a=1}^r\left(M_a\prod_{i_a<j_a}\frac{\Gamma[\tau_{i_a\,j_a}]\Gamma[\delta_{i_a\,j_a}-\tau_{i_a\,j_a}]}{\Gamma[\delta_{i_a\,j_a}]}\right)
\\
&\times\Gamma[{\textstyle\frac{\Delta_{A_0}+(r-2)h+\sum_{a=1}^rc_{(a)}}{2}}+\sum_{a=1}^{r+q}w_a]\prod_{a=1}^r\frac{\Gamma[-w_a]\Gamma[-c_{(a)}-w_a]\Gamma[\frac{h+c_{(a)}-s_{(a)}}{2}+w_a]}{\Gamma[\frac{h-c_{(a)}-s_{(a)}}{2}]}\\
&\times\prod_{a=r+1}^{r+q}\frac{\Gamma[\frac{\Delta_{A_{a}}-\Delta_a}{2}]\Gamma[\frac{\Delta_{A_a}+\Delta_a}{2}-h]}{\Gamma[\frac{\Delta_{A_{a}}-s_{(a)}}{2}]}
\Gamma[-w_a]\Gamma[h-\Delta_a-w_a]\Gamma[{\textstyle\frac{\Delta_a-s_{(a)}}{2}+w_a}],
\end{split}
\end{equation}
where $\Delta_{A_0}=\sum_{a=r+1}^{r+q}\Delta_a$. Hence in computing the residue at this $s_{(a)}$ we necessarily localize the $w_a$ contour to be the residue contour around $w_a=0$ (More detailed account on residue computations is provided in Appendix \ref{app:sec:residues}). In the end we obtain an expression for the desired new Mellin pre-amplitude,
\begin{equation}\label{eq:forest2treespecialmellingeneral}
\begin{split}\checked{}
M[\delta]=&\int[\mathrm{d}\tau][\mathrm{d}w]\prod_{a=1}^r\left(M_a\prod_{i_a<j_a}\frac{\Gamma[\tau_{i_a\,j_a}]\Gamma[\delta_{i_a\,j_a}-\tau_{i_a\,j_a}]}{\Gamma[\delta_{i_a\,j_a}]}\right)
\\
&\times{\color{ProcessBlue}\Gamma[{\textstyle\frac{\Delta_{A_0}+(r-2)h+\sum_{a=1}^{r}c_{(a)}}{2}}+\sum_{a=1}^{r}w_a]}\prod_{a=1}^r\frac{\Gamma[-w_a]\Gamma[-c_{(a)}-w_a]\Gamma[\frac{h+c_{(a)}-s_{(a)}}{2}+w_a]}{\Gamma[\frac{h-c_{(a)}-s_{(a)}}{2}]},
\end{split}
\end{equation}
which is the same as the one for the special case \eqref{eq:forest2treespecialmellin} except for the highlighted factor above.

From our result \eqref{eq:forest2treespecialmellin} and \eqref{eq:forest2treespecialmellingeneral} it is obvious that the only dependence on the $\tau_{i_a\,j_a}$'s ($i_a,j_a\in A_{(a)}$ $\forall a$) in the kernel is restricted to the factor $\prod_{i_a<j_a}\frac{\Gamma[\tau_{i_a\,j_a}]\Gamma[\delta_{i_a\,j_a}-\tau_{i_a\,j_a}]}{\Gamma[\delta_{i_a\,j_a}]}$, under the single constraint
\begin{equation}\checked{}
\sum_{i_a<j_a}\tau_{i_a\,j_a}=\frac{\Delta_{A_{(a)}}-h+c_{(a)}}{2}.
\end{equation}
That is to say, every connected component of the original diagram  contributes to the new Mellin pre-amplitude in exactly the same way as a single connected diagram in the operation of vertex insertion that we discussed in Section \ref{sec:vertexinsertiongeneral}.

As an important application of the above observation, let us consider first constructing a connected diagram from a disconnected one with $r$ components using the above procedure, and then further insert another vertex to the vertex which we used in the gluing. It is not hard to see that the same argument that we provided in Section \ref{sec:localityofinsertion} holds. This immediately implies that the insertion of the new vertex in the second step is not going to modify at all the contribution from any of the original connected component, but only locally modifies the contribution from the vertex to which it is connected (i.e., the vertex added in the first step).

As a result, we conclude that the Mellin amplitude for any tree-level diagram satisfies a set of universal diagrammatic rules. Furthermore, the general rule for any bulk vertex can be read off from any available expression for the satellite diagrams, by chopping off the contributions from each satellite vertex together with the bulk-to-bulk propagator attached to it.

\subsection{Correction function for the bulk vertices}\label{app:sec:vertexcorrection}

We now apply the above results to work out the general formula for the correction function $C_A$ for a bulk vertex $A$, which comes as part of the tree-level diagrammatic rules.  As was noted in Section \ref{sec:treediagrammaticrules} it suffices that we restrict our scope to the satellite diagrams, where every original component is a contact diagram, which becomes an end vertex of the new tree. 
Hence proceeding from \eqref{eq:forest2treespecialmellingeneral}
the remaining $\tau$'s can be directly integrated, leading to the substitution \eqref{eq:contactsimplification}
\begin{equation}\label{app:eq:endleafsubstitution}
\checked{}
M_a\prod_{i_a<j_a}\frac{\Gamma[\tau_{i_a\,j_a}]\,\Gamma[\delta_{i_a\,j_a}-\tau_{i_a\,j_a}]}{\Gamma[\delta_{i_a\,j_a}]}\longrightarrow
\frac{\Gamma[\frac{\Delta_{A_{(a)}}-h\pm c_{(a)}}{2}]\,\Gamma[\frac{h-c_{(a)}-s_{(a)}}{2}]}{\Gamma[\frac{\Delta_{A_{(a)}}-s_{(a)}}{2}]}.
\end{equation}
Furthermore, each end vertex together with the propagator attached to it contribute to the Mellin amplitude by
\begin{equation}\label{app:eq:endleafcontribution}
\checked{}
\frac{\Gamma[\frac{\Delta_{A_{(a)}}-h\pm c_{(a)}}{2}]}{\Gamma[\frac{\Delta_{A_{(a)}}-s_{(a)}}{2}]}\Gamma[{\textstyle\frac{h\pm c_{(a)}-s_{(a)}}{2}}].
\end{equation}
Chopping off these contributinos, therefore the contribution coming from the central vertex in the satellite diagram is
\begin{equation}\label{app:eq:vertexcontributionspecialmellin}
\checked{}
\int[\mathrm{d}w]\,\Gamma[{\textstyle\frac{\Delta_{A_0}+(r-2)h+\sum_{a=1}^rc_{(a)}}{2}}+\sum_{a=1}^rw_a]\prod_{a=1}^r\frac{\Gamma[-w_a]\Gamma[-c_{(a)}-w_a]\Gamma[\frac{h+c_{(a)}-s_{(a)}}{2}+w_a]}{\Gamma[\frac{h\pm c_{(a)}-s_{(a)}}{2}]}.
\end{equation}
This is the entire contribution from a generic bulk scalar vertex to a tree diagram. The actual poles of this integral can be easily analyzed, following the general discussions in Appendix \ref{app:sec:mellinintegrals}. Firstly note that for each $a$ if we contract $\Gamma[-w_a]$ or $\Gamma[-c_{(a)}-w_a]$ with $\Gamma[\frac{h+c_{(a)}-s_{(a)}}{2}+w_a]$, the result is exactly $\Gamma[\frac{h\pm c_{(a)}-s_{(a)}}{2}]$, which already appear in the denominator. This immediately indicates that all the actual poles arise from contracting the overall factor $\Gamma[\frac{\Delta_{A_0}+(r-2)h+\sum_{a=1}^rc_{(a)}}{2}+\sum_{a=1}^2w_a]$ with the $\Gamma[-w_a]$'s or $\Gamma[-c_{(a)}-w_a]$'s, such that the combination is free of any $w$ variables. The only difference between the two group of factors is that one preserves the sign of the corresponding $c_a$ in the overall factor, while the other switches it to the opposite. We thus conclude that the poles \eqref{app:eq:vertexcontributionspecialmellin} in terms of $\Gamma$ functions as
\begin{equation}\label{app:eq:vertexpolesexplicit}
\checked{}
\Gamma[{\textstyle\frac{\Delta_{A_0}+(r-2)h\pm c_{(1)}\pm c_{(2)}\pm\cdots\pm c_{(r)}}{2}}].
\end{equation}
Taking the ratio of \eqref{app:eq:vertexcontributionspecialmellin} and \eqref{app:eq:vertexpolesexplicit} thus results in the formula for the correction function \eqref{eq:vertexcorrectionspecial}, which is free of poles.

\subsection{Relation to the tree diagrammatic rules in previous literature}\label{app:sec:proofofequivalence}

Diagrammatic rules for scalar Witten diagrams at tree level were previously observed in \cite{Fitzpatrick:2011ia,Paulos:2011ie} and proven in \cite{Fitzpatrick:2011ia,Fitzpatrick:2011hu,Nandan:2011wc}. In particular \cite{Paulos:2011ie} summarized a compact expression for the vertex rules in terms of generlized Lauricella hypergeometric functions. To be explicit, let us consider again a vertex that is purely internal, with $r$ bulk-to-bulk propagators attached to it, whose conformal dimensions are $\{\underline\Delta_1,\underline\Delta_2,\ldots,\underline\Delta_r\}$. Then with $r$ non-negative integers $\{m_1,m_2,\ldots,m_r\}$ we assign to this vertex a function
\begin{equation}\label{app:eq:Paulosresult}
\begin{split}
V_{[n_1,\ldots,m_r]}^{\underline\Delta_1,\ldots,\underline\Delta_r}=&\Gamma[{\textstyle\frac{\sum_{a=1}^r\underline\Delta_a}{2}-h}]\left(\prod_{a=1}^r(\underline\Delta_a-h+1)_{n_a}\right)\\
&\times F_{\rm A}^{(r)}[{\textstyle\frac{\sum_{a=1}^r\underline\Delta_a}{2}-h},\{-n_1,\ldots,-n_r\},\{\underline\Delta_1-h+1,\ldots,\underline\Delta_r-h+1\};1,\ldots,1],
\end{split}
\end{equation}
where $F_{\rm A}^{(r)}$ is the generlized Lauricella hypergeometric function, defined as
\begin{equation}
F_{\rm A}^{(r)}[a,\{b_1,\ldots,b_r\},\{c_1,\ldots,c_r\};x_1,\ldots,x_r]:=
\sum_{i_1,\ldots,i_r=0}^\infty\frac{(a)_{i_1+\cdots+i_r}(b_1)_{i_1}\cdots(b_r)_{i_r}}{(c_1)_{i_1}\cdots(c_r)_{i_r}\,i_1!\cdots i_r!}\,x_1^{i_1}\cdots x_r^{i_r}.
\end{equation}

In this subsection we show the equivalence between our rules for the Mellin amplitude integrand $M$ and the above existing result, by deriving \eqref{app:eq:Paulosresult} from \eqref{app:eq:satellitediagramspecial}. For this purpose we can in fact follow the treatment in (4.21) of \cite{Paulos:2011ie} and expand
\begin{equation}
(1+\sum_{a=1}^r\beta_a)^{-\frac{(r-2)h+\sum_{a=1}^rc_{(a)}}{2}}=\sum_{m_1,\ldots,m_r=0}^\infty({\textstyle\frac{(r-2)h+\sum_{a=1}^rc_{(a)}}{2}})_{m_1+m_2+\cdots+m_r}\,\prod_{a=1}^r\frac{(-\beta_a)^{m_a}}{m_a!}.
\end{equation}
Then all the $\beta$ integrals can be straightforwardly performed. Applying the substitution \eqref{app:eq:endleafsubstitution} and further chopping off the contribution \eqref{app:eq:endleafcontribution} for each end leaf, \eqref{app:eq:satellitediagramspecial} leads to the contribution from such vertex as expressed by
\begin{equation}\label{app:eq:derivinglauricellaintermediate}
\begin{split}
\checked{}
&\Gamma[{\textstyle\frac{(r-2)h+\sum_{a=1}^rc_{(a)}}{2}}]\sum_{m_1,\ldots,m_r=0}^\infty({\textstyle\frac{(r-2)h+\sum_{a=1}^rc_{(a)}}{2}})_{m_1+m_2+\cdots+m_r}\prod_{a=1}^r\frac{(-1)^{m_a}(\frac{h+c_{(a)}-s_{(a)}}{2})_{m_a}\Gamma[-c_{(a)}-m_a]}{m_a!\,\Gamma[\frac{h-c_{(a)}-s_{(a)}}{2}]}\\
&=\Gamma[{\textstyle\frac{(r-2)h+\sum_{a=1}^rc_{(a)}}{2}}]\prod_{a=1}^r\frac{\Gamma[-c_{(a)}]}{\Gamma[\frac{h-c_{(a)}-s_{(a)}}{2}]}\\
&\quad\times F_{\rm A}^{(r)}[{\textstyle\frac{(r-2)h+\sum_{a=1}^rc_{(a)}}{2}},\{{\textstyle\frac{h+c_{(1)}-s_{(1)}}{2}},\ldots,{\textstyle\frac{h+c_{(r)}-s_{(r)}}{2}}\},\{c_{(1)}+1,\ldots,c_{(r)}+1\};1,\ldots,1].
\end{split}
\end{equation}

In matching to the resulting in previous literature, we are supposed to sum over all possible poles in the Mandelstam variables
\begin{equation}
\checked{}
\sum_{n_1,\ldots,n_p=0}^\infty\frac{R_{n_1,\ldots,n_p}}{(s_{(1)}-\underline\Delta_1-2n_1)\cdots(s_{(p)}-\underline\Delta_p-2n_p)},
\end{equation}
where $p$ refers to the total number of bulk-to-bulk propagators. The previous diagrammatic rules are then about the structure of the residue $R$ for any set of non-negative integers $\{n_1,\ldots,n_p\}$. From out previous discussion in Section \ref{sec:treediagrammaticrules} we know that the poles of $M$ in the Mandelstam variables only arise in the form $\Gamma[\frac{h\pm c-s}{2}]$ for each propagator, hence it is not hard to see that the physical poles can only be produced from the spectrum integrations by colliding each such pole with either of the two poles $\frac{1}{(\underline\Delta-h)^2-c^2}$ from the spectrum measure. Due to this, the contribution of a given bulk vertex to the residue $R$ is nothing but to evaluate \eqref{app:eq:derivinglauricellaintermediate} at\footnote{Similar to what we encounter frequently in other computations, each physical pole receives contributions from both $\frac{1}{(\underline\Delta-h)-c}$ and $\frac{1}{(\underline\Delta-h)+c}$, but they lead to identical contributions, and so here we only consider one of them, keeping in mind an overall constant.}
\begin{equation}
\checked{}
c_{(a)}=\underline\Delta_a-h,\quad
s_{(a)}=\underline\Delta_a+2n_a,\quad
\forall a,
\end{equation}
Consequently yielding the vertex rule \eqref{app:eq:Paulosresult}.

Furthermore, when taking the residue at the physical pole there is an extra factor for each spectrum variable $c_a$
\begin{equation}
\checked{}
\frac{1}{n_a!}\frac{-1}{\Gamma[\underline\Delta_a-h+n_a+1]}
\end{equation}
coming from the measure and the contribution $\Gamma[\frac{h\pm c_{(a)}-s_{(a)}}{2}]$ from the corresponding propagator to $M$, which is exactly the propagator rule summarized in (1.11) of \cite{Paulos:2011ie}.

As a result we fully verified that our rules for the Mellin integrand $M$ leads to the known diagrammatic rule for the Mellin amplitude $\mathcal{M}$ in \cite{Paulos:2011ie}.





\newpage

\section{Multi-dimensional Mellin Integrals}\label{app:sec:mellinintegrals}

\temp{Fixed.}

Our recursive construction of Mellin amplitudes is based on a special class of integrals, the Mellin integrals. In order that this construction be practical it is crucial to have efficient tools for either evaluating them or probing their analytic properties.  In this appendix we provide a self-contained and detailed discussion with this regard. 


We only aim at a pedagogical description with intuitive arguments and simple examples that illustrates the main points, which should be sufficient for most physics applications. 

\subsection{Basics}

We refer an integral as a \emph{Mellin integral} if its integrand is meromorphic in the integration variables and its poles are separated into \emph{left poles} and \emph{right poles} by its contour, which stretches between $\pm i\infty$. The notion of ``left'' and ``right'' is up to our choice, and once they are specified the contour is unique up to homotopy. 

In most applications the poles of the integrand group into several evenly spaced semi-infinite families that can be treated as arising from $\Gamma$ functions, and (in the case of one variable) can be written into
\begin{equation}\label{app:eq:mellinintegralgeneric}
\int_{-i\infty}^{+i\infty}\frac{\mathrm{d}z}{2\pi i}\,f[z]\,\prod_{i=1}^{n_L}\Gamma[a_i+p_iz]\prod_{j=1}^{n_R}\Gamma[b_j-q_jz],
\end{equation}
where all $p$ and $q$ coefficients are some fixed positive numbers. $f[z]$ is an extra factor that is free of poles in $z$, e.g., $1/\Gamma[z]$. In this case the contour is conventionally defined to be left to all the poles from $\prod_{j=1}^{n_R}\Gamma[b_j-q_jz]$ (right poles) and right to all those from $\prod_{i=1}^{n_L}\Gamma[a_i+p_iz]$ (left poles), if not specified otherwise.  From now on we abbreviate $\int_{-i\infty}^{+i\infty}$ to $\int$.

Generalizing to multi-dimensional integrals, we require that the arguments inside the $\Gamma$ functions all have linear dependence on the integration variables, and for each individual variable the division of left/right poles follows the sign of its coefficients as in \eqref{app:eq:mellinintegralgeneric}. This notion of contour is invariant under arbitrary non-degenerate affine transformation of the integration variables: the result remains invariant as long as the new contour follows the same prescription wrst the new variables.

Two well-known examples are the Barnes' first lemma
\begin{equation}\label{app:eq:barneslemma1}
\int\frac{\mathrm{d}z}{2\pi i}\,\Gamma[a_1+z]\Gamma[a_2+z]\Gamma[b_1-z]\Gamma[b_2-z]=\frac{\prod_{i=1}^2\prod_{j=1}^2\Gamma[a_i+b_j]}{\Gamma[a_1+a_2+b_1+b_2]},
\end{equation}
and the Barnes' second lemma
\begin{equation}\label{app:eq:barneslemma2}
\begin{split}
\int\frac{\mathrm{d}z}{2\pi i}\,&\frac{\Gamma[a_1+z]\Gamma[a_2+z]\Gamma[b_1-z]\Gamma[b_2-z]\Gamma[b_3-z]}{\Gamma[c-z]}
=\frac{\prod_{i=1}^2\prod_{j=1}^3\Gamma[a_i+b_j]}{\Gamma[c-b_1]\Gamma[c-b_2]\Gamma[c-b_3]},
\end{split}
\end{equation}
where $c=a_1+a_2+b_1+b_2+b_3$.

Also, the following useful identity in flat space loop momentum integrations \cite{Smirnov:2006ry}
\begin{equation}\label{app:eq:smirnovmellin}
\frac{1}{(A+B)^\lambda}=\int\frac{\mathrm{d}z}{2\pi i}\,\frac{\Gamma[\lambda+z]\Gamma[-z]}{\Gamma[\lambda]}A^{-\lambda-z}B^z
\end{equation}
belongs to the same type.

As a further example, for an integral with a single left family of poles with the following form, one can deform the contour to the left picking up all the poles in the left family $\Gamma[a_1+z]$, and the residues resum into a hypergeometric function
\begin{equation}\label{app:eq:poleresummation}
\int\frac{\mathrm{d}z}{2\pi i}\,\frac{\Gamma[a_1+z]\prod_{j=1}^{n}\Gamma[b_j-z]}{\prod_{k=1}^{n-1}\Gamma[c_k-z]}=\frac{\prod_{j=1}^{n}\Gamma[a_1+b_j]}{\prod_{k=1}^{n-1}\Gamma[a_1+c_k]}\,{}_{n}F_{n-1}\left[\substack{a_1+b_1,\cdots,a_1+b_{n}\\a_1+c_1,\cdots,a_1+c_{n-1}},-1\right].
\end{equation}
Note that although there are poles arising from the arguments $a_1+c_k$ in the hypergeometric function, they are canceled by the zeros from $1/\Gamma[a_1+c_k]$, and so the only poles come from the prefactor $\prod_{j=1}^{n}\Gamma[a_1+b_j]$.

Mellin integral representation of a function is very useful if the two Barnes' lemmas can be applied repeatedly. This is the basic ingredient that leads to the recursion relations with the simplified kernel in Section \ref{sec:treediagramsrevisited}, \ref{sec:oneloopdiagrams} and \ref{sec:arbitraryloops}. Otherwise one often has to either perform numerical evaluations \cite{Czakon:2005rk} or, when analytic result is in need, deform the contour and resum infinite series of residues. \eqref{app:eq:poleresummation} is a simple manifestation of this latter strategy; for a more general discussion, e.g., see Appendix A of \cite{Gerhardus:2015sla} \footnote{This approach is, however, not optimal for the understanding analytic structures, because it usually yields a linear combination of complicated expressions, which term by term is poluted by significant amoung of spurious poles. This can be easily observed by, e.g., anaylyzing Barnes' first lemma \eqref{app:eq:barneslemma1} in this way.}.

Since the computation of Mellin amplitudes involve certain amount of integrals that remain non-trivial, there is always the question regarding what class of standard functions are best-tailored as analytic representations of the final result, which is not generally answered\footnote{Here it is useful to draw the analogous issue with the flat space scattering amplitudes for comparison. There a fairly large class of Feynman diagrams turn out to be linear combination of interated integrals (though this fact itself is already non-trivial), hence can at least be represented using Goncharov polylogarithms, and their analytic structure can be easily handled using the recently introduced ``symbology'' \cite{Goncharov:2009,Goncharov:2010jf}. Despite of these, even the simplest sunrise diagram with massive propagators in two dimensions are essentially elliptic \cite{Laporta:2004rb,Adams:2013kgc} (see \cite{Bourjaily:2017bsb} for a more non-trivial example). 
While at least for generic one-loop diagrams there has been a clear understanding \cite{Arkani-Hamed:2017ahv}, it is in general an open question what type of function a Feynman diagram belongs to.}. We do not attempt to address this question either in this paper.

In fact, in some sense this question may not even be necessary: we know amplitudes have to possess delicate structures, as they encode non-trivial physical information regarding unitary evolution, causality, etc, thus deserving to be treated as certain standard functions on their own. The crucial question is whether representations exist such that on one hand their analytic properties are manifest and that on the other hand the result can be efficiently evaluated numerically upon any given data.

For Mellin integrals arising in this paper, a systematic study of their analytic structures can indeed be carried out. On the one hand, there exists an efficient algorithm for detecting all their singularities (which turn out to be poles only) (Appendix \ref{app:sec:detectionofpoles}), and on the other hand, the residue at each pole can be directly represented using the same integrand but assigning different integration contours (Appendix \ref{app:sec:residues}). These pave the way for our analysis in Section \ref{sec:polestructure} and \ref{sec:residuecomputation}.

\subsection{Detection of poles}\label{app:sec:detectionofpoles}

We now discussion a general strategy in detecting the singularities of Mellin integrals, which results in an explicit algorithm presented in Appendix \ref{app:sec:polealgoritm}.

\subsubsection{Emergence of poles}\label{app:sec:poleemergence}

Singularity of an integral arises whenever its contour is pinched by the singularities of its integrand such that the contour cannot be deformed away to avoid the pinching (for a careful discussion of the general situation, see Section 2.1 of \cite{Eden:1966dnq}). The nature of the singularity can be learned by examining the behavior of the contour in the neighborhood of the pinching point in the parameter space. 

A very first qualitative property to be drawn in general is that, for Mellin integrals where the integrand is meromorphic in both the integration variables and the parameters (hence integrals like \eqref{app:eq:smirnovmellin} are excluded), the result is always meromorphic as well. 

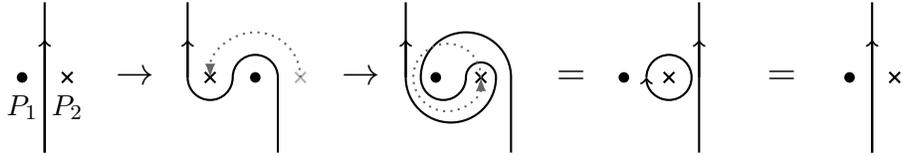
\begin{figure}[ht]
\captionsetup{margin=4em}
\begin{center}
\begin{tikzpicture}
\begin{scope}
\draw [black,thick] (0,-1) -- (0,1);
\draw [black,thick,->] (0,-1) -- (0,.5);
\draw [black,thick] ($(.3,0)+(-135:.1)$) -- ++(45:.2); 
\draw [black,thick] ($(.3,0)+(-45:.1)$) -- ++(135:.2); 
\fill [black] (-.3,0) circle [radius=2pt];
\node [anchor=north] at (-.3,-.1) {$P_1$};
\node [anchor=north] at (.3,-.1) {$P_2$};
\end{scope}
\node [anchor=center] at (1.2,0) {\Large $\rightarrow$};
\begin{scope}[xshift=2.5cm]
\draw [black,thick] (.6,-1) -- (.6,0) arc [start angle=0,end angle=180,radius=.3] arc [start angle=0,end angle=-180,radius=.3] -- (-.6,1);
\draw [black,thick,->] (-.6,0) -- (-.6,.5);
\draw [black,thick] ($(-.3,0)+(-135:.1)$) -- ++(45:.2); 
\draw [black,thick] ($(-.3,0)+(-45:.1)$) -- ++(135:.2); 
\draw [black!30!white,thick] ($(.9,0)+(-135:.1)$) -- ++(45:.2); 
\draw [black!30!white,thick] ($(.9,0)+(-45:.1)$) -- ++(135:.2); 
\draw [black!60!white,thick,dotted,-latex] (.9,0) arc [start angle=0,end angle=180,radius=.6];
\fill [black] (.3,0) circle [radius=2pt];
\end{scope}
\node [anchor=center] at (4.2,0) {\Large $\rightarrow$};
\begin{scope}[xshift=5.5cm]
\draw [black,thick] ($(.3,0)+(-135:.1)$) -- ++(45:.2); 
\draw [black,thick] ($(.3,0)+(-45:.1)$) -- ++(135:.2); 
\fill [black] (-.3,0) circle [radius=2pt];
\draw [black,thick] (.7,-1) -- (.7,0) arc [start angle=0,end angle=180,radius=.6] arc [start angle=180,end angle=360,radius=.3] arc [start angle=180,end angle=0,radius=.2] arc [start angle=0,end angle=-180,radius=.6] -- (-.7,1);
\draw [black,thick,->] (-.7,0) -- +(0,.5);
\draw [black!60!white,thick,dotted,-latex] (.3,0) arc [start angle=0,end angle=360,radius=.45];
\end{scope}
\node [anchor=center] at (7,0) {\Large $=$};
\begin{scope}[xshift=8cm]
\draw [black,thick] (.7,-1) -- (.7,1);
\draw [black,thick,->] (.7,0) -- +(0,.5);
\draw [black,thick,->] (0,0) arc [start angle=180,end angle=-180,radius=.3];
\draw [black,thick] ($(.3,0)+(-135:.1)$) -- ++(45:.2); 
\draw [black,thick] ($(.3,0)+(-45:.1)$) -- ++(135:.2); 
\fill [black] (-.3,0) circle [radius=2pt];
\end{scope}
\node [anchor=center] at (9.8,0) {\Large $=$};
\begin{scope}[xshift=11cm]
\draw [black,thick] (0,-1) -- (0,1);
\draw [black,thick,->] (0,-1) -- (0,.5);
\draw [black,thick] ($(.3,0)+(-135:.1)$) -- ++(45:.2); 
\draw [black,thick] ($(.3,0)+(-45:.1)$) -- ++(135:.2); 
\fill [black] (-.3,0) circle [radius=2pt];
\end{scope}
\end{tikzpicture}
\end{center}
\vspace{-1.5em}\caption{Move the right pole around the left pole to detect the nature of the singularity that arises from pinching the contour by this pair of poles.}
\label{fig:poleofMellinintegral}
\end{figure}

It suffices to illustrate this in the case of a single integral. Without loss of generality let us assume the singularity arises from the collision between a left pole $P_1$ and a right pole $P_2$. We keep an infinitesimal distance between them and deform parameters so that $P_2$ moves around $P_1$ and back to the original position, as shown in Figure \ref{fig:poleofMellinintegral}. The contour is forced to deform accordingly. However, the resulting contour turns out to be homotopic to the original one, indicating trivial monodromy. Since the integrand clearly stays the same, any singularity of such Mellin integrals is necessarily a pole.

It is also simple to book-keep the specific location of the poles that emerge from Mellin integrals.
Following the notation in \eqref{app:eq:mellinintegralgeneric}, every left pole of the integrand is located at $z=\frac{-a_i-m_L}{p_i}$ for some $i$ and $m_L\in\mathbb{N}$, and every right pole at $z=\frac{b_j+m_R}{q_j}$ similarly. Hence for each pair of left family $\Gamma[a_i+p_iz]$ and right family $\Gamma[b_j-q_jz]$ in the integrand the $z$ integral produces a semi-infinite family of poles at the locations
\begin{equation}
\frac{a_i}{p_i}+\frac{b_j}{q_j}+\frac{m_L}{p_i}+\frac{m_R}{q_j}=0,\qquad m_L,m_R\in\mathbb{N}.
\end{equation}
In particular, if $p_i$ is a multiple of $q_j$ (or vice versa) then this family reduces to an evenly spaced series again and can be treated as poles in $\Gamma[\frac{q_j}{p_i}a_i+b_j]$ (or $\Gamma[a_i+\frac{p_i}{q_j}b_j]$). Even though the actual residues at these poles differ from $\Gamma$ functions in general, this is nevertheless a convenient notation for the poles.

For simplicity let us focus on the case when the coefficients $p$ and $q$ are just 1, which is what we mostly encounter in this paper (after rescaling the variables). Then the function resulting from \eqref{app:eq:mellinintegralgeneric} can schematically be written into the form
\begin{equation}
F[\{a\},\{b\}]\prod_{i=1}^{n_L}\prod_{j=1}^{n_R}\Gamma[a_i+b_j],
\end{equation}
where $F[\{a\},\{b\}]$ is again a function without additional poles, which is only responsible for the residues at each pole. This structure is in particular manifest in the two Barnes' lemmas \eqref{app:eq:barneslemma1} \eqref{app:eq:barneslemma2}. Since the result share similar structure with the original integrand, this fact continues to hold for multi-dimensional Mellin integrals.

While the above discussion in principle provides a systematic way to learn about the complete pole structure of Mellin integrals, subtlety may arise in actual applications due to fake poles. 
Before we discuss how to cure this problem, let us illustrate this phenomenon with a simple yet non-trivial example.

\subsubsection{Fake poles from the na\"ive contour analysis: an example}

Let us think about the following 2-fold integral
\begin{equation}\label{app:eq:mellinexample1}
\int\frac{\mathrm{d}x\,\mathrm{d}y}{(2\pi i)^2}\,\Gamma[a_1+x]\Gamma[a_2-x]\Gamma[b_1+x-y]\Gamma[b_2-x+y]\Gamma[c_1+y]\Gamma[c_2-y].
\end{equation}
This integral can be done using Barnes' first lemma \eqref{app:eq:barneslemma1}, yielding
\begin{equation}
\frac{\Gamma[a_1+a_2]\Gamma[b_1+b_2]\Gamma[c_1+c_2]\Gamma[a_1+b_2+c_2]\Gamma[a_2+b_1+c_1]}{\Gamma[a_1+a_2+b_1+b_2+c_1+c_2]},
\end{equation}
which does not depend on our choice of the ordering in integrating $x$ and $y$. Altogether we have five families of poles, as indicated by the resulting $\Gamma$ functions.

However, if we na\"ively apply the arguments of pinching contours, we will encounter poles that are not present. Specifically, if we first perform such analysis on the $x$ integral and then on the $y$ integral, we will conclude with an additional family of poles
\begin{equation}
\Gamma[a_1+a_2+b_1+b_2]
\end{equation}
apart from the ones in the correct result; while in the other ordering we will conclude with a different additional family
\begin{equation}
\Gamma[b_1+b_2+c_1+c_2].
\end{equation}
Hence by doing the analysis in different orderings, we result in different na\"ive conclusions, each containing one family of fake poles. In this specific case the reason that these are fake poles is obvious from the Barnes' first lemma: the first integral produces exactly the same factor in the denominator, so that in the second integral these na\"ive poles are killed by extra zeros. However, this only becomes obvious after actually performing the first integral.

\subsubsection{Strategy to avoid fake poles}\label{app:sec:poleglobalview}

In generic Mellin integrals where Barnes' lemmas do not directly apply, it is hard to extract detailed information about the zeros of an integral, and even worse, we cannot expect that the zeros always form evenly spaced families at generic values of the parameters $\{a,b\}$ as the poles do (although they may get evenly aligned as the parameters approach the values such that a pole is potentially encountered). Due to this, it is hopeless to apply the cancellation argument to identify the fake poles as we did for \eqref{app:eq:mellinexample1}.


In the previous example we noticed that, while the actual poles always arise from the contour analysis for whatever ordering of integrations we choose, the fake poles may not. This suggests that we can perform the contour analysis for every ordering of integrations and then keep their common poles. However, the amount of computations involved is daunting.

Nevertheless, there is one important fact in this observation: (up to a sign) the result of Mellin integrals should not depend on the specific ordering we choose to perform the integrals. Due to this, a much better strategy is to characterize the poles of multi-dimensional Mellin integrals from a global point of view, instead of the local view that they emerge step by step from the composition of a specific sequence of individual Mellin integrals. This should give us a better understanding of how fake poles are different from actual poles, which will lead to tremendous simplification for the analysis.

To start with, let us index the $\Gamma$ factor in the integrand to keep track of the original families of poles, so for example in \eqref{app:eq:mellinexample1} we have altogether six $\Gamma$ factors
\begin{equation}
\Gamma_1[a_1+x]\Gamma_2[a_2-x]\Gamma_3[b_1+x-y]\Gamma_4[b_2-x+y]\Gamma_5[c_1+y]\Gamma_6[c_2-y].
\end{equation}
Based on this, for a $\Gamma$ function corresponding to a family of poles emerged from an integration, we attach to it the union of labels of the original $\Gamma$'s from which they originate, e.g., $\Gamma_{\{2,3\}}=\Gamma[(a_2-x)+(b_1+x-y)]\equiv\Gamma[a_2+b_1-y]$ arises from the pinching between $\Gamma_2$ and $\Gamma_3$ during the $x$ integration. In this simple case the new argument is literally the summation of the original ones, such that the integration variable cancel away. Regardless of how we perform the integrals, it is obvious that the poles at the end should associate to $\Gamma$ functions whose arguments are free of any integration variables. The complete list of such emergent $\Gamma$'s are
\begin{equation}
\begin{split}
&\Gamma_{\{1,2\}}=\Gamma[a_1+a_2],\qquad\Gamma_{\{3,4\}}=\Gamma[b_1+b_2],\qquad\Gamma_{\{5,6\}}=\Gamma[c_1+c_2],\\
&\Gamma_{\{1,4,6\}}=\Gamma[a_1+b_2+c_2],\qquad\Gamma_{\{2,3,5\}}=\Gamma[a_2+b_1+c_1],\\
&\Gamma_{\{1,2,3,4\}}=\Gamma[a_1+a_2+b_1+b_2],\qquad\Gamma_{\{1,2,5,6\}}=\Gamma[a_1+a_2+c_1+c_2],\\
&\qquad\Gamma_{\{3,4,5,6\}}=\Gamma[b_1+b_2+c_1+c_2],\\
&\Gamma_{\{1,2,3,4,5,6\}}=\Gamma[a_1+a_2+b_1+b_2+c_1+c_2],
\end{split}
\end{equation}
where we graded these functions according to the length of their subscripts.

With the above notation it is easy to observe that there exists a natural notion of compositeness in these emergent $\Gamma$ functions: a $\Gamma$ function (or its corresponding poles) is ``\emph{composite}'' if its subscripts contain a genuine subset which itself corresponds to a $\Gamma$ function free of the integration variables. For example, $\Gamma_{\{1,2,3,4\}}$ is composite because it contains both $\Gamma_{\{1,2\}}$ and $\Gamma_{\{3,4\}}$. So in the above example we have the classification
\begin{align}
\text{non-composite:}&\qquad\Gamma_{\{1,2\}},\Gamma_{\{3,4\}},\Gamma_{\{5,6\}},\Gamma_{\{1,4,6\}},\Gamma_{\{2,3,5\}},\\
\text{composite:}&\qquad\Gamma_{\{1,2,3,4\}},\Gamma_{\{1,2,5,6\}},\Gamma_{\{3,4,5,6\}},\Gamma_{\{1,2,3,4,5,6\}}.
\end{align}

For a $\Gamma$ function in the above, if it corresponds to genuine poles of the integral, it must be able to arise in every possible way of performing the integrations as a sequence of label accumulations. Take $\Gamma_{\{1,4,6\}}$ as an example, it can arise in two ways
\begin{equation}
\begin{split}
\Gamma_1\Gamma_4\Gamma_6\xrightarrow{\int\mathrm{d}x}\Gamma_{\{1,4\}}[a_1+b_2+y]\Gamma_6[c_2-y]\xrightarrow{\int\mathrm{d}y}\Gamma_{\{1,4,6\}}[a_1+b_2+c_2],&\\
\Gamma_1\Gamma_4\Gamma_6\xrightarrow{\int\mathrm{d}y}\Gamma_1[a_1+x]\Gamma_{\{4,6\}}[b_2+c_2-x]\xrightarrow{\int\mathrm{d}x}\Gamma_{\{1,4,6\}}[a_1+b_2+c_2].&
\end{split}
\end{equation}
In order that such accumulation can proceed from the beginning to the end, it is crucial that the arguments in the $\Gamma$ functions in each intermediate step must depend on at least one integration variable. Obviously, while this is obeyed by the non-composite $\Gamma$ functions, the composite ones always admit of an accumulation procedure that violates it. In consequence, we draw a proposition that underlies the algorithm in the next subsubsection
\begin{proposition}\label{app:prop:disconnectedpoles}
All composite poles are fake.
\end{proposition}

In actual practice it is not necessary and not economic to begin by enumerating all the $\Gamma$ functions free of integration variables, simply because of the exponential growth with the number of $\Gamma$'s to begin with. Instead, it suffices to work through the contour analysis following an arbitrarily chosen ordering of integration variables. Along the way, we label each emergent $\Gamma$ function by merging the two sets of labels associated to the two original $\Gamma$ functions responsible for it via the integration. For simple integrals we just identify the composite $\Gamma$ functions at the end of this contour analysis and throw them away.  When the number of integrations is high, a more efficient way is to identify $\Gamma$ functions whose arguments are already free of integration variables after each contour analysis of a single integral, and throw away $\Gamma$ functions whose subscript contains either of the former as a proper subset, and then use the remaining $\Gamma$ functions for the contour analysis in the next. 

To further illustrate Proposition \ref{app:prop:disconnectedpoles} and the resulting method, let us slightly modify our example in the previous subsubsection and study
\begin{equation}\label{app:eq:eqforprop1}
\int\frac{\mathrm{d}x\,\mathrm{d}y}{(2\pi i)^2}\,\frac{\Gamma_1[a_1+x]\Gamma_2[a_2-x]\Gamma_3[b_1+x-y]\Gamma_4[b_2-x+y]\Gamma_5[c_1+y]\Gamma_6[c_2-y]\Gamma_7[d+y]}{\Gamma[a_1+a_2+b_1+b_2+c_1+c_2+d+y]}.
\end{equation}
The result can again be explicitly worked out, by first integrating $x$ using Barnes' first lemma and then integrating $y$ using Barnes' second lemma
\begin{equation}\label{app:eq:eqforprop1result}
\begin{split}
&\Gamma_{\{1,2\}}[a_1+a_2]\Gamma_{\{3,4\}}[b_1+b_2]\Gamma_{\{2,3,5\}}[a_2+b_1+c_1]\Gamma_{\{1,4,6\}}[a_1+b_2+c_2]\Gamma_{\{5,6\}}[c_1+c_2]\\
&\times\Gamma_{\{2,3,7\}}[a_2+b_1+d]\Gamma_{\{6,7\}}[c_2+d]/\\
&(\Gamma[a_1+a_2+b_1+b_2+c_1+c_2]\Gamma[a_1+a_2+b_1+b_2+c_2+d]\Gamma[a_2+b_1+c_1+c_2+d]).
\end{split}
\end{equation}
All the $\Gamma$ functions here are obviously non-composite. While in principle we could obtain the same result by doing the integrations in the other ordering, this is very non-obvious. Let us now insist to analyze the poles using the contour arguments, first on $y$ and then on $x$, then apart from the ones already shown in the correct result na\"ively we will conclude with two additional families of poles
\begin{equation}
\Gamma_{\{3,4,5,6\}}[b_1+b_2+c_1+c_2]\Gamma_{\{3,4,6,7\}}[b_1+b_2+c_2+d].
\end{equation}
Since we cannot apply the Barnes' second lemma when integrating $y$ first, we are not able to argue that there are correspondingly zeros in the same pattern before integrating $x$ (and in fact this is not true, as can be checked numerically). However, both families of poles are composite. Hence following our logic they can be directly ruled out.

\subsubsection{Algorithm for pole detection in multi-dimensional Mellin integrals}\label{app:sec:polealgoritm}

Let us now turn the discussion so far into an explicit algorithm in detecting the complete set of poles in generic multi-dimensional Mellin integrals.

In the previous subsubsection we specialized in cases where the emerged poles are coincident to poles from a single $\Gamma$ function. As mentioned at the beginning this may not always happen. Although the labeling of the emergent pole basically remains the same, to take care of how the poles in a family are aligned in each step causes slight complication. In fact, this is not even necessary, because all what we care about are the locations of the poles at the end. When detecting the presence of poles the only crucial point is to identify their origin through multiple integrations, and as we are going to discuss in detail later, with this information at hand it is straightforward to work out the detailed location of each pole.

With this in mind we can temporarily put off the detailed contents of the poles and introduce an abstract notation for them. For each factor in the original integrand generically of the form $\Gamma_i[a_0+\sum_{j=1}^ra_jz_j]$ (where all $z$'s are integration variables), we can denote its corresponding family of poles as $\gamma_i[\sum_{j=1}^ra_jz_j]$, where we ignore the constant term $a_0$ and understood the argument as being equivalent up to an arbitrary \emph{positive} overall rescaling
\begin{equation}
\gamma_i[{\textstyle\sum_{j=1}^ra_jz_j}]\sim\gamma_i[{\textstyle c\sum_{j=1}^ra_jz_j}],\quad c>0.
\end{equation}
In every step of integration, say $z_1$, for every pair of $\gamma$'s where this variables appear with coefficients of opposite signs, there should be a new $\gamma$ family of poles emerging from them, whose argument is obtained by eliminating $z_1$, i.e.,
\begin{equation}\label{app:eq:gammaemergence}
\gamma_{A}\!\left[\sum_{i=1}^ra_iz_i\right]\,\gamma_{B}\!\left[\sum_{i=1}^rb_iz_i\right]\xrightarrow{\int\mathrm{d}z_1}\gamma_{A\cup B}\!\left[\frac{1}{|a_1|}\sum_{i=2}^ra_iz_i+\frac{1}{|b_1|}\sum_{i=2}^rb_iz_i\right],
\end{equation}
where we assume $a_1$ and $b_1$ have opposite signs. Here $A$ and $B$ each can be some set of original labels (where possibly $A\cap B\neq\varnothing$), and we label the emergent $\gamma$ by their \emph{union}. Even when these pole are not evenly align as if coming from a single $\Gamma$ function, we still collect them in the same $\gamma$ family, because they share the same origin as indicated by the new subscript $A\cup B$. It is then apparent that a $\gamma$ family will stop to affect later integrations once its argument becomes zero, $\gamma[0]$. In particular, after all integrations are performed, every $\gamma$ family showing up in the end has to be of this type.

With this setup, for a given multi-dimensional Mellin integral we now provide the following algorithm in working out its poles:
\begin{enumerate}[noitemsep,nolistsep]
\item[(a).] Turn every $\Gamma$ function in the (numerator of the) original integrand into its corresponding $\gamma$ notation and label them. (We temporarily put aside the remaining part of the integrand.)
\item[(b).] Pick up an arbitrary integration variable $z$, and replace the $\gamma$'s containing $z$ by all $\gamma$'s that can arise from the $z$ integration following \eqref{app:eq:gammaemergence}.
\item[(c).] In the resulting set of $\gamma$'s, collect the subscripts of those in the form $\gamma[0]$.
\item[(d).] Inspect each $\gamma$: if its corresponding subscript contains any subscript collected in the previous step as a proper subset, we delete this $\gamma$ (i.e., these are composite poles, following Proposition \ref{app:prop:disconnectedpoles}).
\item[(e).] Collecting the remaining $\gamma$'s. If there are any other integration variables, pick up an arbitrary one, and repeat the analysis from (b) to (d). After all the integration variables are analyzed, the resulting $\gamma$'s are all in the form $\gamma_A[0]$ for some set of labels $A$. This terminates the algorithm.
\end{enumerate}
This algorithm returns a list of $\gamma$'s recording the families of poles that may exist (apart from those that directly descend from the integrand, which we put aside in step (a)).


Although the algorithm above starts with a specific choice of ordering in doing the integrations, attentive readers may already find in practice that the set of $\gamma$ families in the output are independent of this choice.  The reason for this independence will be clarified later on.

\subsection{Local properties of poles}\label{app:sec:residues}

We now analyze the local properties of poles emerged from Mellin integrals. This includes several aspects: (1) the location of the pole, (2) the order of the pole, (3) the residue at the pole. As we are going to see, these are encoded in the detailed geometry of the pinching, for which the $\gamma$ notation worked out by Algorithm \ref{app:sec:polealgoritm} provides a particularly convenient starting point.

For simplicity, in this subsection we always assume that the poles in the original integrand are all simple poles.


\subsubsection{Revisiting a single Mellin integral}\label{app:sec:revisitsingleintegral}

Before diving into generic situations, let us quickly review how the analysis is performed for a single Mellin integral of the form
\begin{equation}\label{app:eq:singleintegraleg}
\int\frac{\mathrm{d}z}{2\pi i}\,\Gamma[a+z]\,\Gamma[b-z]\,H[z].
\end{equation}
Here the function $H[z]$ can contain other poles.

Obviously there is potentially a family of poles corresponding to $\Gamma[a+b]$, from the pinching of $\Gamma[a+z]$ and $\Gamma[b-z]$. In the simplest situation we assume that there is no pole nor zero from $H[z]$ getting to the pinching point when the contour is pinched. Then it is not hard to verify that these poles are simple poles. To check this, e.g., for the leading pole $a+b=0$, we can turn on an infinitesimal value $a+b=\epsilon>0$. As $\epsilon\to0$, the whole integrand scales as $\epsilon^{-2}$ in the neighborhood of $z=\frac{b-a}{2}$, but the size of the part of integration contour affected by this diverging behavior only scales as $\epsilon$. Hence the integral itself scales as $\epsilon^{-1}$, which indicates a simple pole.

Similar argument applies to situations when other poles or zero in $H[z]$ get to the same pinching point.  Most generally we can imagine altogether $r$ poles and $r'$ zeros that have this behavior (including those from $\Gamma[a+z]$ and $\Gamma[b-z]$). Regardless of how they are distributed on the left and right, as long as the contour gets pinched, then the corresponding emergent pole has to be of order $r-r'-1$.

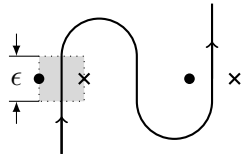
\begin{figure}[ht]
\captionsetup{margin=4em}
\begin{center}
\begin{tikzpicture}
\begin{scope}
\draw [black,thin,dotted,fill=black!15!white] (-.3,-.3) rectangle (.3,.3);
\draw [black] (-.3,.3) -- ++(-.4,0);
\draw [black] (-.3,-.3) -- ++(-.4,0);
\draw [black,-latex] (-.6,.6) -- (-.6,.3);
\draw [black,-latex] (-.6,-.6) -- (-.6,-.3);
\node [anchor=center] at (-.6,0) {$\epsilon$};
\draw [black,thick] (0,-1) -- (0,.3) arc [start angle=180,end angle=0,radius=.5] -- ++(0,-.6) arc [start angle=-180,end angle=0,radius=.5] -- (2,1);
\draw [black,thick,->] (0,-1) -- ++(0,.5);
\draw [black,thick,->] (2,0) -- ++(0,.5);
\draw [black,thick] ($(.3,0)+(-135:.1)$) -- ++(45:.2); 
\draw [black,thick] ($(.3,0)+(-45:.1)$) -- ++(135:.2); 
\fill [black] (-.3,0) circle [radius=2pt];
\draw [black,thick] ($(2.3,0)+(-135:.1)$) -- ++(45:.2); 
\draw [black,thick] ($(2.3,0)+(-45:.1)$) -- ++(135:.2); 
\fill [black] (1.7,0) circle [radius=2pt];
\end{scope}
\end{tikzpicture}
\end{center}
\vspace{-1.5em}\caption{The same pole from distinct pinching.}
\label{app:fig:distinctpinching}
\end{figure}

Another lesson to be drawn from this picture is that the pinching of a contour is \emph{local}. Already as we go on to the next pole $a+b=1$, the contour is pinched at two points, $z=-a=b+1$, and $z=-a-1=b$. Since the two points are distinct, as shown in Figure \ref{app:fig:distinctpinching}, the $\epsilon^{-1}$ divergence arising locally from each neighborhood only add up together, and hence the subleading poles are still simple poles. In more non-trivial integrals, the same pole could emerge from pinching between different pairs of families. For instance, consider the integral
\begin{equation}\label{app:eq:multiplepinchinglocationseg}
\int\frac{\mathrm{d}z}{2\pi i}\Gamma_1[a+z]\,\Gamma_2[b-z]\,\Gamma_3[a+b+z]\,\Gamma_4[-z]=\frac{\Gamma[a+b]^2\,\Gamma_{\{1,4\}}[a]\,\Gamma_{\{2,3\}}[a+2b]}{\Gamma[2a+2b]}.
\end{equation}
Here $\Gamma[a+b]$ emerges both from $\Gamma_1\Gamma_2$ and from $\Gamma_3\Gamma_4$. However, in the first case the pinching occurs at $z=-a=b$, while in the other at $z=-a-b=0$. As long as $a\neq b$ these two pinching points are distinct. As a result, the poles in the family $\Gamma[a+b]$ are again simple poles for this integral, as can also be easily seen by the presence of $\Gamma[2a+2b]$ in the denominator on RHS.

Now let us return to the case \eqref{app:eq:singleintegraleg}  when poles in $\Gamma[a+b]$ are simple. We begin by focusing on the leading pole $a+b=0$. For the computation of its residue, we can first push the contour to the right, so that the integral is re-written as
\begin{equation}
-\residue{z=b}\,\Gamma[a+z]\,\Gamma[b-z]\,H[z]+\int'\frac{\mathrm{d}z}{2\pi i}\Gamma[a+z]\,\Gamma[b-z]\,H[z].
\end{equation}
The contour in the second term is again anchored at $\pm i\infty$ but now it is to the right of all left poles as well as the leading right pole $z=b$, and to the left of all the remaining right poles. Since this term remains regular in the pinching limit, it cannot have any non-vanishing contribution to the residue. Hence we conclude that 
\begin{equation}\label{app:eq:singleresidueright}
\residue{b=-a}\int\frac{\mathrm{d}z}{2\pi i}\,\Gamma[a+z]\,\Gamma[b-z]\,H[z]=-\residue{b=-a}\,\residue{z=b}\,\Gamma[a+z]\,\Gamma[b-z]\,H[z].
\end{equation}
Alternatively we can also think about pushing the original contour to the left, so as to express the same residue as
\begin{equation}\label{app:eq:singleresidueleft}
\residue{b=-a}\,\residue{z=-a}\,\Gamma[a+z]\,\Gamma[b-z]\,H[z],
\end{equation}
where the difference in the overall sign is due to the orientation of the contours.  

In brief, when computing the residue of the original integral, the original contour responsible for the emergent pole has to be replaced by a $T^1$ residue contour locally wrapping around either of the original poles in the pinching. In other words, the contour for the integration that is relevant for an emergent pole has to be localized for the residue at this pole.

Similar logic applies to residue at subleading poles of $\Gamma[a+b]$ as well. The only difference is that we need to push the original contour even further, as it is now pinched at several locations. If we are interested in the pole at $b=-a-m$ for some positive integer $m$, then we first need to decompose the $z$ contour into residue contours at each $z=b+k$ with $k\leq m$ plus a remaining contour that is irrelevant, and the residue is thus expressed as
\begin{equation}\label{app:eq:subleadingresidue1d}
\residue{b=-a-m}\int\frac{\mathrm{d}z}{2\pi i}\,\Gamma[a+z]\,\Gamma[b-z]\,H[z]=-\residue{b=-a-m}\sum_{k=0}^m\residue{z=b+k}\,\Gamma[a+z]\,\Gamma[b-z]\,H[z].
\end{equation}
Again an equivalent expression can be obtained by pushing the contour to the left.

More generally when the pole arises from different pinching configurations as in \eqref{app:eq:multiplepinchinglocationseg}, the residue computation is jus a summation of contributions from each pinching. For instance, in that example the residue at the leading pole is thus computed as
\begin{equation}
\left(-\residue{b=-a}\,\residue{z=b}-\residue{b=-a}\,\residue{z=0}\right)\Gamma[a+z]\,\Gamma[b-z]\,\Gamma[a+b+z]\,\Gamma[-z]=2\,\Gamma[a]\,\Gamma[-a].
\end{equation}
We mind the reader that while in one-dimensional integrals the above case is basically the same as the computation for a subleading pole \eqref{app:eq:subleadingresidue1d}, in multi-dimensional integrals there is some essential difference between the two, as will be discussed in Section \ref{app:sec:higherpoles}.

\subsubsection{Geometry of generic pinching}
\label{app:sec:pinchinggeometry}

We now move on to inspect the local properties of individual emergent pole in the multi-dimensional Mellin integral. For simplicity we restrict to leading poles in each $\gamma$ family for the time being.

To motivate the general picture, let us put the one-dimensional integral \eqref{app:eq:singleintegraleg} discussed above in a wider context by treating $H[z]$ itself as arising from some other integrations. In other words, we consider a multi-dimensional integral that contains poles in the family $\Gamma[a+b]$ but these poles emerge from the $z$ integration only.  For Mellin integrals we can always tune the value of the parameters such that the integration contour is simply defined: it has a \emph{fixed} real part and the integration is performed over the entire imaginary space for each integration variable.  In particular for this example, the existence of such simple contour requires that $\mathfrak{Re}(a+z)>0$ and $\mathfrak{Re}(b-z)>0$, which is possible as long as $a+b>0$ (for this discussion we also restrict all parameters to be real).  When such simple contour is well-defined, it suffices that we restrict our focus on the real slice of the entire space spanned by the integration variables, and by definition it is guaranteed that the simple contour intersects it at a unique point. For convenience, later on when we say ``integration contour'' we equally refer to this intersection point in the real slice.

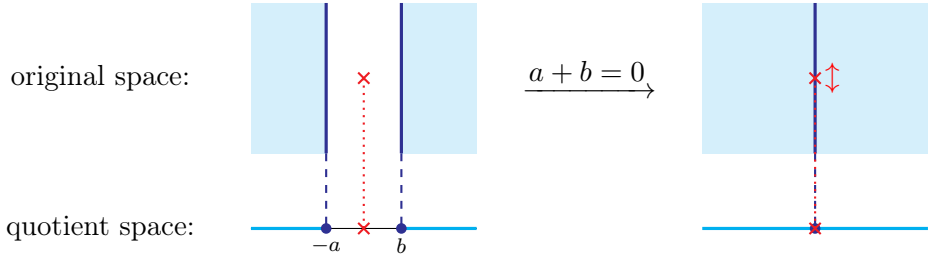
\begin{figure}[ht]
\captionsetup{margin=4em}
\begin{center}
\begin{tikzpicture}
\begin{scope}[xshift=-3cm]
\fill [ProcessBlue!15!white] (-.5,-1) -- (-.5,1) -- (-1.5,1) -- (-1.5,-1) -- cycle;
\draw [Blue,very thick] (-.5,-1) -- (-.5,1);
\draw [Blue,thick,dashed] (-.5,-1) -- (-.5,-2);
\fill [ProcessBlue!15!white] (.5,-1) -- (.5,1) -- (1.5,1) -- (1.5,-1) -- cycle;
\draw [Blue,very thick] (.5,-1) -- (.5,1);
\draw [Blue,thick,dashed] (.5,-1) -- (.5,-2);
\draw [black] (-1.5,-2) -- (1.5,-2);
\draw [ProcessBlue,very thick] (-1.5,-2) -- (-.5,-2);
\draw [ProcessBlue,very thick] (.5,-2) -- (1.5,-2);
\fill [Blue] (-.5,-2) circle [radius=2pt];
\fill [Blue] (.5,-2) circle [radius=2pt];
\node [anchor=north] at (-.5,-2) {$\scriptstyle -a$};
\node [anchor=north] at (.5,-2) {$\scriptstyle b$};
\draw [Red,thick] ($(-45:3pt)$) -- +(135:6pt);
\draw [Red,thick] ($(45:3pt)$) -- +(-135:6pt);
\draw [Red,thick,dotted] (0,0) -- (0,-2);
\draw [Red,thick] ($(0,-2)+(-45:3pt)$) -- +(135:6pt);
\draw [Red,thick] ($(0,-2)+(45:3pt)$) -- +(-135:6pt);
\end{scope}
\node [anchor=center] at (0,0) {$\xrightarrow{\textstyle a+b=0}$};
\begin{scope}[xshift=3cm]
\fill [ProcessBlue!15!white] (-1.5,-1) -- (-1.5,1) -- (1.5,1) -- (1.5,-1) -- cycle;
\draw [Blue,very thick] (0,-1) -- (0,1);
\draw [Blue,thick,dashed] (0,-1) -- (0,-2);
\draw [ProcessBlue,very thick] (-1.5,-2) -- (1.5,-2);
\fill [Blue] (0,-2) circle [radius=2pt];
\draw [Red,thick] ($(-45:3pt)$) -- +(135:6pt);
\draw [Red,thick] ($(45:3pt)$) -- +(-135:6pt);
\draw [Red,thick,dotted] (0,0) -- (0,-2);
\draw [Red,thick] ($(0,-2)+(-45:3pt)$) -- +(135:6pt);
\draw [Red,thick] ($(0,-2)+(45:3pt)$) -- +(-135:6pt);
\node [anchor=west] at (0,0) {\color{red}$\updownarrow$};
\end{scope}
\node [anchor=center] at (-6.5,0) {original space:};
\node [anchor=center] at (-6.5,-2) {quotient space:};
\end{tikzpicture}
\end{center}
\vspace{-1.5em}\caption{1d pinching in a multi-dimensional integral. The red cross represents the intersection point of the integration contour and the real slice, which is forbidden to sit in the blue regions.}
\end{figure}

In this case the two hyperplanes $a+z=0$ and $b-z=0$ border a strip region, in which the contour is allowed to live.  Since the other integrations are irrelevant for the pole under study, we could equally quotient away those directions and view the whole configuration in the resulting one-dimensional space, which reduces back to the example \eqref{app:eq:singleintegraleg}. In this quotient space the intersection point stays inside a 1-simplex $(-a,b)$.  Now as we hit the pole $a+b=0$, the strip region crashes down to a hyperplane $z=-a=b$, where its two original boundaries collide.  Note that while in the quotient picture the contour is uniquely fixed as the entire simplex shrinks to a point, in the original space it is still allowed to move within the hyperplane since it is pinched only in one direction. For later convenience we call this hyperplane the \emph{pinching plane} for the pole under study. More generally, the pinching plane is defined to be the set of points allowed for the contour when a pinching occurs\footnote{Of course in general there will be additional constraints from the integration, such that the actual region for the allowed contour (for the remaining integrals that are not relevant for the pole) is resitricted to a finite region within the pinching plane. But this is not important for our current discussion.}. These two geometric pictures, one in the original space, the other in the quotient space, are important for different purposes.

\paragraph{Shrinking simplices.}

Generalizing to arbitrary multi-dimensional Mellin integrals, let us assume that there are in total $n$ integrations and that for the pole under study only $r$ of the integrations are relavent. This means that in the real slice the contour intersection has to be restricted in $r$ independent directions but not in the rest. In the simplest senario, the original region of allowed intersection has to have the shape $\Delta_{(r)}\times\mathbb{R}^{n-r}$, where $\Delta_{(r)}$ is a certain $r$-simplex. This is bordered by $r+1$ hyperplanes parallel to each other in the $n-r$ un-pinched directions, which come from $r+1$ leading poles of the integrand. Switching to the quotient space, as we hit the emergent pole again, the entire simplex $\Delta_{(r)}$ has to shrink to a single poin, or in other words, the $r+1$ hyperplanes has to simultaneous pass through a common point.

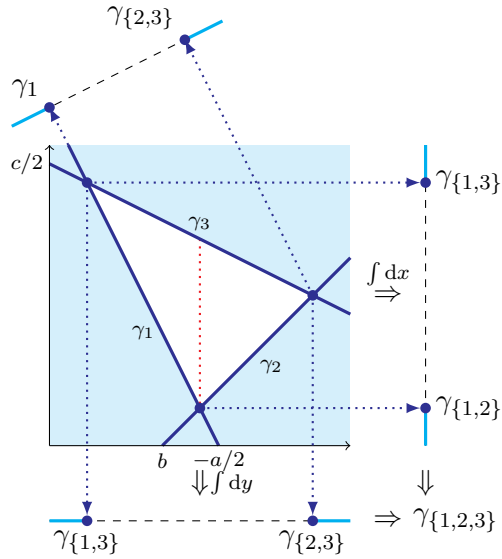
\begin{figure}[ht]
\captionsetup{margin=4em}
\begin{center}
\begin{tikzpicture}
\begin{scope}
\fill [ProcessBlue!15!white] (-2,-.5) -- (-2,3.5) -- (2,3.5) -- (2,-.5) -- cycle;
\coordinate (i31) at (-1.5,3);
\coordinate (i12) at (0,0);
\coordinate (i23) at (1.5,1.5);
\fill [white] (i31) -- (i12) -- (i23);
\draw [Red,thick,dotted] (0,2.25) -- (0,0);
\draw [Blue,very thick] (.25,-.5) -- (-1.75,3.5);
\draw [Blue,very thick] (-.5,-.5) -- (2,2);
\draw [Blue,very thick] (-2,3.25) -- (2,1.25);
\fill [Blue] (i31) circle [radius=2pt];
\fill [Blue] (i12) circle [radius=2pt];
\fill [Blue] (i23) circle [radius=2pt];
\node [anchor=center] at ($(i31)!.5!(i12)+(0,-.5)$) {$\scriptstyle\gamma_1$};
\node [anchor=center] at ($(i12)!.5!(i23)+(.2,-.2)$) {$\scriptstyle\gamma_2$};
\node [anchor=center] at ($(i23)!.5!(i31)+(0,.2)$) {$\scriptstyle\gamma_3$};
\draw [black,->] (-2,-.5) -- (2,-.5);
\draw [black,->] (-2,-.5) -- (-2,3.5);
\node [anchor=center] at (.25,-.7) {$\scriptstyle-a/2$};
\node [anchor=center] at (-.5,-.7) {$\scriptstyle b$};
\node [anchor=center] at (-2.3,3.25) {$\scriptstyle c/2$};
\draw [black,dashed] (-2,-1.5) -- (2,-1.5);
\node [anchor=center] at (0,-1) {$\Downarrow$};
\node [anchor=west] at (0,-1) {$\scriptstyle\int\mathrm{d}y$};
\draw [Blue,thick,dotted,-latex] (i31) -- ($(-1.5,-1.5)+(0,2pt)$);
\draw [Blue,thick,dotted,-latex] (i23) --  ($(1.5,-1.5)+(0,2pt)$);
\draw [ProcessBlue,very thick] (-2,-1.5) -- (-1.5,-1.5);
\draw [ProcessBlue,very thick] (2,-1.5) -- (1.5,-1.5);
\fill [Blue] (-1.5,-1.5) circle [radius=2pt];
\fill [Blue] (1.5,-1.5) circle [radius=2pt];
\node [anchor=north] at (-1.5,-1.5) {$\gamma_{\{1,3\}}$};
\node [anchor=north] at (1.5,-1.5) {$\gamma_{\{2,3\}}$};
\draw [black,dashed] (3,-.5) -- (3,3.5);
\node [anchor=center] at (2.5,1.5) {$\Rightarrow$};
\node [anchor=south] at (2.5,1.5) {$\scriptstyle\int\mathrm{d}x$};
\draw [Blue,thick,dotted,-latex] (i31) -- ($(3,3)+(-2pt,0)$);
\draw [Blue,thick,dotted,-latex] (i12) -- ($(3,0)+(-2pt,0)$);
\draw [ProcessBlue,very thick] (3,-.5) -- (3,0);
\draw [ProcessBlue,very thick] (3,3) -- (3,3.5);
\fill [Blue] (3,0) circle [radius=2pt];
\fill [Blue] (3,3) circle [radius=2pt];
\node [anchor=west] at (3,0) {$\gamma_{\{1,2\}}$};
\node [anchor=west] at (3,3) {$\gamma_{\{1,3\}}$};
\node [anchor=center] at (3.4,-1.5) {$\gamma_{\{1,2,3\}}$};
\node [anchor=center] at (2.5,-1.5) {$\Rightarrow$};
\node [anchor=center] at (3,-1) {$\Downarrow$};
\draw [Blue,thick,dotted,-latex] (-1.75,3.5) -- (-2,4);
\draw [Blue,thick,dotted,-latex] (1.5,1.5) -- (-.2,4.9);
\draw [black,dashed] (-2.5,3.75) -- (.3,5.15);
\draw [ProcessBlue,very thick] (-2.5,3.75) -- (-2,4);
\draw [ProcessBlue,very thick] (.3,5.15) -- (-.2,4.9);
\fill [Blue] (-2,4) circle [radius=2pt];
\fill [Blue] (-.2,4.9) circle [radius=2pt];
\node [anchor=south east] at (-2,4) {$\gamma_1$};
\node [anchor=south east] at (-.2,4.9) {$\gamma_{\{2,3\}}$};
\end{scope}
\end{tikzpicture}
\end{center}
\vspace{-1.5em}\caption{$\gamma$ poles induced by a simplex. Any integration projects the simplex down to one lower dimenisonal space.}
\label{app:fig:gammageometry}
\end{figure}

Let us more closely inspect how such configuration is captured by Algorithm \ref{app:sec:polealgoritm} in the case of two relevant integrals. To be specific let us assume this 2-simplex comes from the leading poles of $\Gamma_1[a+2x+y]\,\Gamma_2[b-x+y]\,\Gamma_3[c-x-2y]$ in the integral. When $a+b+c>0$ the contour can be simply defined, as depicted in Figure \ref{app:fig:gammageometry}, sitting anywhere in the white triangle region.
Assume that we first integrate $y$ and then $x$. As we immediately observe, there is an ambiguity in how the contour has to be pinched during the $y$ integration: if it is to the left of the red line then it is pinched by the singularity hyperplane from $\gamma_1$ and $\gamma_3$, otherwise from $\gamma_2$ and $\gamma_3$. These two possibilities produce $\gamma_{\{1,3\}}$ and $\gamma_{\{2,3\}}$ respectively. Geometrically this can be interpreted as projecting the 2-simplex along the $y$ direction. In the resulting 1d space parametrized by $x$ we thus obtain a 1-simplex, with its one boundary point $\gamma_{\{1,3\}}$ being the image of the intersection $\gamma_1\cap\gamma_3$ under the projection, while the other boundary point $\gamma_{\{2,3\}}$ being the image of $\gamma_2\cap\gamma_3$. The next integration $x$ then merges $\gamma_{\{1,3\}}$ and $\gamma_{\{2,3\}}$ into $\gamma_{\{1,2,3\}}$, which can be understood as a trivial projection, closing the 1d simplex further into a point.

We can equally choose the other ordering of integrations, for which the 1d simplex has its two boundaries originating from different set of preimages, $\gamma_1\cap\gamma_3$ and $\gamma_1\cap\gamma_2$. Nevertheless the two projections yields the same $\gamma_{\{1,2,3\}}$. We can even make the configuration a bit more special by rotating the frame, so that, e.g., $\gamma_1$ becomes parallel to the direction of the first projection (as shown in Figure \ref{app:fig:gammageometry}). In this case, in the first integration there is only a unique choice for the pinching. Correspondingly, in the first projection the whole $\gamma_1$ collaps to one boundary point of the new 1d simplex, which we still denote as $\gamma_1$, while the other boundary point $\gamma_{\{2,3\}}$ originates from $\gamma_2\cap\gamma_3$. In the end this again produces $\gamma_{\{1,2,3\}}$. As a consequence we see that the family of poles $\gamma_{\{1,2,3\}}$ from the shrinking of this 2d simplex does not rely on how we apply Algorithm \ref{app:sec:polealgoritm}. 

As we consider an $r$-simplex with generic $r$, it is obvious that there always exist a frame such that in every step of projection the resulting image maintains to be a simplex. This indicates that
\begin{proposition}\label{app:prop:simplextogamma}
Every simplex uniquely specifies one $\gamma$ family, whose subscript is the set of codim-1 boundaries of the simplex.
\end{proposition}

\paragraph{Shrinking polytopes.}

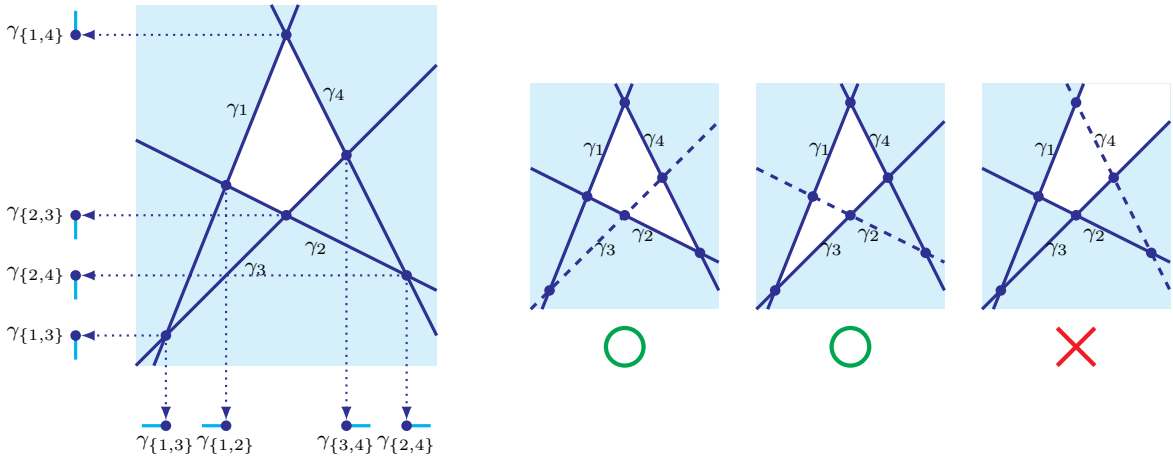
\begin{figure}[ht]
\captionsetup{margin=4em}
\begin{center}
\begin{tikzpicture}
\begin{scope}[scale=.8]
\fill [ProcessBlue!15!white] (-2.5,-2.5) -- (-2.5,3.5) -- (2.5,3.5) -- (2.5,-2.5) -- cycle;
\coordinate (iA) at (0,3);
\coordinate (iB) at (-1,.5);
\coordinate (iC) at (-2,-2);
\coordinate (iD) at (1,1);
\coordinate (iE) at (2,-1);
\coordinate (iF) at (0,0);
\fill [white] (iA) -- (iB) -- (iF) -- (iD) -- cycle;
\draw [Blue,very thick] (-2.2,-2.5) -- (.2,3.5);
\draw [Blue,very thick] (-2.5,1.25) -- (2.5,-1.25);
\draw [Blue,very thick] (-2.5,-2.5) -- (2.5,2.5);
\draw [Blue,very thick] (-.25,3.5) -- (2.5,-2);
\fill [Blue] (iA) circle [radius=2.5pt]; 
\fill [Blue] (iB) circle [radius=2.5pt]; 
\fill [Blue] (iC) circle [radius=2.5pt]; 
\fill [Blue] (iD) circle [radius=2.5pt]; 
\fill [Blue] (iE) circle [radius=2.5pt]; 
\fill [Blue] (iF) circle [radius=2.5pt]; 
\node [anchor=center] at ($(iA)!.5!(iB)+(-.3,0)$) {$\scriptstyle\gamma_1$};
\node [anchor=center] at ($(iB)!.5!(iE)+(0,-.3)$) {$\scriptstyle\gamma_2$};
\node [anchor=center] at ($(iC)!.5!(iD)+(0,-.4)$) {$\scriptstyle\gamma_3$};
\node [anchor=center] at ($(iA)!.5!(iD)+(.3,0)$) {$\scriptstyle\gamma_4$};
\draw [Blue,thick,dotted,-latex] (iC) -- ($(-2,-3.5)+(0,2.5pt)$);
\draw [Blue,thick,dotted,-latex] (iB) -- ($(-1,-3.5)+(0,2.5pt)$);
\draw [Blue,thick,dotted,-latex] (iD) -- ($(1,-3.5)+(0,2.5pt)$);
\draw [Blue,thick,dotted,-latex] (iE) -- ($(2,-3.5)+(0,2.5pt)$);
\draw [ProcessBlue,very thick] (-2,-3.5) -- +(-.4,0);
\fill [Blue] (-2,-3.5) circle [radius=2.5pt];
\node [anchor=north] at (-2,-3.5) {$\scriptstyle\gamma_{\{1,3\}}$};
\draw [ProcessBlue,very thick] (-1,-3.5) -- +(-.4,0);
\fill [Blue] (-1,-3.5) circle [radius=2.5pt];
\node [anchor=north] at (-1,-3.5) {$\scriptstyle\gamma_{\{1,2\}}$};
\draw [ProcessBlue,very thick] (1,-3.5) -- +(.4,0);
\fill [Blue] (1,-3.5) circle [radius=2.5pt];
\node [anchor=north] at (1,-3.5) {$\scriptstyle\gamma_{\{3,4\}}$};
\draw [ProcessBlue,very thick] (2,-3.5) -- +(.4,0);
\fill [Blue] (2,-3.5) circle [radius=2.5pt];
\node [anchor=north] at (2,-3.5) {$\scriptstyle\gamma_{\{2,4\}}$};
\draw [Blue,thick,dotted,-latex] (iA) -- ($(-3.5,3)+(2.5pt,0)$);
\draw [Blue,thick,dotted,-latex] (iF) -- ($(-3.5,0)+(2.5pt,0)$);
\draw [Blue,thick,dotted,-latex] (iE) -- ($(-3.5,-1)+(2.5pt,0)$);
\draw [Blue,thick,dotted,-latex] (iC) -- ($(-3.5,-2)+(2.5pt,0)$);
\draw [ProcessBlue,very thick] (-3.5,3) -- +(0,.4);
\fill [Blue] (-3.5,3) circle [radius=2.5pt];
\node [anchor=east] at (-3.5,3) {$\scriptstyle\gamma_{\{1,4\}}$};
\draw [ProcessBlue,very thick] (-3.5,0) -- +(0,-.4);
\fill [Blue] (-3.5,0) circle [radius=2.5pt];
\node [anchor=east] at (-3.5,0) {$\scriptstyle\gamma_{\{2,3\}}$};
\draw [ProcessBlue,very thick] (-3.5,-1) -- +(0,-.4);
\fill [Blue] (-3.5,-1) circle [radius=2.5pt];
\node [anchor=east] at (-3.5,-1) {$\scriptstyle\gamma_{\{2,4\}}$};
\draw [ProcessBlue,very thick] (-3.5,-2) -- +(0,-.4);
\fill [Blue] (-3.5,-2) circle [radius=2.5pt];
\node [anchor=east] at (-3.5,-2) {$\scriptstyle\gamma_{\{1,3\}}$};
\end{scope}
\begin{scope}[scale=.5,xshift=9cm]
\fill [ProcessBlue!15!white] (-2.5,-2.5) -- (-2.5,3.5) -- (2.5,3.5) -- (2.5,-2.5) -- cycle;
\coordinate (iA) at (0,3);
\coordinate (iB) at (-1,.5);
\coordinate (iC) at (-2,-2);
\coordinate (iD) at (1,1);
\coordinate (iE) at (2,-1);
\coordinate (iF) at (0,0);
\fill [white] (iA) -- (iB) -- (iE) -- cycle;
\draw [Blue,very thick] (-2.2,-2.5) -- (.2,3.5);
\draw [Blue,very thick] (-2.5,1.25) -- (2.5,-1.25);
\draw [Blue,very thick,dashed] (-2.5,-2.5) -- (2.5,2.5);
\draw [Blue,very thick] (-.25,3.5) -- (2.5,-2);
\fill [Blue] (iA) circle [radius=4pt]; 
\fill [Blue] (iB) circle [radius=4pt]; 
\fill [Blue] (iC) circle [radius=4pt]; 
\fill [Blue] (iD) circle [radius=4pt]; 
\fill [Blue] (iE) circle [radius=4pt]; 
\fill [Blue] (iF) circle [radius=4pt]; 
\node [anchor=center] at ($(iA)!.5!(iB)+(-.3,0)$) {$\scriptstyle\gamma_1$};
\node [anchor=center] at ($(iB)!.5!(iE)+(0,-.3)$) {$\scriptstyle\gamma_2$};
\node [anchor=center] at ($(iC)!.5!(iD)+(0,-.4)$) {$\scriptstyle\gamma_3$};
\node [anchor=center] at ($(iA)!.5!(iD)+(.3,0)$) {$\scriptstyle\gamma_4$};
\draw [Green,ultra thick] (0,-3.5) circle [radius=.5];
\end{scope}
\begin{scope}[scale=.5,xshift=15cm]
\fill [ProcessBlue!15!white] (-2.5,-2.5) -- (-2.5,3.5) -- (2.5,3.5) -- (2.5,-2.5) -- cycle;
\coordinate (iA) at (0,3);
\coordinate (iB) at (-1,.5);
\coordinate (iC) at (-2,-2);
\coordinate (iD) at (1,1);
\coordinate (iE) at (2,-1);
\coordinate (iF) at (0,0);
\fill [white] (iA) -- (iC) -- (iD) -- cycle;
\draw [Blue,very thick] (-2.2,-2.5) -- (.2,3.5);
\draw [Blue,very thick,dashed] (-2.5,1.25) -- (2.5,-1.25);
\draw [Blue,very thick] (-2.5,-2.5) -- (2.5,2.5);
\draw [Blue,very thick] (-.25,3.5) -- (2.5,-2);
\fill [Blue] (iA) circle [radius=4pt]; 
\fill [Blue] (iB) circle [radius=4pt]; 
\fill [Blue] (iC) circle [radius=4pt]; 
\fill [Blue] (iD) circle [radius=4pt]; 
\fill [Blue] (iE) circle [radius=4pt]; 
\fill [Blue] (iF) circle [radius=4pt]; 
\node [anchor=center] at ($(iA)!.5!(iB)+(-.3,0)$) {$\scriptstyle\gamma_1$};
\node [anchor=center] at ($(iB)!.5!(iE)+(0,-.3)$) {$\scriptstyle\gamma_2$};
\node [anchor=center] at ($(iC)!.5!(iD)+(0,-.4)$) {$\scriptstyle\gamma_3$};
\node [anchor=center] at ($(iA)!.5!(iD)+(.3,0)$) {$\scriptstyle\gamma_4$};
\draw [Green,ultra thick] (0,-3.5) circle [radius=.5];
\end{scope}
\begin{scope}[scale=.5,xshift=21cm]
\fill [ProcessBlue!15!white] (-2.5,-2.5) -- (-2.5,3.5) -- (2.5,3.5) -- (2.5,-2.5) -- cycle;
\coordinate (iA) at (0,3);
\coordinate (iB) at (-1,.5);
\coordinate (iC) at (-2,-2);
\coordinate (iD) at (1,1);
\coordinate (iE) at (2,-1);
\coordinate (iF) at (0,0);
\fill [white] (.2,3.5) -- (iB) -- (iF) -- (2.5,2.5) -- (2.5,3.5) -- cycle;
\draw [Blue,very thick] (-2.2,-2.5) -- (.2,3.5);
\draw [Blue,very thick] (-2.5,1.25) -- (2.5,-1.25);
\draw [Blue,very thick] (-2.5,-2.5) -- (2.5,2.5);
\draw [Blue,very thick,dashed] (-.25,3.5) -- (2.5,-2);
\fill [Blue] (iA) circle [radius=4pt]; 
\fill [Blue] (iB) circle [radius=4pt]; 
\fill [Blue] (iC) circle [radius=4pt]; 
\fill [Blue] (iD) circle [radius=4pt]; 
\fill [Blue] (iE) circle [radius=4pt]; 
\fill [Blue] (iF) circle [radius=4pt]; 
\node [anchor=center] at ($(iA)!.5!(iB)+(-.3,0)$) {$\scriptstyle\gamma_1$};
\node [anchor=center] at ($(iB)!.5!(iE)+(0,-.3)$) {$\scriptstyle\gamma_2$};
\node [anchor=center] at ($(iC)!.5!(iD)+(0,-.4)$) {$\scriptstyle\gamma_3$};
\node [anchor=center] at ($(iA)!.5!(iD)+(.3,0)$) {$\scriptstyle\gamma_4$};
\draw [Red,ultra thick] (-.5,-4) -- +(1,1);
\draw [Red,ultra thick] (-.5,-3) -- +(1,-1);
\end{scope}
\end{tikzpicture}
\end{center}
\vspace{-1.5em}\caption{Shrinking of a polytope. On the right we show two 2-simplices induced by the polytope on the left, and a third one that is forbidden.}
\label{app:fig:shrinkpolytope}
\end{figure}

Most generally we have to consider as well situations when the integration contour is constrained within some polytope region in the quotient space. It suffices to inspect a 2d case, where the polytope is a 4-gon, in order to understand the general feature. An explicit example is illustrated in Figure \ref{app:fig:shrinkpolytope}.

Depending on how we deform the parameters the shape of the polytope may even change when we shrink them, which already causes certain ambiguity. Hence when we shrink it to a point we may not always expect that all its codim-1 boundary get to the pinching point simultaneously. Nevertheless, the set of codim-1 boundaries of a polytope always contains some subset that carves out a simplex in the space, and it is obvious that when the contour is finally pinched it has to be pinched by at least one such simplex. For instance, in Figure \ref{app:fig:shrinkpolytope} the boundary set allows two simplices, bounded by $\{\gamma_1,\gamma_2,\gamma_4\}$ and $\{\gamma_1,\gamma_3,\gamma_4\}$ respectively. However, not all subsets are allowed, e.g., $\{\gamma_1,\gamma_2,\gamma_3\}$, as in such case (if we ignore $\gamma_4$) the contour is never pinched.

Let us see what Algorithm \ref{app:sec:polealgoritm} dictates in this situation. Say we first perform the integral associated to the vertical direction. Note here we need to take into consideration four possible types of pinching, as all of them may possibly happen depending on specific deformation of the parameters. For each pinching we track the intersection of the pair of relevant hyperplanes and project it down to one lower dimension, in the same way as we did for a simplex. Although in this situation the resulting image is no longer a polytope, it remains true that the image of each intersection carves out a hyperplane in the new space and that the contour can only sit on one side relative to it, which is determined by the relative position of the allowed region before projection. As a result from this integration we obtain two families of left poles $\gamma_{\{1,3\}}$ and $\gamma_{\{1,2\}}$, and two families of right poles $\gamma_{\{3,4\}}$ and $\gamma_{\{2,4\}}$. As we further perform another integration we obtain three families
\begin{equation}
\gamma_{\{1,2,4\}},\quad\gamma_{\{1,3,4\}},\quad\gamma_{\{1,2,3,4\}}.
\end{equation}
Observe that the first two families, with subscripts of length three, exactly corresponds to the two different simplices we mentioned before. Apart from these there is an extra family containing all the four labels. We could also do the integrations in the other way, as shown in Figure \ref{app:fig:shrinkpolytope}; where although the intermediate $\gamma$'s differ we still land on the same conclusion.

Now recall that due to Proposition \ref{app:prop:disconnectedpoles} $\gamma_{\{1,2,3,4\}}$ is composite and so we rule it out in the algorithm\footnote{In fact, the appearance of such composite poles (when ignoring Proposition \ref{app:prop:disconnectedpoles}) is not even invariant. To understand this, for instance we can again rotation the example in Figure \ref{app:fig:shrinkpolytope} such that the direction corresponding to the first integration parallels $\gamma_3$. Then obviously from the first integration we obtain $\gamma_{\{1,4\}}$, $\gamma_3$ and $\gamma_{\{2,4\}}$. The second integration then yields $\gamma_{\{1,2,4\}}$ and $\gamma_{\{1,3,4\}}$ only. This further confirms the necessity of incorporating Proposition \ref{app:prop:disconnectedpoles} into Algorithm \ref{app:sec:polealgoritm}.}. This indicates that in general if we start with certain polytopes, then the algorithm always returns to us a set of $\gamma$ poles precisely corresponding to all the simplices induced by this polytope.

In consequence, we land on the following general feature of Algorithm \ref{app:sec:polealgoritm}
\begin{proposition}
Every $\gamma$ family from Algorithm \ref{app:sec:polealgoritm} originates from a shrinking simplex.
\end{proposition}

Note that even when we analyze a simplex, if we start in higher dimensions we could still encounter general polytopes from the projection (though not always). In the simplest case, for example, we can in principle obtain the same configuration as in Figure \ref{app:fig:shrinkpolytope} by projecting from a 3-simplex. Here the four boundaries, instead of being $\{\gamma_1,\gamma_2,\gamma_3,\gamma_4\}$, are relabled as, e.g., $\{\gamma_{\{1,2\}},\gamma_{\{2,3\}},\gamma_{\{3,4\}},\gamma_{\{1,4\}}\}$. To see the difference in the result as compared to that of Figure \ref{app:fig:shrinkpolytope}, after projecting along the $y$ direction the four original boundary points switch to
\begin{equation}
\gamma_{\{1,3\}}\mapsto\gamma_{\{1,2,3,4\}},\quad
\gamma_{\{1,2\}}\mapsto\gamma_{\{1,2,3\}},\quad
\gamma_{\{3,4\}}\mapsto\gamma_{\{1,3,4\}},\quad
\gamma_{\{2,4\}}\mapsto\gamma_{\{1,2,3,4\}}.
\end{equation}
Attention that although we obtain two $\gamma_{\{1,2,3,4\}}$'s, the argument in each still depends on the remaining variable, where its coefficient comes with different signs (as one can check by working with a specific 3-simplex). Hence these are in fact two distinct boundaries. Then the last projection only generates a $\gamma_{\{1,2,3,4\}}[0]$. This is consistent with Proposition \ref{app:prop:simplextogamma} that a shrinking simplex always generates a unique $\gamma$ family.

\paragraph{Subleading poles.}

So far we have been focusing on a leading pole, which comes from some shrinking simplex whose boundaries are the leading poles of original $\Gamma$ functions. The subleading poles arise similarly, where the only difference is that the contour is now constrained by singularity hyperplanes corresponding to some subleading pole of each revelant $\Gamma$ function in the original integrand. We do not illustate this further.

\subsubsection{Locations of poles}

Since we have concluded that every $\gamma$ family emerges from a shrinking simplex, it is then simple to work out the explicit locations of the pole for any $\gamma$. Without loss of generality let us assume an $r$-simplex with boundaries $\{\Gamma_0[E_0],\Gamma_1[E_1],\ldots,\Gamma_r[E_r]\}$. There are altogether $r$ integrations relevant for the corresponding $\gamma$, which we denote as $\{z_1,\ldots,z_r\}$. For this $\gamma_{\{1,2,\ldots,r\}}$ we collect the $r+1$ equations
\begin{equation}
E_i+m_i=0,\quad m_i\in\mathbb{N},\quad i=0,1,\ldots,r,
\end{equation}
and eliminate the $z$ variables. Obviously this yields a unique equation
\begin{equation}\label{app:eq:gammapolelocations}
E+\sum_{i=0}^rp_im_i=0,\quad m_0,\ldots,m_r\in\mathrm{N},
\end{equation}
for some expression $E$ and numerical coefficients $\{p\}$ (up to an overall rescaling). It is guaranteed that $p_i>0$ ($\forall i$). \eqref{app:eq:gammapolelocations} thus enumerates locations of all the poles in $\gamma_{\{1,2,\ldots,r\}}$. In particular the leading pole locates at $E=0$. Let us further assume an overall rescaling such that $\min(p_0,\ldots,p_r)=1$. In this case, if all $p_i$'s turn out to be positive integers, then this $\gamma$ family coincide with the poles of $\Gamma[E]$. 

Here note that each subleading pole $E+m=0$ (for some $m>0$) in general receives contributions from several distinct pinching, corresponding to nodes in a finite lattice
\begin{equation}\label{app:eq:pinchinglattice}
m_i\geq 0\quad (i=0,\ldots,r),\quad\sum_{i=0}^rp_im_i= m.
\end{equation}

\subsubsection{Intersection of pinching planes and higher-order poles}
\label{app:sec:higherpoles}

For the poles in each individual $\gamma$ family emerged from the shrinking of an $r$-simplex, if we just focus on the $r+1$ $\Gamma$ functions responsible for them but ignore the rest of the original integrand (say by assuming it is always finite), then by a simple counting of divergence (as in Section \ref{app:sec:revisitsingleintegral}) we easily observe that the poles are simple poles. This is true for both the leading pole and the subleading poles, because although the subleading poles may come from different pinching sites as shown in \eqref{app:eq:pinchinglattice} all these sites are distinct from each other.

If it turns out that the locations of poles in any $\gamma$ family is different from any other $\gamma$ family, then all the poles indeed simply emerges from some shrinking simplex and is thus simple. However, in practice it may occur that the locations of poles as worked out in \eqref{app:eq:gammapolelocations} coincide between different $\gamma$ families. Here let us just focus on the leading poles, as there is no difficulty in its generalization to subleading ones.

Without loss of generality we assume that the integral in total contains $n$ integrations. When we hit a pole in some $\gamma$ family, the contour has to be restricted to a codim-$r$ hyperplane, which we defined as its corresponding pinching plane. Now let us consider two families $\Gamma_A[E]$ and $\Gamma_B[E]$ \footnote{Here for simplicity we assume both families are encoded in some $\Gamma$ functions for the clarify of presentation, but this condition is not essentially needed.}. We denote their corresponding pinching planes as $P_A$ and $P_B$ respectively. In general $P_A$ and $P_B$ can have different codimensions.

If it occurs that $P_A\cap P_B=\varnothing$, this just indicates that $P_A$ and $P_B$ contributes \emph{additively} to the local properties of the emergent pole $\Gamma[E]$. Since in the original integration space they are separated by finite distance, the local pinching pattern at each is not modified, and the order of the pole remains the same. This is in analogy to the one-dimensional case as represented in \eqref{app:eq:multiplepinchinglocationseg}.

\begin{figure}[ht]
\captionsetup{margin=4em}
\begin{center}
\begin{tikzpicture}
\begin{scope}
\fill [ProcessBlue,opacity=.15] (1.5,1.5) -- ++(0,-1) -- ++(-3,0) -- ++(0,1) -- cycle;
\fill [ProcessBlue,opacity=.15] (1.5,-1.5) -- ++(0,1) -- ++(-3,0) -- ++(0,-1) -- cycle;
\fill [ProcessBlue,opacity=.15] (1.5,1.5) -- ++(-1,0) -- ++(0,-3) -- ++(1,0) -- cycle;
\fill [ProcessBlue,opacity=.15] (-1.5,1.5) -- ++(1,0) -- ++(0,-3) -- ++(-1,0) -- cycle;
\draw [Blue,very thick] (-1.5,.5) -- (1.5,.5);
\draw [Blue,very thick] (-1.5,-.5) -- (1.5,-.5);
\draw [Blue,very thick] (.5,-1.5) -- (.5,1.5);
\draw [Blue,very thick] (-.5,-1.5) -- (-.5,1.5);
\node [anchor=east] at (-.5,0) {$\scriptsize \gamma_1$};
\node [anchor=north] at (0,-.5) {$\scriptsize \gamma_2$};
\node [anchor=west] at (.5,0) {$\scriptsize \gamma_3$};
\node [anchor=south] at (0,.5) {$\scriptsize \gamma_4$};
\end{scope}
\begin{scope}[xshift=5cm]
\fill [ProcessBlue,opacity=.15] (1.5,1.5) -- ++(0,-1.5) -- ++(-3,0) -- ++(0,1.5) -- cycle;
\fill [ProcessBlue,opacity=.15] (1.5,-1.5) -- ++(0,.75) -- ++(-3,0) -- ++(0,-.75) -- cycle;
\fill [ProcessBlue,opacity=.15] (-.5,1.5) -- (-.5,-.75) -- (.5,-.75) -- (-.5,.75) -- (-.5,1.5) -- (1.5,1.5) -- ++(0,-3) -- ++(-3,0) -- ++(0,3) -- cycle;
\draw [Blue,very thick] (-1.5,0) -- (1.5,0);
\draw [Blue,very thick] (-1.5,-.75) -- (1.5,-.75);
\draw [Blue,very thick] (1,-1.5) -- (-1,1.5);
\draw [Blue,very thick] (-.5,-1.5) -- (-.5,1.5);
\node [anchor=east] at (-.5,-.375) {$\scriptsize \gamma_1$};
\node [anchor=north] at (0,-.75) {$\scriptsize \gamma_2$};
\node [anchor=west] at (.25,-.375) {$\scriptsize \gamma_3$};
\node [anchor=south] at (-.25,0) {$\scriptsize \gamma_4$};
\end{scope}
\node [anchor=center] at (0,-2) {\Large$\Downarrow$};
\begin{scope}[yshift=-3.5cm]
\draw [Blue,ultra thick] (-1.5,0) -- (1.5,0);
\node [anchor=south west] at (-1.5,0) {$\scriptsize P_{\{2,4\}}$};
\draw [Blue,ultra thick] (0,-1) -- (0,1);
\node [anchor=south west] at (0,-1) {$\scriptsize P_{\{1,3\}}$};
\end{scope}
\node [anchor=center] at (5,-2) {\Large$\Downarrow$};
\begin{scope}[xshift=5cm,yshift=-3.5cm]
\draw [Blue,ultra thick] (-1.5,0) -- (1.5,0);
\node [anchor=south west] at (-1.5,0) {$\scriptsize P_{\{2,4\}}$};
\fill [Blue] (0,0) circle [radius=3pt];
\node [anchor=north] at (0,0) {$\scriptsize P_{\{1,2,3\}}$};
\end{scope}
\end{tikzpicture}
\end{center}
\vspace{-1.5em}\caption{Higher-order poles from the intersection of pinching planes.}
\label{app:fig:emergehigherpoles}
\end{figure}
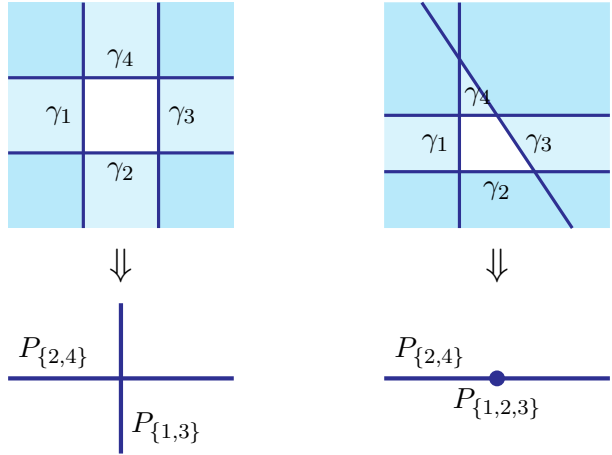
On the other hand, if $P_A$ and $P_B$ indeed have a non-trivial intersection, then the contour in fact has to be further restricted to $P_A\cap P_B$. This is essentially because as we hit the pole in $\gamma$ the contour is forced to become incident to $\Gamma_i$ singularities for both $i\in A$ and $i\in B$. For instance, in Figure \ref{app:fig:emergehigherpoles} we present two examples of 2d integrals that are qualitatively different. In the first case the algorithm generates $\gamma_{\{1,3\}}$ and $\gamma_{\{2,4\}}$ that are assumed to share the same poles, both emerged from some shrinking $1$-simplex. Their pinching planes $P_{\{1,3\}}$ and $P_{\{2,4\}}$ are both codim-1 but intersect at a codim-2 plane, and so for the poles under consideration the contour is restricted to $P_{\{1,3\}}\cap P_{\{2,4\}}$. In the second case, the algorithm generates $\gamma_{\{2,4\}}$ and $\gamma_{\{1,2,3\}}$. The pinching plane of the former is codim-1 while that of the latter is codim-2. Here we further have $P_{\{1,2,3\}}$ being embedded in $P_{\{2,4\}}$. So the contour is restricted to $P_{\{1,2,3\}}\equiv P_{\{1,2,3\}}\cap P_{\{2,4\}}$.

In both cases the resulting poles are double poles, as they emerge from four $\Gamma$'s and two integrations, or equivalently, as they come as the overlapping of two families of simple poles. 

As a natural generalization of the above discussion, assume we have in total $t$ families of poles $\gamma_{A_1}\gamma_{A_2}\cdots\gamma_{A_t}$ such that all of them coincide. If $P_{A_1}\cap P_{A_2}\cap\cdots\cap P_{A_t}\neq\varnothing$, then the actual contour has to be restricted to this common intersection, which is always a hyperplane of certain codimension, and the resulting poles are of order $t$. More generally we may encounter $P_{A_1}\cap P_{A_2}\cap\cdots\cap P_{A_t}=\varnothing$ as well. In this case, we look for all possible intersection of some subset of them, such that it does not further intersect with any other element in the set. This returns a disconnected set of hyperplanes (of possibly different codimensions), such that the actual contour has to be pinched by each connected component once. Correspondingly the order of the resulting pole is determined by the maximal number of $P$'s whose common intersection is non-empty (thus being one of the connected components). 

To name one such example, consider three families $\gamma_1\gamma_2\gamma_3$ whose poles coincide. Let us assume that $P_1$ and $P_2$ are planes while $P_3$ is a line, and any two of them have a non-trivial intersection but $P_1\cap P_2\cap P_3=\varnothing$. Then the contour has to be pinching in $P_1\cap P_2$ (a line), $P_1\cap P_3$ and $P_2\cap P_3$ (points). In this case the emergent pole is a double pole.

\temp{More subtle care for the role of the zeros.}

\paragraph{Further comments on the zeros.}

In practice we also need to take into consideration zeros of the integrand and the possibility of their intersection or overlap with the pinching plane. In that case the pole has to have a lower order or even vanish. In particular, if the pole is only a simple pole then any overlapping zero will kill it. 

This is particularly important when the zeros arise from some Beta functions in the integrand (as we frequently see in the integration kernels under study). Here we can identify the reciprocal $\Gamma$ function therein with two subscripts, from the two Gamma functions in the same Beta, i.e., of the pattern $\frac{\Gamma_1[x]\Gamma_2[y]}{\Gamma_{\{1,2\}}[x+y]}$. After working out the non-composite $\Gamma$ functions, we should further delete those that contains both subscripts. This holds regardless of whether the reciprocal $\Gamma$ function itself depends on some integration variables or not.

It suffices to illustrate this point by the following two-fold integral
\begin{equation}
\begin{split}
&\int\frac{\mathrm{d}x\,\mathrm{d}y}{(2\pi i)^2}\,\frac{\Gamma_1[a_1+x]\Gamma_{2}[b_2-x+y]}{\Gamma_{\{1,2\}}[a_1+b_2+y]}\Gamma_3[a_2-x]\Gamma_{4}[b_1+x-y]\Gamma_5[c_1+y]\Gamma_6[c_2+y]\Gamma_7[c_3-y]\\
&=\frac{\Gamma_{\{1,3\}}[a_1+a_2]\Gamma_{\{2,4\}}[b_1+b_2]\Gamma_{\{3,4,5\}}[a_2+b_1+c_1]\Gamma_{\{3,4,6\}}[a_2+b_1+c_2]\Gamma_{\{5,7\}}[c_1+c_3]\Gamma_{\{6,7\}}[c_2+c_3]}{\Gamma_{\{1,2,3,4\}}[a_1+a_2+b_1+b_2]\Gamma_{\{3,4,5,6,7\}}[a_2+b_1+c_1+c_2+c_3]}.
\end{split}
\end{equation}
By ignoring the factor $1/\Gamma_{\{1,2\}}$ first we conclude with the elementary Gamma's
\begin{equation}
\{\Gamma_{\{1,3\}},\Gamma_{\{2,4\}},\Gamma_{\{5,7\}},\Gamma_{\{6,7\}},\Gamma_{\{1,2,7\}},\Gamma_{\{3,4,5\}},\Gamma_{\{3,4,6\}}\}.
\end{equation}
However, $\Gamma_{\{1,2,7\}}$ contains $\{1,2\}$ in its subscripts, so it has to be excluded.

\subsubsection{Residue contours}
\label{app:sec:genericresiduecontour}

Based on the geometric picture for the pinching in Section \ref{app:sec:pinchinggeometry}, we now provide a general prescription for the integration contour that computes the residue at a specified pole, which generalizes the practice in one-dimensional integrals discussed in Appendix \ref{app:sec:revisitsingleintegral}.

\paragraph{Simple pole.}

First focus on the case of a simple pole. As discussed before, this uniquely emerges from a shrinking simplex in the pinching. Without loss of generality let us again consider $\gamma_{\{0,1,\ldots,r\}}$ that emerges from $\{\Gamma_{i}[E_{i}]\}$ ($i=0,1,\ldots,r$) via the $r$-fold integrations of variables $\{z_1,\ldots,z_r\}$. Here the arguments inside the $\Gamma$ in general have the form
\begin{equation}
E_i=a_i+\sum_{j=1}^rp_{i\,j}\,z_j,
\end{equation}
where the $p$ coefficients can be either positive or negative.

Since the original integral is invariant under arbitrary non-degenerate affine transformation, it is in particular convenient to transform $\{z_1,\ldots,z_r\}$ into a standard frame associated to these poles. This is done by, e.g., assigning $E_i=e_i$ ($i=1,\ldots,r$) and take $\{e_1,\ldots,e_r\}$ as our new integration variables
\begin{equation}
\int[\mathrm{d}z]\,H[z]\prod_{i=0}^r\Gamma_i[E_i]=\det\Big(\frac{\partial E}{\partial z}\Big)^{-1}\int[\mathrm{d}e]\,H[z[e]]\,\Gamma_0[E_0]\prod_{i=1}^r\Gamma_i[e_i].
\end{equation}
In the $e$ frame it must happens that $E_0$ has the form
\begin{equation}
E_0=E-\sum_{j=1}^rp'_{0\,j}\,e_j,
\end{equation}
with all $\{p'\}$ positive (since otherwise the pinching cannot occur). Then the residue at the pole $E+m=0$ ($k\in\mathbb{N}$) is simply computed by
\begin{equation}
\begin{split}
&\det\Big(\frac{\partial E}{\partial z}\Big)^{-1}\residue{E+m=0}\prod_{i=1}^r(-\!\!\residue{e_i=-m_i})\,H\,\Gamma_0[E_0]\prod_{i=1}^r\Gamma_i[e_i]\\
&\qquad=\frac{(-1)^{r+\sum_{i=1}^rm_i}}{\prod_{i=1}^rm_i!}\det\Big(\frac{\partial E}{\partial z}\Big)^{-1}\residue{E+m=0}\sum_{m_1,\ldots,m_r=0}^{\sum_{i=1}^rp'_{0\,i}m_i\leq m}H\,\Gamma_0[E+\sum_{i=1}^rp'_{0\,i}m_i].
\end{split}
\end{equation}
Here the summation is induced from the lattice \eqref{app:eq:pinchinglattice}.

\paragraph{Higher-order pole.}

When a higher-order pole is emerged, the direct prescription for its corresponding residue is involved, due to the complicated pinching configuration as discussed before. Nevertheless, one can start by deforming the original integral with some infinitesimal parameter $\epsilon$, such that the higher-order pole desolves into a bunch of simple poles. At this point, we compute the residue contributed by each of the pinching planes and sum them up. Then in the $\epsilon\to0$ limit, although each piece has to diverge, the divergent part cancel away and the remaining finite part is guaranteed to be the correct residue at the original higher-order pole.

Let us illustrate this by one simple example
\begin{equation}\label{app:eq:doublepoleexample}
\int\frac{\mathrm{d}x\,\mathrm{d}y}{(2\pi i)^2}\,\Gamma_1[a+y]\,\Gamma_2[-y],\Gamma_3[-x]\,\Gamma_4[a+x+y]\,\Gamma_5[b-y-x]\,\Gamma_6[c+x]=\frac{\Gamma[a]^2\,\Gamma[a+b]\,\Gamma[c]\,\Gamma[a+b+c]}{\Gamma[2a+b+c]}.
\end{equation}
Here the double poles $\Gamma[a]$ emerge both as $\gamma_{\{1,2\}}$ and as $\gamma_{\{2,3,4\}}$. To compute the residue, say at the leading pole $a=0$, we start by deforming $\Gamma_1[a+y]\mapsto\Gamma_1[a+\epsilon+y]$. The the two families become $\Gamma_{\{1,2\}}[a+\epsilon]$ and $\Gamma_{\{2,3,4\}}[a]$. Computing the residue at $a=-\epsilon$ for $\gamma_{\{1,2\}}$ we obtain
\begin{equation}
-\residue{a=-\epsilon}\,\residue{y=0}\,(\text{same integrand})=\frac{\Gamma[c]\,\Gamma[b+c]\,\Gamma[b-\epsilon]\,\Gamma[-\epsilon]}{\Gamma[b+c-\epsilon]},
\end{equation}
and computing the residue at $a=0$ for $\gamma_{\{2,3,4\}}$ yields
\begin{equation}
\residue{a=0}\,\residue{y=0}\,\residue{x=0}\,(\text{same integrand})=\Gamma[b]\,\Gamma[c]\,\Gamma[\epsilon].
\end{equation}
Each piece is divergent as $\epsilon\mapsto0$, which is expected. However, the summation of these two pieces has a finite limit that matches the correct residue
\begin{equation}
-\Gamma[b]\,\Gamma[c]\,(2\gamma_{\rm E}-\psi[b]+\psi[b+c]),
\end{equation}
where $\gamma_{\rm E}$ denotes the Euler constant and $\psi[x]\equiv\Gamma'[x]/\Gamma[x]$.

\newpage

\section{Selected Verifications of the Fake Poles of Pre-amplitudes}\label{sec:checkfakepoles}

\temp{This appendix can be completely fixed.}

In this Appendix we verify that some of the poles (marked green) obtained in the examples in Section \ref{sec:polestructure} are absent. These occurred in
\begin{itemize}
\item 4-point triangle diagram with only cubic vertices:
\begin{equation}
\checked{}
\underset{\{3,4,6,7,9,10\}}{\Gamma[{\textstyle\frac{2h-S-c_3-c_4}{2}}]}\,
\underset{\{3,4,6,8,9,10\}}{\Gamma[{\textstyle\frac{2h-S-c_3+c_4}{2}}]}\,
\underset{\{3,5,6,7,9,10\}}{\Gamma[{\textstyle\frac{2h-S+c_3-c_4}{2}}]}\,
\underset{\{3,5,6,8,9,10\}}{\Gamma[{\textstyle\frac{2h-S+c_3+c_4}{2}}]};
\end{equation}
\item 4-point box diagram: 
\begin{equation}
\begin{split}\checked{}
&\underset{\{4,8,9,12,17,19,20,24\}}{\cancel{\Gamma}[{\textstyle\frac{\Delta_{34}-c_1-c_3}{2}}]}\,
\underset{\{4,11,12,13,17,18,19,20\}}{\cancel{\Gamma}[{\textstyle\frac{\Delta_{12}+c_1-c_3}{2}}]}\,
\underset{\{5,8,9,12,17,19,20,24\}}{\cancel{\Gamma}[{\textstyle\frac{\Delta_{34}-c_1+c_3}{2}}]}\,
\underset{\{5,11,12,13,17,18,19,20\}}{\cancel{\Gamma}[{\textstyle\frac{\Delta_{12}+c_1+c_3}{2}}]}\\
&\underset{\{4,8,9,11,12,13,17,19,20\}}{\cancel{\Gamma}[{\textstyle h-c_3}]}\,
\underset{\{5,8,9,11,12,13,17,19,20\}}{\cancel{\Gamma}[{\textstyle h+c_3}]}.
\end{split}
\end{equation}
\end{itemize}
Their absence are all due to the vanishing of the vertex correction functions (from the original tree diagram) when the contour is pinched. Note that all these families by themselves are simple poles, and so they have to be absent as long as we find a zero from any part of the remaining integrand. In the following we provide a detailed check.

\subsection{4-point triangle with only cubic vertices}\label{app:sec:checkfakepoletriangle}

For this diagram it suffices to focus on the first family $\Gamma_{\{3,4,6,7,9,10\}}$, as the analysis on the other three families are similar. The original $\Gamma$ functions relevant for these poles are
\begin{equation}
\begin{split}\checked{}
\Gamma_3[{\textstyle\frac{t_1-S}{2}}]\,
\Gamma_4[{\textstyle\frac{h-c_3-t_3}{2}}]\,
\Gamma_6[{\textstyle\frac{h-c_2-t_2+t_3}{2}}]\,
\Gamma_7[{\textstyle\frac{h-c_4-t_4}{2}}]\,
\Gamma_9[{\textstyle\frac{h+c_2+t_4-t_5}{}}]\,
\Gamma_{10}[{\textstyle\frac{-2h-t_1+t_2+t_5}{2}}].
\end{split}
\end{equation}
It is easily seen that all the integrals are pinched in order to produce this $\gamma$ family. In particular, at the leading pole we have
\begin{equation}\label{app:eq:4ptcubicpinchvaluesleading}
t_1=2h-c_3-c_4,\quad t_3=h-c_3,\quad t_4=h-c_4.
\end{equation}
The remaining part of the integrand  \eqref{eq:remainingintegrand3ptcubic} stays finite, except for the vertex correction $C_{A_2}$, which depends on $\{t_1,t_3,t_4,c_1,c_3,c_4\}$ via \eqref{eq:vertexcorrectionspecial}
\begin{equation}\label{eq:Corr3}
\checked{}
\int\!\!\prod_{a\in\{1,3,4\}}\!\!\frac{\mathrm{d}w_a}{2\pi i}\,\underbrace{\frac{\Gamma[\frac{h+c_1+c_3+c_4}{2}+\underset{a\in\{1,3,4\}}{\sum}w_a]}{\Gamma[\frac{h\pm c_1\pm c_3\pm c_4}{2}]}\!\!\prod_{a\in\{1,3,4\}}\!\!\!\!\frac{\Gamma[-w_a]\Gamma[-c_a-w_a]\Gamma[\frac{h+c_a-t_a}{2}+w_a]}{\Gamma[\frac{h\pm c_a-t_a}{2}]}}_{c_{A_2}}.
\end{equation}
Evaluating on the values \eqref{app:eq:4ptcubicpinchvaluesleading}, obviously the integrand $c_{A_2}$ produces a double zero at the above values of $t_3$ and $t_4$. Moreover, poles at the same location also emerges from the $w_3$ integration due to the factors $\Gamma[-c_3-w_3]$ and $\Gamma[\frac{h+c_3-t_3}{2}+w_3]$, and from the $w_4$ integration due to $\Gamma[-c_4-w_4]$ and $\Gamma[\frac{h+c_4-t_4}{2}+w_4]$. Hence the computation involves a $\frac{\infty}{\infty}$ effect, which necessarily forces to further localize the $w_3$ and $w_4$ contours, resulting in
\begin{equation}
\begin{split}\checked{}
C_{A_2}=&\frac{1}{4}\int\frac{\mathrm{d}w_1}{2\pi i}\,\residue{t_4=h-c_4}\,\residue{w_4=-c_4}\,\residue{t_3=h-c_3}\,\residue{w_3=-c_3}\,\left(c_{A_2}\,\Gamma[{\textstyle\frac{h-c_3-t_3}{2}}]\,\Gamma[{\textstyle\frac{h-c_4-t_4}{2}}]\right)\\
=&\int\frac{\mathrm{d}w_1}{2\pi i}\,\frac{\Gamma[-c_1-w_1]\Gamma[-w_1]\Gamma[\frac{-h+c_1+c_3+c_4}{2}+w_1]\Gamma[\frac{h+c_1-c_3-c_4}{2}+w_1]}{\Gamma[\frac{h\pm c_1\pm c_3\pm c_4}{2}]\Gamma[\frac{-h\pm c_1+c_3+c_4}{2}]},
\end{split}
\end{equation}
where we have already substituted $t_1=2h-c_3-c_4$. The remaining $w_1$ integral can then be performed by Barnes' first lemma, which however produces a $1/\Gamma[0]$, thus forcing $C_{A_2}$ to vanish. 

Let us move on to the subleading poles, at $S=2h-c_3-c_4+2m$ for some positive integer $m$. In general the contour is pinched at several different sites, where their $\{t_1,t_3,t_4\}$ coordinates are
\begin{equation}
\checked{}
t_1=2h-c_3-c_4+2n,\qquad
t_3=h-c_3+2p,\qquad
t_4=h-c_4+2q,\qquad
m\geq n\geq p+q,
\end{equation}
for some non-negative integers $n,p,q$. As we evaluate $C_{A_2}$ at each specific pinching site $(n,p,q)$, the proper contour prescription is
\begin{equation}
\checked{}
\frac{(-1)^{p+q}p!q!}{4}\int\frac{\mathrm{d}w_1}{2\pi i}\,\residue{t_4=h-c_4+2q}\left(\sum_{j=0}^q\residue{w_4=-c_4+2j}\right)\residue{t_3=h-c_3+2p}\left(\sum_{i=0}^p\residue{w_3=-c_3+2i}\right)\left(c_{A_2}\,\Gamma[{\textstyle\frac{h-c_3-t_3}{2}}]\,\Gamma[{\textstyle\frac{h-c_4-t_4}{2}}]\right),
\end{equation}
which then is further evaluated at $t_1=2h-c_3-c_4+2n$. It is not hard to explicitly verify that term by term in this double summation always integrates to zero in similar way as what we observed for the leading pole. Hence again effectively there is no pinching at all.

\subsection{The 4-point box}\label{app:sec:checkfakepolesbox}

For the 4-point box it suffices to focus on $\Gamma_{\{4,8,9,12,17,19,20,24\}}$ and $\Gamma_{\{4,8,9,11,12,13,17,19,20\}}$. As mentioned in Section \ref{sec:boxpoles} these can already be ruled out by the symmetries of the box diagram, nevertheless it is good to inspect explicitly how they are absent. Here we only work with the leading poles; the subleading ones follow similarly as what we discussed above.

\subsubsection*{$\Gamma_{\{4,8,9,12,17,19,20,24\}}[\frac{\Delta_{34}-c_1-c_3}{2}]$.} 

This family of poles originate from the factors
\begin{equation}
\begin{split}\checked{}
&\Gamma_4[{\textstyle\frac{h-c_3-t_3}{2}}]\,
\Gamma_8[{\textstyle\frac{h-c_1+t_4-t_5}{2}}]\,
\Gamma_9[{\textstyle\frac{-t_1+t_3-t_4+t_5}{2}}]\,
\Gamma_{12}[{\textstyle\frac{-2h+t_5-t_6+t_7}{2}}]\,
\Gamma_{17}[{\textstyle\frac{t_1-t_5+t_6-\Delta_2}{2}}]\\
&\Gamma_{19}[{\textstyle\frac{t_6-t_7+t_8-\Delta_3}{2}}]\,
\Gamma_{20}[{\textstyle\frac{\Delta_{23}-t_6}{2}}]\,
\Gamma_{24}[{\textstyle\frac{\Delta_{34}-t_8}{2}}].
\end{split}
\end{equation}
Except for $t_2$, the integrals in all other directions are pinched. In particular we have $t_3=h-c_3$ and $t_4=h-c_3+\Delta_3$. Recall the original Mellin integrand contains two additional correction functions $C_{A_2}$ and $C_{A_3}$. Let us first focus on $C_{A_2}$, which in this case explicitly is
\begin{equation}\label{app:eq:ca2box}
\checked{}
\int\frac{\mathrm{d}w_2\mathrm{d}w_3}{(2\pi i)^2}\,\underbrace{\frac{\Gamma[\frac{\Delta_2+c_2+c_3}{2}+w_2+w_3]}{\Gamma[\frac{\Delta_2\pm c_3\pm c_3}{2}]}\!\!\prod_{a\in\{2,3\}}\!\!\!\!\frac{\Gamma[-w_a]\Gamma[-c_a-w_a]\Gamma[\frac{h+c_a-t_a}{2}+w_a]}{\Gamma[\frac{h\pm c_a-t_a}{2}]}}_{c_{A_2}}.
\end{equation}
This function only depend on $\{t_2,t_3\}$ and no other integration variables. Again, as we assign $t_3=h-c_3$, $\Gamma[\frac{h-c_3-t_3}{2}]$ in the denominator diverges, but a pole at the same location is also produced by the $w_3$ integral due to the factors $\Gamma[-c_3-w_3]$ and $\Gamma[\frac{h+c_3-t_3}{2}+w_3]$ in the numerator. Hence the value of $C_{A_2}$ is computed as
\begin{equation}
\checked{}
C_{A_2}=\frac{1}{2}\int\frac{\mathrm{d}w_2}{2\pi i}\residue{t_3=h-c_3}\,\residue{w_3=-c_3}\,\left(c_{A_2}\,\Gamma[{\textstyle\frac{h-c_3-t_3}{2}}]\right)=\frac{1}{\Gamma[\frac{\Delta_2\pm c_2+t_3}{2}]\Gamma[\frac{h-c_3-t_2+\Delta_2}{2}]},
\end{equation}
which is finite. Similar computation needs to be carried out for the evaluation of $C_{A_3}$, which upon the localized values of $t_3$ and $t_4$ turns out to be zero. Hence the leading pole of $\Gamma[\frac{\Delta_{34}-c_1-c_3}{2}]$ is absent.

\subsubsection*{$\Gamma_{\{4,8,9,11,12,13,17,19,20\}}[h-c_3]$.}

This family of poles emerge from
\begin{equation}
\begin{split}\checked{}
&\Gamma_{4}[{\textstyle{\frac{h-c_3-t_3}{2}}}]\,
\Gamma_{8}[{\textstyle{\frac{h-c_1+t_4-t_5}{2}}}]\,
\Gamma_{9}[{\textstyle{\frac{-t_1+t_3-t_4+t_5}{2}}}]\,
\Gamma_{11}[{\textstyle{\frac{h+c_1+t_2-t_7}{2}}}]\,
\Gamma_{12}[{\textstyle{\frac{-2h+t_5-t_6+t_7}{2}}}]\\
&\Gamma_{13}[{\textstyle{\frac{-t_2+t_3+t_7-t_8}{2}}}]\,
\Gamma_{17}[{\textstyle{\frac{t_1-t_5+t_6-\Delta_2}{2}}}]\,
\Gamma_{19}[{\textstyle{\frac{t_6-t_7+t_8-\Delta_3}{2}}}]\,
\Gamma_{20}[{\textstyle{\frac{\Delta_{23}-t_6}{2}}}].
\end{split}
\end{equation}
At the leading pole the contour is pinched at the locations where $t_2=\Delta_2$, $t_3=0$ and $c_3=h$. Similar evaluations of $C_{A_2}$ as before shows that this factor vanishes. Hence this leading pole is absent.

\newpage

\section{Poles of Pre-amplitudes for Generalized Bubbles}\label{app:sec:polegbubbles}

\temp{No major revision further needed in this appendix.}

In this appendix we list out seven examples of generalized bubble diagrams. The pole structure of their pre-amplitudes are worked out from the effective bulk-to-bulk propagator following Appendix \ref{app:sec:bbpropagators}. It can be explicitly checked that these results are consistent with Conjecture \ref{sec:enhancedrules}. And for this purpose we also present the families of poles that arise from the four diagrammatic rules but are absent due to their compositeness (note that since all the diagrams are generalized bubbles presumably every bulk-to-bulk propagator should contribute a $\Gamma[h\pm c]$ according to the generalized bubble rule in Section \ref{sec:Mgbubbles}).

Furthermore, we collect here the explicit expressions for residues at their leading non-minimal poles (if they allow non-minimal cuts), which is an evidence that we have to exclude the generalized bubbles from our Conjecture \ref{sec:conjectureonamplitude} regarding the absence of non-minimal poles. For this purpose also study an additional diagram that contains a generalized bubble as a proper sub-diagram at the end.

\begin{enumerate}[noitemsep,nolistsep]
\item[\textbf{(A)}] For the diagram (We denote the total boundary conformal dimension on the left as $\Delta_{A_0}$ and that on the right as $\Delta_{A_1}$. The same applies to other diagrams.)
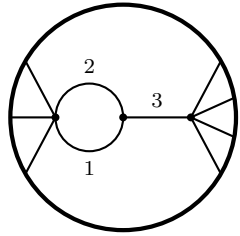
\begin{center}
\begin{tikzpicture}
\begin{scope}[scale=.6]
\draw [black,ultra thick] (0,0) circle [radius=2.5];
\draw [black,thick] (-1.5,0) -- (150:2.5);
\draw [black,thick] (-1.5,0) -- (180:2.5);
\draw [black,thick] (-1.5,0) -- (210:2.5);
\draw [black,thick] (1.5,0) -- (30:2.5);
\draw [black,thick] (1.5,0) -- (10:2.5);
\draw [black,thick] (1.5,0) -- (-10:2.5);
\draw [black,thick] (1.5,0) -- (-30:2.5);
\draw [black,thick] (-.75,0) circle [radius=.75];
\draw [black,thick] (0,0) -- (1.5,0);
\fill [black] (-1.5,0) circle [radius=2.5pt];
\fill [black] (0,0) circle [radius=2.5pt];
\fill [black] (1.5,0) circle [radius=2.5pt];
\node [anchor=north] at (-.75,-.75) {\scriptsize $1$};
\node [anchor=south] at (-.75,.75) {\scriptsize $2$};
\node [anchor=south] at (.75,0) {\scriptsize $3$};
\end{scope}
\end{tikzpicture}
\end{center}
the pre-amplitude has poles
\begin{equation}
\checked{}
\Gamma[{\textstyle\frac{h\pm c_1\pm c_2\pm c_3}{2}}]\,
\Gamma[{\textstyle\frac{\Delta_{A_1}-h\pm c_3}{2}}]\,
\Gamma[{\textstyle\frac{\Delta_{A_0}-h\pm c_3}{2}}]\,
\Gamma[{\textstyle\frac{h\pm c_3-S}{2}}].
\end{equation}
In addition, the poles ruled out by compositeness are
\begin{equation}\checked{}
{\color{ForestGreen}
\cancel{\Gamma}[h\pm c_1]\,
\cancel{\Gamma}[h\pm c_2]\,
\cancel{\Gamma}[h\pm c_3]\,
\cancel{\Gamma}[{\textstyle\frac{\Delta_{A_0}\pm c_1\pm c_2}{2}}]\,
\cancel{\Gamma}[{\textstyle\frac{2h\pm c_1\pm c_2-S}{2}}]}.
\end{equation}
This diagram allows a non-minimal cut $(12)$, hence the non-minimal poles $\Gamma[\frac{\underline\Delta_{12}-S}{2}]$. The residue at its leading pole is
\begin{equation}
\residue{S=\underline\Delta_{12}}\mathcal{M}=
-\prod_i\frac{\mathcal{C}_{\Delta_i}}{\Gamma[\Delta_i]}\,\frac{\Gamma[\underline\Delta_1]\,\Gamma[\underline\Delta_2]\,\Gamma[\frac{\underline\Delta_{12}+\Delta_{A_0}}{2}-h]\,\Gamma[\frac{\underline\Delta_{12}+\Delta_{A_1}}{2}-h]}{8\prod_{a=1}^2\!\Gamma[\underline\Delta_a-h+1]\;\Gamma[\underline\Delta_{12}]\,(\underline\Delta_{12}-\underline\Delta_3)\,(2h-\underline\Delta_{123})}.
\end{equation}
\item[\textbf{(B)}] For the diagram
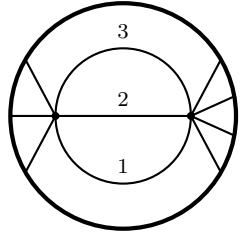
\begin{center}
\begin{tikzpicture}
\begin{scope}[scale=.6]
\draw [black,ultra thick] (0,0) circle [radius=2.5];
\draw [black,thick] (-1.5,0) -- (150:2.5);
\draw [black,thick] (-1.5,0) -- (180:2.5);
\draw [black,thick] (-1.5,0) -- (210:2.5);
\draw [black,thick] (1.5,0) -- (30:2.5);
\draw [black,thick] (1.5,0) -- (10:2.5);
\draw [black,thick] (1.5,0) -- (-10:2.5);
\draw [black,thick] (1.5,0) -- (-30:2.5);
\draw [black,thick] (0,0) circle [radius=1.5];
\draw [black,thick] (-1.5,0) -- (1.5,0);
\fill [black] (-1.5,0) circle [radius=2.5pt];
\fill [black] (1.5,0) circle [radius=2.5pt];
\node [anchor=south] at (0,-1.5) {\scriptsize $1$};
\node [anchor=south] at (0,0) {\scriptsize $2$};
\node [anchor=south] at (0,1.5) {\scriptsize $3$};
\end{scope}
\end{tikzpicture}
\end{center}
the pre-amplitude has poles
\begin{equation}
\begin{split}\checked{}
&\Gamma[{\textstyle h\pm c_1}]\,
\Gamma[{\textstyle h\pm c_2}]\,
\Gamma[{\textstyle h\pm c_3}]\,
\Gamma[{\textstyle\frac{\Delta_{A_0}+h\pm c_1\pm c_2\pm c_3}{2}}]\,
\Gamma[{\textstyle\frac{\Delta_{A_1}+h\pm c_1\pm c_2\pm c_3}{2}}]\\
&\Gamma[{\textstyle\frac{\Sigma\Delta}{2}-h}]\,
\Gamma[{\textstyle\frac{3h\pm c_1\pm c_2\pm c_3-S}{2}}].
\end{split}
\end{equation}
There are no composite poles from this diagram, hence no non-minimal cuts as well.
\item[\textbf{(C)}] For the diagram
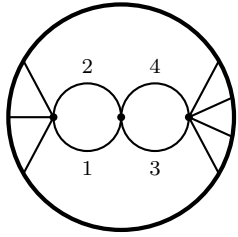
\begin{center}
\begin{tikzpicture}
\begin{scope}[scale=.6]
\draw [black,ultra thick] (0,0) circle [radius=2.5];
\draw [black,thick] (-1.5,0) -- (150:2.5);
\draw [black,thick] (-1.5,0) -- (180:2.5);
\draw [black,thick] (-1.5,0) -- (210:2.5);
\draw [black,thick] (1.5,0) -- (30:2.5);
\draw [black,thick] (1.5,0) -- (10:2.5);
\draw [black,thick] (1.5,0) -- (-10:2.5);
\draw [black,thick] (1.5,0) -- (-30:2.5);
\draw [black,thick] (-.75,0) circle [radius=.75];
\draw [black,thick] (.75,0) circle [radius=.75];
\fill [black] (-1.5,0) circle [radius=2.5pt];
\fill [black] (0,0) circle [radius=2.5pt];
\fill [black] (1.5,0) circle [radius=2.5pt];
\node [anchor=north] at (-.75,-.75) {\scriptsize $1$};
\node [anchor=south] at (-.75,.75) {\scriptsize $2$};
\node [anchor=north] at (.75,-.75) {\scriptsize $3$};
\node [anchor=south] at (.75,.75) {\scriptsize $4$};
\end{scope}
\end{tikzpicture}
\end{center}
the pre-amplitude has poles
\begin{equation}
\begin{split}\checked{}
&\Gamma[{\textstyle h\pm c_1}]\,
\Gamma[{\textstyle h\pm c_2}]\,
\Gamma[{\textstyle h\pm c_3}]\,
\Gamma[{\textstyle h\pm c_4}]\,
\Gamma[{\textstyle\frac{\Delta_{A_0}\pm c_1\pm c_2}{2}}]\,
\Gamma[{\textstyle\frac{2h\pm c_1\pm c_2\pm c_3\pm c_4}{2}}]\\
&\Gamma[{\textstyle\frac{\Delta_{A_1}\pm c_3\pm c_4}{2}}]\,
\Gamma[{\textstyle\frac{\Delta_{A_0}\pm c_3\pm c_4}{2}}]\,
\Gamma[{\textstyle\frac{\Delta_{A_1}\pm c_1\pm c_2}{2}}]\,
\Gamma[{\textstyle\frac{\Sigma\Delta}{2}-h}]\\
&\Gamma[{\textstyle\frac{2h\pm c_1\pm c_2-S}{2}}]\,
\Gamma[{\textstyle\frac{2h\pm c_3\pm c_4-S}{2}}].
\end{split}
\end{equation}
There are no composite poles from this diagram, hence no non-minimal cuts as well.
\item[\textbf{(D)}] For the diagram
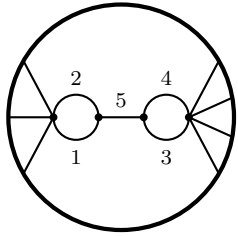
\begin{center}
\begin{tikzpicture}
\begin{scope}[scale=.6]
\draw [black,ultra thick] (0,0) circle [radius=2.5];
\draw [black,thick] (-1.5,0) -- (150:2.5);
\draw [black,thick] (-1.5,0) -- (180:2.5);
\draw [black,thick] (-1.5,0) -- (210:2.5);
\draw [black,thick] (1.5,0) -- (30:2.5);
\draw [black,thick] (1.5,0) -- (10:2.5);
\draw [black,thick] (1.5,0) -- (-10:2.5);
\draw [black,thick] (1.5,0) -- (-30:2.5);
\draw [black,thick] (-1,0) circle [radius=.5];
\draw [black,thick] (1,0) circle [radius=.5];
\draw [black,thick] (-.5,0) -- (.5,0);
\fill [black] (-1.5,0) circle [radius=2.5pt];
\fill [black] (-.5,0) circle [radius=2.5pt];
\fill [black] (.5,0) circle [radius=2.5pt];
\fill [black] (1.5,0) circle [radius=2.5pt];
\node [anchor=north] at (-1,-.5) {\scriptsize $1$};
\node [anchor=south] at (-1,.5) {\scriptsize $2$};
\node [anchor=north] at (1,-.5) {\scriptsize $3$};
\node [anchor=south] at (1,.5) {\scriptsize $4$};
\node [anchor=south] at (0,0) {\scriptsize $5$};
\end{scope}
\end{tikzpicture}
\end{center}
the pre-amplitude has poles
\begin{equation}
\begin{split}\checked{}
&\Gamma[{\textstyle\frac{h\pm c_1\pm c_2\pm c_5}{2}}]\,
\Gamma[{\textstyle\frac{h\pm c_3\pm c_4\pm c_5}{2}}]\,
\Gamma[{\textstyle\frac{\Delta_{A_0}-h\pm c_5}{2}}]\,
\Gamma[{\textstyle\frac{\Delta_{A_1}-h\pm c_5}{2}}]\,
\Gamma[{\textstyle\frac{h\pm c_5-S}{2}}].
\end{split}
\end{equation}
In addition, the poles ruled out by compositeness are
\begin{equation}
\begin{split}\checked{}
&{\color{ForestGreen}
\cancel{\Gamma}[h\pm c_1]\,
\cancel{\Gamma}[h\pm c_2]\,
\cancel{\Gamma}[h\pm c_3]\,
\cancel{\Gamma}[h\pm c_4]\,
\cancel{\Gamma}[h\pm c_5]\,
\cancel{\Gamma}[{\textstyle\frac{\Delta_{A_0}\pm c_1\pm c_2}{2}}]\,
\cancel{\Gamma}[{\textstyle\frac{\Delta_{A_1}\pm c_3\pm c_4}{2}}]}\\
&{\color{ForestGreen}\cancel{\Gamma}[{\textstyle\frac{2h\pm c_1\pm c_2-S}{2}}]\,
\cancel{\Gamma}[{\textstyle\frac{2h\pm c_3\pm c_4-S}{2}}]}.
\end{split}
\end{equation}
This diagram allows two non-minimal cuts, hence two families of non-minimal poles $\Gamma[\frac{\underline\Delta_{12}-S}{2}]$ and $\Gamma[\frac{\underline\Delta_{34}-S}{2}]$. Due to the symmetry of this diagram we only check the leading pole of the first family, whose residue is
\begin{equation}
\begin{split}
\residue{S=\underline\Delta_{12}}\!\mathcal{M}\!=
&\frac{-1}{2^6\pi^h}\prod_i\frac{\mathcal{C}_{\Delta_i}}{\Gamma[\Delta_i]}\int\prod_{a=3}^4\frac{[\mathrm{d}c_a]_{\underline\Delta_a}}{\Gamma[\pm c_a]}\,\Gamma[{\textstyle\frac{2h\pm c_3\pm c_4-\underline\Delta_{12}}{2}}]\,\Gamma[{\textstyle\frac{\underline\Delta_{34}\pm c_3\pm c_4}{2}}]\\
&\times\frac{\Gamma[\underline\Delta_1]\,\Gamma[\underline\Delta_2]\,\Gamma[\frac{\underline\Delta_{12}+\Delta_{A_0}}{2}-h]\,\Gamma[\frac{\underline\Delta_{12}+\Delta_{A_1}}{2}-h]}{\prod_{a=1}^2\!\Gamma[\underline\Delta_a\!-\!h\!+\!1]\;\Gamma[\underline\Delta_{12}]^2\,\Gamma[2h\!-\!\underline\Delta_{12}]\,\Gamma[h]\,(\underline\Delta_{12}\!-\!\underline\Delta_5)\,(2h\!-\!\underline\Delta_{125})}.
\end{split}
\end{equation}
\item[\textbf{(E)}] For the diagram
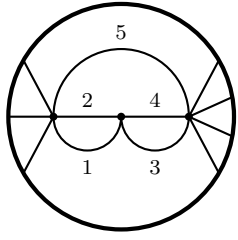
\begin{center}
\begin{tikzpicture}
\begin{scope}[scale=.6]
\draw [black,ultra thick] (0,0) circle [radius=2.5];
\draw [black,thick] (-1.5,0) -- (150:2.5);
\draw [black,thick] (-1.5,0) -- (180:2.5);
\draw [black,thick] (-1.5,0) -- (210:2.5);
\draw [black,thick] (1.5,0) -- (30:2.5);
\draw [black,thick] (1.5,0) -- (10:2.5);
\draw [black,thick] (1.5,0) -- (-10:2.5);
\draw [black,thick] (1.5,0) -- (-30:2.5);
\draw [black,thick] (-1.5,0) -- (1.5,0);
\draw [black,thick] (1.5,0) arc [start angle=0,end angle=180,radius=1.5] arc [start angle=180,end angle=360,radius=.75] arc [start angle=180,end angle=360,radius=.75];
\fill [black] (-1.5,0) circle [radius=2.5pt];
\fill [black] (0,0) circle [radius=2.5pt];
\fill [black] (1.5,0) circle [radius=2.5pt];
\node [anchor=north] at (-.75,-.75) {\scriptsize $1$};
\node [anchor=south] at (-.75,0) {\scriptsize $2$};
\node [anchor=north] at (.75,-.75) {\scriptsize $3$};
\node [anchor=south] at (.75,0) {\scriptsize $4$};
\node [anchor=south] at (0,1.5) {\scriptsize $5$};
\end{scope}
\end{tikzpicture}
\end{center}
the pre-amplitude has poles
\begin{equation}
\begin{split}\checked{}
&\Gamma[{\textstyle h\pm c_1}]\,
\Gamma[{\textstyle h\pm c_2}]\,
\Gamma[{\textstyle h\pm c_3}]\,
\Gamma[{\textstyle h\pm c_4}]\,
\Gamma[{\textstyle h\pm c_5}]\,
\Gamma[{\textstyle\frac{\Delta_{A_0}+h\pm c_1\pm c_2\pm c_5}{2}}]\\
&\Gamma[{\textstyle\frac{2h\pm c_1\pm c_2\pm c_3\pm c_4}{2}}]\,
\Gamma[{\textstyle\frac{\Delta_{A_1}+h\pm c_3\pm c_4\pm c_5}{2}}]\,
\Gamma[{\textstyle\frac{\Delta_{A_0}+h\pm c_3\pm c_4\pm c_5}{2}}]\,
\Gamma[{\textstyle\frac{\Delta_{A_1}+h\pm c_1\pm c_2\pm c_5}{2}}]\\
&\Gamma[{\textstyle\frac{\Sigma\Delta}{2}-h}]\,
\Gamma[{\textstyle\frac{3h\pm c_1\pm c_2\pm c_5-S}{2}}]\,
\Gamma[{\textstyle\frac{3h\pm c_3\pm c_4\pm c_5-S}{2}}].
\end{split}
\end{equation}
There are no composite poles from this diagram, hence no non-minimal cuts as well.
\item[\textbf{(F)}] For the diagram
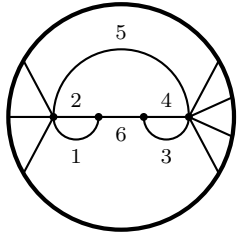
\begin{center}
\begin{tikzpicture}
\begin{scope}[scale=.6]
\draw [black,ultra thick] (0,0) circle [radius=2.5];
\draw [black,thick] (-1.5,0) -- (150:2.5);
\draw [black,thick] (-1.5,0) -- (180:2.5);
\draw [black,thick] (-1.5,0) -- (210:2.5);
\draw [black,thick] (1.5,0) -- (30:2.5);
\draw [black,thick] (1.5,0) -- (10:2.5);
\draw [black,thick] (1.5,0) -- (-10:2.5);
\draw [black,thick] (1.5,0) -- (-30:2.5);
\draw [black,thick] (-1.5,0) -- (1.5,0);
\draw [black,thick] (.5,0) arc [start angle=-180,end angle=0,radius=.5] arc [start angle=0,end angle=180,radius=1.5] arc [start angle=180,end angle=360,radius=.5];
\fill [black] (-1.5,0) circle [radius=2.5pt];
\fill [black] (-.5,0) circle [radius=2.5pt];
\fill [black] (.5,0) circle [radius=2.5pt];
\fill [black] (1.5,0) circle [radius=2.5pt];
\node [anchor=north] at (-1,-.5) {\scriptsize $1$};
\node [anchor=south] at (-1,0) {\scriptsize $2$};
\node [anchor=north] at (1,-.5) {\scriptsize $3$};
\node [anchor=south] at (1,0) {\scriptsize $4$};
\node [anchor=south] at (0,1.5) {\scriptsize $5$};
\node [anchor=north] at (0,0) {\scriptsize $6$};
\end{scope}
\end{tikzpicture}
\end{center}
the pre-amplitude has poles
\begin{equation}
\begin{split}\checked{}
&\Gamma[{\textstyle h\pm c_5}]\,
\Gamma[{\textstyle\frac{h\pm c_1\pm c_2\pm c_6}{2}}]\,
\Gamma[{\textstyle\frac{h\pm c_3\pm c_4\pm c_6}{2}}]\,
\Gamma[{\textstyle\frac{\Delta_{A_0}\pm c_5\pm c_6}{2}}]\,
\Gamma[{\textstyle\frac{\Delta_{A_1}\pm c_5\pm c_6}{2}}]\,
\Gamma[{\textstyle\frac{\Sigma\Delta}{2}-h}]\\
&\Gamma[{\textstyle\frac{2h\pm c_5\pm c_6-S}{2}}].
\end{split}
\end{equation}
In addition, the poles ruled out by compositeness are
\begin{equation}
\begin{split}\checked{}
&{\color{ForestGreen}
\cancel{\Gamma}[h\pm c_1]\,
\cancel{\Gamma}[h\pm c_2]\,
\cancel{\Gamma}[h\pm c_3]\,
\cancel{\Gamma}[h\pm c_4]\,
\cancel{\Gamma}[{\textstyle\frac{\Delta_{A_0}+h\pm c_1\pm c_2\pm c_5}{2}}]\,
\cancel{\Gamma}[{\textstyle\frac{\Delta_{A_1}+h\pm c_3\pm c_4\pm c_5}{2}}]}\\
&{\color{ForestGreen}\cancel{\Gamma}[{\textstyle\frac{3h\pm c_1\pm c_2\pm c_5-S}{2}}]\,
\cancel{\Gamma}[{\textstyle\frac{3h\pm c_3\pm c_4\pm c_5-S}{2}}]}.
\end{split}
\end{equation}
This diagram allows two non-minimal cuts, hence two families of non-minimal poles $\Gamma[\frac{\underline\Delta_{125}-S}{2}]$ and $\Gamma[\frac{\underline\Delta_{345}-S}{2}]$. Due to the symmetry of this diagram we only check the leading pole of the first family, whose residue is
\begin{equation}
\begin{split}
\residue{S=\underline\Delta_{125}}\!\mathcal{M}\!=
&\frac{1}{2^7\pi^{2h}}\prod_i\frac{\mathcal{C}_{\Delta_i}}{\Gamma[\Delta_i]}\int\prod_{a=3}^4\frac{[\mathrm{d}c_a]_{\underline\Delta_a}}{\Gamma[\pm c_a]}\,\Gamma[{\textstyle\frac{2h\pm c_3\pm c_4-\underline\Delta_{12}}{2}}]\,\Gamma[{\textstyle\frac{\underline\Delta_{34}\pm c_3\pm c_4}{2}}]\\
&\times\frac{\Gamma[\underline\Delta_1]\,\Gamma[\underline\Delta_2]\,\Gamma[\underline\Delta_5]\,\Gamma[\frac{\underline\Delta_{125}+\Delta_{A_0}}{2}-h]\,\Gamma[\frac{\underline\Delta_{125}+\Delta_{A_1}}{2}-h]}{\!\!\!\!\!\!\underset{a\in\{1,2,5\}}{\prod}\!\!\!\!\Gamma[\underline\Delta_a\!-\!h\!+\!1]\,\Gamma[\underline\Delta_{12}]\Gamma[\underline\Delta_{125}]\Gamma[2h\!-\!\underline\Delta_{12}]\Gamma[h](\underline\Delta_{12}\!-\!\underline\Delta_6)(2h\!-\!\underline\Delta_{126})}.
\end{split}
\end{equation}
\item[\textbf{(G)}] For the diagram
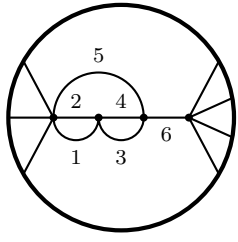
\begin{center}
\begin{tikzpicture}
\begin{scope}[scale=.6]
\draw [black,ultra thick] (0,0) circle [radius=2.5];
\draw [black,thick] (-1.5,0) -- (150:2.5);
\draw [black,thick] (-1.5,0) -- (180:2.5);
\draw [black,thick] (-1.5,0) -- (210:2.5);
\draw [black,thick] (1.5,0) -- (30:2.5);
\draw [black,thick] (1.5,0) -- (10:2.5);
\draw [black,thick] (1.5,0) -- (-10:2.5);
\draw [black,thick] (1.5,0) -- (-30:2.5);
\draw [black,thick] (-1.5,0) -- (1.5,0);
\draw [black,thick] (-.5,0) arc [start angle=-180,end angle=0,radius=.5] arc [start angle=0,end angle=180,radius=1] arc [start angle=180,end angle=360,radius=.5];
\fill [black] (-1.5,0) circle [radius=2.5pt];
\fill [black] (-.5,0) circle [radius=2.5pt];
\fill [black] (.5,0) circle [radius=2.5pt];
\fill [black] (1.5,0) circle [radius=2.5pt];
\node [anchor=north] at (-1,-.5) {\scriptsize $1$};
\node [anchor=south] at (-1,0) {\scriptsize $2$};
\node [anchor=north] at (0,-.5) {\scriptsize $3$};
\node [anchor=south] at (0,0) {\scriptsize $4$};
\node [anchor=south] at (-.5,1) {\scriptsize $5$};
\node [anchor=north] at (1,0) {\scriptsize $6$};
\end{scope}
\end{tikzpicture}
\end{center}
the pre-amplitude has poles
\begin{equation}
\begin{split}\checked{}
&\Gamma[{\textstyle h\pm c_1}]\,
\Gamma[{\textstyle h\pm c_2}]\,
\Gamma[{\textstyle h\pm c_3}]\,
\Gamma[{\textstyle h\pm c_4}]\,
\Gamma[{\textstyle h\pm c_5}]\,
\Gamma[{\textstyle\frac{2h\pm c_1\pm c_2\pm c_3\pm c_4}{2}}]\\
&\Gamma[{\textstyle\frac{2h\pm c_3\pm c_4\pm c_5\pm c_6}{2}}]\,
\Gamma[{\textstyle\frac{2h\pm c_1\pm c_2\pm c_5\pm c_6}{2}}]\,
\Gamma[{\textstyle\frac{\Delta_{A_0}-h\pm c_6}{2}}]\,
\Gamma[{\textstyle\frac{\Delta_{A_1}-h\pm c_6}{2}}]\,
\Gamma[{\textstyle\frac{h\pm c_6-S}{2}}].
\end{split}
\end{equation}
In addition, the poles ruled out by compositeness are
\begin{equation}
\begin{split}\checked{}
{\color{ForestGreen}
\cancel{\Gamma}[{\textstyle\frac{\Delta_{A_0}+h\pm c_1\pm c_2\pm c_5}{2}}]\,
\cancel{\Gamma}[{\textstyle\frac{\Delta_{A_0}+h\pm c_3\pm c_4\pm c_5}{2}}]\,
\cancel{\Gamma}[{\textstyle\frac{3h\pm c_1\pm c_2\pm c_5-S}{2}}]\,
\cancel{\Gamma}[{\textstyle\frac{3h\pm c_3\pm c_4\pm c_5-S}{2}}]}.
\end{split}
\end{equation}
This diagram again allows two non-minimal cuts, hence two families of non-minimal poles $\Gamma[\frac{\underline\Delta_{125}-S}{2}]$ and $\Gamma[\frac{\underline\Delta_{345}-S}{2}]$. The Mellin amplitude still enjoys the invariance under exchange of labels $(12)\leftrightarrow(34)$, and so we only check the leading pole of the first family, whose residue is
\begin{equation}
\begin{split}
\residue{S=\underline\Delta_{125}}\!\mathcal{M}\!=
&\frac{1}{2^7\pi^{2h}}\prod_i\frac{\mathcal{C}_{\Delta_i}}{\Gamma[\Delta_i]}\int\prod_{a=3}^4\frac{[\mathrm{d}c_a]_{\underline\Delta_a}}{\Gamma[\pm c_a]}\,\Gamma[{\textstyle\frac{2h\pm c_3\pm c_4-\underline\Delta_{12}}{2}}]\,\Gamma[{\textstyle\frac{\underline\Delta_{34}\pm c_3\pm c_4}{2}}]\\
&\times\frac{\Gamma[\underline\Delta_1]\,\Gamma[\underline\Delta_2]\,\Gamma[\underline\Delta_5]\,\Gamma[\frac{\underline\Delta_{125}+\Delta_{A_0}}{2}-h]\,\Gamma[\frac{\underline\Delta_{125}+\Delta_{A_1}}{2}-h]}{\!\!\!\!\!\!\underset{a\in\{1,2,5\}}{\prod}\!\!\!\!\Gamma[\underline\Delta_a\!-\!h\!+\!1]\,\Gamma[\underline\Delta_{12}]\Gamma[\underline\Delta_{125}]\Gamma[2h\!-\!\underline\Delta_{12}]\Gamma[h](\underline\Delta_{12}\!-\!\underline\Delta_6)(2h\!-\!\underline\Delta_{126})}.
\end{split}
\end{equation}
This result turns out to be exactly the same as that of diagram (F).
\item[\textbf{(H)}] Finally let us look at a 4-point diagram
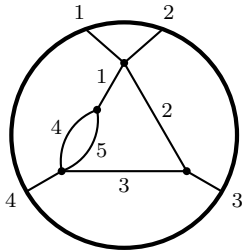
\begin{center}
\begin{tikzpicture}
\begin{scope}[scale=.6]
\draw [black,ultra thick] (0,0) circle [radius=2.5];
\draw [black,thick] (210:1.6) -- (-30:1.6) -- (90:1.6) -- ++(240:1.2);
\draw [black,thick] (70:2.5) -- (90:1.6) -- (110:2.5);
\draw [black,thick] (-30:1.6) -- (-30:2.5);
\draw [black,thick] (210:1.6) -- (210:2.5);
\draw [black,thick] ($(90:1.6)+(240:1.2)$) .. controls ($(90:1.6)+(240:1.2)+(280:.6)$) and ($(210:1.6)+(20:.6)$) .. (210:1.6);
\draw [black,thick] ($(90:1.6)+(240:1.2)$) .. controls ($(90:1.6)+(240:1.2)+(200:.6)$) and ($(210:1.6)+(100:.6)$) .. (210:1.6);
\fill [black] (210:1.6) circle [radius=2.5pt];
\fill [black] (-30:1.6) circle [radius=2.5pt];
\fill [black] (90:1.6) circle [radius=2.5pt];
\fill [black] ($(90:1.6)+(240:1.2)$) circle [radius=2.5pt];
\node [anchor=center] at (110:2.9) {\scriptsize $1$};
\node [anchor=center] at (70:2.9) {\scriptsize $2$};
\node [anchor=center] at (-30:2.9) {\scriptsize $3$};
\node [anchor=center] at (210:2.9) {\scriptsize $4$};
\node [anchor=center] at (-.5,1.3) {\scriptsize $1$};
\node [anchor=center] at (30:1.1) {\scriptsize $2$};
\node [anchor=center] at (0,-1.1) {\scriptsize $3$};
\node [anchor=south] at (-1.5,-.2) {\scriptsize $4$};
\node [anchor=south] at (-.5,-.7) {\scriptsize $5$};
\end{scope}
\end{tikzpicture}
\end{center}
which contains a generalized bubble (in fact diagram (A)) as a proper sub-diagram. For this diagram we can directly apply the result \eqref{eq:preMtriangle4ptpolesall} for the 4-point triangle diagram with a quartic vertex constructed in Figure \ref{fig:triangle} (B). Together with the effective spectrum function corrponding to the generalized bubble sub-diagram (formed by propagator 1,4,5), we observe that its pre-amplitude has poles
\begin{equation}
\begin{split}\checked{}
&\Gamma[{\textstyle\frac{\Delta_{12}\pm c_1\pm c_2}{2}}]\,
\Gamma[{\textstyle\frac{\Delta_3\pm c_2\pm c_3}{2}}]\,
\Gamma[{\textstyle\frac{h\pm c_1\pm c_4\pm c_5}{2}}]\,
\Gamma[{\textstyle\frac{\Delta_4\pm c_1\pm c_3}{2}}]\,
\Gamma[{\textstyle\frac{\Sigma\Delta}{2}-h}]\\&
\Gamma[{\textstyle\frac{2h\pm c_1\pm c_2-S}{2}}]\,
\Gamma[{\textstyle\frac{2h\pm c_2\pm c_3-\Delta_3}{2}}]\,
\Gamma[{\textstyle\frac{2h\pm c_1\pm c_3-\Delta_4}{2}}].
\end{split}
\end{equation}
In the above, apart from $\Gamma[{\textstyle\frac{h\pm c_1\pm c_4\pm c_5}{2}}]$ all the rest families directly descend from \eqref{eq:preMtriangle4ptpolesall}, but now as viewed by our diagrammatic rules the family $\Gamma[{\textstyle\frac{\Delta_4\pm c_1\pm c_3}{2}}]$ arises from the contraction of the bubble instead of from an original bulk vertex. Again there are poles absent due to compositeness
\begin{equation}
{\color{ForestGreen}\Gamma[h\pm c_1]\,
\Gamma[h\pm c_4]\,
\Gamma[h\pm c_5]\,
\Gamma[{\textstyle\frac{\Delta_4+h\pm c_3\pm c_4\pm c_5}{2}}]\,
\Gamma[{\textstyle\frac{3h\pm c_3\pm c_4\pm c_5-\Delta_4}{2}}]\,
\Gamma[{\textstyle\frac{3h\pm c_2\pm c_4\pm c_5-S}{2}}]}.
\end{equation}
This diagram allows one non-minimal cut in the $S$ channel, in correspondence to the non-minimal poles $\Gamma[{\textstyle\frac{\underline\Delta_{245}-S}{2}}]$. Explicit computation shows that these are indeed present in the Mellin amplitude, and the residue at the leading pole is
\begin{equation}\label{app:eq:resgbsubdiagram}
\begin{split}
\residue{S=\frac{\underline\Delta_{245}}{2}}\!\!\mathcal{M}\!=&-\frac{1}{2^6\pi^h}\prod_{i=1}^4\frac{\mathcal{C}_{\Delta_i}}{\Gamma[\Delta_i]}\int\frac{[\mathrm{d}c_3]_{\underline\Delta_3}}{\Gamma[\pm c_3]}\\
&\frac{\Gamma[\underline\Delta_4]\,\Gamma[\underline\Delta_5]\,\Gamma[\frac{\underline\Delta_2\pm c_3\pm(\Delta_3-h)}{2}]\,\Gamma[\frac{\underline\Delta_{45}\pm c_3\pm(\Delta_4-h)}{2}]\,\Gamma[\frac{\underline\Delta_{245}+\Delta_{12}}{2}-h]}{\!\!\!\!\underset{a\in\{2,4,5\}}{\prod}\!\!\!\!\Gamma[\underline\Delta_a\!-\!h\!+\!1]\,\Gamma[\underline\Delta_{45}]\,\Gamma[\frac{\underline\Delta_{245}\pm(\Delta_3-\Delta_4)}{2}]\,\Gamma[\frac{\underline\Delta_{245}-\Delta_{34}}{2}\!+\!h]\,(\underline\Delta_{45}\!-\!\underline\Delta_1)\,(2h\!-\!\underline\Delta_{145})}.
\end{split}
\end{equation}
\end{enumerate}

\newpage

\section{Poles of $\mathcal{M}$}\label{app:sec:resultpolesofamplitudes}

\temp{Need serious checks.}

In this section we collect results on the pole structure of the Mellin amplitude $\mathcal{M}$ for various diagrams.

\subsection{4-point bubble}

In the integrand $\mathcal{N}M_0K$ for the construction of the 4-point bubble in Figure \ref{fig:bubble4pt}, in addition to the poles already listed in \eqref{eq:bubblelabeledintegrand} we further have
\begin{equation}\label{app:eq:bubble4tremainpoles}
\begin{split}\checked{}
&\Gamma_{13}[{\textstyle\frac{\Delta_{12}-c_1-c_2}{2}}]\,
\Gamma_{14}[{\textstyle\frac{\Delta_{12}+c_1-c_2}{2}}]\,
\Gamma_{15}[{\textstyle\frac{\Delta_{34}-c_1+c_2}{2}}]\,
\Gamma_{16}[{\textstyle\frac{\Delta_{34}+c_1+c_2}{2}}]\\
&\gamma_{17}[(\underline\Delta_1-h)+c_1]\,
\gamma_{18}[(\underline\Delta_1-h)-c_1]\,
\gamma_{19}[(\underline\Delta_2-h)+c_2]\,
\gamma_{20}[(\underline\Delta_2-h)-c_2].
\end{split}
\end{equation}
Recall that here the notation $\gamma_a[E]$ refers to a single pole at $E=0$ in the family. The poles of $\mathcal{M}$ are
\begin{equation}\label{app:eq:bubble4ptcalM}\checked{}
\underset{\{10,11,12\}}{\Gamma[{\textstyle\frac{\Sigma\Delta}{2}-h}]}\,
\underset{\substack{\{1,3,4,17,19\}\\\{1,6,8,17,20\}\\\{2,3,4,18,19\}\\\{2,6,8,18,20\}}}{\Gamma[{\textstyle\frac{\underline\Delta_{12}-S}{2}}]}\,
\underset{\substack{\{13,17,19\}\\\{14,18,19\}\\\{1,6,10,11,17,20\}\\\{2,6,10,11,18,20\}}}{\Gamma[{\textstyle\frac{\underline\Delta_{12}+\Delta_{12}}{2}-h}]}\,
\underset{\substack{\{15,17,20\}\\\{16,18,20\}\\\{1,3,10,12,17,19\}\\\{2,3,10,12,18,19\}}}{\Gamma[{\textstyle\frac{\underline\Delta_{12}+\Delta_{34}}{2}-h}]}.
\end{equation}
In addition, there are poles
\begin{equation}\checked{}
\underset{\substack{\{1, 3, 6, 10, 17\}\\\{2, 3, 6, 10, 18\}}}{\Gamma[\underline\Delta_1]}\,
\underset{\substack{\{3, 5, 19\}\\\{6, 9, 20\}}}{\Gamma[\underline\Delta_2]}
\end{equation}
if $h$ is not an integer (i.e., not in even boundary spacetime dimensions).

\subsection{3-point triangle}

In the integrand $\mathcal{N}M_0K$ for the construction of the 3-point triangle in Figure \ref{fig:triangle} (A), in addition to the poles already listed in \eqref{eq:labeledpolesintegrandtriangle3pt} we further have
\begin{equation}\label{app:eq:triangl3ptremainpoles}
\begin{split}\checked{}
&\Gamma_{15}[{\textstyle\frac{\Delta_1-c_2-c_3}{2}}]\,
\Gamma_{16}[{\textstyle\frac{\Delta_1-c_2+c_3}{2}}]\,
\Gamma_{17}[{\textstyle\frac{\Delta_2-c_1-c_3}{2}}]\,
\Gamma_{18}[{\textstyle\frac{\Delta_2+c_1-c_3}{2}}]\,
\Gamma_{19}[{\textstyle\frac{\Delta_2-c_1+c_3}{2}}]\,
\Gamma_{20}[{\textstyle\frac{\Delta_2+c_1+c_3}{2}}]\\
&\Gamma_{21}[{\textstyle\frac{\Delta_3-c_1+c_2}{2}}]\,
\Gamma_{22}[{\textstyle\frac{\Delta_3+c_1+c_2}{2}}]\,
\gamma_{23}[(\underline\Delta_1-h)+c_1]\,
\gamma_{24}[(\underline\Delta_1-h)-c_1]\\
&\gamma_{25}[(\underline\Delta_2-h)+c_2]\,
\gamma_{26}[(\underline\Delta_2-h)-c_2]\,
\gamma_{27}[(\underline\Delta_3-h)+c_3]\,
\gamma_{28}[(\underline\Delta_3-h)-c_3].
\end{split}
\end{equation}
The poles of $\mathcal{M}$ are
\begin{equation}\label{app:eq:triangle3ptcalMpoles}\checked{}
\begin{split}
&\underset{\{9,10,14\}}{\Gamma[{\textstyle\frac{\Sigma\Delta}{2}-h}]}\,
\underset{\substack{\{21,23,26\}\\\{22,24,26\}\\\{1,3,9,14,23,25\}\\\{2,3,9,14,24,25\}}}{\Gamma[{\textstyle\frac{\underline\Delta_{12}+\Delta_3}{2}-h}]}\,
\underset{\substack{\{15,25,27\}\\\{16,25,28\}\\\{4,6,9,10,26,27\}\\\{5,6,9,10,26,28\}}}{\Gamma[{\textstyle\frac{\underline\Delta_{23}+\Delta_1}{2}-h}]}\,
\underset{\substack{\{17,13,27\}\\\{18,24,27\}\\\{19,23,28\}\\\{20,24,28\}}}{\Gamma[{\textstyle\frac{\underline\Delta_{31}+\Delta_2}{2}-h}]}\\
&\underset{\substack{\{1,3,11,23,25\}\\\{1,6,12,23,26\}\\\{2,3,11,24,25\}\\\{2,6,12,24,26\}}}{\Gamma[{\textstyle\frac{\underline\Delta_{12}-\Delta_3}{2}}]}\,
\underset{\substack{\{3,4,7,25,27\}\\\{3,5,7,25,28\}\\\{4,6,8,26,27\}\\\{5,6,8,26,28\}}}{\Gamma[{\textstyle\frac{\underline\Delta_{23}-\Delta_1}{2}}]}\,
\underset{\substack{\{1,3,4,6,9,23,27\}\\\{1,3,5,6,9,23,28\}\\\{2,3,4,6,9,24,27\}\\\{2,3,5,6,9,24,28\}}}{\Gamma[{\textstyle\frac{\underline\Delta_{31}-\Delta_2}{2}}]}.
\end{split}
\end{equation}
All the poles above are simple poles. Apart from these, when $h$ is not an integer, there are also three families of poles
\begin{equation}\checked{}
{\color{Magenta}\underset{\substack{\{1,3,4,6,9,11,19,21,23\}\\\{1,3,4,6,9,12,14,19,23\}\\\{1,3,5,6,9,11,17,21,23\}\\\{1,3,5,6,9,12,17,21,23\}\\\{2,3,4,6,9,11,14,22,24\}\\\{2,3,4,6,9,12,14,20,24\}\\\{2,3,5,6,9,11,18,22,24\}\\\{2,3,5,6,9,12,14,18,24\}}}{\Gamma[\underline\Delta_{1}]}}\,
\underset{\substack{\{1,6,12,22,26\}\\\{2,6,12,21,26\}\\\{3,4,7,16,25\}\\\{3,5,7,15,25\}}}{\Gamma[\underline\Delta_{2}]}\,
{\color{Magenta}\underset{\substack{\{1,3,4,6,7,9,10,18,27\}\\\{1,3,4,6,8,9,15,18,27\}\\\{1,3,5,6,7,9,10,20,28\}\\\{1,3,5,6,8,9,16,20,28\}\\\{2,3,4,6,7,9,10,17,27\}\\\{2,3,4,6,8,9,15,17,27\}\\\{2,3,5,6,7,9,10,19,28\}\\\{2,3,5,6,8,9,16,19,28\}}}{\Gamma[\underline\Delta_{3}]}}.
\end{equation}
Here for the two families marked magenta, each pinching pattern indicated are actually the overlap of two simplices, hence the poles by themselves are double poles. However, the order is reduced by one by an extra zero from the vertex correction $C_{A_2}$.


\subsection{4-point triangle with a quartic vertex}

In the integrand $\mathcal{N}M_0K$ for the construction of the 4-point triangle with a quartic vertex in Figure \ref{fig:triangle} (B), in addition to the poles already listed in \eqref{eq:labeledpolesintegrandtriangle4pt} we further have
\begin{equation}\label{eq:triangle4ptremainpoles}
\checked{}
\begin{split}
&\Gamma_{17}[{\textstyle\frac{\Delta_3-c_2+c_3}{2}}]\,
\Gamma_{18}[{\textstyle\frac{\Delta_3+c_2+c_3}{2}}]\,
\Gamma_{19}[{\textstyle\frac{\Delta_4-c_1-c_3}{2}}]\,
\Gamma_{20}[{\textstyle\frac{\Delta_4+c_1-c_3}{2}}]\,
\Gamma_{21}[{\textstyle\frac{\Delta_{12}-c_1-c_2}{2}}]\,
\Gamma_{22}[{\textstyle\frac{\Delta_{12}+c_1-c_2}{2}}]\\
&\Gamma_{23}[{\textstyle\frac{\Delta_{12}-c_1+c_2}{2}}]\,
\Gamma_{24}[{\textstyle\frac{\Delta_{12}+c_1+c_2}{2}}]\,
\gamma_{25}[(\underline\Delta_1-h)+c_1]\,
\gamma_{26}[(\underline\Delta_1-h)-c_1]\\
&\gamma_{27}[(\underline\Delta_2-h)+c_2]\,
\gamma_{28}[(\underline\Delta_2-h)-c_2]\,
\gamma_{29}[(\underline\Delta_3-h)+c_3]\,
\gamma_{30}[(\underline\Delta_3-h)-c_3].
\end{split}
\end{equation}
First of all, there is one family of poles in $S$
\begin{equation}\label{app:eq:triangle4ptQpolesS}\checked{}
\underset{\substack{\{1,3,5,6,7,8,25,27\}\\\{1,4,5,6,7,8,25,28\}\\\{2,3,5,6,7,8,26,27\}\\\{2,4,5,6,7,8,26,28\}}}{\Gamma[{\textstyle\frac{\underline\Delta_{12}-S}{2}}]}.
\end{equation}
Apart from these, there are also poles
\begin{equation}\label{app:eq:triangle4ptQpolesall}\checked{}
\begin{split}
&\underset{\{8,11,14,16\}}{\Gamma[{\textstyle\frac{\Sigma\Delta}{2}-h}]}\,
\underset{\substack{\{3,7,10,27,30\}\\\{4,7,10,28,30\}\\\{3,5,6,9,27,29\}\\\{4,5,6,9,28,29\}}}{\Gamma[{\textstyle\frac{\underline\Delta_{23}-\Delta_3}{2}}]}\,
\underset{\substack{\{1,5,12,25,29\}\\\{2,5,12,26,29\}\\\{1,6,7,13,25,30\}\\\{2,6,7,13,26,30\}}}{\Gamma[{\textstyle\frac{\underline\Delta_{13}-\Delta_4}{2}}]}\,
\underset{\substack{\{21,25,27\}\\\{22,26,27\}\\\{23,25,28\}\\\{24,26,28\}}}{\Gamma[{\textstyle\frac{\underline\Delta_{12}+\Delta_{12}}{2}-h}]}\,
\underset{\substack{\{17,27,30\}\\\{18,28,30\}\\\{3,5,8,11,27,29\}\\\{4,5,8,11,28,29\}}}{\Gamma[{\textstyle\frac{\underline\Delta_{23}+\Delta_3}{2}-h}]}\,
\underset{\substack{\{19,25,29\}\\\{20,26,29\}\\\{1,7,8,14,25,30\}\\\{2,7,8,14,26,30\}}}{\Gamma[{\textstyle\frac{\underline\Delta_{13}+\Delta_4}{2}-h}]}\\
&\underset{\substack{\{1,5,12,17,25,27\}\\\{1,5,12,18,25,28\}\\\{2,5,12,17,26,27\}\\\{2,5,12,18,26,28\}}}{\Gamma[{\textstyle\frac{\underline\Delta_{12}+\Delta_3-\Delta_4}{2}}]}\,
\underset{\substack{\{3,7,10,19,25,27\}\\\{3,7,10,20,26,27\}\\\{4,7,10,19,25,28\}\\\{4,7,10,20,26,28\}}}{\Gamma[{\textstyle\frac{\underline\Delta_{12}-\Delta_3+\Delta_4}{2}}]}\,
\underset{\substack{\{17,19,25,27\}\\\{17,20,26,27\}\\\{18,19,25,28\}\\\{18,20,26,28\}\\\{1,3,5,7,8,11,14,25,27\}\\\{1,4,5,7,8,11,14,25,28\}\\\{2,3,5,7,8,11,14,26,27\}\\\{2,4,5,7,8,11,14,26,28\}}}{\Gamma[{\textstyle\frac{\underline\Delta_{12}+\Delta_{34}}{2}-h}]}
\end{split}
\end{equation}
Then again there are also three families when $h$ is not an integer
\begin{equation}\checked{}
\underset{\substack{\{1,6,7,13,19,25\}\\\{2,6,7,13,20,26\}\\\{1,5,7,8,12,14,25\}\\\{2,5,7,8,12,14,26\}}}{\Gamma[\underline\Delta_1]}\,
\underset{\substack{\{3,5,6,9,17,27\}\\\{4,5,6,9,18,28\}\\\{3,5,7,8,10,11,27\}\\\{4,5,7,8,10,11,28\}}}{\Gamma[\underline\Delta_2]}\,
\underset{\substack{\{1,5,12,20,29\}\\\{2,5,12,19,29\}\\\{3,7,10,18,30\}\\\{4,7,10,17,30\}}}{\Gamma[\underline\Delta_3]}
\end{equation}

\subsection{4-point triangle with cubic vertices}

In the integrand $\mathcal{N}M_0K$ for the construction of the 4-point triangle with only cubic vertices in Figure \ref{fig:triangle4ptC}, in addition to the poles already listed in \eqref{eq:labeledpolesintegrandtriangle4ptC} we further have
\begin{equation}\label{app:eq:triangle4ptCremainpoles}
\begin{split}\checked{}
&\Gamma_{18}[{\textstyle\frac{h-c_1-c_3-c_4}{2}}]\,
\Gamma_{19}[{\textstyle\frac{h+c_1-c_3-c_4}{2}}]\,
\Gamma_{20}[{\textstyle\frac{h-c_1+c_3-c_4}{2}}]\,
\Gamma_{21}[{\textstyle\frac{h+c_1+c_3-c_4}{2}}]\,
\Gamma_{22}[{\textstyle\frac{h-c_1-c_3+c_4}{2}}]\\
&\Gamma_{23}[{\textstyle\frac{h+c_1-c_3+c_4}{2}}]\,
\Gamma_{24}[{\textstyle\frac{h-c_1+c_3+c_4}{2}}]\,
\Gamma_{25}[{\textstyle\frac{h+c_1+c_3+c_4}{2}}]\,
\Gamma_{26}[{\textstyle\frac{\Delta_3+c_2-c_3}{2}}]\,
\Gamma_{27}[{\textstyle\frac{\Delta_3+c_2+c_3}{2}}]\\
&\Gamma_{28}[{\textstyle\frac{\Delta_4-c_2-c_4}{2}}]\,
\Gamma_{29}[{\textstyle\frac{\Delta_4-c_2+c_4}{2}}]\,
\Gamma_{30}[{\textstyle\frac{\Delta_{12}-h-c_1}{2}}]\,
\Gamma_{31}[{\textstyle\frac{\Delta_{12}-h+c_1}{2}}]\,
\gamma_{32}[(\underline\Delta_1-h)+c_1]\\
&\gamma_{33}[(\underline\Delta_1-h)-c_1]\,
\gamma_{34}[(\underline\Delta_2-h)+c_2]\,
\gamma_{35}[(\underline\Delta_2-h)-c_2]\,
\gamma_{36}[(\underline\Delta_3-h)+c_3]\\
&\gamma_{37}[(\underline\Delta_3-h)-c_3]\,
\gamma_{38}[(\underline\Delta_4-h)+c_4]\,
\gamma_{39}[(\underline\Delta_4-h)-c_4].
\end{split}
\end{equation}
First of all, there is only one family of poles in $S$
\begin{equation}\label{app:eq:triangle4ptCpoleS}\checked{}
\underset{\substack{\{1,3,32\}\\\{2,3,33\}}}{\Gamma[{\textstyle\frac{\underline\Delta_{1}-S}{2}}]}.
\end{equation}
Other poles of $\mathcal{M}$ are
\begin{equation}\label{app:eq:triangle4ptCpolesall}\checked{}
\begin{split}
&\underset{\substack{\{1,10,13,16,31\}\\\{2,10,13,16,30\}}}{\Gamma[{\textstyle\frac{\Sigma\Delta}{2}-h}]}\,
\underset{\substack{\{4,9,12,35,36\}\\\{5,9,12,35,37\}\\\{3,4,6,11,34,36\}\\\{3,5,6,11,34,37\}}}{\Gamma[{\textstyle\frac{\underline\Delta_{23}-\Delta_3}{2}}]}\,
\underset{\substack{\{6,7,14,34,38\}\\\{6,8,14,34,39\}\\\{3,7,9,15,35,38\}\\\{3,8,9,15,35,39\}}}{\Gamma[{\textstyle\frac{\underline\Delta_{24}-\Delta_4}{2}}]}\,
\underset{\substack{\{20,32\}\\\{31,33\}}}{\Gamma[{\textstyle\frac{\underline\Delta_1+\Delta_{12}}{2}-h}]}\,
\underset{\substack{\{1, 10, 13, 16, 32\}\\\{2, 10, 13, 16, 33\}}}{\Gamma[{\textstyle\frac{\underline\Delta_1+\Delta_{34}}{2}-h}]}\\
&\underset{\substack{\{26,35,36\}\\\{27,35,37\}\\\{4,6,10,13,34,36\}\\\{5,6,10,13,34,37\}}}{\Gamma[{\textstyle\frac{\underline\Delta_{23}+\Delta_3}{2}-h}]}\,
\underset{\substack{\{28,34,38\}\\\{29,34,39\}\\\{7,9,10,16,35,38\}\\\{8,9,10,16,35,39\}}}{\Gamma[{\textstyle\frac{\underline\Delta_{24}+\Delta_4}{2}-h}]}\,
\underset{\substack{\{18,31,36,38\}\\\{19,30,36,38\}\\\{20,31,37,38\}\\\{21,30,37,38\}\\\{22,31,36,39\}\\\{23,30,36,39\}\\\{24,31,37,39\}\\\{25,30,37,39\}}}{\Gamma[{\textstyle\frac{\underline\Delta_{34}+\Delta_{12}}{2}-h}]}\,
\underset{\substack{\{26, 28, 36, 38\}\\\{26, 29, 36, 39\}\\\{27, 28, 37, 38\}\\\{27, 29, 37, 39\}}}{\Gamma[{\textstyle\frac{\underline\Delta_{34}+\Delta_{34}}{2}-h}]}\,
\underset{\substack{\{18,32,36,38\}\\\{19,33,36,38\}\\\{20,32,37,38\}\\\{21,33,37,38\}\\\{22,32,36,39\}\\\{23,33,36,39\}\\\{24,32,37,39\}\\\{25,33,37,39\}}}{\Gamma[{\textstyle\frac{\underline\Delta_{134}}{2}-h}]}.
\end{split}
\end{equation}
In addition, when $h$ is not an integer we also have
\begin{equation}\checked{}
\underset{\substack{\{4, 9, 12, 27, 35\}\\\{5, 9, 12, 26, 35\}\\\{6, 7, 14, 29, 34\}\\\{6, 8, 14, 28, 34\}}}{\Gamma[\Sigma_2]}.
\end{equation}

\subsection{4-point box}

In the integrand $\mathcal{N}M_0K$ for the construction of the 4-point box in Figure \ref{fig:box4pt}, in addition to the poles already listed in \eqref{eq:labeledpolesintgrandbox} we further have
\begin{equation}\label{app:eq:box4ptremainpoles}
\begin{split}\checked{}
&\Gamma_{25}[{\textstyle\frac{\Delta_1-c_1-c_2}{2}}]\,
\Gamma_{26}[{\textstyle\frac{\Delta_1-c_1+c_2}{2}}]\,
\Gamma_{27}[{\textstyle\frac{\Delta_2-c_2-c_3}{2}}]\,
\Gamma_{28}[{\textstyle\frac{\Delta_2+c_2-c_3}{2}}]\,
\Gamma_{29}[{\textstyle\frac{\Delta_2-c_2+c_3}{2}}]\\
&\Gamma_{30}[{\textstyle\frac{\Delta_2+c_2+c_3}{2}}]\,
\Gamma_{31}[{\textstyle\frac{\Delta_3-c_3-c_4}{2}}]\,
\Gamma_{32}[{\textstyle\frac{\Delta_3+c_3-c_4}{2}}]\,
\Gamma_{33}[{\textstyle\frac{\Delta_3-c_3+c_4}{2}}]\,
\Gamma_{34}[{\textstyle\frac{\Delta_3+c_3+c_4}{2}}]\\
&\Gamma_{35}[{\textstyle\frac{\Delta_4+c_1-c_4}{2}}]\,
\Gamma_{36}[{\textstyle\frac{\Delta_4+c_1+c_4}{2}}]\,
\gamma_{37}[(\underline\Delta_1-h)+c_1]\,
\gamma_{38}[(\underline\Delta_1-h)-c_1]\\
&\gamma_{39}[(\underline\Delta_2-h)+c_2]\,
\gamma_{40}[(\underline\Delta_2-h)-c_2]\,
\gamma_{41}[(\underline\Delta_3-h)+c_3]\\
&\gamma_{42}[(\underline\Delta_3-h)-c_3]\,
\gamma_{43}[(\underline\Delta_4-h)+c_4]\,
\gamma_{44}[(\underline\Delta_4-h)-c_4].
\end{split}
\end{equation}
Firstly the poles of $\mathcal{M}$ in the Mandelstam variables are
\begin{equation}\label{app:eq:box4ptpoleS}\checked{}
\underset{\substack{\{1,4,8,9,37,41\}\\\{1,5,8,9,37,42\}\\\{4,11,13,14,38,41\}\\\{5,11,13,14,38,42\}}}{\Gamma[{\textstyle\frac{\underline\Delta_{13}-S}{2}}]}\,
\underset{\substack{\{2,6,8,10,11,12,39,43\}\\\{2,7,8,10,11,12,39,44\}\\\{3,6,8,10,11,12,40,43\}\\\{3,7,8,10,11,12,40,44\}}}{\Gamma[{\textstyle\frac{\underline\Delta_{24}-T}{2}}]}.
\end{equation}
Other poles are
\begin{equation}\label{app:eq:box4ptpolesall}\checked{}
\begin{split}
&
\underset{\substack{\{2,8,9,15,37,39\}\\\{3,8,9,15,37,40\}\\\{2,10,11,14,16,38,39\}\\\{3,10,11,14,16,38,40\}}}{\Gamma[{\textstyle\frac{\underline\Delta_{12}-\Delta_1}{2}}]}\,
\underset{\substack{\{2,4,8,9,11,12,17,39,41\}\\\{2,5,8,9,11,12,17,39,42\}\\\{3,4,8,9,11,12,17,40,41\}\\\{3,5,8,9,11,12,17,40,42\}}}{\Gamma[{\textstyle\frac{\underline\Delta_{23}-\Delta_2}{2}}]}\,
\underset{\substack{\{4,6,8,11,12,13,19,41,43\}\\\{4,7,8,11,12,13,19,41,44\}\\\{5,6,8,11,12,13,19,42,43\}\\\{5,7,8,11,12,13,19,42,44\}}}{\Gamma[{\textstyle\frac{\underline\Delta_{34}-\Delta_3}{2}}]}\,
\underset{\substack{\{6,11,13,22,38,43\}\\\{7,11,13,22,38,44\}\\\{1,6,8,10,21,37,43\}\\\{1,7,8,10,21,37,44\}}}{\Gamma[{\textstyle\frac{\underline\Delta_{41}-\Delta_4}{2}}]}\\
&\underset{\substack{\{25,37,39\}\\\{26,37,40\}\\\{2,11,12,17,18,38,39\}\\\{3,11,12,17,18,38,40\}}}{\Gamma[{\textstyle\frac{\underline\Delta_{12}+\Delta_1}{2}-h}]}\,
\underset{\substack{\{27,39,41\}\\\{28,40,41\}\\\{29,39,42\}\\\{30,40,42\}}}{\Gamma[{\textstyle\frac{\underline\Delta_{23}+\Delta_2}{2}-h}]}\,
\underset{\substack{\{31,41,43\}\\\{32,42,43\}\\\{33,41,44\}\\\{34,42,44\}}}{\Gamma[{\textstyle\frac{\underline\Delta_{34}+\Delta_3}{2}-h}]}\,
\underset{\substack{\{35,38,43\}\\\{36,38,44\}\\\{6,8,12,19,24,37,43\}\\\{7,8,12,19,24,37,44\}}}{\Gamma[{\textstyle\frac{\underline\Delta_{41}+\Delta_4}{2}-h}]}\\
&\underset{\substack{\{25,28,37,41\}\\\{25,30,37,42\}\\\{26,27,37,41\}\\\{26,29,37,42\}\\\{2,11,12,17,18,28,38,41\}\\\{2,11,12,17,18,30,38,42\}\\\{3,11,12,17,18,27,38,41\}\\\{3,11,12,17,18,29,38,42\}}}{\Gamma[\textstyle\frac{\underline\Delta_{13}+\Delta_{12}}{2}-h]}\,
\underset{\substack{\{27,32,39,43\}\\\{27,34,39,44\}\\\{28,32,40,43\}\\\{28,34,40,44\}\\\{29,31,39,43\}\\\{29,33,39,44\}\\\{30,31,40,43\}\\\{30,33,40,44\}}}{\Gamma[{\textstyle\frac{\underline\Delta_{24}+\Delta_{23}}{2}-h}]}\,
\underset{\substack{\{31,36,38,41\}\\\{32,36,38,42\}\\\{33,35,38,41\}\\\{34,35,38,42\}\\\{6,8,12,19,24,33,37,41\}\\\{6,8,12,19,24,34,37,42\}\\\{7,8,12,19,24,31,37,41\}\\\{7,8,12,19,24,32,37,42\}}}{\Gamma[{\textstyle\frac{\underline\Delta_{13}+\Delta_{34}}{2}-h}]}\,
\underset{\substack{\{25,35,39,43\}\\\{25,36,39,44\}\\\{26,35,40,43\}\\\{26,36,40,44\}\\\{2,6,8,11,12,17,18,19,24,39,43\}\\\{2,7,8,11,12,17,18,19,24,39,44\}\\\{3,6,8,11,12,17,18,19,24,40,43\}\\\{3,7,8,11,12,17,18,19,24,40,44\}}}{\Gamma[{\textstyle\frac{\underline\Delta_{24}+\Delta_{41}}{2}-h}]}\\
&\underset{\{12,17,18,19,20,24\}}{\Gamma[{\textstyle\frac{\Sigma\Delta}{2}-h}]}.
\end{split}
\end{equation}

\newpage

\section{Non-planar Diagrams at Two Loops}\label{app:sec:nonplanar2loop}

In this appendix we collect some results on the simplest 4-point non-planar diagrams, at two loops. The main purpose is of two folds. One is to further illustrate the new features associated to the choice of Mandelstam variables that occur in the construction of higher-loop diagrams, as was discussed in Section \ref{sec:Mandelstamindependence}. The other is to provide more non-trivial evidence for Conjecture \ref{sec:enhancedrules} and Conjecture \ref{sec:conjectureonamplitude}, regarding the pole structure of $M$ and $\mathcal{M}$ respectively.

\subsection{Example with a unique channel}

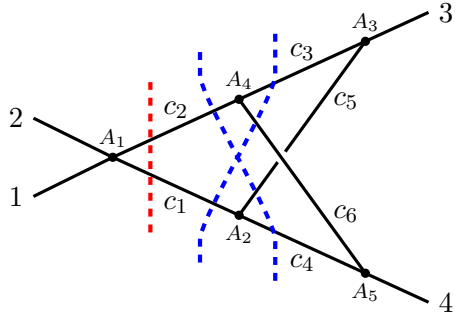
\begin{figure}[ht]
\captionsetup{margin=2em}
\begin{center}
\begin{tikzpicture}
\coordinate (tip) at (-1.9,0);
\draw [black,very thick] (170:3) -- (tip) -- (-170:3);
\draw [black,very thick] (40:3) -- (tip) -- (-40:3);
\draw [black,very thick] ($(tip)!.8!(40:3)$) -- ($(tip)!.4!(-40:3)$);
\fill [white] (.35,0) circle [radius=2.5pt];
\draw [black,very thick] ($(tip)!.8!(-40:3)$) -- ($(tip)!.4!(40:3)$);
\fill [black] (tip) circle [radius=1.75pt];
\node [anchor=south] at (tip) {\scriptsize $A_1$};
\fill [black] ($(tip)!.8!(40:3)$) circle [radius=1.75pt];
\node [anchor=south] at ($(tip)!.8!(40:3)$) {\scriptsize $A_3$};
\fill [black] ($(tip)!.4!(-40:3)$) circle [radius=1.75pt];
\node [anchor=north] at ($(tip)!.4!(-40:3)$) {\scriptsize $A_2$};
\fill [black] ($(tip)!.8!(-40:3)$) circle [radius=1.75pt];
\node [anchor=north] at ($(tip)!.8!(-40:3)$) {\scriptsize $A_5$};
\fill [black] ($(tip)!.4!(40:3)$) circle [radius=1.75pt];
\node [anchor=south] at ($(tip)!.4!(40:3)$) {\scriptsize $A_4$};
\node [anchor=east] at (-170:3) {$1$};
\node [anchor=east] at (170:3) {$2$};
\node [anchor=west] at (40:3) {$3$};
\node [anchor=west] at (-40:3) {$4$};
\node [anchor=north] at ($(tip)!.2!(-40:3)$) {$c_1$};
\node [anchor=south] at ($(tip)!.2!(40:3)$) {$c_2$};
\node [anchor=south] at ($(tip)!.6!(40:3)$) {$c_3$};
\node [anchor=north] at ($(tip)!.6!(-40:3)$) {$c_4$};
\node [anchor=center] at (1.2,.75) {$c_5$};
\node [anchor=center] at (1.2,-.75) {$c_6$};
\draw [red,ultra thick,dashed] (-1.4,-1) -- ++(0,2);
\begin{scope}[xshift=-.24cm]
\draw [blue,ultra thick,dashed] (-.5,1.4) -- (-.5,1.1) .. controls (-.5,.8) and (.5,-.8) .. (.5,-1.1) -- (.5,-1.7);
\end{scope}
\begin{scope}[xshift=-.24cm]
\draw [blue,ultra thick,dashed] (-.5,-1.4) -- (-.5,-1.1) .. controls (-.5,-.8) and (.5,.8) .. (.5,1.1) -- (.5,1.7);
\end{scope}
\end{tikzpicture}
\end{center}
\vspace{-1.5em}\caption{A 4-point two-loop non-planar diagram. The red dashed line indicates the unique minimal cut. The blue dashed lines indicate the two non-minimal cuts.}
\label{app:fig:nonplanarA}
\end{figure}

We first study the two-loop diagram shown in Figure \ref{app:fig:nonplanarA}.  Its Mellin (pre-)amplitude obviously depends only on the Mandelstam variable $S$,  but this channel allows a unique minimal cut as well as two non-minimal cuts, as indicated in the figure.

\begin{figure}[ht]
\captionsetup{margin=2em}
\begin{center}
\begin{tikzpicture}
\begin{scope}
\node [anchor=center] at (0,-3.5) {(A)};
\fill [black!15!white] (180:2) circle [radius=.4];
\fill [black!15!white] ($(120:2)+(210:.4)$) arc [start angle=210,end angle=390,radius=.4] -- ++(120:1) arc [start angle=30,end angle=210,radius=.4] -- cycle;
\fill [black!15!white] (60:2) circle [radius=.4];
\fill [black!15!white] (0:2) circle [radius=.4];
\draw [black,very thick,dotted] (0,0) circle [radius=2];
\draw [black,very thick] (-60:2) arc [start angle=-60,end angle=240,radius=2];
\coordinate (p1) at ($(180:2)+(210:1)$);
\node [anchor=center] at ($(0,0)!1.1!(p1)$) {$1$};
\coordinate (p2) at ($(180:2)+(150:1)$);
\node [anchor=center] at ($(0,0)!1.1!(p2)$) {$2$};
\draw [black,very thick] (p1) -- (180:2) -- (p2);
\fill [black] (180:2) circle [radius=2pt];
\node [anchor=center] at (180:1.7) {\scriptsize $A_1$};
\coordinate (A5) at (120:3);
\coordinate (p4) at ($(A5)+(150:1)$);
\node [anchor=center] at ($(0,0)!1.1!(p4)$) {$4$};
\coordinate (c6p) at ($(A5)+(90:1)$);
\node [anchor=center] at ($(0,0)!1.1!(c6p)$) {$c_6^+$};
\draw [black,very thick] (p4) -- (A5) -- (c6p);
\draw [black,very thick] (A5) -- (120:2);
\fill [black] (120:2) circle [radius=2pt];
\fill [black] (A5) circle [radius=2pt];
\node [anchor=center] at (120:1.7) {\scriptsize $A_2$};
\node [anchor=west] at (120:3) {\scriptsize $A_5$};
\coordinate (p3) at (60:3);
\node [anchor=center] at ($(0,0)!1.1!(p3)$) {$3$};
\draw [black,very thick] (p3) -- (60:2);
\fill [black] (60:2) circle [radius=2pt];
\node [anchor=center] at (60:1.7) {\scriptsize $A_3$};
\coordinate (c6m) at (0:3);
\node [anchor=center] at ($(0,0)!1.1!(c6m)$) {$c_6^-$};
\draw [black,very thick] (c6m) -- (0:2);
\fill [black] (0:2) circle [radius=2pt];
\node [anchor=center] at (0:1.7) {\scriptsize $A_4$};
\node [anchor=center] at ($(0,0)!1.2!(240:2)$) {$c_2^-$};
\node [anchor=center] at ($(0,0)!1.2!(-60:2)$) {$c_2^+$};
\draw [black,fill=white] (240:2) circle [radius=2.5pt];
\draw [black,fill=white] (-60:2) circle [radius=2.5pt];
\draw [red,very thick,dashed] (158:3) -- (158:2) .. controls (158:.5) and (255:.5) .. (255:2) -- (255:2.5);
\node [anchor=center] at ($(0,0)!1.1!(255:2.5)$) {\color{red}$u_1$};
\draw [red,very thick,dashed] (90:3) -- (270:2.5);
\node [anchor=center] at ($(0,0)!1.1!(270:2.5)$) {\color{red}$u_5$};
\draw [red,very thick,dashed] (15:3) -- (15:2) .. controls (15:.5) and (-75:.5) .. (-75:2) -- (-75:2.5);
\node [anchor=center] at ($(0,0)!1.1!(-75:2.5)$) {\color{red}$u_3$};
\draw [orange,very thick,dashed] (195:3) -- (195:2) .. controls (195:1) and (165:1) .. (165:2) -- (165:3);
\node [anchor=center] at ($(0,0)!1.15!(195:3)$) {\color{orange}$u_2|t_1$};
\draw [orange,very thick,dashed] (210:3) -- (210:2) .. controls (210:.5) and (100:.5) .. (100:2) -- (100:3);
\node [anchor=center] at ($(0,0)!1.15!(210:3)$) {\color{orange}$u_6|t_2$};
\draw [orange,very thick,dashed] (136:4) .. controls ($(120:2.5)+(210:.35)$) .. (120:2.5) .. controls ($(120:2.5)+(30:.35)$) .. (104:4);
\node [anchor=center] at ($(0,0)!1.1!(136:4)$) {\color{orange}$u_4|t_4$};
\draw [orange,very thick,dashed] (136:3) -- (136:2) .. controls (136:.5) and (45:.5) .. (45:2) -- (45:3);
\node [anchor=center] at ($(0,0)!1.1!(45:3)$) {\color{orange}$u_{8}|t_3$};
\draw [orange,very thick,dashed] (225:3) -- (225:2) .. controls (225:.5) and (30:.5) .. (30:2) -- (30:3);
\node [anchor=center] at ($(0,0)!1.1!(225:3)$) {\color{orange}$u_{7}|h-c_6$};
\draw [orange,very thick,dashed] (75:3) -- (75:2) .. controls (75:.5) and (-20:.5) .. (-20:2) -- (-20:3);
\node [anchor=center] at ($(0,0)!1.1!(75:3)$) {\color{orange}$u_{10}|t_2$};
\draw [orange,very thick,dashed] (147:3) -- (147:2) .. controls (147:.5) and (-40:.5) .. (-40:2) -- (-40:3);
\node [anchor=center] at ($(0,0)!1.1!(-40:3)$) {\color{orange}$u_{9}|t_1$};
\end{scope}
\begin{scope}[xshift=7.5cm]
\node [anchor=center] at (0,-3.5) {(B)};
\fill [black!15!white] (162:2) circle [radius=.4];
\fill [black!15!white] (90:2) circle [radius=.4];
\fill [black!15!white] (18:2) circle [radius=.4];
\draw [black,very thick,dotted] (-180:2) arc [start angle=-180,end angle=0,radius=2];
\draw [black,very thick] (-54:2) arc [start angle=-54,end angle=90,radius=2];
\draw [black,very thick] (162:2) arc [start angle=162,end angle=234,radius=2];
\draw [black,very thick] (198:2) .. controls (198:.75) and (90:.75) .. (90:2);
\fill [white] (126:.66) circle [radius=3pt];
\draw [black,very thick] (162:2) .. controls (162:.75) and (54:.75) .. (54:2);
\fill [black] (198:2) circle [radius=2pt];
\fill [black] (54:2) circle [radius=2pt];
\node [anchor=center] at (198:2.3) {\scriptsize $A_4$};
\node [anchor=center] at (54:2.3) {\scriptsize $A_2$};
\coordinate (p1) at ($(162:2)+(192:1)$);
\node [anchor=center] at ($(0,0)!1.1!(p1)$) {$1$};
\coordinate (p2) at ($(162:2)+(132:1)$);
\node [anchor=center] at ($(0,0)!1.1!(p2)$) {$2$};
\draw [black,very thick] (p1) -- (162:2) -- (p2);
\fill [black] (162:2) circle [radius=2pt];
\node [anchor=center] at (153:2) {\scriptsize $A_1$};
\coordinate (p3) at (90:3);
\node [anchor=center] at ($(0,0)!1.1!(p3)$) {$3$};
\draw [black,very thick] (p3) -- (90:2);
\fill [black] (90:2) circle [radius=2pt];
\node [anchor=center] at (99:2) {\scriptsize $A_3$};
\coordinate (p4) at (18:3);
\node [anchor=center] at ($(0,0)!1.1!(p4)$) {$4$};
\draw [black,very thick] (p4) -- (18:2);
\fill [black] (18:2) circle [radius=2pt];
\node [anchor=center] at (18:1.7) {\scriptsize $A_5$};
\node [anchor=center] at ($(0,0)!1.2!(234:2)$) {$c_6^-$};
\node [anchor=center] at ($(0,0)!1.2!(-54:2)$) {$c_6^+$};
\draw [black,fill=white] (234:2) circle [radius=2.5pt];
\draw [black,fill=white] (-54:2) circle [radius=2.5pt];
\draw [red,very thick,dashed] (126:3) -- (126:2) .. controls (126:.5) and (258:.5) .. (258:2) -- (258:2.5);
\node [anchor=center] at ($(0,0)!1.1!(258:2.5)$) {\color{red}$t_3$};
\draw [red,very thick,dashed] (42:3) -- (42:2) .. controls (42:.5) and (282:.5) .. (282:2) -- (282:2.5);
\node [anchor=center] at ($(0,0)!1.1!(282:2.5)$) {\color{red}$t_4$};
\draw [orange,very thick,dashed] (138:3) -- (138:2) .. controls (138:1) and (186:1) .. (186:2) -- (186:3);
\node [anchor=center] at ($(0,0)!1.1!(186:3)$) {\color{orange}$t_1|S$};
\draw [ProcessBlue,very thick,dashed] (210:3) -- (210:2) .. controls (210:1) and (66:1) .. (66:2) -- (66:3);
\node [anchor=center] at ($(0,0)!1.1!(210:3)$) {\color{ProcessBlue}$t_5|\Delta_4$};
\draw [orange,very thick,dashed] (114:3) -- (114:2) .. controls (114:.5) and (-18:.5) .. (-18:2) -- (-18:3);
\node [anchor=center] at ($(0,0)!1.1!(-18:3)$) {\color{orange}$t_6|S$};
\end{scope}
\end{tikzpicture}
\end{center}
\vspace{-1.5em}\caption{Construction of the non-planar diagram in Figure \ref{app:fig:nonplanarA}. (A) constructs the first loop. (B) constructs the second loop. Shaded areas denote different $A_{(a)}$ regions.}
\label{app:fig:constructnonplanarA}
\end{figure}
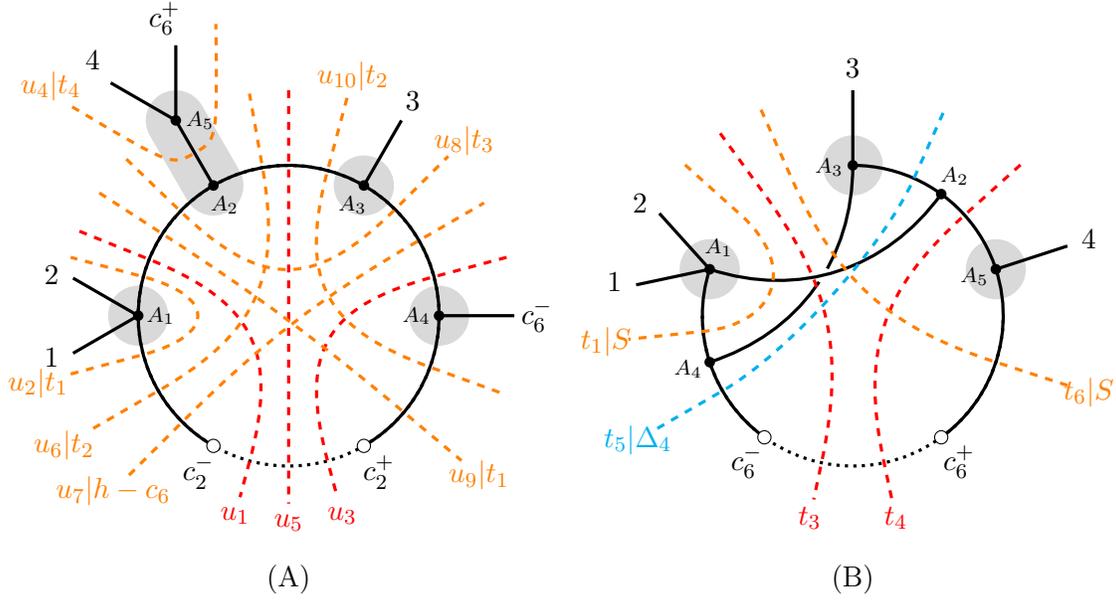

Starting with a tree with bulk-to-bulk propagators $\{1,3,4,5\}$, we first create the propagator 2 to form a one-loop box diagram, and then further create the propagator 6 to obtain this two-loop diagram, as shown in Figure \ref{app:fig:constructnonplanarA} (we use $u$ variables for the tree and $t$ for the intermediate one-loop diagram). The expressions for the pre-amplitude of the tree diagram as well as the kernels in the two steps can be directly read off the figure, and we skip writing them out explicitly.

Similar to the analysis in Section \ref{sec:doubletrianglediagram}, here let us again list the necessary labels for poles in the integrand.  In the one-loop construction the poles relevant to the $u$ integrals are
\begin{equation}\label{app:eq:nonplanarAintegrandpoles1}\checked{}
\begin{split}
&\Gamma_{1}[{\textstyle\frac{h-c_1-u_1}{2}}]\,
\Gamma_{2}[{\textstyle\frac{h+c_1-u_1}{2}}]\,
\Gamma_{3}[{\textstyle\frac{u_2-t_1}{2}}]\,
\Gamma_{4}[{\textstyle\frac{h-c_2-u_1+u_2}{2}}]\,
\Gamma_{5}[{\textstyle\frac{h-c_3-u_3}{2}}]\,
\Gamma_{6}[{\textstyle\frac{h+c_3-u_3}{2}}]\\
&\Gamma_{7}[{\textstyle\frac{h-c_4-u_4}{2}}]\,
\Gamma_{8}[{\textstyle\frac{h+c_4-u_4}{2}}]\,
\Gamma_{9}[{\textstyle\frac{u_4-t_4}{2}}]\,
\Gamma_{10}[{\textstyle\frac{h-c_5-u_5}{2}}]\,
\Gamma_{11}[{\textstyle\frac{h+c_5-u_5}{2}}]\,
\Gamma_{12}[{\textstyle\frac{u_2+u_4-u_6}{2}}]\\
&\Gamma_{13}[{\textstyle\frac{t_1-t_2+t_4-u_2-u_4+u_6}{2}}]\,
\Gamma_{14}[{\textstyle\frac{u_1-u_2-u_5+u_6}{2}}]\,
\Gamma_{15}[{\textstyle\frac{h-c_2+u_3-u_7}{2}}]\,
\Gamma_{16}[{\textstyle\frac{-u_3+u_5-u_6+u_7}{2}}]\\
&\Gamma_{17}[{\textstyle\frac{-h+c_6+t_2+t_3-t_4+u_4-u_6+u_7-u_8}{2}}]\,
\Gamma_{18}[{\textstyle\frac{-t_3+t_4-u_4+u_8}{2}}]\,
\Gamma_{19}[{\textstyle\frac{-u_4+u_6-u_7+u_8}{2}}]\\
&\Gamma_{20}[{\textstyle\frac{h+c_2+u_1-u_9}{2}}]\,
\Gamma_{21}[{\textstyle\frac{3h-c_6+t_1-t_3-u_7+u_8-u_9}{2}}]\,
\Gamma_{22}[{\textstyle\frac{-2h+u_7-u_8+u_9}{2}}]\,
\Gamma_{23}[{\textstyle\frac{-u_1+u_5+u_9-u_{10}}{2}}]\\
&\Gamma_{24}[{\textstyle\frac{-t_1+t_2+t_3-u_8+u_9-u_{01}}{2}}]\,
\Gamma_{25}[{\textstyle\frac{u_{10}-t_2}{2}}]\,
\Gamma_{26}[{\textstyle\frac{-h+c_6+u_3-u_5+u_{10}}{2}}]\\
&\Gamma_{27}[{\textstyle\frac{u_8-u_9+u_{10}-\Delta_3}{2}}]\,
\Gamma_{28}[{\textstyle\frac{\Delta_3+u_4-u_8}{2}}]\,
\Gamma_{29}[{\textstyle\frac{\Delta_3+h-c_6-u_{10}}{2}}]\,
\Gamma_{30}[{\textstyle\frac{\Delta_{12}-u_2}{2}}].
\end{split}
\end{equation}
The 10 $u$ integrals then give rise to the following families of emergent poles 
\begin{equation}\label{app:eq:nonplanarAemergentpolesonloop}\checked{}
\begin{split}
&\Gamma[{\textstyle\frac{h-c_2\pm c_3-c_6}{2}}]\,
\Gamma[{\textstyle\frac{h\pm c_2\pm c_3+c_6}{2}}]\,
\Gamma[{\textstyle\frac{2h\pm c_1\pm c_2-t_1}{2}}]\,
\Gamma[{\textstyle\frac{2h\pm c_2\pm c_5-t_2}{2}}]\,
\Gamma[{\textstyle\frac{2h\pm c_1\pm c_3-t_3}{2}}]\\
&\Gamma[{\textstyle\frac{h\pm c_4-t_4}{2}}]\,
\Gamma[{\textstyle\frac{2h\pm c_3\pm c_5-\Delta_3}{2}}]\,
\Gamma[{\textstyle\frac{\Delta_{12}\pm c_1+c_2}{2}}]\,
\Gamma[{\textstyle\frac{\Delta_{123}\pm c_4-c_6}{2}}].
\end{split}
\end{equation}
Apart from these there are also poles directly passing to the one loop, which we label as
\begin{equation}\label{app:eq:nonplanarArestpolesoneloop}\checked{}
\begin{split}
&\Gamma_{31}[{\textstyle\frac{h-c_1-c_4-c_5}{2}}]\,
\Gamma_{32}[{\textstyle\frac{h+c_1-c_4-c_5}{2}}]\,
\Gamma_{33}[{\textstyle\frac{h-c_1+c_4-c_5}{2}}]\,
\Gamma_{34}[{\textstyle\frac{h+c_1+c_4-c_5}{2}}]\,
\Gamma_{35}[{\textstyle\frac{h-c_1-c_4+c_5}{2}}]\\
&\Gamma_{36}[{\textstyle\frac{h+c_1-c_4+c_5}{2}}]\,
\Gamma_{37}[{\textstyle\frac{h-c_1+c_4+c_5}{2}}]\,
\Gamma_{38}[{\textstyle\frac{h+c_1+c_4+c_5}{2}}]\,
\Gamma_{39}[{\textstyle\frac{h+c_2-c_3-c_6}{2}}]\,
\Gamma_{40}[{\textstyle\frac{h+c_2+c_3-c_6}{2}}]\\
&\Gamma_{41}[{\textstyle\frac{\Delta_3-c_3-c_5}{2}}]\,
\Gamma_{42}[{\textstyle\frac{\Delta_3+c_3-c_5}{2}}]\,
\Gamma_{43}[{\textstyle\frac{\Delta_3-c_3+c_5}{2}}]\,
\Gamma_{44}[{\textstyle\frac{\Delta_3+c_3+c_5}{2}}]\,
\Gamma_{45}[{\textstyle\frac{\Delta_4-c_4+c_6}{2}}]\\
&\Gamma_{46}[{\textstyle\frac{\Delta_4+c_4+c_6}{2}}]\,
\Gamma_{47}[{\textstyle\frac{\Delta_{12}-c_1-c_2}{2}}]\,
\Gamma_{48}[{\textstyle\frac{\Delta_{12}+c_1-c_2}{2}}].
\end{split}
\end{equation}
\eqref{app:eq:nonplanarAemergentpolesonloop} together with \eqref{app:eq:nonplanarArestpolesoneloop} constitute the entire set of poles of the intermediate one-loop diagram constructed in Figure \ref{app:fig:constructnonplanarA} (A), which is consistent with Conjecture \ref{sec:enhancedrules}.  In particular, note that there are no poles corresponding to $\Gamma[\frac{2h\pm c_1\pm c_5-t_4}{2}]$, which are composite.  This one-loop diagram depends one Mandelstam variables $\{t_1,t_2,t_3,t_4\}$, which are independent from each other.

Based on the above resulting one-loop pole structure we then move on to the two-loop construction. Attention here that the Mandelstam variable $t_2$ does not explicitly enter the corresponding kernel. In this case, as instructed in Section \ref{sec:Mandelstamindependence} there should exist linear relations between $t_2$ and the extra variables introduced in the kernel, i.e., $\{t_5,t_6\}$. Indeed, we can find out that
\begin{equation}\checked{}
\Delta_3+h-c_6+t_1-t_2-t_3+t_4-t_5=0.
\end{equation}
Consequently we should use the same kernel but substitute one of these $t$ variables in terms of the rest using the above relation. For convenience we choose $t_5$ to be dependent. Hence the construction of this two-loop diagram involves in total 10 $u$ integrals and 5 $t$ integrals.

After this substitution the additional poles from the two-loop kernel are explicitly
\begin{equation}\label{app:eq:nonplanarAintegrandpoles2}\checked{}
\begin{split}
&\Gamma_{77}[{\textstyle\frac{t_1-S}{2}}]\,
\Gamma_{78}[{\textstyle\frac{h-c_6+t_1-t_3}{2}}]\,
\Gamma_{79}[{\textstyle\frac{-h+c_6+t_2+t_3-t_4}{2}}]\,
\Gamma_{80}[{\textstyle\frac{h+c_6+t_3-t_6}{2}}]\,
\Gamma_{81}[{\textstyle\frac{t_6-S}{2}}]\\
&\Gamma_{82}[{\textstyle\frac{-h-c_6+t_1-t_2-t_3+t_4+t_6}{2}}]\,
\Gamma_{83}[{\textstyle\frac{-t_1+t_2+t_3-\Delta_3}{2}}]\,
\Gamma_{84}[{\textstyle\frac{h-c_6-t_2+\Delta_3}{2}}]\,
\Gamma_{85}[{\textstyle\frac{-t_3+t_4+t_6-\Delta_4}{2}}]\\
&\Gamma_{86}[{\textstyle\frac{h+S-c_6-t_2-t_3+t_4+\Delta_3-\Delta_4}{2}}]\,
\Gamma_{87}[{\textstyle\frac{h+c_6-t_4+\Delta_4}{2}}]\,
\Gamma_{88}[{\textstyle\frac{h+S+c_6-t_1+t_2+t_3-t_4-t_6-\Delta_3+\Delta_4}{2}}]\\
&\Gamma_{89}[{\textstyle\frac{\Delta_{34}-t_6}{2}}]\,
\Gamma_{90}[{\textstyle\frac{\Delta_{12}-t_1}{2}}].
\end{split}
\end{equation}

\subsubsection{Poles of $M$}

With the above setup, we can conveniently work out the pole structure of the pre-amplitude of this non-planar diagram, which is
\begin{equation}\label{app:eq:nonplanarAprepoles}\checked{}
\begin{split}
&\underbrace{\Gamma[{\textstyle\frac{\Delta_{12}\pm c_1\pm c_2}{2}}]\,
\Gamma[{\textstyle\frac{h\pm c_1\pm c_4\pm c_5}{2}}]\,
\Gamma[{\textstyle\frac{h\pm c_2\pm c_3\pm c_6}{2}}]\,
\Gamma[{\textstyle\frac{\Delta_3\pm c_3\pm c_5}{2}}]\,
\Gamma[{\textstyle\frac{\Delta_4\pm c_4\pm c_6}{2}}]}_{\text{vertex}}\\
&\underbrace{\Gamma[{\textstyle\frac{2h-S\pm c_1\pm c_2}{2}}]\,
\Gamma[{\textstyle\frac{2h\pm c_3\pm c_5-\Delta_3}{2}}]\,
\Gamma[{\textstyle\frac{2h\pm c_4\pm c_6-\Delta_4}{2}}]}_{\text{channel}}\,
\underbrace{\Gamma[{\textstyle\frac{\Delta_{34}\pm c_1\pm c_2}{2}}]\,
\Gamma[{\textstyle\frac{\Sigma\Delta}{2}-h}]}_{\text{loop}}.
\end{split}
\end{equation}
Sources of the poles are indicated above as well.  While the contributions from the original vertices all exist, there are missing ones in correspondence to the channel rule and the loop contraction rule
\begin{equation}\label{app:eq:nonplanarAprepolescomposite}\checked{}
\begin{split}
&{\color{ForestGreen}\underbrace{\cancel{\Gamma}[{\textstyle\frac{3h\pm c_1\pm c_3\pm c_4-\Delta_3}{2}}]\,
\cancel{\Gamma}[{\textstyle\frac{3h\pm c_2\pm c_5\pm c_6-\Delta_3}{2}}]\,
\cancel{\Gamma}[{\textstyle\frac{3h\pm c_1\pm c_5\pm c_6-\Delta_4}{2}}]\,
\cancel{\Gamma}[{\textstyle\frac{3h\pm c_2\pm c_3\pm c_4-\Delta_4}{2}}]}_{\text{channel}}}\\
&{\color{ForestGreen}\underbrace{\cancel{\Gamma}[{\textstyle\frac{3h\pm c_1\pm c_3\pm c_6-S}{2}}]\,
\cancel{\Gamma}[{\textstyle\frac{3h\pm c_2\pm c_4\pm c_5-S}{2}}]}_{\text{channel}}\,
\underbrace{\color{ForestGreen}\cancel{\Gamma}[{\textstyle\frac{\Delta_{123}\pm c_4\pm c_6}{2}}]\,
\cancel{\Gamma}[{\textstyle\frac{\Delta_{124}\pm c_3\pm c_5}{2}}]}_{\text{loop}}}.
\end{split}
\end{equation}
It is simple to observe that all of these are composite, in the sense of the ability to emerge from the genuine poles \eqref{app:eq:nonplanarAprepoles} via $c$ integrals or $\Delta $ integrals. Hence we have the first verification of Conjecture \ref{sec:enhancedrules} in a non-planar diagram.

\subsubsection{Poles of $\mathcal{M}$}

Regarding the properties of the Mellin amplitude of the diagram under study, we only focus on checking the absence of non-minimal poles in the unique Mandelstam variable $S$. For this discussion let us also label the extra poles from the normalization $\mathcal{N}$
\begin{equation}\label{app:eq:nonplanarAintegrandpoles3}\checked{}
\begin{split}
&\gamma_{91}[(\underline\Delta_1-h)+c_1]\,
\gamma_{92}[(\underline\Delta_1-h)-c_1]\,
\gamma_{93}[(\underline\Delta_2-h)+c_2]\,
\gamma_{94}[(\underline\Delta_2-h)-c_2]\\
&\gamma_{95}[(\underline\Delta_3-h)+c_3]\,
\gamma_{96}[(\underline\Delta_3-h)-c_3]\,
\gamma_{97}[(\underline\Delta_4-h)+c_4]\,
\gamma_{98}[(\underline\Delta_4-h)-c_4]\\
&\gamma_{99}[(\underline\Delta_5-h)+c_5]\,
\gamma_{100}[(\underline\Delta_5-h)-c_5]\,
\gamma_{101}[(\underline\Delta_6-h)+c_6]\,
\gamma_{102}[(\underline\Delta_6-h)-c_6].
\end{split}
\end{equation}

There are two families of non-minimal poles in correspondence to the two channel families in \eqref{app:eq:nonplanarAprepolescomposite}. We discuss $\Gamma[\frac{\underline\Delta_{245}-S}{2}]$ only, as the other family is related by reflection of the diagram. These emerge via the $\{c_2,c_4,c_5\}$ integrals from the intermediate poles (observed in the two-loop pre-amplitude)
\begin{equation}\checked{}
\Gamma[{\textstyle\frac{2h\pm c_1\pm c_2-S}{2}}]\,
\Gamma[{\textstyle\frac{h\pm c_1\pm c_4\pm c_5}{2}}],
\end{equation}
together with additional poles from the normalization $\mathcal{N}$. They receive contribution from altogether 16 $\gamma$ families, which are
\begin{equation*}\checked{}
\begin{split}
&\gamma_{\{1, 3, 14, 15, 16, 32, 77, 93, 97, 99\}}\,
\gamma_{\{1, 3, 14, 15, 16, 34, 77, 93, 98, 99\}}\,
\gamma_{\{1, 3, 14, 15, 16, 36, 77, 93, 97, 100\}}\\
&\gamma_{\{1, 3, 14, 15, 16, 38, 77, 93, 98, 100\}}\,
\gamma_{\{2, 3, 14, 15, 16, 31, 77, 93, 97, 99\}}\,
\gamma_{\{2, 3, 14, 15, 16, 33, 77, 93, 98, 99\}}\\
&\gamma_{\{2, 3, 14, 15, 16, 35, 77, 93, 97, 100\}}\,
\gamma_{\{2, 3, 14, 15, 16, 37, 77, 93, 98, 100\}}\,
\gamma_{\{1, 9, 18, 20, 24, 25, 32, 77, 94, 97, 99\}}\\
&\gamma_{\{1, 9, 18, 20, 24, 25, 34, 77, 94, 98, 99\}}\,
\gamma_{\{1, 9, 18, 20, 24, 25, 36, 77, 94, 97, 100\}}\,
\gamma_{\{1, 9, 18, 20, 24, 25, 38, 77, 94, 98, 100\}}\\
&\gamma_{\{2, 9, 18, 20, 24, 25, 31, 77, 94, 97, 99\}}\,
\gamma_{\{2, 9, 18, 20, 24, 25, 33, 77, 94, 98, 99\}}\,
\gamma_{\{2, 9, 18, 20, 24, 25, 35, 77, 94, 97, 100\}}\\
&\gamma_{\{2, 9, 18, 20, 24, 25, 37, 77, 94, 98, 100\}}.
\end{split}
\end{equation*}
From the localized values of the relevant $c$ variables, it is obvious that the pinching planes for these $\gamma$ families do not have any intersection. Hence the poles are all simple poles, and when computing the residue it suffices to consider additively the contribution from each $\gamma$ pinching. This computation is straightforward and we do not explicitly write out here. It turns out that the contribution from each $\gamma$ family vanishes individually (even before we perform the remaining $u$, $t$ and $c$ integrals that are not localized by the pinching). So we conclude that the non-minimal poles corresponding to the three cuts are completely absent.

\subsection{Example with multiple channels}

\begin{figure}[ht]
\captionsetup{margin=2em}
\begin{center}
\begin{tikzpicture}
\begin{scope}
\coordinate (up) at (-.3,.7);
\coordinate (down) at (-.3,-.7);
\draw [black,very thick] (150:3) -- ($(up)+(-1.4,0)$) -- ($(down)+(-1.4,0)$) -- (-150:3);
\draw [black,very thick] (40:3) -- (up) -- ($(up)+(-1.4,0)$);
\draw [black,very thick] ($(down)+(-1.4,0)$) -- (down) -- (-40:3);
\draw [black,very thick] (up) -- ($(down)!.6!(-40:3)$);
\fill [white] (.21,0) circle [radius=2.5pt];
\draw [black,very thick] (down) -- ($(up)!.6!(40:3)$);
\fill [black] (up) circle [radius=1.75pt];
\fill [black] (down) circle [radius=1.75pt];
\fill [black] ($(up)+(-1.4,0)$) circle [radius=1.75pt];
\fill [black] ($(down)+(-1.4,0)$) circle [radius=1.75pt];
\fill [black] ($(up)!.6!(40:3)$) circle [radius=1.75pt];
\fill [black] ($(down)!.6!(-40:3)$) circle [radius=1.75pt];
\node [anchor=east] at (-150:3) {$1$};
\node [anchor=east] at (150:3) {$2$};
\node [anchor=west] at (40:3) {$3$};
\node [anchor=west] at (-40:3) {$4$};
\node [anchor=north] at ($(down)+(-.7,0)$) {$c_1$};
\node [anchor=south] at ($(up)+(-.7,0)$) {$c_2$};
\node [anchor=north] at ($(down)!.3!(-40:3)$) {$c_4$};
\node [anchor=south] at ($(up)!.3!(40:3)$) {$c_3$};
\node [anchor=center] at (1,.5) {$c_5$};
\node [anchor=center] at (1,-.5) {$c_6$};
\node [anchor=east] at (-1.7,0) {$c_7$};
\end{scope}
\begin{scope}[scale=.5,xshift=13cm,yshift=5cm]
\node [anchor=east] at (-4,0) {$S$:};
\coordinate (up) at (-.3,.7);
\coordinate (down) at (-.3,-.7);
\draw [black,very thick] (150:3) -- ($(up)+(-1.4,0)$) -- ($(down)+(-1.4,0)$) -- (-150:3);
\draw [black,very thick] (40:3) -- (up) -- ($(up)+(-1.4,0)$);
\draw [black,very thick] ($(down)+(-1.4,0)$) -- (down) -- (-40:3);
\draw [black,very thick] (up) -- ($(down)!.6!(-40:3)$);
\fill [white] (.21,0) circle [radius=4pt];
\draw [black,very thick] (down) -- ($(up)!.6!(40:3)$);
\draw [red,ultra thick] (up) -- ($(up)+(-1.4,0)$);
\draw [red,ultra thick] ($(down)+(-1.4,0)$) -- (down);
\fill [black] (up) circle [radius=1.75pt];
\fill [black] (down) circle [radius=1.75pt];
\fill [black] ($(up)+(-1.4,0)$) circle [radius=1.75pt];
\fill [black] ($(down)+(-1.4,0)$) circle [radius=1.75pt];
\fill [black] ($(up)!.6!(40:3)$) circle [radius=1.75pt];
\fill [black] ($(down)!.6!(-40:3)$) circle [radius=1.75pt];
\node [anchor=east] at (-150:3) {\scriptsize $1$};
\node [anchor=east] at (150:3) {\scriptsize $2$};
\node [anchor=west] at (40:3) {\scriptsize $3$};
\node [anchor=west] at (-40:3) {\scriptsize $4$};
\end{scope}
\begin{scope}[scale=.5,xshift=13cm]
\node [anchor=east] at (-4,0) {$T$:};
\coordinate (up) at (-.3,.7);
\coordinate (down) at (-.3,-.7);
\draw [black,very thick] (150:3) -- ($(up)+(-1.4,0)$) -- ($(down)+(-1.4,0)$) -- (-150:3);
\draw [black,very thick] (40:3) -- (up) -- ($(up)+(-1.4,0)$);
\draw [black,very thick] ($(down)+(-1.4,0)$) -- (down) -- (-40:3);
\draw [red,ultra thick] (up) -- ($(down)!.6!(-40:3)$);
\fill [white] (.21,0) circle [radius=4pt];
\draw [red,ultra thick] (down) -- ($(up)!.6!(40:3)$);
\draw [red,ultra thick] ($(up)+(-1.4,0)$) -- ($(down)+(-1.4,0)$);
\fill [black] (up) circle [radius=1.75pt];
\fill [black] (down) circle [radius=1.75pt];
\fill [black] ($(up)+(-1.4,0)$) circle [radius=1.75pt];
\fill [black] ($(down)+(-1.4,0)$) circle [radius=1.75pt];
\fill [black] ($(up)!.6!(40:3)$) circle [radius=1.75pt];
\fill [black] ($(down)!.6!(-40:3)$) circle [radius=1.75pt];
\node [anchor=east] at (-150:3) {\scriptsize $1$};
\node [anchor=east] at (150:3) {\scriptsize $2$};
\node [anchor=west] at (40:3) {\scriptsize $3$};
\node [anchor=west] at (-40:3) {\scriptsize $4$};
\end{scope}
\begin{scope}[scale=.5,xshift=13cm,yshift=-5cm]
\node [anchor=east] at (-4,0) {$U$:};
\coordinate (up) at (-.3,.7);
\coordinate (down) at (-.3,-.7);
\draw [black,very thick] (150:3) -- ($(up)+(-1.4,0)$) -- ($(down)+(-1.4,0)$) -- (-150:3);
\draw [black,very thick] (40:3) -- (up) -- ($(up)+(-1.4,0)$);
\draw [black,very thick] ($(down)+(-1.4,0)$) -- (down) -- (-40:3);
\draw [black,very thick] (up) -- ($(down)!.6!(-40:3)$);
\fill [white] (.21,0) circle [radius=4pt];
\draw [black,very thick] (down) -- ($(up)!.6!(40:3)$);
\draw [red,ultra thick] ($(up)+(-1.4,0)$) -- ($(down)+(-1.4,0)$);
\draw [red,ultra thick] (up) -- ($(up)!.6!(40:3)$);
\draw [red,ultra thick] ($(down)!.6!(-40:3)$) -- (down);
\fill [black] (up) circle [radius=1.75pt];
\fill [black] (down) circle [radius=1.75pt];
\fill [black] ($(up)+(-1.4,0)$) circle [radius=1.75pt];
\fill [black] ($(down)+(-1.4,0)$) circle [radius=1.75pt];
\fill [black] ($(up)!.6!(40:3)$) circle [radius=1.75pt];
\fill [black] ($(down)!.6!(-40:3)$) circle [radius=1.75pt];
\node [anchor=east] at (-150:3) {\scriptsize $1$};
\node [anchor=east] at (150:3) {\scriptsize $2$};
\node [anchor=west] at (40:3) {\scriptsize $3$};
\node [anchor=west] at (-40:3) {\scriptsize $4$};
\end{scope}
\end{tikzpicture}
\end{center}
\vspace{-1.5em}\caption{Another 4-point two-loop non-planar diagram. Red lines indicate the propagators in the unique minimal cut of the corresponding OPE channels.}
\label{app:fig:nonplanarB}
\end{figure}
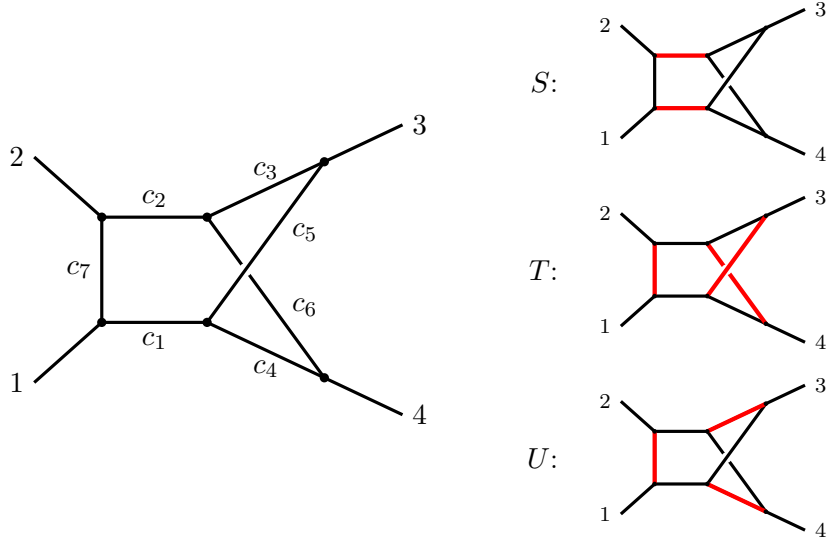

As an additional example, we slightly modify the diagram in Figure \ref{app:fig:nonplanarA} by splitting the vertex $A_1$ therein into two bulk vertices, connected by an extra propagator 7, as shown in Figure \ref{app:fig:nonplanarB}. Diagrammatically it is easy to see that its (pre-)amplitude has non-trivial dependence on all the three Mandestam variables $\{S,T,U\}$ of a 4-point diagram.
This involves a new feature that has not been captured by our previous examples, due to the linear dependence among these variables
\begin{equation}\checked{}
S+T+U=\Delta_1+\Delta_2+\Delta_3+\Delta_4.
\end{equation}
Recall that in our construction everytime when we create a new loop we are treating the original lower-loop diagram effectively as a tree, such that the construction inherently specifies a planar ordering, according to which an independent set of Mandelstam variables of the new diagram are assigned to the integral kernel. Particularly for this diagram, say if we choose to construct it by forming propagator 7, in which we use the planar ordering as literally shown in Figure \ref{app:fig:nonplanarB}. Then in the resulting integral representation only $\{S,T\}$ explicitly enter. It is a non-trivial outcome of the integrals that poles in all the three channels should correctly emerge. This diagram contains a unique minimal cut in each of the three channels, as indicated by the three diagrams on the right. Indeed, we explicitly verified that the pre-amplitude possesses the three corresponding types of poles
\begin{equation}\label{app:eq:nonplanarBMpoleSTU}\checked{}
\Gamma[{\textstyle\frac{2h\pm c_1\pm c_2-S}{2}}]\,
\Gamma[{\textstyle\frac{3h\pm c_5\pm c_6\pm c_7-T}{2}}]\,
\Gamma[{\textstyle\frac{3h\pm c_3\pm c_4\pm c_7+S+T-\Sigma\Delta}{2}}],
\end{equation}
which provides an extra consistency check of our construction. We do not provide the explicit integral representation here, but just to list out the other poles of the pre-amplitude thus resulted, which are
\begin{equation}\label{app:eq:nonplanarBpolesrest}\checked{}
\begin{split}
&\Gamma[{\textstyle\frac{\Delta_1\pm c_1\pm c_7}{2}}]\,
\Gamma[{\textstyle\frac{\Delta_2\pm c_2\pm c_7}{2}}]\,
\Gamma[{\textstyle\frac{h\pm c_1\pm c_4\pm c_5}{2}}]\,
\Gamma[{\textstyle\frac{h\pm c_2\pm c_3\pm c_6}{2}}]\,
\Gamma[{\textstyle\frac{\Delta_3\pm c_3\pm c_5}{2}}]\,
\Gamma[{\textstyle\frac{\Delta_4\pm c_4\pm c_6}{2}}]\\
&\Gamma[{\textstyle\frac{2h\pm c_1\pm c_7-\Delta_1}{2}}]\,
\Gamma[{\textstyle\frac{2h\pm c_2\pm c_7-\Delta_2}{2}}]\,
\Gamma[{\textstyle\frac{2h\pm c_3\pm c_5-\Delta_3}{2}}]\,
\Gamma[{\textstyle\frac{2h\pm c_4\pm c_6-\Delta_4}{2}}]\\
&\Gamma[{\textstyle\frac{\Delta_{34}\pm c_1\pm c_2}{2}}]\,
\Gamma[{\textstyle\frac{\Sigma\Delta}{2}-h}].
\end{split}
\end{equation}
Once again, the computation rules out the poles that follow the diagrammatic rules in Section \ref{sec:preliminaryobservations} but are composite, in complete agreement with Conjecture \ref{sec:enhancedrules}.

As for poles of the Mellin amplitude, we also explicitly confirmed that the potential non-minimal poles in the $S$ channel are absent as well (there are obviously no non-minimal cuts in the other two channels).

\newpage


\bibliographystyle{JHEP}
\bibliography{wdiagrams}

\end{document}